\documentclass{aastex63}

\usepackage{amssymb}
\usepackage{amsmath}
\usepackage{wrapfig}
\usepackage{rotating}
\usepackage{multirow}

\shorttitle{$N_{\rm GC}$ and $M_\bullet$ in spiral galaxies}
\shortauthors{Gonz\'alez-L\'opezlira et al.}

\newcommand\uiks{$u^*i^\prime K_s$}
\newcommand\ks{$K_s$}
\newcommand\ip{$i^\prime$}
\newcommand\ust{$u^*$}
\newcommand\gp{$g^\prime$}
\newcommand\rp{$r^\prime$}
\newcommand\samename{\vrule height0.4pt depth0.0pt width1.0in \thinspace.}

\received{2021 August 9}
\revised{2022 September 15}
\accepted{2022 October 3}
\submitjournal{ApJ}

\begin{document}

\title{The relation between globular cluster systems and supermassive black holes in spiral galaxies III. The link to the $M_\bullet - M_\ast$ correlation.}

\correspondingauthor{Rosa A.\ Gonz\'alez-L\'opezlira}
\email{r.gonzalez@irya.unam.mx}

\author[0000-0003-1557-4931]{Rosa A.\ Gonz\'alez-L\'opezlira}
\affiliation{Instituto de Radioastronomia y Astrofisica, UNAM, Campus Morelia,
     Michoacan, Mexico, C.P.\ 58089
}

\author[0000-0003-2127-2841]{Luis Lomel\'{\i}-N\'u\~nez}
\affiliation{Instituto Nacional de Astrof\'{\i}sica, \'Optica y Electr\'onica, Luis Enrique Erro 1, Tonantzintla, Puebla, C.P. 72840, Mexico}
\affiliation{Instituto de Radioastronomia y Astrofisica, UNAM, Campus Morelia,
     Michoacan, Mexico, C.P.\ 58089}

\author[0000-0001-7966-7606]{Yasna \'Ordenes-Brice\~no}
\affiliation{Instituto de Astrof\'{\i}sica, Pontificia Universidad Cat\'olica de Chile, Av.\ Vicu\~na Mackenna 4860, 7820436 Macul, Santiago, Chile}

\author[0000-0002-5635-3345]{Laurent Loinard}
\affiliation{Instituto de Radioastronomia y Astrofisica, UNAM, Campus Morelia,
     Michoacan, Mexico, C.P.\ 58089}

\author[0000-0001-8221-8406]{Stephen Gwyn}
\affiliation{Herzberg Institute of Astrophysics, National Research Council of Canada, Victoria, BC V9E 2E7, Canada}

\author[0000-0002-5897-7813]{Karla \'Alamo-Mart\'{\i}nez}
\affiliation{Departamento de Astronom\'{\i}a, Universidad de Guanajuato, Apdo.\ Postal 144, Guanajuato 36000, Mexico}

\author[0000-0002-6971-5755]{Gustavo Bruzual}
\affiliation{Instituto de Radioastronomia y Astrofisica, UNAM, Campus Morelia,
     Michoacan, Mexico, C.P.\ 58089}

\author[0000-0002-7214-8296]{Ariane Lan\c con}
\affiliation{Observatoire Astronomique de Strasbourg, Universit\'e de Strasbourg, CNRS, UMR 7550, 11 rue de l'Universit\'e,  67000 Strasbourg, France}

\author[0000-0003-0350-7061]{Thomas H.\ Puzia}
\affiliation{Instituto de Astrof\'{\i}sica, Pontificia Universidad Cat\'olica de Chile, Av.\ Vicu\~na Mackenna 4860, 7820436 Macul, Santiago, Chile}

\begin{abstract}

We continue to explore the relationship between globular cluster total number,
$N_{\rm GC}$, and central black hole mass, $M_\bullet$, in spiral galaxies. We
present here results for the Sab galaxies NGC\,3368, NGC\,4736 (M\,94) and
NGC\,4826 (M\,64), and the Sm galaxy NGC\,4395.  The globular cluster (GC)
candidate selection is based on the (\ust\ - \ip)  versus (\ip\ - \ks)
color-color diagram, and \ip-band shape parameters.  We determine the
$M_\bullet$ versus $N_{\rm GC}$ correlation for these spirals, plus NGC\,4258,
NGC\,253, M\,104, M\,81, M\,31, and the Milky Way. We also redetermine the
correlation for the elliptical sample in \citet{harr14}, with updated galaxy
types from \citet{sahu19b}.  Additionally, we derive total stellar galaxy mass,
$M_\ast$, from its two-slope correlation with $N_{\rm GC}$ \citep{huds14}, and
fit $M_\bullet$ versus $M_\ast$ for both spirals and ellipticals. We obtain log
$M_\bullet \propto$ (1.01 $\pm$ 0.13) log $N_{\rm GC}$ for ellipticals, and log
$M_\bullet \propto$ (1.64 $\pm$ 0.24) log $N_{\rm GC}$ for late type galaxies
(LTG).  The linear $M_\bullet$ versus $N_{\rm GC}$ correlation in ellipticals
could be due to statistical convergence through mergers, but not the much
steeper correlation for LTG.  However, in the $M_\bullet$ versus total stellar
mass ($M_\ast$) parameter space, {\it with $M_\ast$ derived from its
correlation with $N_{\rm GC}$}, $M_\bullet \propto$ (1.48 $\pm$ 0.18) log
$M_\ast$ for ellipticals, and $M_\bullet \propto$ (1.21 $\pm$ 0.16) log
$M_\ast$ for LTG.  The observed agreement between ellipticals and LTG in this
parameter space may imply that black holes and galaxies co-evolve through
``calm” accretion, AGN feedback, and other secular processes.

\end{abstract}


\section{Introduction} \label{sec:intro}

It is widely accepted that all massive galaxies contain a supermassive black hole (SMBH). In
spheroidal systems, the masses of the SMBH, $M_{\bullet}$, correlate with other
properties of their host galaxies: the bulge luminosity \citep[the $M_\bullet
$--$L_{\rm bulge}$ relation, e.g.,][]{korm93,korm95}, the bulge mass \citep[the
$M_\bullet$--$M_{\rm bulge}$ relation, e.g.,][]{dres89,mago98}, the bulge
stellar velocity dispersion \citep[the $M_\bullet$--$\sigma_\ast$ relation,
e.g.,][]{ferr00,gebh00}, the dark matter halo mass \citep{spit09}.  Today, the
$M_\bullet$--$\sigma_\ast$ relation is known to have a $\approx$ 0.6 dex
scatter and depend on the galaxy's merger history \citep{bogd18,sahu19b}, and
the $M_\bullet$--$M_{\rm bulge}$ relation depends on the morphology of the
galaxy \citep{savo16,vand16,sahu19a}.  \citet{grah12,grah13,scot13,davi19a}
find that there are correlations between $M_\bullet$, and the luminosity and
mass of the bulge, but that the slope is steeper for spiral bulges with a
Sersic profile than for early type galaxies (ETG), irrespective of whether the
latter are better fitted by a Sersic or a core-Sersic profile. 

Exciting as these correlations are because of the clues that they may provide
regarding galaxy formation and assembly, and about their possible co-evolution with
the central black hole, they are not without controversy.  While
\citet{ferr02,baes03,davi19b} state that $M_\bullet$ correlates with the disk
maximum rotation velocity and hence with the dark matter halo mass,
\citet{korm11b} retort that black holes do not correlate directly with dark
matter haloes, and that a correlation appears only if the galaxy has a bulge.
Likewise, \citet{korm11a} contend that black holes do not correlate with either
disks or pseudobulges. \citet{davi18} argue that $M_\bullet$ actually
correlates with disk mass, albeit with a low correlation coefficient, $r = $
0.3.

Investigating the limits and deviations of these scaling relations, i.e., at
high and low masses, may be particularly useful to uncover the 
processes that are involved in the formation and evolution of galaxies and their black holes.
For example, the log $M_\bullet$ versus log $M_\ast$ correlation, with $M_\ast$
the total galaxy stellar mass, has been explored in both ETG and late type
galaxies (LTG).  \citet{vand16,davi19a,sahu19a} find that
the slope for LTG is between 70\% and 100\% steeper than the nearly linear one
followed by ETG. \citet{davi19a} rules out that such a steep correlation in LTG could
be due to statistical convergence through hierarchical galaxy formation
\citep{peng07,jahn11}; in addition, since the LTG in their sample sit on the same
correlation, regardless of whether they contain or not an active galactic
nucleus (AGN), they call into question the relevance of AGN feedback to the
relation. \citet{rein15} analyze two samples, one of ETG and classical bulges,
mostly without active AGN, and another of LTG with active AGN. They find
similar slopes (respectively, 1.4 and 1.0, i.e., less steep for LTGs), but with
an offset of more than one order of magnitude at log $M_\ast/M_\odot \sim$ 10.
In the same parameter space, \citet{simm17} determine that the slope for a
sample of AGN in disk-dominated, merger-free, galaxies is 1.2, roughly equal
to that of the bulge sample of \citet{hari04}, and with a small offset of
$\sim$ 0.2 dex at log $M_\ast/M_\odot =$ 10; they hence conclude that mergers
are inconsequential for the co-evolution of black holes and galaxies, and that
secular processes, like ``calm" accretion and AGN feedback, are more
fundamental for galaxy assembly and black hole growth. 

An intriguing correlation, given the extremely disparate scales, is the one
between central black hole mass and the total number of globular clusters
($N_{\rm GC}$) \citep{burk10,harr11,harr14}. The correlation can be expressed
as $N_{\rm GC} \propto M_\bullet^{1.02\pm0.10}$, and spans over 3 orders of
magnitude.  Possible causal links have been proposed, e.g., feedback by the
jets of AGN \citep[e.g.,][]{silk98,fabi12} and
cannibalization of GCs by black holes
\citep[e.g.,][]{capu01,capu05,capu09,gned14}; since it is roughly linear for
ellipticals, it has been argued that it could be due to statistical convergence through
merging.  On the other hand, GC systems are known to be very good tracers of
galaxy virial mass \citep[e.g.,][]{blak97,spit09,geor10,huds14,harr17}. This
correlation is good over 7 orders of magnitude in halo mass, from dwarf
galaxies to galaxy clusters, with a strongly declining dispersion with
increasing mass, $\delta N_{\rm GC}$/$N_{\rm GC} \propto$ $M_{\rm vir}^{1/2}$
\citep{forb18,burk20}. 

The study of GC systems in spiral galaxies has been hampered by the sparsity of
their globular clusters, compared to ellipticals, and has been mostly limited
to edge-on objects, given the difficulty of detecting individual GCs projected
on a spiral disk.  Until relatively recently, there were only 5 spiral galaxies
with precise measurements of both  $N_{\rm GC}$ and $M_\bullet$: the Milky Way
(MW, Sbc), M\,104 (Sa), M\,81 (Sab), M\,31 (Sb) and NGC\,253 (Sc).
\citet{burk10} only included the first four, and readily noticed that, while
M\,31, M\,81, and M\,104 fall right on the $N_{\rm GC} - M_\bullet$ correlation
for ellipticals, our Galaxy has a black hole that is about one order of
magnitude lighter than expected from its $N_{\rm GC}$. They also observed that,
compared to M\,31, M\,81 and M\,104, the MW has both a later Hubble type and
possibly a pseudobulge, and hypothesized these factors could be related to its
non-compliance with the correlation.

We are in the process of studying the $N_{\rm GC}$ versus $M_\bullet$
correlation in all nine spirals with precise black hole measurements within 16
Mpc: NGC\,1068, NGC\,3368, NGC\,4051, NGC\,4151, NGC\,4258, NGC\,4395,
NGC\,4736, NGC\,4826, and NGC\,5055, although so far we have full data sets
only for NGC\,3368, NGC\,4258, NGC\,4395, NGC\,4736, and NGC\,4826.  Bulge type
was not considered when selecting the sample, but with one exception all of
them seem to have a pseudobulge \citep{korm10,korm13a,korm13b}, while
NGC\,4258, NGC\,4826, and NGC\,5055 may also have a classical bulge. The single
exception, NGC\,4395, is a pure disk, bulgeless, spiral \citep{fili03,korm13a}.
For completeness, we should mention that M\,31 and M\,81 have classical bulges,
NGC\,253 and the  MW have pseudobulges, and M\,104 has both a bulge and a
pseudobulge \citep{korm13a}.  Pseudobulges are widely considered the product of
secular evolution with few major mergers \citep[e.g.,][]{korm04,atha05},
although according to, for example, \citet{sauv18}, they can also be formed by
major mergers of gas-rich progenitors in the distant universe. 

In this paper, we present new measurements and analyses of the globular cluster
systems of the Sab galaxies NGC\,3368 (M\,96), NGC\,4736 (M\,94) and NGC\,4826
(M\,64), and the Sm galaxy NGC\,4395.  Their surface brightness fluctuations
(SBF) distances are 10.4$^{+1.1}_{-1.0}$ Mpc for NGC\,3368, 5.0$\pm$ 0.4 Mpc
for NGC\,4736, and 7.3$\pm$ 0.7 Mpc for NGC\,4826 \citep{tonr01,blak10}.
NGC\,4395 is completely devoid of a bulge, and hence lacks an SBF measurement;
we adopt a Cepheid distance of 4.3$\pm$ 0.4 Mpc \citep{thim04}.  The masses of
their central supermassive black holes are (3.6 $\pm$ 1.1)$\times 10^5\
M_\odot$ for NGC\,4395 \citep{pete05}; (7.5 $\pm$ 1.5)$\times 10^6\ M_\odot$
for NGC\,3368; 6.8$^{+1.5}_{-1.6} \times 10^6\ M_\odot$ for NGC\,4736, and
(1.6$\pm$0.4)$\times 10^6\ M_\odot$ for NGC\,4826 \citep{korm13a}. Their
inclinations to the line of sight are, respectively, 46$\fdg$2 (NGC\,3368),
33$\fdg$7 (NGC\,4395), 35$\fdg$6 (NGC\,4736) and  57$\fdg$5 (NGC\,4826)
\citep{rc3}. We take the following position angles (PA; North through East):
5$\degr$ (NGC\,3368), 147$\degr$ (NGC\,4395), 115$\degr$ (NGC\,4736 and
NGC\,4826). 

In order to identify globular cluster candidates (GCCs) in these galaxies, we
apply the ($u^*-i'$) vs.\ ($i'-K_s$) diagram technique (\uiks\ hereafter). It
is extremely powerful to isolate, in different regions and according to their
stellar populations and star formation histories, foreground Galactic stars,
background galaxies, young clusters in the disk, and GCCs. In addition, objects
in such separate areas of the plot also differ in their light concentration
parameters, at least up to a distance of $\sim$ 16 Mpc \citep{muno14}. The
efficacy of the procedure has been corroborated by \citet{powa16} in M87, and
by \citet{gonz19}, with OSIRIS observations of the GC candidates in the
megamaser prototype NGC\,4258. The method allows for a very low contamination
of the samples from foreground stars, background galaxies, and young stellar
clusters, even in disk galaxies that are not completely edge-on, and in the
absence of radial velocity measurements. 

\section{Data} \label{sec:data}

All the data for this work were obtained with the
Canada-France-Hawaii-Telescope (CFHT), the optical images with MegaCam
\citep{boul03}, and the \ks-band ones with the Wide-field InfraRed Camera
\citep[WIRCam;][]{puge04}. MegaCam has a field-of-view (FOV) of 0$\fdg96 \times
0\fdg$94, with a plate scale of 0$\farcs$186 pixel$^{-1}$, while the FOV of the
WIRCam is $\sim 21^\prime \times 21^\prime$, with a plate scale of 0$\farcs$307
pixel$^{-1}$.  Brief descriptions of the detectors' layouts, cameras'
performances, and data reduction procedures are given in \citet{gonz17}.
Table~\ref{tab:tabobs} gives a summary of the observations.  

Except for NGC\,4395, the \gp, \ip, and \rp\ data are archival, and were
originally secured through programs 11AC08 (P.I.\ G.\ Harris), and 13AS03
(P.I.\ Zhao-Yu Li). The \ust\ and \ks-band images, as well as all the NGC\,4395
data, are ours, and were acquired through programs 17AF12, 18AF99 (P.I.\ K.\
\'Alamo-Mart\'{\i}nez), and 19AF03 (P.I.\ Y.\ \'Ordenes-Brice\~no). 

To avoid saturation, individual \ks\ exposures of the galaxies were only 10\,s
long. For NGC\,3368, one image was taken in each of the four quadrants of the
WIRCam, before nodding the telescope to obtain a sky exposure 2$\fdg$5 apart.
The other three galaxies were observed in an approximately circular pattern
with a radius $\sim$ 1$\farcm$6. Given their large diameters compared to the
WIRCam FOV \citep[i.e., $R_{25} = 6\farcm6$, $5\farcm6$, and  5$^{\prime}$,
respectively, for NGC\,4395, NGC\,4736, and NGC\, 4826;][]{rc3}, separate (also
10\, s long) sky frames were taken, respectively, 2$\fdg$5 (NGC\,4395),
1$\fdg$5 (NGC\,4736) and 3$\fdg$5 (NGC\,4826) away, with the same circular
pattern, using the target(T)-sky(S) sequence STTSSTTSS...TTS.

\begin{figure}[ht!]
\plottwo{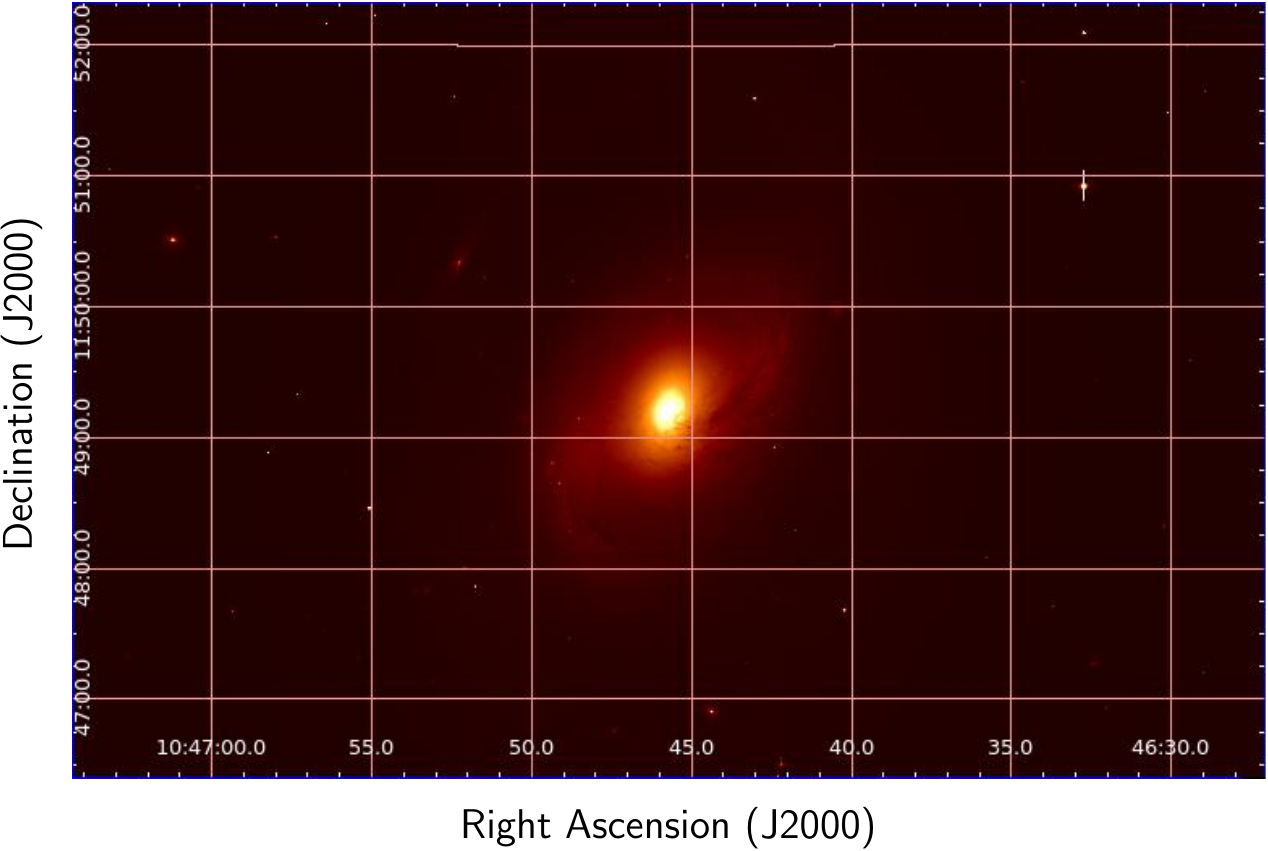}{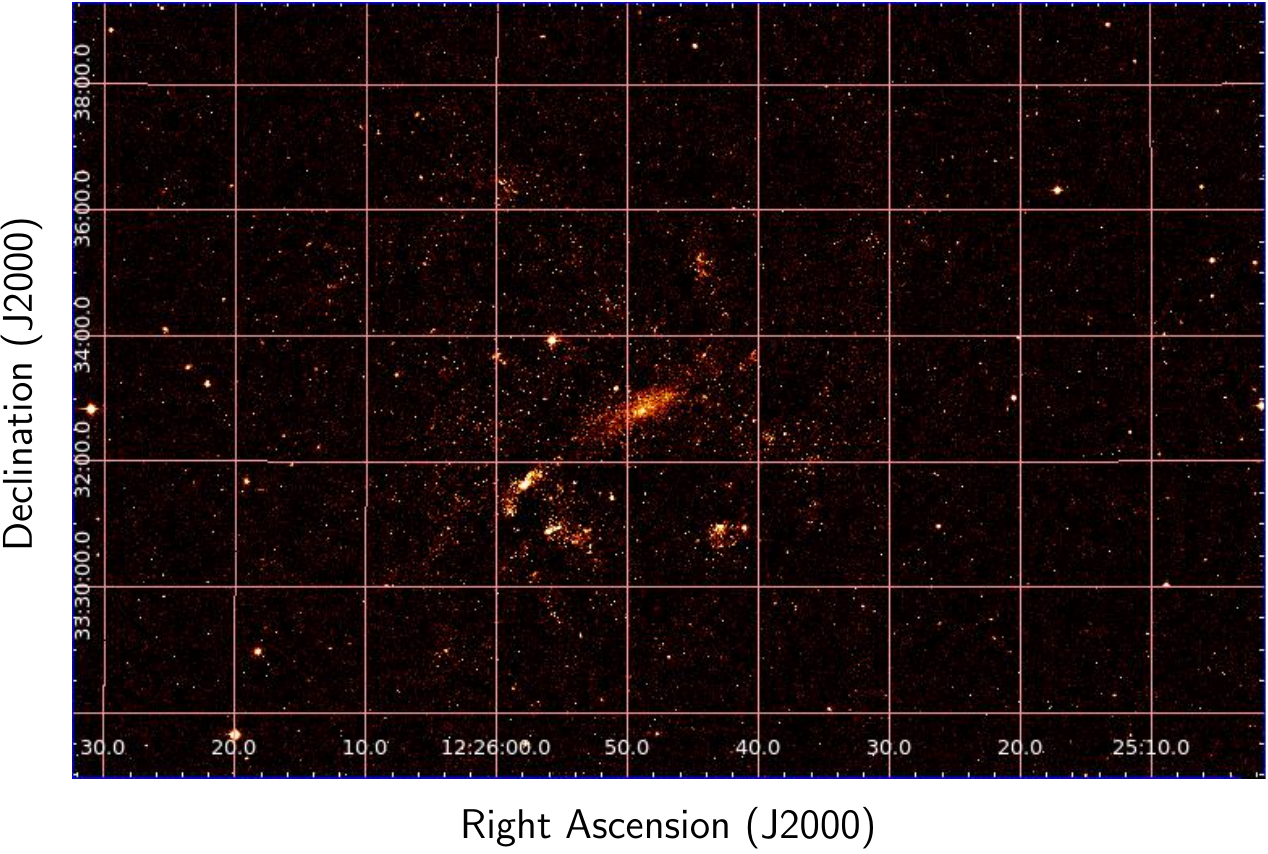}
\plottwo{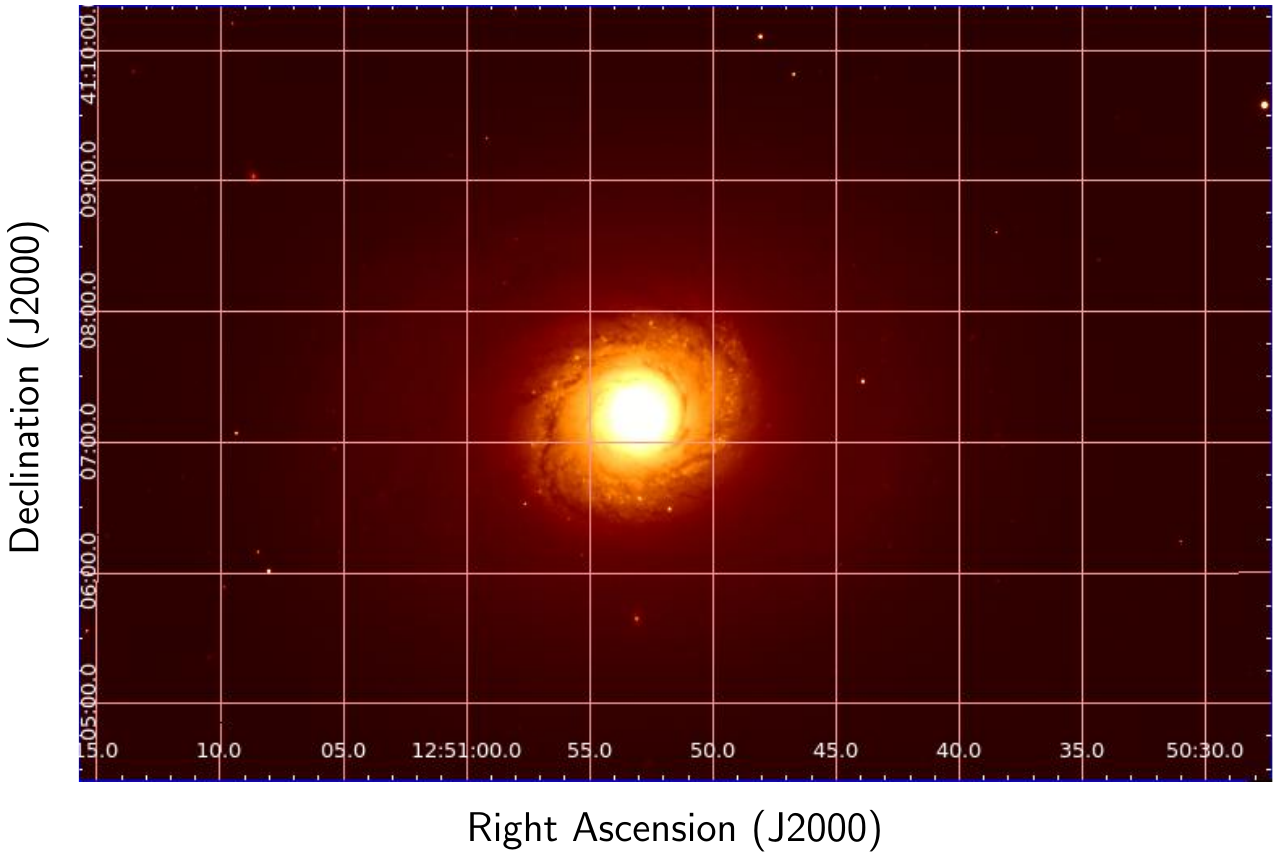}{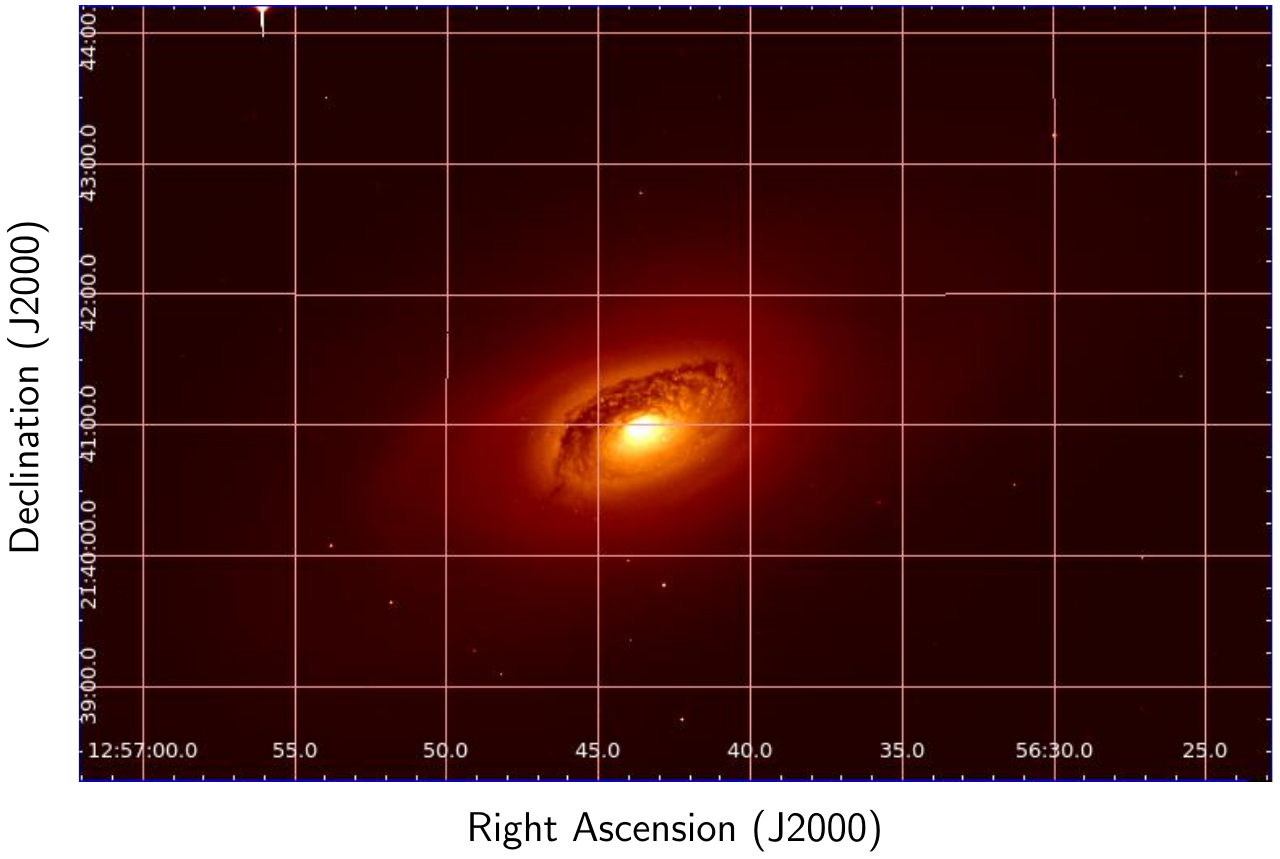}
\caption{CFHT \ip-band images, asinh scale. {\it Top left:} NGC\,3368; {\it top
right:} NGC\,4395; {\it bottom left:} NGC\,4736; {\it bottom right:} NGC\,4826.
\label{fig:lasgalaxias}}
\end{figure}

Before combining them into mosaics, the individual optical images were
recalibrated with the MegaPipe pipeline \citep{gwyn08}, in order to achieve
astrometric internal and external accuracies of 0$\farcs$04 and 0$\farcs$15,
respectively, and a photometric accuracy of 0.03 mag.  The \ks-band mosaics
were produced with the WIRwolf pipeline \citep{gwyn14}, which sky-subtracts
individual images and achieves an internal astrometric accuracy of 0$\farcs$1.
Both MegaPipe and WIRwolf render images with photometry in the AB system
\citep{oke74}, with zero points ($zp$) set to 30 mag.
Figure~\ref{fig:lasgalaxias} shows the images of the four galaxies in the
\ip-band.

\medskip

\floattable
\begin{deluxetable}{cCRRrlccl}
\tablecaption{ Observation Log\label{tab:obsdata}}
\tablewidth{0pt}
\tablehead{
\colhead{Galaxy} &\colhead{\rm Filter} & \colhead{$\lambda_{\rm cen}^a$} & \colhead{ {\rm FWHM}$^b$} & \colhead{Exposure} & \colhead{Camera} &\colhead{Pixel size} &\colhead{ Program }  &\colhead{Date} \\
\colhead{} & \colhead{       }  & \colhead{                 }  & \colhead{          } & \colhead{s} &
\colhead{} & \colhead{$\arcsec$} & \colhead{} & \colhead{UT}
}
\startdata
NGC\,3368 &  u^*     (302)  & 3537\  \text{\AA} & 867\  \text{\AA} & 5200   &Megacam & 0.186  &18AF99  & 2018 June 16         \\               
          &                 &                   &                  &        &        &        &19AF03  & 2019 March 04         \\               
          & g^\prime (401)  & 4873\  \text{\AA}  & 1455\  \text{\AA}&1000   &Megacam & 0.186  &11AC08  & 2011 March 01         \\
          & i^\prime (702)  & 7776\  \text{\AA}&  1508\  \text{\AA} &1600   &Megacam & 0.186  &11AC08  & 2011 March 01         \\
          & K_s             & 2.15 \mu\text{m} & 0.33\ \mu\text{m}  & 1660   &WIRCam  & 0.307  &17AF12  & 2017 April 18         \\ 
NGC\,4395 & u^*     (302)   & 3537\  \text{\AA} & 867\  \text{\AA} &  1500  &Megacam & 0.186  &18AF99  & 2018 June 14      \\
          & g^\prime (402)  & 4720\  \text{\AA} & 1520\ \text{\AA} &  400   & Megacam& 0.186  &18AF04  & 2018 May 23         \\
          &  i^\prime (703) & 7764\  \text{\AA} & 1554\ \text{\AA} & 200    &Megacam & 0.186  &19AF03  & 2019 April 03       \\
          & K_s             & 2.15 \mu\text{m} & 0.33\ \mu\text{m}  &460    &WIRCam  & 0.307  &18AF98  & 2018 July  12       \\ 
NGC\,4736 &  u^*     (302)  &  3537\  \text{\AA} & 867\  \text{\AA} &  6900 & Megacam & 0.186  & 18AF99 & 2018 May 23, June 12 \\
         &                  &           &          &        &         &        & 19AF03 & 2019 March 04 \\
         &  g^\prime (401)  &  4873\  \text{\AA}  & 1455\  \text{\AA}       &   1645 & Megacam & 0.186  & 13AS03 & 2013 February 15\\
         &  r^\prime (601)  &  6282\  \text{\AA}  & 1219\  \text{\AA}       &  3304  & Megacam & 0.186     & 13AS03 & 2013 February 13, 15\\
         &  i^\prime (702)  &  7776\  \text{\AA}&  1508\  \text{\AA}        &  1127  & Megacam & 0.186     & 13AS03 & 2013 February 12  \\
         & K_s  &  2.15\ \mu\text{m} & 0.33\ \mu\text{m} & 760 &  WIRCam  &   0.307 &  17AF12 & 2017 April 08  \\
NGC\,4826 &  u^*     (302)  &  3537\  \text{\AA} & 867\  \text{\AA} &  9600 & Megacam & 0.186  & 18AF99  & 2018 May 23, June 13 \\
         &                  &           &          &        &         &        & 19AF03 & 2019 April 03 \\
         &  g^\prime (401)  &  4873\  \text{\AA}  & 1455\  \text{\AA}       &   2345 & Megacam & 0.186  & 11AC08  & 2011 March 01\\
         &                  &             &                 &        &         &        & 13AS03  & 2013 February 13\\
         &  r^\prime (601)  &  6282\  \text{\AA}  & 1219\  \text{\AA}       &  4963  & Megacam & 0.186     & 13AS03 & 2013 February 15, 17; May 03\\
         &  i^\prime (702)  &  7776\  \text{\AA}&  1508\  \text{\AA}        &  1827  & Megacam & 0.186     & 11AC08 & 2011  March 01  \\
         &                  &           &                   &        &         &           & 13AS03 & 2013  May 02  \\
         & K_s  &  2.15\ \mu\text{m} & 0.33\ \mu\text{m} & 880 &  WIRCam  &   0.307 &  17AF12  & 2017 April 13 
\enddata

\tablecomments{\ $^a$The central wavelength between the two points defining FWMH
(\url{http://svo2.cab.inta-csic.es/svo/theory/fps3/index.php?id=CFHT/}). \\
$^b$ {\it Ibid.}
}
\label{tab:tabobs}
\end{deluxetable}


\section{Detection and photometry}\label{sec:detection}

Source detection and photometric measurements in all the stacked images were
carried out with SExtractor \citep{bert96} and PSFEx \citep{bert11}. The
detection was performed on images from which their median-filtered version was
subtracted; faint sources are detected more easily in a median-subtracted
image. All median images were produced with a 31 pix$\times$31 pix median
filter. Conversely, the photometry measurements were made on the original
stacked images.  PSFEx employs point sources detected in a first pass of
SExtractor to build a point-spread function (PSF) model that SExtractor can
then apply in a second pass to obtain PSF magnitudes of sources.  We chose
adequate PSF stars manually, based on their brightness versus compactness, as
measured by SExtractor parameters MAG\_AUTO \citep[a Kron-like elliptical
aperture magnitude,][]{kron80} and FLUX\_RADIUS (similar to the effective
radius); their values for each filter and galaxy are given in
Table~\ref{tab:psfcrit}. A detailed description of this procedure is presented
in \citet{gonz17}. The spatial variations of the PSF were modeled with
polynomials of degree 3.  To create the PSF, the flux of each star was measured
in an 18 pixel aperture in all bands (equivalent to $3\farcs3$ in the optical
and $5\farcs5$ at $K_s$); such aperture, determined through the growth-curve
method for each passband, is large enough to measure the total flux of the
stars, but small enough to reduce the likelihood of contamination by extraneous
objects.

\floattable
\begin{deluxetable}{cCccccr}
\tablecaption{PSF stars parameters\label{tab:psfcrit}}
\tablewidth{0pt}
\tablehead{
\colhead{      }& \colhead{      }& \multicolumn2c{FLUX\_RADIUS}& \multicolumn2c{MAG\_AUTO}& \colhead {}\\
\colhead{Galaxy}& \colhead{Filter}& \colhead{Min}&  \colhead{Max}& \colhead{Min}& \colhead{Max}& \colhead{$N$ PSF stars} \\
\colhead{}& \colhead {} &           \multicolumn2c{pixel} & \multicolumn2c{mag} & \colhead{}
}
\decimals
\startdata
NGC\,3368 & u^*       &   2.10  & 2.60   & 16.0    & 23.0    &  1998 \\
         & g^\prime   &   1.60  & 2.00   & 17.5    & 25.5    &  3876 \\
         & i^\prime   &   1.30  & 1.60   & 17.0    & 23.0    &  5306 \\
         & K_s        &   1.10  & 1.60   & 15.5    & 20.0    &  1162 \\
NGC\,4395 & u^*       &   1.80  & 2.40   & 17.0    & 23.0    &  2379 \\
         & g^\prime   &   1.70  & 2.20   & 17.0    & 23.0    &  3937 \\
         & i^\prime   &   1.60  & 2.10   & 15.0    & 21.5    &  3741 \\
         & K_s        &   1.10  & 1.60   & 16.0    & 20.0    &   449 \\
NGC\,4736 & u^*       &   2.00  & 2.70   & 15.5    & 22.0    &  1296 \\
         & g^\prime   &   3.10  & 3.70   & 16.0    & 22.9    &  3112 \\
         & r^\prime   &   3.20  & 3.70   & 15.5    & 22.0    &  4302 \\
         & i^\prime   &   2.50  & 2.80   & 16.0    & 22.0    &  5047 \\
         & K_s        &   1.30  & 1.70   & 15.0    & 19.5    &   320 \\
NGC 4826 &  u^*       &   2.50  & 3.10   & 15.0    & 23.0    &  2300 \\
         & g^\prime   &   2.00  & 2.40   & 16.0    & 23.0    &  2379 \\
         & r^\prime   &   3.00  & 3.60   & 16.0    & 22.0    &  5115 \\
         & i^\prime   &   1.80  & 2.20   & 16.0    & 22.0    &  4805 \\
         & K_s        &   1.10  & 1.40   & 15.5    & 20.5    &   519 \\
\enddata
\end{deluxetable}

\section{Completeness} \label{sec:completeness}

Completeness tests were only performed on the \ks-band mosaics; these are the
smallest and shallowest images, and in fact set our object detection limits. In
order to determine the GC detection completeness as a function of magnitude,
65,000 artificial point sources were generated for each galaxy, based on their
respective \ks\ PSF model, in the interval  19 mag $< m_{Ks} <$ 24 mag, and
with a uniform, or box-shaped, magnitude distribution.  Due to the WIRCam pixel
size,\footnote{0$\farcs$307 or $\sim$ 6 pc, $\gtrsim$ 7 pc, $\sim$ 11 pc, and
$\sim$ 14.5 pc  at the distances of, respectively, NGC\,4395, NGC\,4736,
NGC\,4826, and NGC\,3368.} GCCs are mostly unresolved in the \ks-band images.
The artificial objects were added only 500 at a time, to avoid creating
artificial crowding; that is, a total of 130 images with simulated sources were
produced for each galaxy.  The positions of the added sources were random, and
assigned  without regard to the positions of other artificial objects or of
real sources.

Artificial sources were then recovered by running SExtractor on each one of the
simulated images, with the same parameters used for the original \ks\ galaxy
images, and cross-matching the positions of all detections with the known input
coordinates of the added objects.  Sources with SExtractor output parameter
FLAGS $\neq$ 0 were eliminated; this operation excludes artificial objects
falling on top of other artificial or real sources, and hence discards
preferentially added objects in crowded regions. The restriction FLAGS=0 was
also applied in the selection of true sources (see Section~\ref{sec:uiks}).

\begin{figure}[ht!]
\plottwo{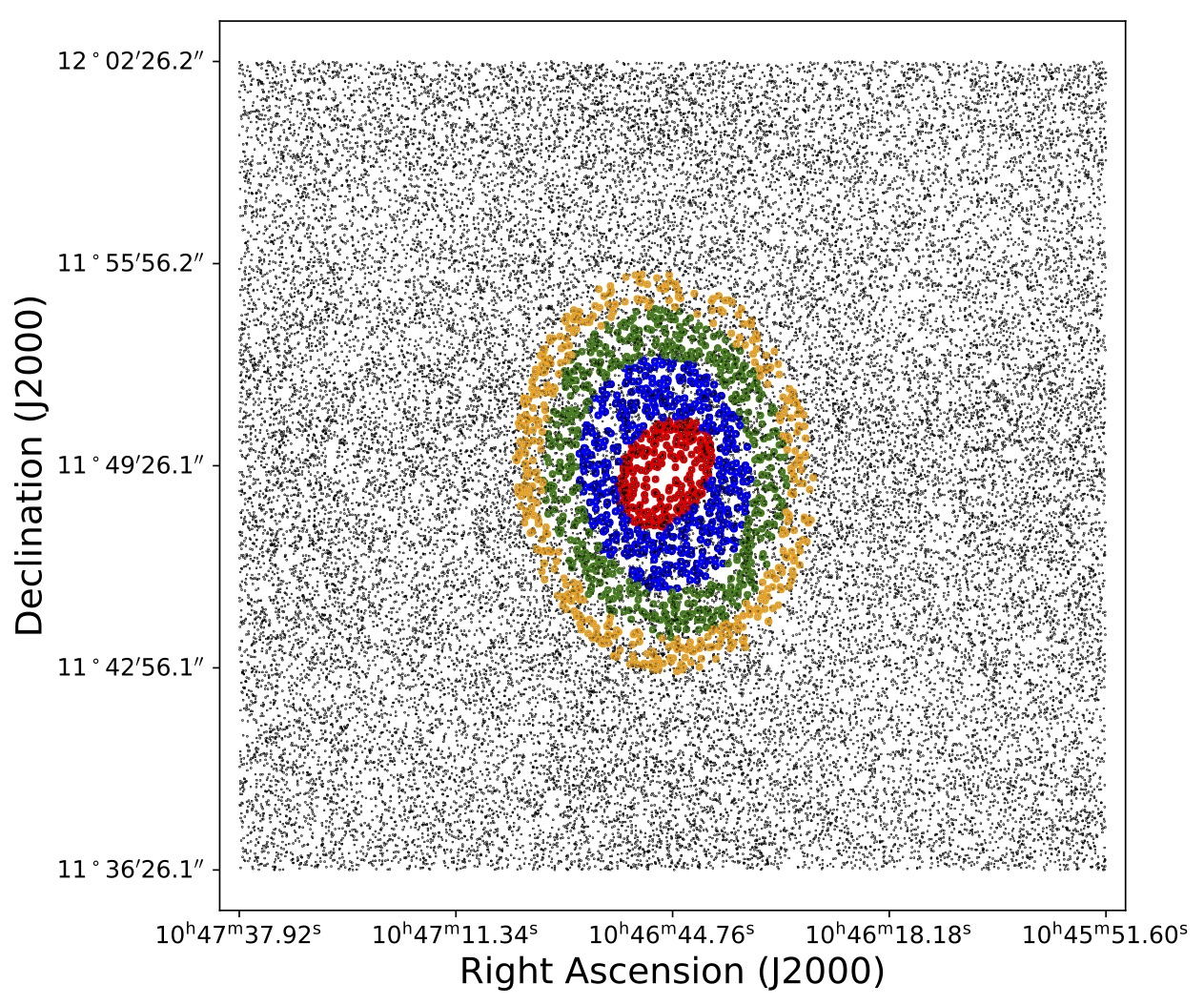}{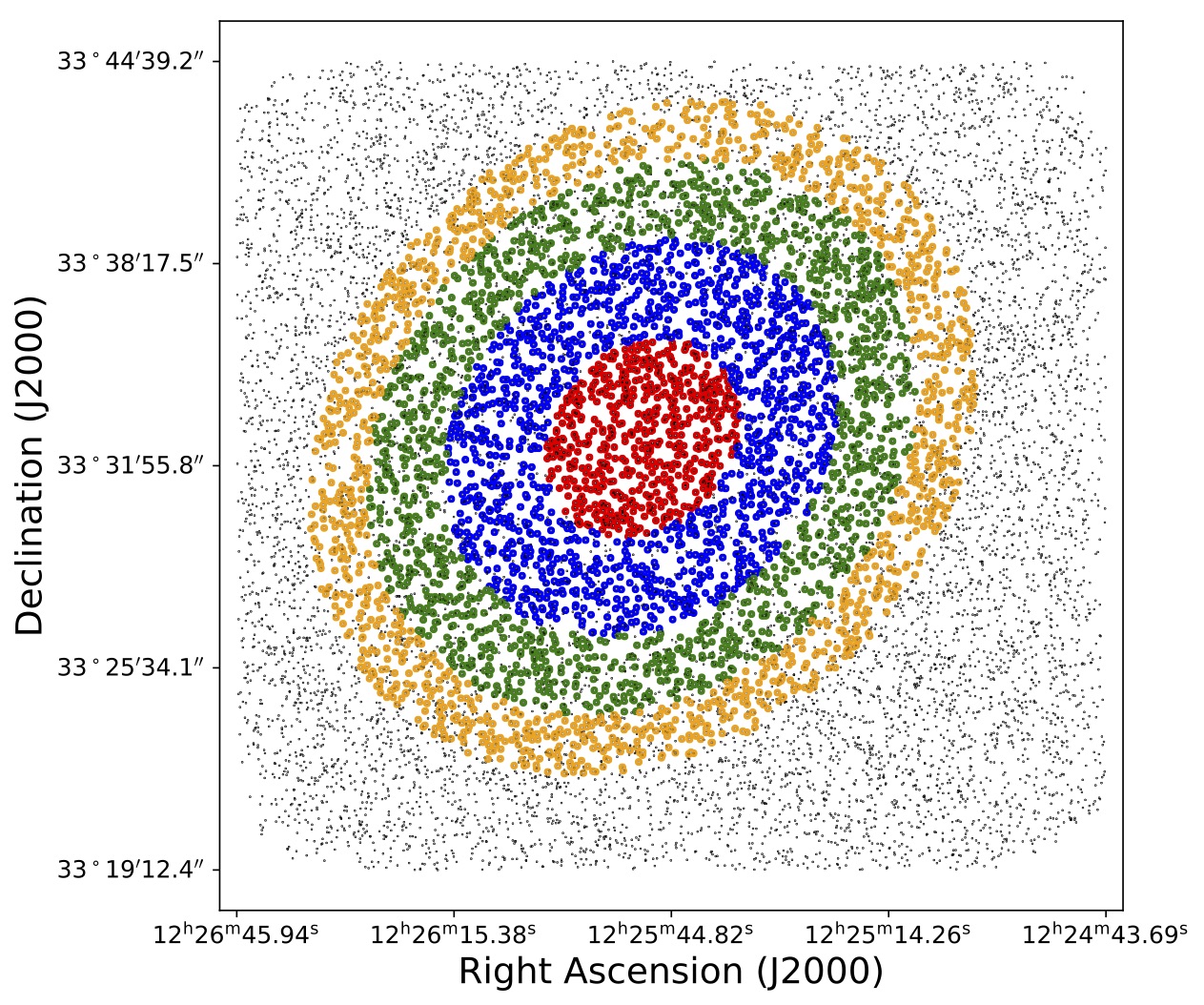}
\plottwo{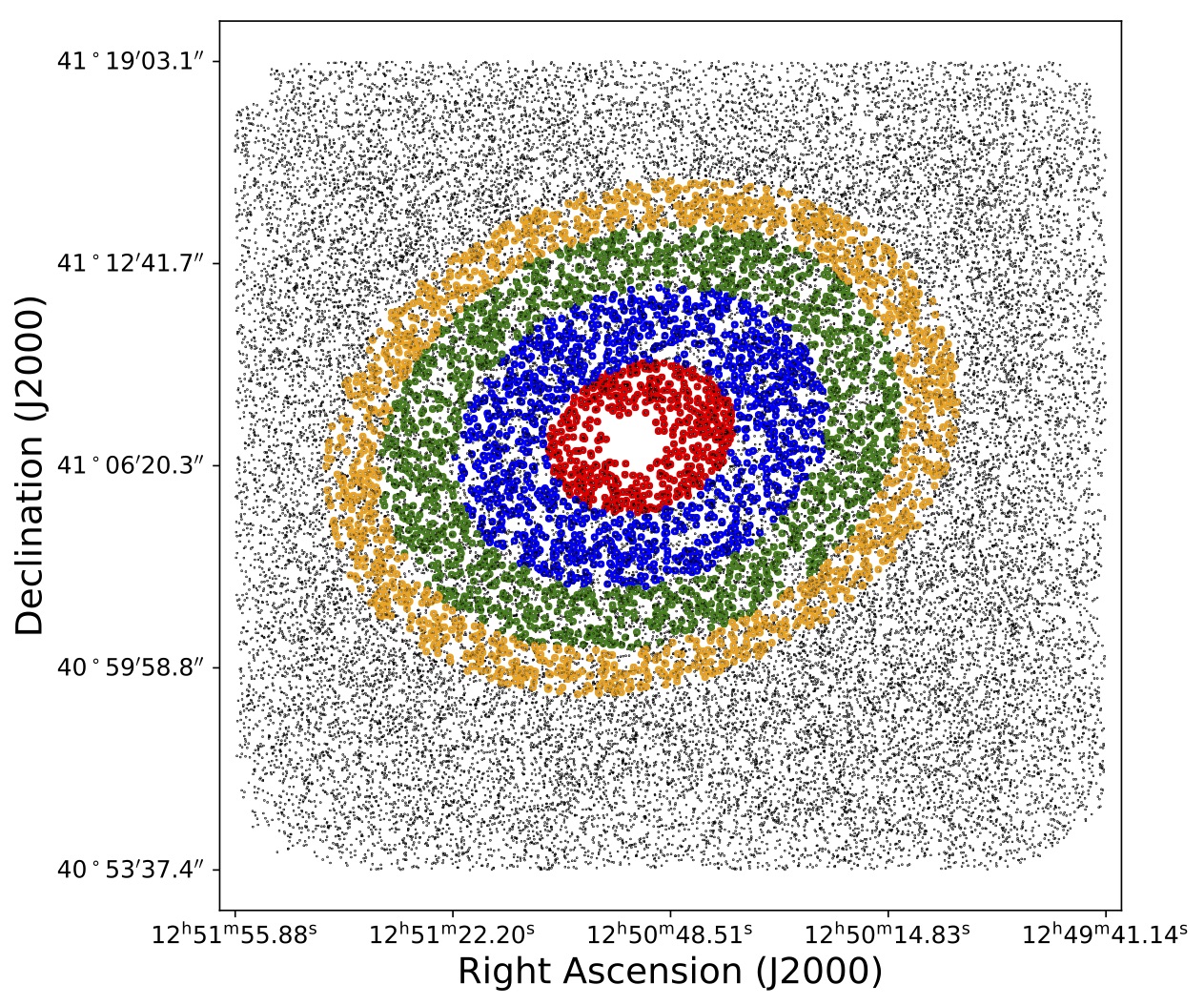}{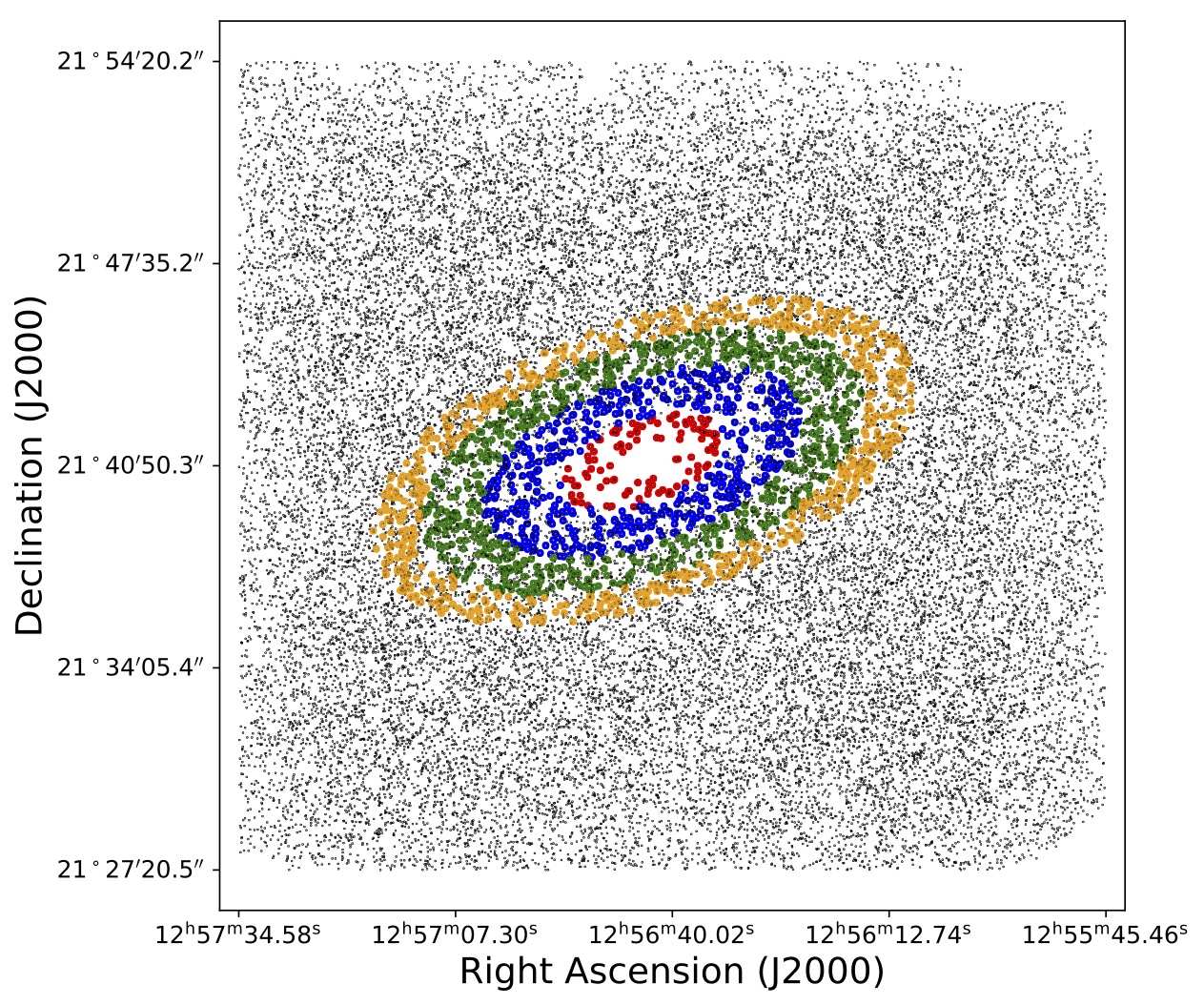}
\caption{Completeness tests, recovered artificial point sources. {\it Top
left:} NGC\,3368; {\it top right:} NGC\,4395; {\it bottom left:} NGC\,4736;
{\it bottom right:} NGC\,4826.  Colors indicate points from artificial images
that were actually used in the fits, to ensure the samples of all regions were
statistically equivalent.
\label{fig:artificial}}
\end{figure}

As done previously with NGC\,4258 by \citet{gonz17}, for all the galaxies we
estimated completeness in four different regions within 1.7 $R_{25}$: an
ellipse with semimajor axis equal to 0.5 $R_{25}$, and three elliptical annuli,
respectively, from 0.5 to 1.0 $R_{25}$, from 1.0 to 1.4 $R_{25}$, and between
1.4 and 1.7 $R_{25}$. With the exception of the innermost ellipse of NGC\,3368,
all regions have the axis ratios and PA assumed for the observed disks of the
galaxies (respectively, 0.69 and 5$\degr$ for NGC\,3368, 0.82 and 147$\degr$
for NGC\,4395, 0.81  and 115$\degr$ for NGC\,4736, and 0.54 and 115$\degr$ for
NGC\,4826); the PA of the central ellipse of NGC\,3368 follows the alignment of
the surface brightness there, and hence differs from the orientation of the
outer isophotes.  With the goal of estimating completeness with statistically
equivalent samples, for each region we considered a number of added sources
inversely proportional to its area; in other words, for the inner ellipses the
artificial sources from all 130 images were taken into account, whereas only
43, 34, and 39 were included for the first, second, and third annuli,
respectively.   On average, for each region we consider $\sim$ 2000 artificial
sources in the cases of NGC\,4395 and NGC\,4736, and $\sim$ 1000 for NGC\,3368
and NGC\,4826.  Figure~\ref{fig:artificial} shows all recovered sources in
black, and those from the artificial images actually considered in colors; in
the innermost ellipse (within 0.5 $R_{25}$), all recovered sources were used,
hence all dots appear colored in both black and red.  

In the case of a box-shaped magnitude distribution, the fraction of recovered
artificial sources as a function of magnitude is often well described by the
Pritchet function \citep[e.g.,][]{mcla94}:

\begin{equation}
f^P(m) = \frac{1}{2}  \left[ 1\ -\ \frac{\alpha_{\rm cutoff} (m - m_{\rm lim})}{\sqrt{1 + \alpha_{\rm cutoff}^2(m - m_{\rm lim})^2}}  \right];
\label{eq:pritchet}
\end{equation}

\noindent
$m_{\rm lim}$ is the magnitude at which completeness is 50\%, and $\alpha_{\rm
cutoff}$ determines the steepness of the cutoff.  However, in the case of
crowded regions, where even at bright magnitudes the recovered fraction is flat
but $< 1$, better fits are obtained with a modified Pritchet function:

\begin{equation}
f^{mP}(m) = \frac{1}{2}  \left[ \gamma - \delta\ \frac{\alpha_{\rm cutoff} (m - m_{\rm lim})}{\sqrt{1 + \alpha_{\rm cutoff}^2(m - m_{\rm lim})^2}}  \right];
\label{eq:pritchmod}
\end{equation}

\noindent here, $\gamma$ is the asymptotic value of the fraction at bright
magnitudes, and $\delta$ is the fraction at the cutoff.

\begin{figure}[ht!]
\hspace*{-0.3cm}\plottwo{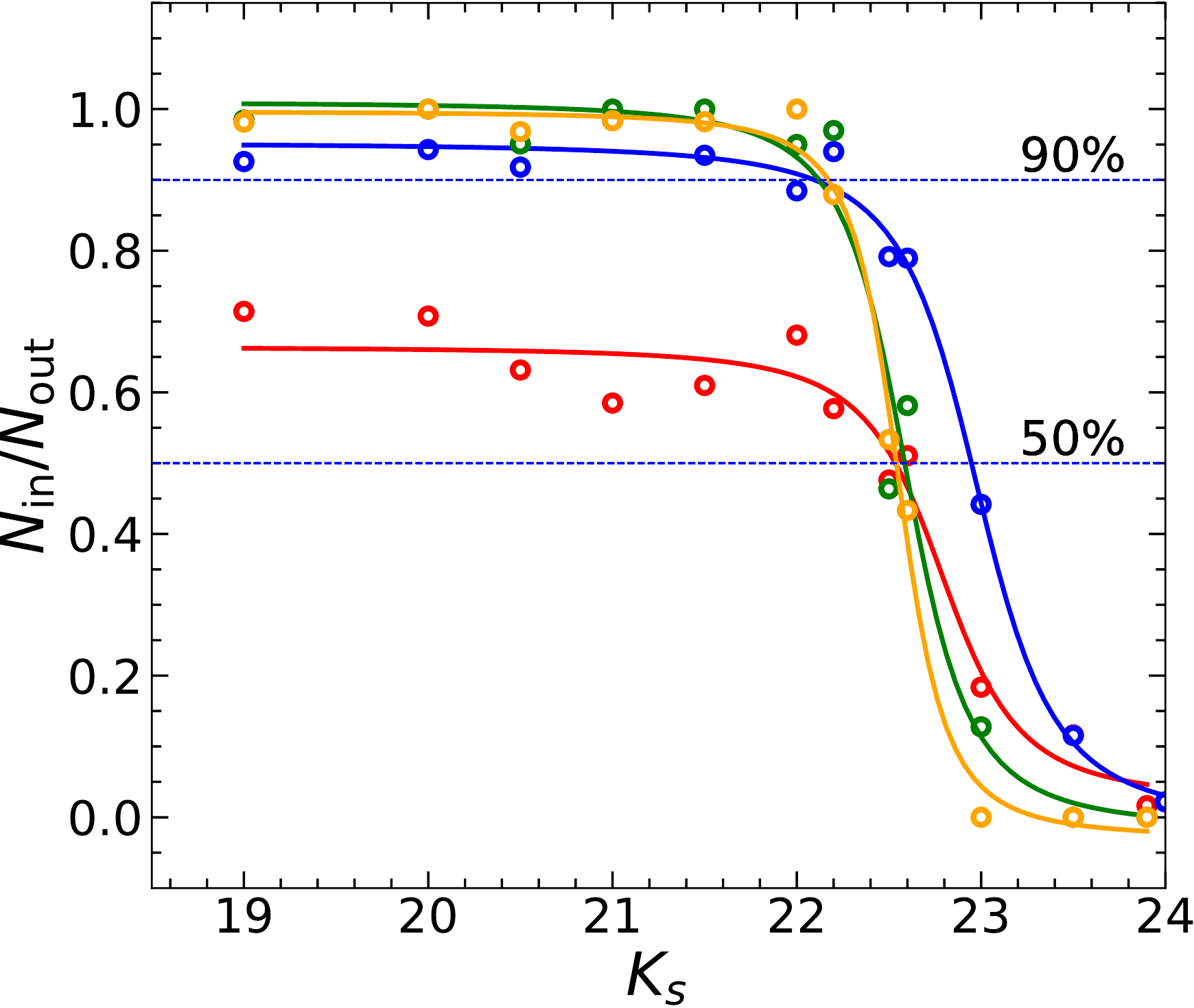}{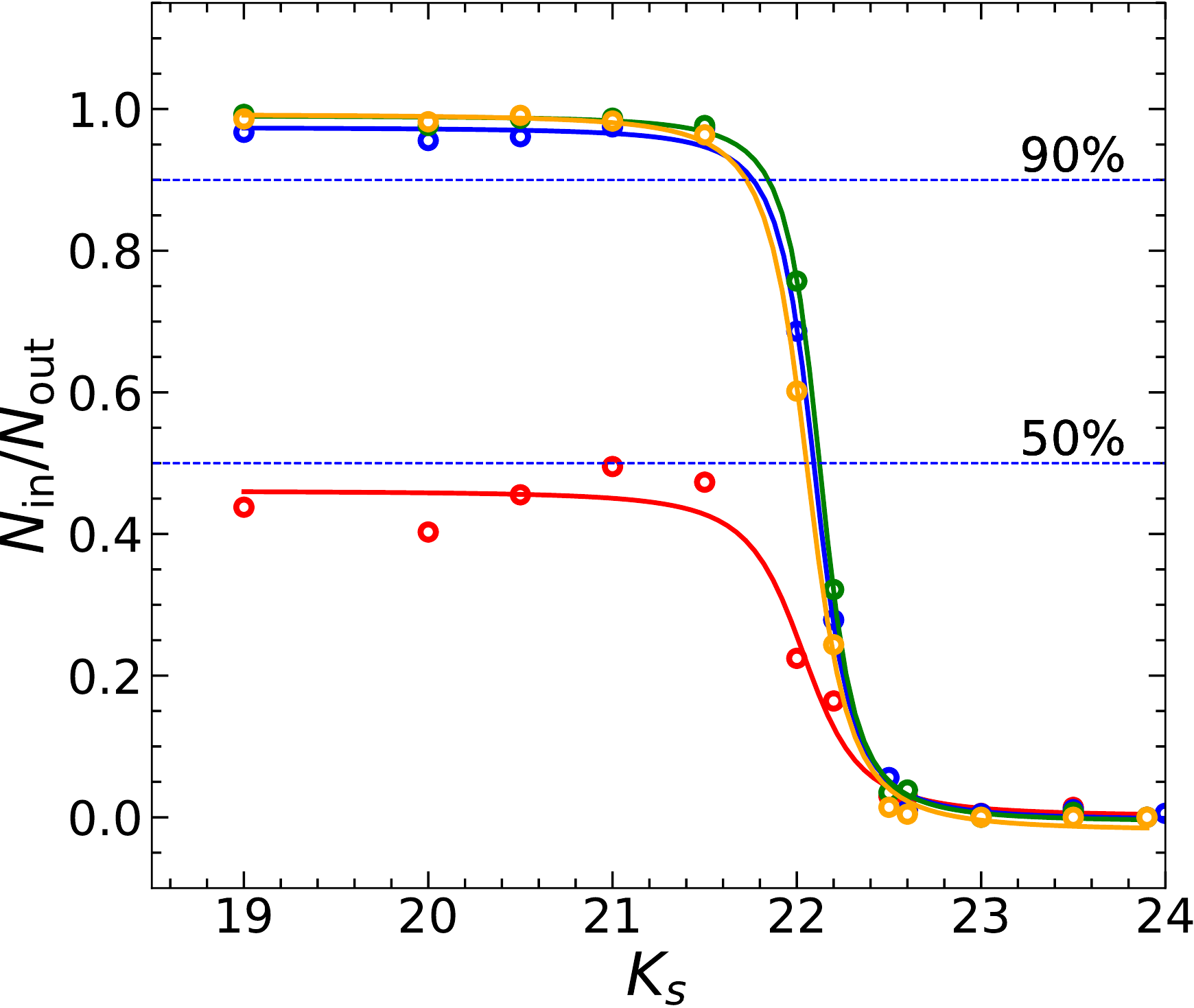}
\hspace*{-0.3cm}\plottwo{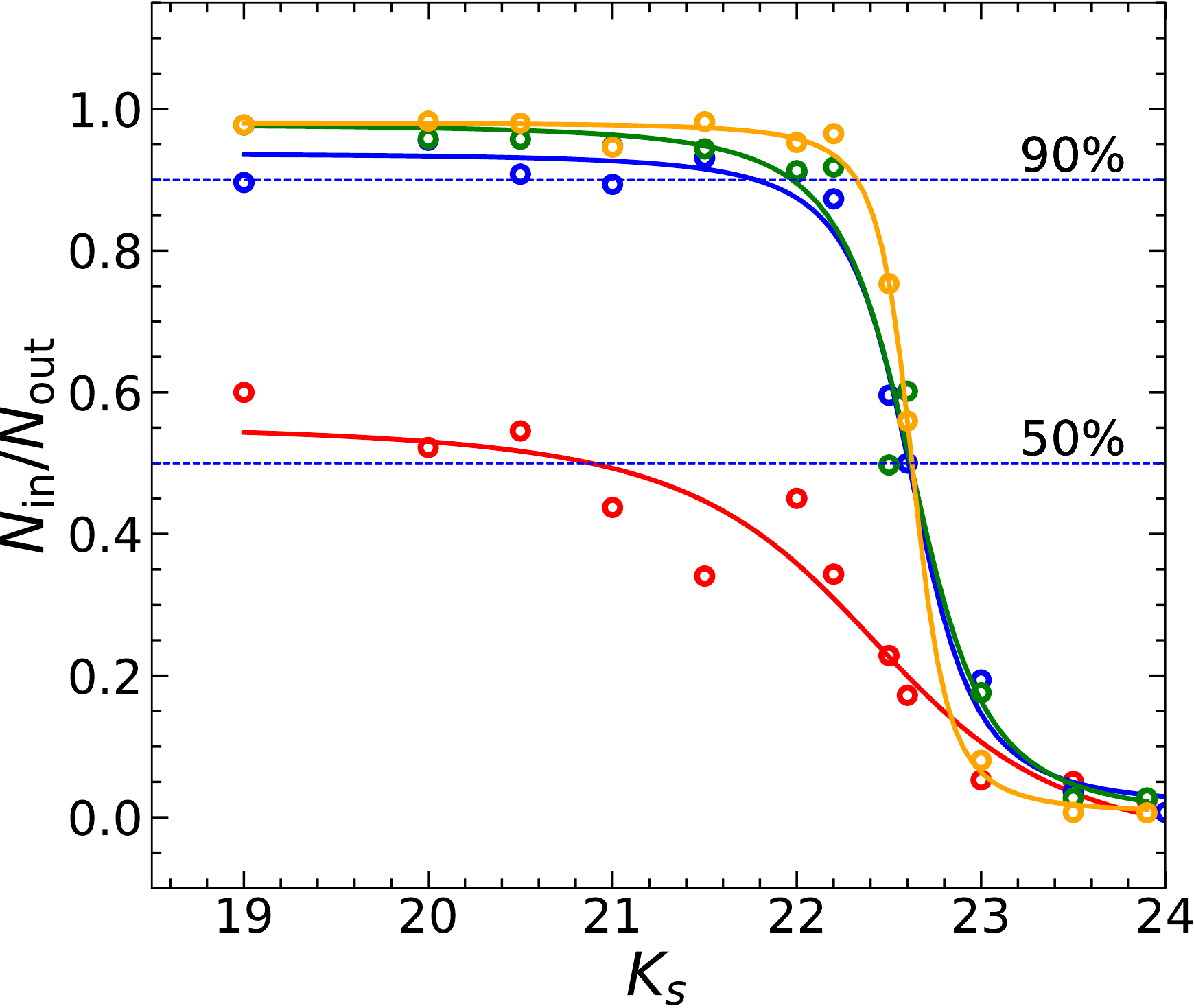}{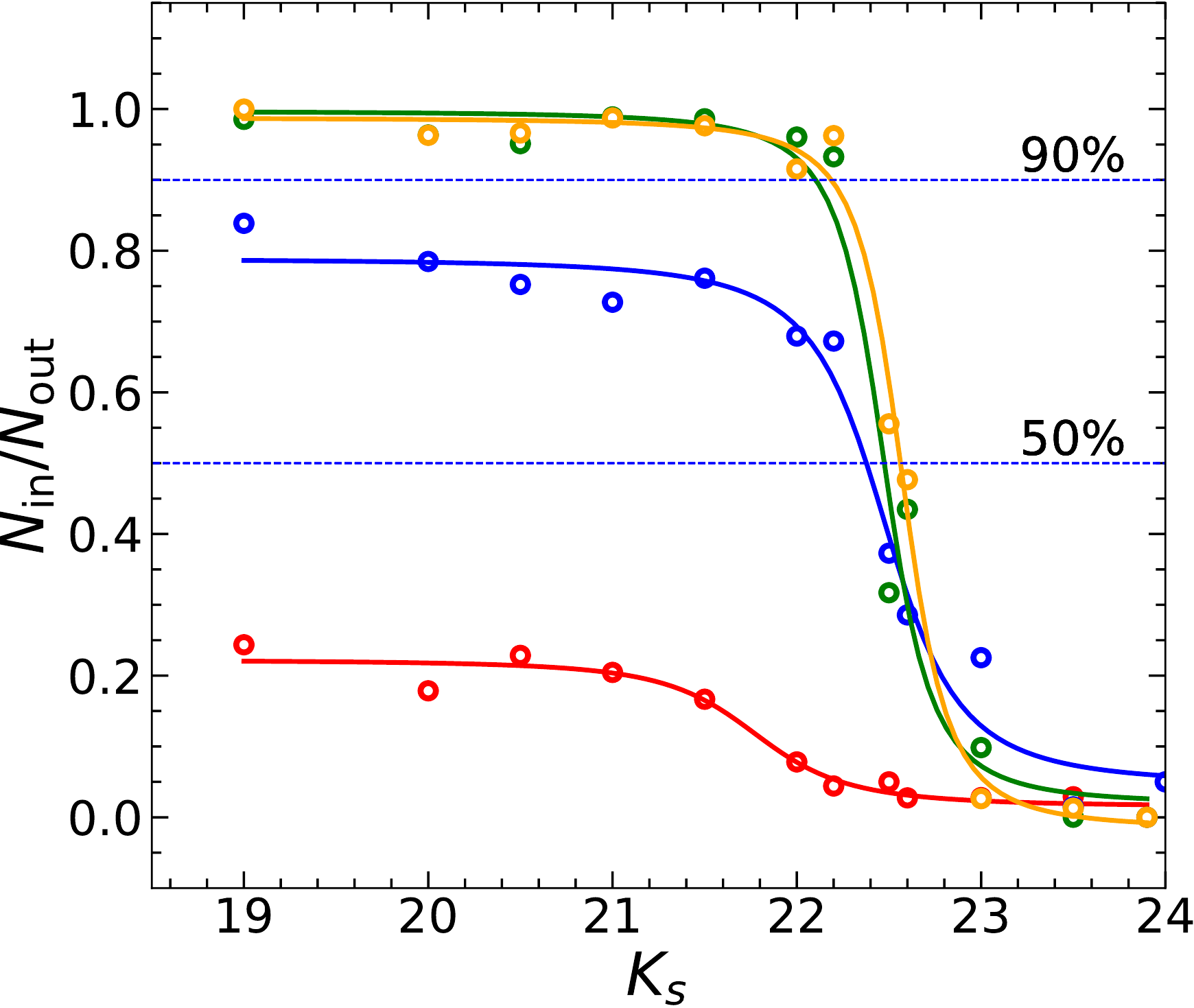}
\caption{Completeness tests. {\it Top left:} NGC\,3368; {\it top right:}
NGC\,4395; {\it bottom left:} NGC\,4736; {\it bottom right:} NGC\,4826.  Fits
({\it solid lines}) to recovered fractions ({\it dots}) versus $K_s$ mag with
eq.~\ref{eq:pritchmod}. Colors ({\it red, blue, green, gold}) refer to
centermost ellipse, and inner, middle, and external annuli, respectively. {\it
Blue dotted lines} indicate the 90\% and 50\% completeness values.
\label{fig:cmpltnss}}
\end{figure}

For each one of the galaxies, the quotient of recovered to added sources as a
function of output PSF magnitude was fit with equation~\ref{eq:pritchmod},
separately for each region. The recovered fractions (dots) and their fits
(solid lines) versus \ks\ mag are displayed in Figure~\ref{fig:cmpltnss}. It is
quite obvious that the regions within 0.5 $R_{25}$ are extremely affected by
crowding and confusion due to partially resolved sources.  This is true also
for NGC\,4395, in spite of the absence of a conspicuous lack of recovered
sources in the central region.  The values of the fit parameters $m_{\rm lim}$,
$\alpha_{\rm cutoff}$, $\gamma$, and $\delta$  are shown in
Table~\ref{tab:prit}.

\floattable
\begin{deluxetable}{cccccc}
\tablecaption{Completeness Fit Parameters\label{tab:prit}}
\tablewidth{0pt}
\tablehead{
\colhead{Galaxy}&\colhead{Region} & \colhead {$m_{\rm lim}$} & \colhead{$\alpha_{\rm cutoff}$} &\colhead{$\gamma$}&\colhead{$\delta$} \\
\colhead{      }&\colhead {$R_{25}$} & \colhead{$K_s$ AB mag} & \colhead{} &\colhead{} & \colhead{} 
}
\decimals
\startdata
NGC\,3368 &   0.0$-$0.5  &  22.89$\pm$0.09   &2.22$\pm$0.71      & 0.69$\pm$0.05 &  0.64$\pm$0.06  \\
         &   0.5$-$1.0  &  22.98$\pm$0.03   &2.24$\pm$0.20      & 0.94$\pm$0.20 &  0.96$\pm$0.20  \\
         &   1.0$-$1.4  &  22.69$\pm$0.05   &2.72$\pm$0.83      & 0.99$\pm$0.83 &  1.03$\pm$0.83  \\
         &   1.4$-$1.7  &  22.65$\pm$0.02   &3.68$\pm$0.66      & 0.97$\pm$0.66 &  1.03$\pm$0.66  \\
NGC\,4395 &  0.0$-$0.5  &  22.14$\pm$0.05   &3.09$\pm$1.15      & 0.46$\pm$0.02 &  0.46$\pm$0.03  \\
         &   0.5$-$1.0  &  22.10$\pm$0.01   &4.77$\pm$0.26      & 0.97$\pm$0.26 &  0.98$\pm$0.26  \\
         &   1.0$-$1.4  &  22.23$\pm$0.01   &5.08$\pm$0.23      & 0.98$\pm$0.23 &  1.00$\pm$0.23  \\
         &   1.4$-$1.7  &  22.16$\pm$0.01   &4.26$\pm$0.36      & 0.97$\pm$0.36 &  1.01$\pm$0.36  \\
NGC 4736 &   0.0$-$0.5  &  22.52$\pm$0.23   &0.90$\pm$0.45      & 0.50$\pm$0.08 &  0.62$\pm$0.14  \\
         &   0.5$-$1.0  &  22.63$\pm$0.03   &2.72$\pm$0.32      & 0.95$\pm$0.32 &  0.92$\pm$0.32  \\
         &   1.0$-$1.4  &  22.73$\pm$0.05   &2.36$\pm$0.43      & 0.98$\pm$0.43 &  0.98$\pm$0.43  \\
         &   1.4$-$1.7  &  22.73$\pm$0.01   &5.01$\pm$0.41      & 0.99$\pm$0.41 &  0.98$\pm$0.41  \\
NGC 4826 &   0.0$-$0.5  &  21.87$\pm$0.11   &1.86$\pm$0.65      & 0.24$\pm$0.01 &  0.21$\pm$0.02  \\
         &   0.5$-$1.0  &  22.48$\pm$0.05   &2.33$\pm$0.19      & 0.83$\pm$0.19 &  0.74$\pm$0.19  \\
         &   1.0$-$1.4  &  22.57$\pm$0.04   &3.64$\pm$0.98      & 1.01$\pm$0.98 &  0.98$\pm$0.98  \\
         &   1.4$-$1.7  &  22.67$\pm$0.02   &3.82$\pm$0.68      & 0.97$\pm$0.68 &  1.01$\pm$0.68  \\
\enddata
\end{deluxetable}

\section{Selection of Globular Cluster Candidates} \label{sec:uiks}

The main tool used for the selection of GCCs is the ($u^*\ -\ i^\prime$) versus
($i^\prime\ -\ K_s$) color-color diagram \citep{muno14}.  The combination of
optical and near infrared (NIR) data in this parameter space constitutes the
most powerful photometric-only method for the selection of a clean sample of GC
candidates.  It probes simultaneously the main-sequence turnoff, and the red
giant and horizontal branches of stellar populations.  In this plane, GCs
occupy a region that is distinct from the areas inhabited by background
galaxies, on the one hand, and foreground Galactic stars, on the other. Objects
in these various regions not only have different stellar populations, but light
concentration parameters that roughly distinguish GCs, which appear marginally
resolved in the optical, from both point and extended sources.  

\begin{figure}[h!]
\plottwo{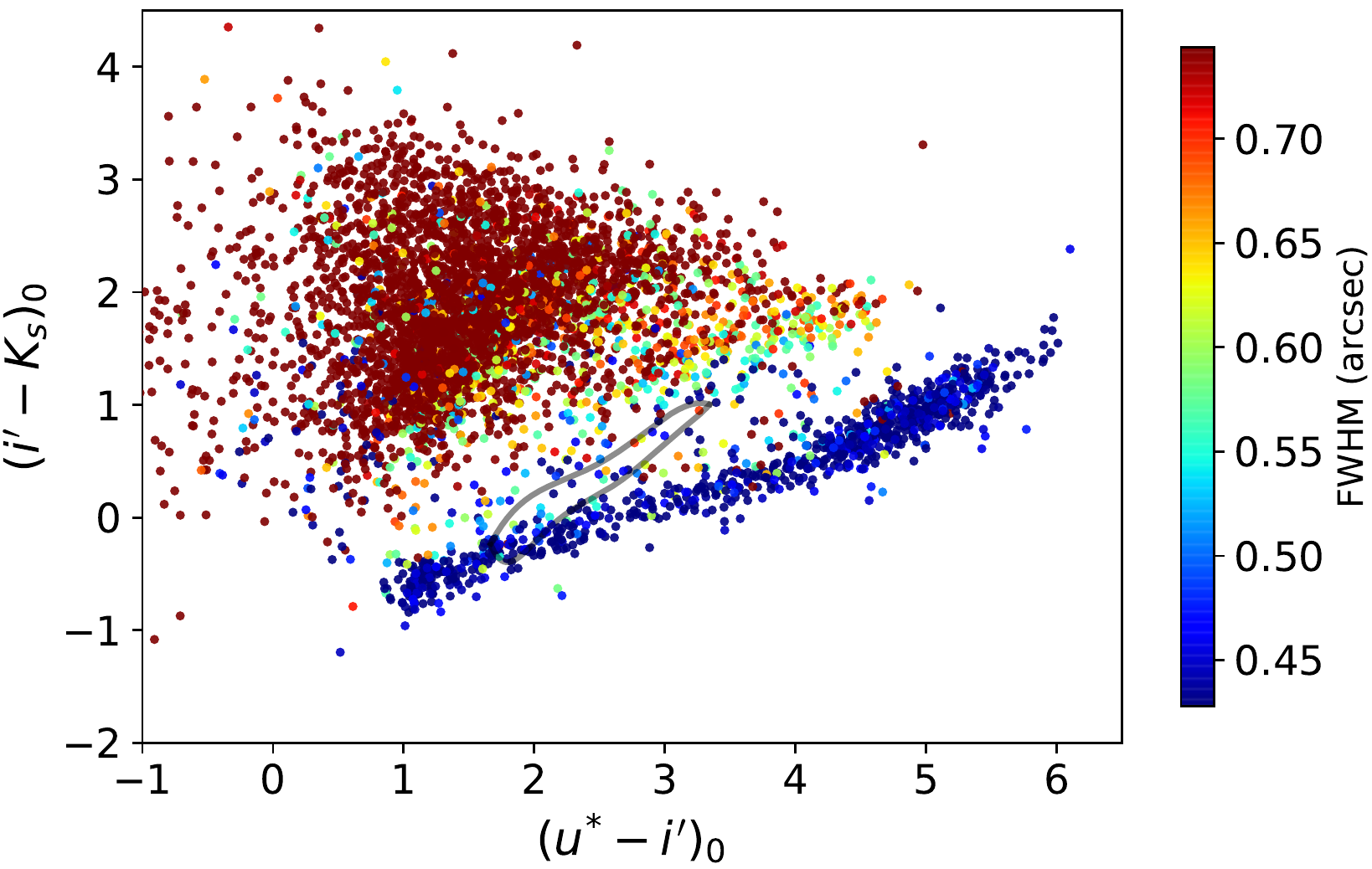}{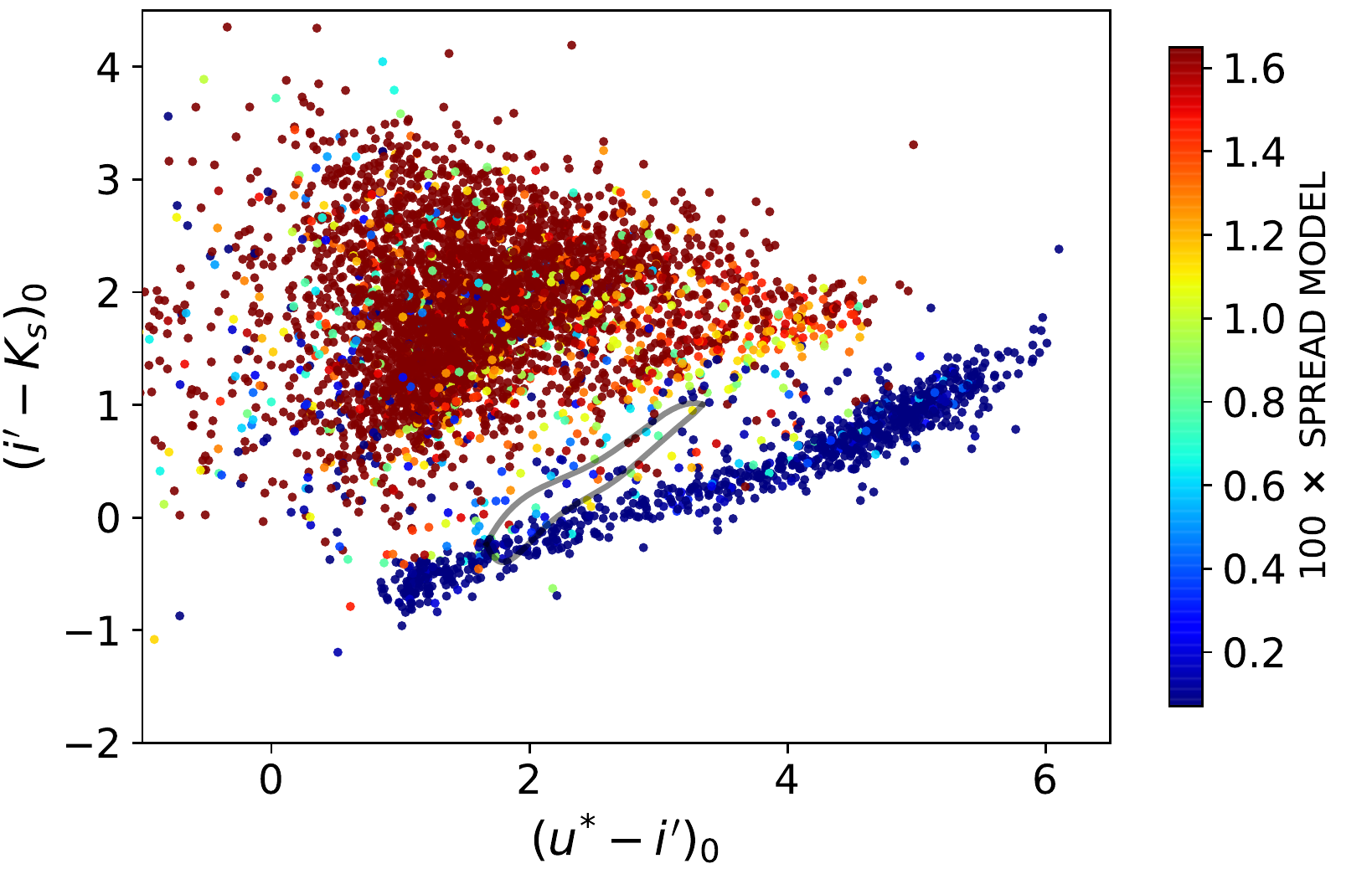}
\plottwo{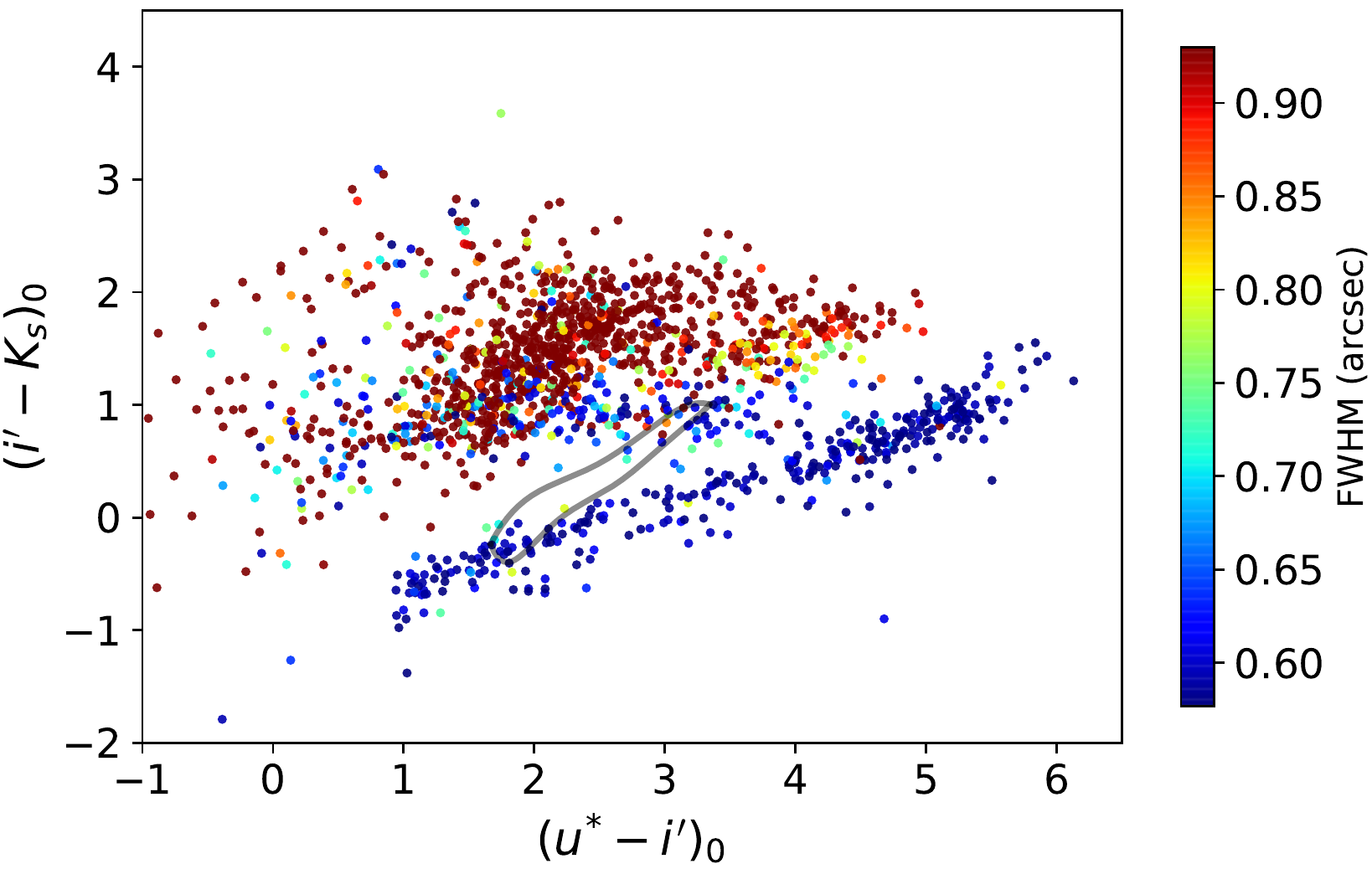}{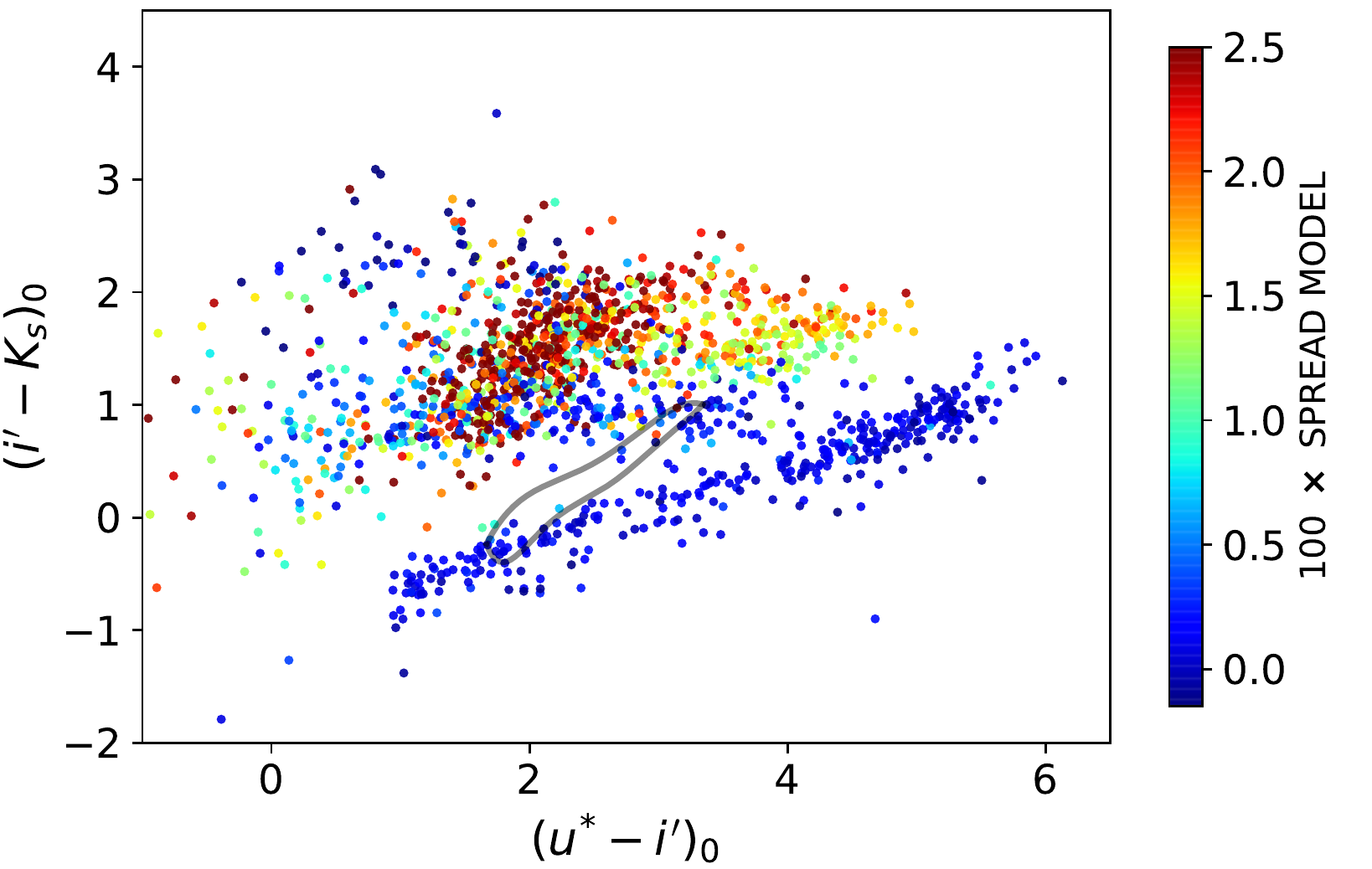}
\plottwo{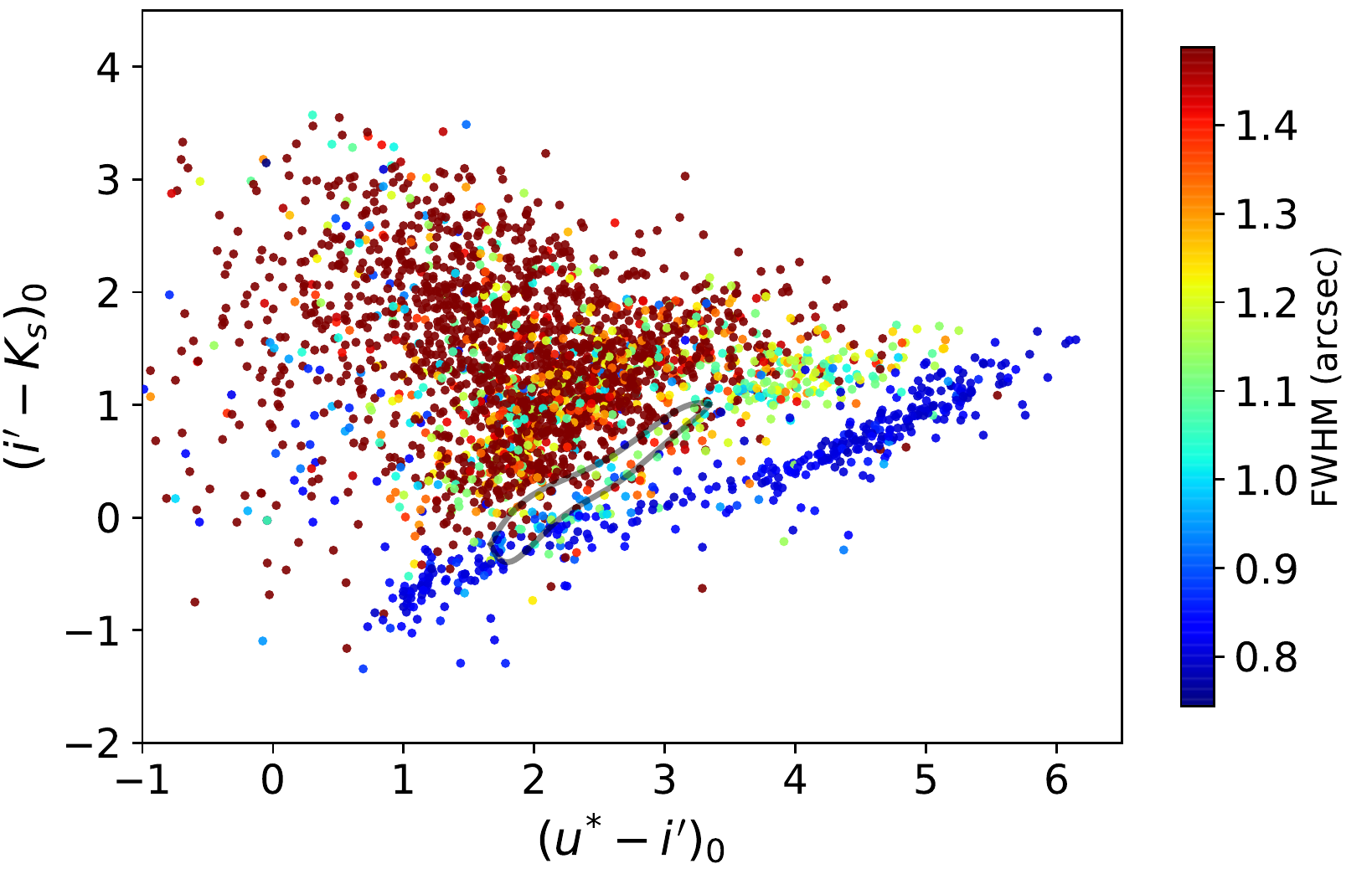}{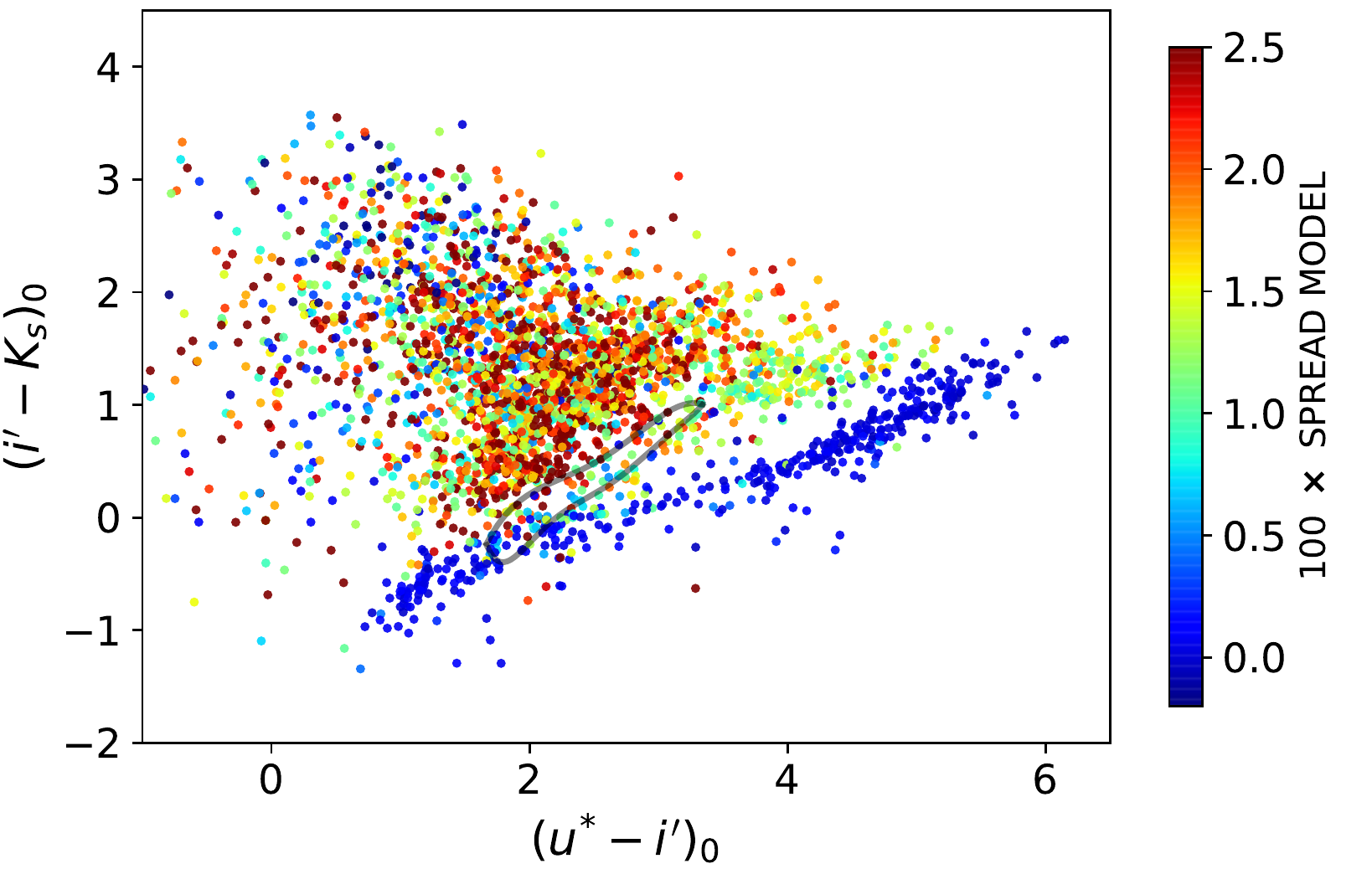}
\plottwo{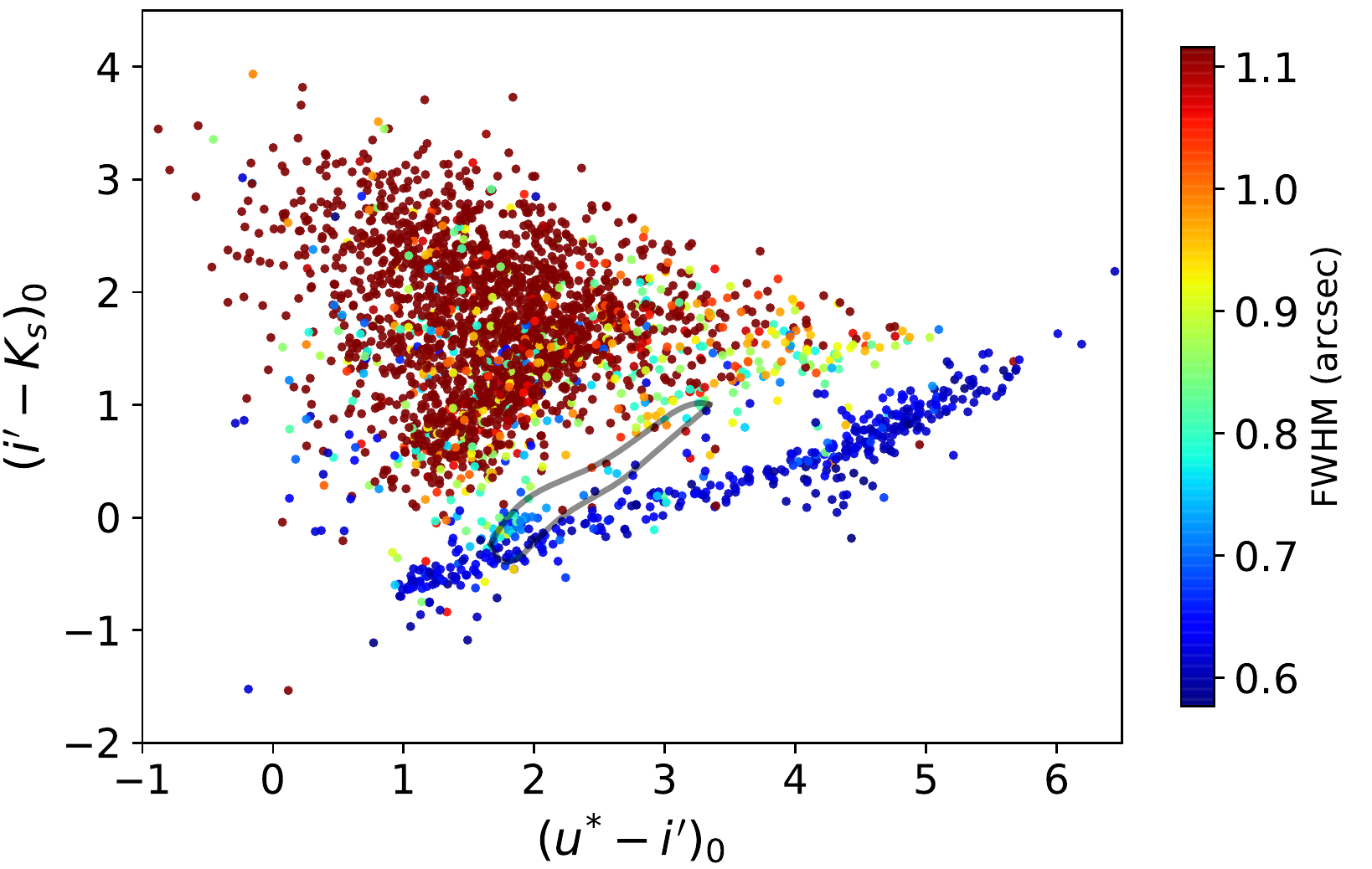}{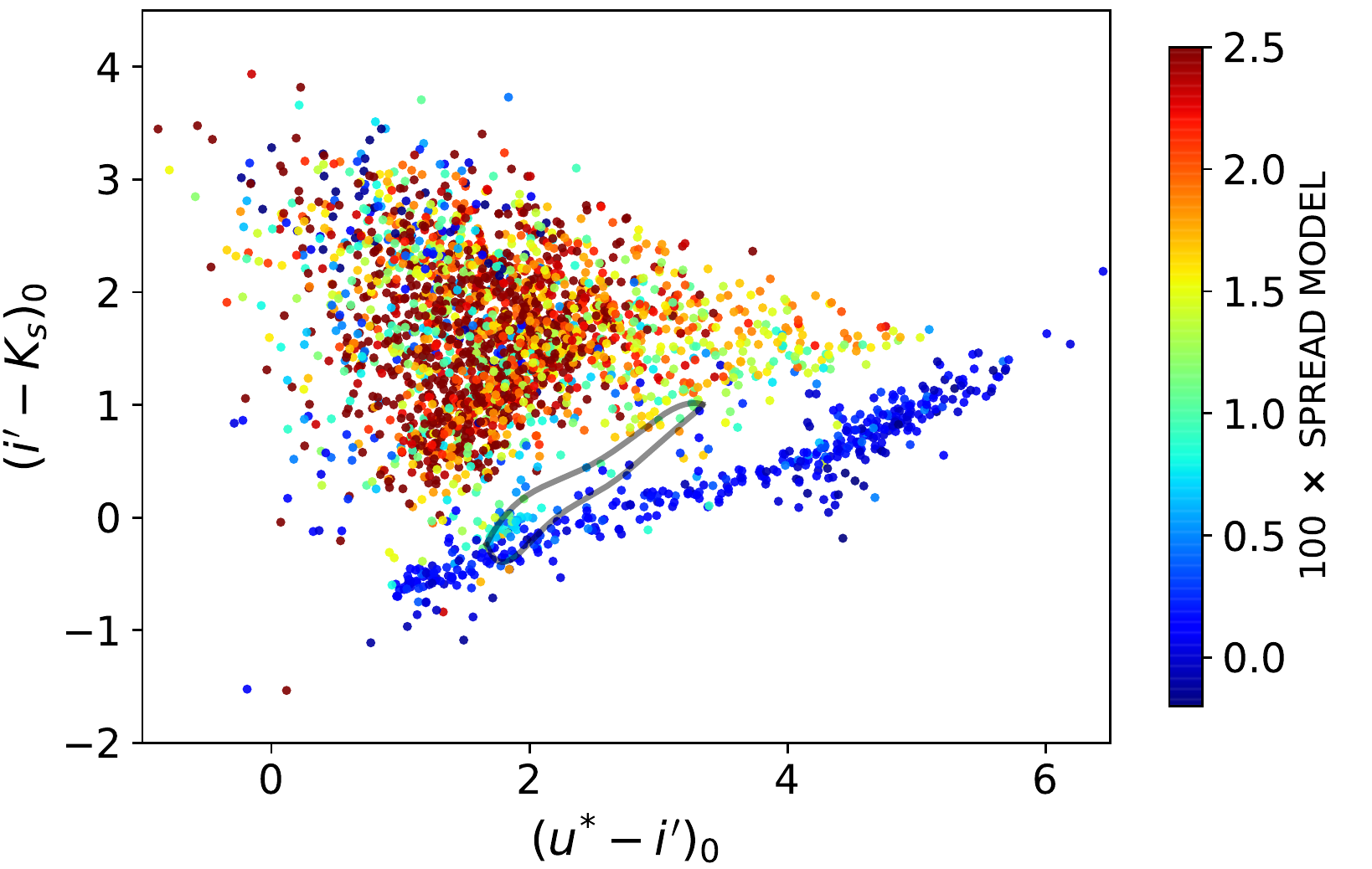}
\caption{Morphology variations across the \uiks\ diagram. {\it Top:} NGC\,3368;
{\it second row:} NGC\,4395; {\it third row:} NGC\,4736; {\it bottom:}
NGC\,4826. Indicators of shape in the \ip-band are coded as shown by the color
bars: FWHM ({\it left}) and 100$\times$SPREAD\_MODEL ({\it right}).
\label{fig:uiks_w_shapes}}
\end{figure}

Figure~\ref{fig:uiks_w_shapes} shows the \uiks\ diagrams for all the sources
detected in the images of NGC\,3368 (top row), NGC\,4395 (second row),
NGC\,4736 (third row) and NGC\,4826 (bottom row).  All the data have been
corrected for Galactic extinction, using the values from \citet{schlaf11} given
for the Sloan Digital Sky Survey (SDSS) filters by the NASA Extragalactic
Database,\footnote{The NASA/IPA Extragalactic Database (NED) is operated by the
Jet Propulsion Laboratory, California Institute of Technology, under contract
with the National Aeronautics and Space Administration.}respectively, 
for NGC\,3368, $A_u = 0.107$, $A_g = 0.083$, $A_i = 0.043$, $A_{Ks} = 0.005$;  
for NGC\,4395, $A_u = 0.073$, $A_g = 0.057$, $A_i = 0.029$, $A_{Ks} = 0.008$; 
for NGC\,4736, $A_u = 0.075$, $A_g = 0.059$, $A_r = 0.041$, $A_i = 0.030$, $A_{Ks} = 0.005$;
for NGC\,4826, $A_u = 0.175$, $A_g = 0.137$, $A_r = 0.095$, $A_i = 0.070$, $A_{Ks} = 0.012$.

The sources are color-coded by their light concentration in the \ip-band, as
gauged by the FWHM (left column), and 100$\times$ the dimensionless SExtractor
parameter SPREAD\_MODEL (right column).\footnote{For each object, the
SPREAD\_MODEL value results from the comparison between its best fitting PSF,
and the convolution of such PSF with an exponential disk whose scale length is
equal to the FWHM of said PSF \citep{desa12}.} In these panels, reddish-brown
sources are extended, blue sources are pointlike, and cyan sources are
marginally resolved. The color limits have been chosen to highlight these
differences. Objects in the reddish-brown cloud are mostly galaxies, while
those in the blue band are main-sequence stars in our Galaxy. 

In between the galaxy cloud and the stellar band, the selection region for GCCs
is indicated by the gray contour.  The selection region was determined using
data of the giant elliptical M\,87 \citep{muno14,powa16}; it has the shape of a
drop or pennon, and contains more than 2000 spectroscopically confirmed GCs in
that galaxy. The technique has already been applied to the spiral megamaser
prototype NGC\,4258 \citep{gonz17}, and its effectiveness in spirals assessed
by \citet{gonz19}.  One difference with the observations of M\,87 and our
previous work on NGC\,4258 is that a new \ust\ filter was used for the
observations presented in this paper.  As a consequence, the \ust\ luminosity
function turn-over (LFTO) is slightly fainter, and the GC selection region
shifted 0.24 mag to the red in (\ust\ - \ip). 

The spectroscopic observations of the NGC\,4258 GCCs reported by \citet{gonz19}
found no false negatives, and a sample contamination of, conservatively, 7/23
or 30\%, fully consistent with the 50--100 contaminants found by \citet{powa16}
for the GCCs of M\,87, in a FOV 9 times larger. Like M\,87 and NGC\,4258, the
fields of NGC\,3368, NGC\,4395, NGC\,4736 and NGC\,4826 lie at high Galactic
latitudes, respectively, $b=57\fdg0$, $b=81\fdg5$, $b=76\fdg0$, and $b
=84\fdg4$.
     
\citet{gonz19} also determined that the contaminants could be reduced by
setting the light concentration parameters SPREAD\_MODEL $\leqslant$ 0.015 and
FWHM $\leqslant$ 29 pc ($0\farcs58$ for NGC\,3368, $1\farcs40$ for NGC\,4395,
$1\farcs20$ for NGC\,4736 and $0\farcs82$ for NGC\,4826).  Moreover, we only
kept objects within $\pm\ 3\sigma$ of the expected GC LFTO magnitude in every
filter, except at $K_s$, since \citet{gonz19} realized that objects more than
1.7$\times \sigma$ brighter than the turnoff at that wavelength were potential
contaminants. Other restrictions were an axis ratio $\geqslant$ 0.7 and an
error MAGERR\_PSF $< 0.2$ mag, both in the \ip-band, as well as FLAGS = 0 in
all bands.  The standard deviation of the GCLF depends on galaxy luminosity
\citep{jord07}; based on their eq.\ 18, we have assumed 0.9 for NGC\,4395, 1.1
for NGC\,4826, and 1.2 for NGC\,3368 and NGC\,4736.\footnote{The MW GCLF has
$\sigma_V =$ 1.15 $\pm$ 0.1 \citep{jord07}, also consistent with their eq.\
18.} Values of the LFTO in the optical were derived by combining the absolute
LFTO magnitude in the $g$ band, $M^0_g = -7.2$ mag \citep{jord07}, with the
(AB) colors given in the MegaCam filter system by \citet[][hereafter
BC03]{bruz03} models for an SSP with $Z = 0.0004$, an age of 12 Gyr, and a
Chabrier initial mass function \citep[IMF;][]{chab03}. The LFTO in the $V$-band
would be $M^0_V = -7.4$ mag.  The LFTO magnitude in the \ks-band was taken from
\citet{wang14}.  TO absolute magnitudes, observed LFTO magnitudes, and
magnitude ranges of the GCLF are shown in Table~\ref{tab:TOdat}.

\floattable
\begin{deluxetable}{cccccccccc}
\tablecaption{Turnover magnitudes and GCLF ranges\label{tab:TOdat}}
\tablewidth{20cm}
\tablehead{\colhead{}& \colhead{} &\multicolumn2c{NGC\,3368}& \multicolumn2c{NGC\,4395}& \multicolumn2c{NGC\,4736}& \multicolumn2c{NGC\,4826}\\
\colhead{Filter} & \colhead {$M^0$} & \colhead{$m^0_{\rm TO}$} & \colhead{$m^0_{\rm TO} \pm\ 3\sigma$} & \colhead{$m^0_{\rm TO}$} & \colhead{$m^0_{\rm TO} \pm\ 3\sigma$}& \colhead{$m^0_{\rm TO}$} & \colhead{$m^0_{\rm TO} \pm\ 3\sigma$} & \colhead{$m^0_{\rm TO}$} & \colhead{$m^0_{\rm TO} \pm\ 3\sigma$} \\
\colhead {} & \colhead{mag} & \colhead{mag} & \colhead{mag} & \colhead{mag} &\colhead{mag}  & \colhead{mag} & \colhead{mag} & \colhead{mag} &\colhead{mag}
}
\decimals
\startdata
 \ust &  -5.96 &  24.12 & $20.52  < m^0_{u^*}      <27.72$ & 22.21&    $19.51 < m^0_{u^*}      < 24.91$&22.53  & $18.93 < m^0_{u^*}      < 26.13$ & 23.36  & $20.06 < m^0_{u^*} < 26.66$\\
 \gp &  -7.20 &   22.88 & $19.28  < m^0_{g^\prime} <26.48$ & 20.97&    $18.27 < m^0_{g^\prime} < 23.67$&21.29  & $17.69 < m^0_{g^\prime} < 24.89$ &22.12 &  $18.82 < m^0_{g^\prime} < 25.42$  \\
 \rp &  -7.69 &   22.40 & $18.80  < m^0_{r^\prime} <26.00$ & 20.48&    $17.78 < m^0_{r^\prime} < 23.18$&20.80  & $17.20 < m^0_{r^\prime} < 24.40$ &21.63 & $18.33 < m^0_{r^\prime} < 24.93$\\
 \ip &  -7.92 &   22.16 & $18.56  < m^0_{i^\prime} <25.76$ & 20.25&    $17.55 < m^0_{i^\prime} < 22.95$&20.57  & $16.97 < m^0_{i^\prime} < 24.17$ &21.40 & $18.10 < m^0_{i^\prime} < 24.70$ \\
 \ks &  -8.1 &    22.0  & $19.9   < m^0_{Ks}       <25.6$ & 20.1&      $18.5  < m^0_{Ks}       < 22.8$&20.4    & $18.4   < m^0_{Ks}       < 24.0$ &21.2 & $19.3 < m^0_{Ks} < 24.5$\\
\enddata
\end{deluxetable}

\begin{figure}[ht!]
\hspace*{1.3cm}\includegraphics[scale=0.45]{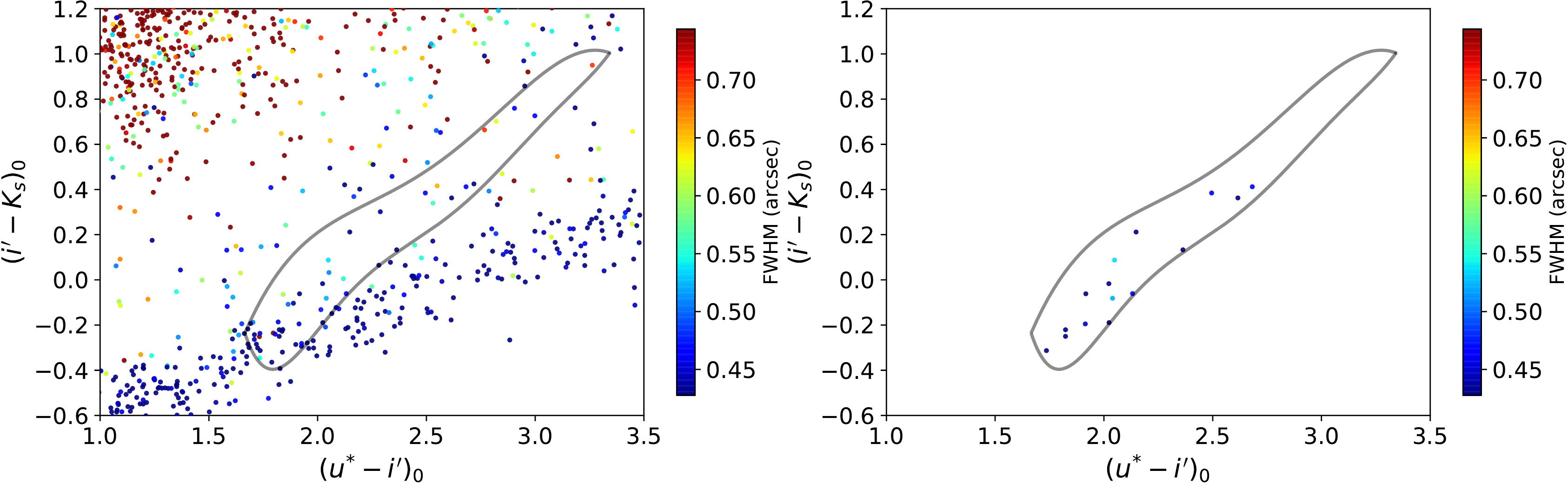}
\hspace*{1.3cm}\includegraphics[scale=0.45]{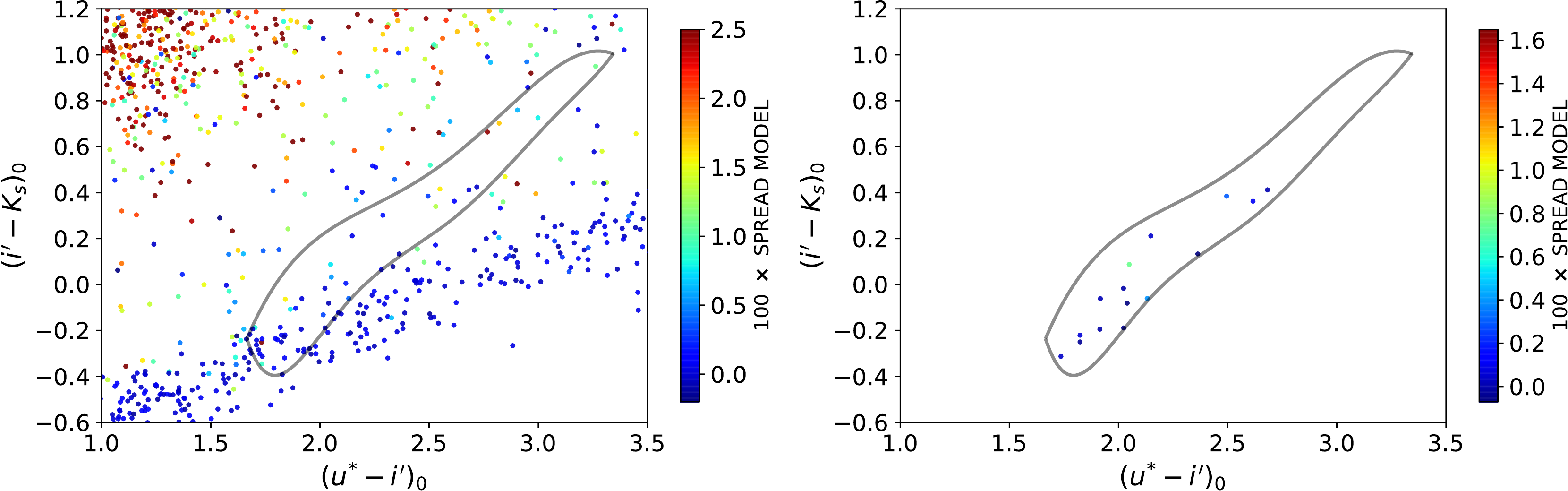}
\caption{(\ust\ - \ip) versus (\ip\ - \ks) colors and structural parameters of
sources in and closely around selection region of NGC\,3368.  {\it Top:} FWHM;
{\it bottom:} 100 $\times$ SPREAD\_MODEL.  {\it Left column:} all sources; {\it
right column:} final sample.  The gray contour outlines the GCC selection
region.  GCCs appear with shades of blue and cyan in both compactness
estimators.
\label{fig:sel_shape_n3368}}
\end{figure}

\begin{figure}[ht!]
\hspace*{1.3cm}\includegraphics[scale=0.45]{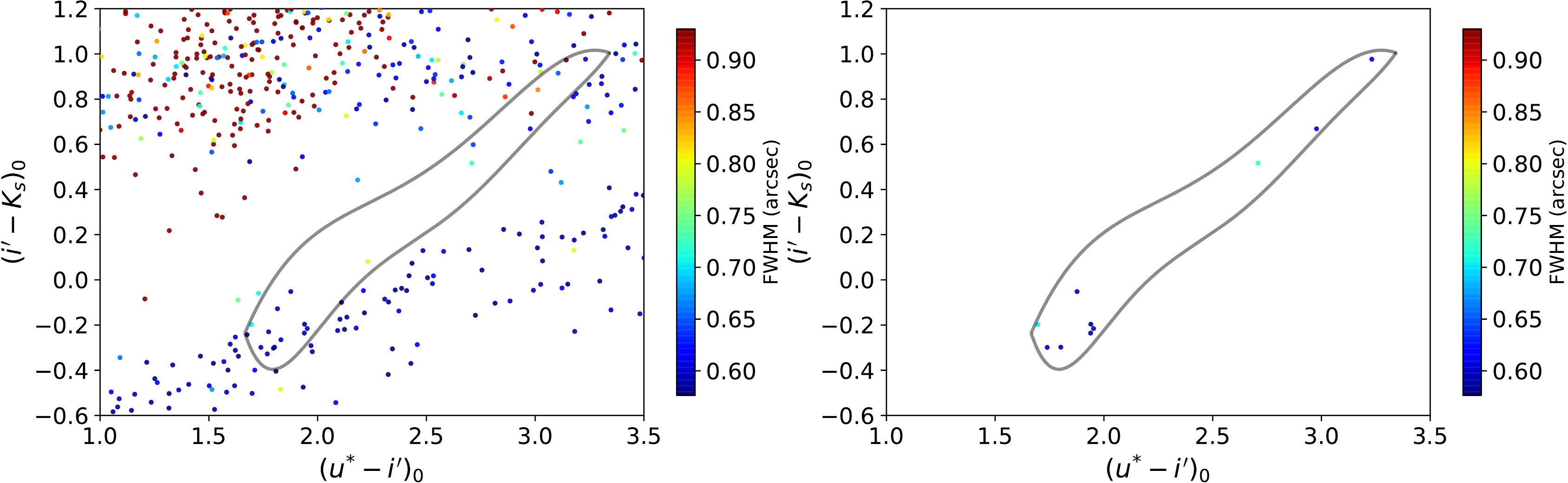}
\hspace*{1.3cm}\includegraphics[scale=0.45]{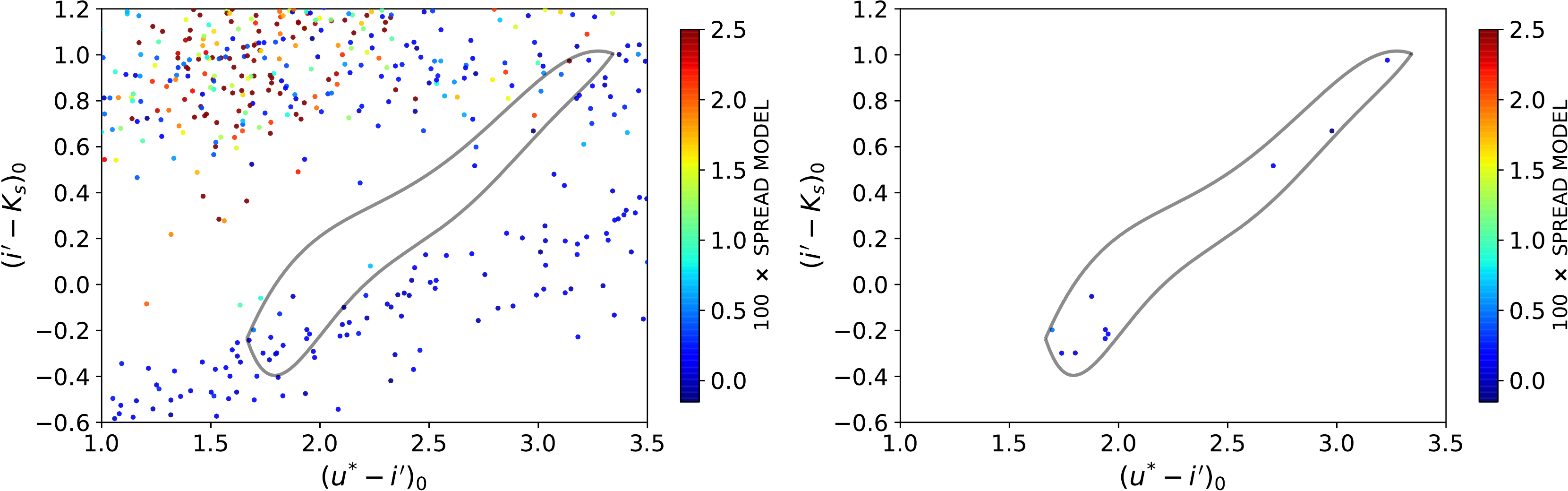}
\caption{(\ust\ - \ip) versus (\ip\ - \ks) colors and structural parameters of
sources in and closely around selection region of NGC\,4395.  {\it Top:} FWHM;
{\it bottom:} 100 $\times$ SPREAD\_MODEL.  {\it Left column:} all sources; {\it
right column:} final sample.  Symbols as in Fig.~\ref{fig:sel_shape_n3368}.
\label{fig:sel_shape_n4395}}
\end{figure}

\begin{figure}[ht!]
\hspace*{1.3cm}\includegraphics[scale=0.45]{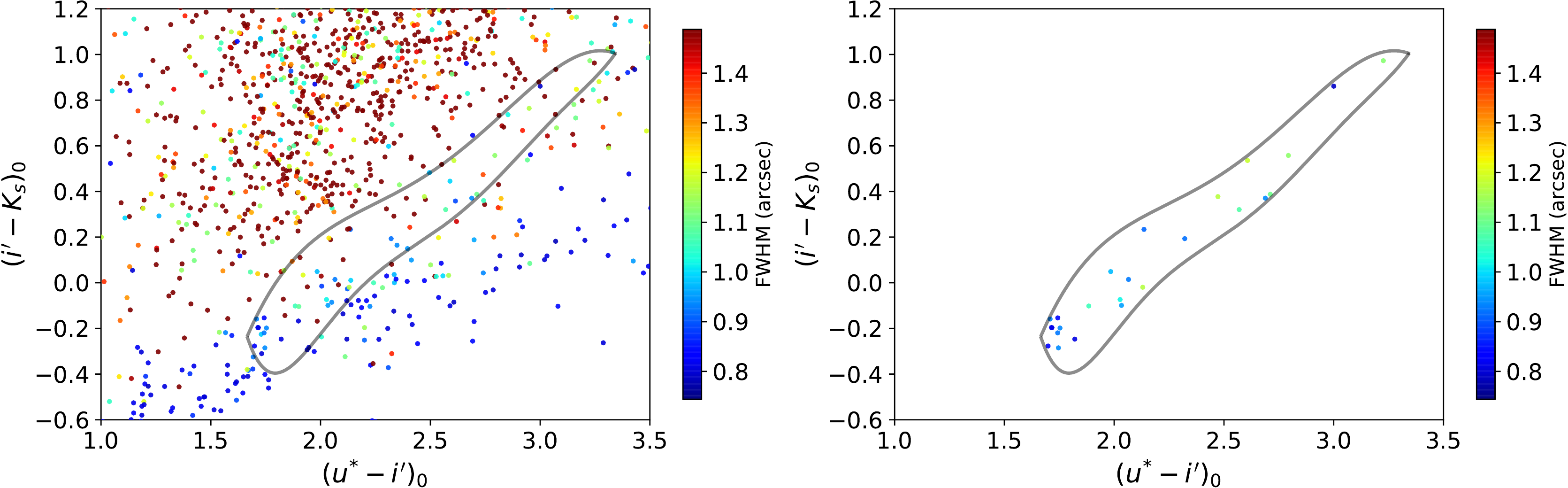}
\hspace*{1.3cm}\includegraphics[scale=0.45]{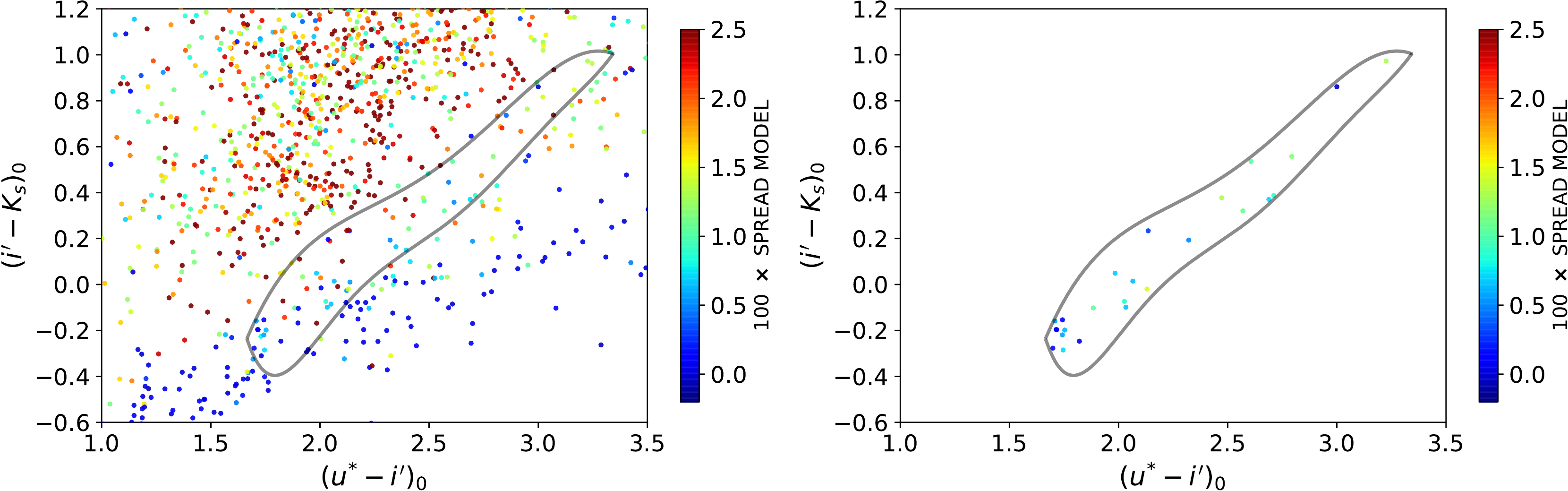}
\caption{(\ust\ - \ip) versus (\ip\ - \ks) colors and structural parameters of
sources in and closely around selection region of NGC\,4736.  {\it Top:} FWHM;
{\it bottom:} 100 $\times$ SPREAD\_MODEL.  {\it Left column:} all sources; {\it
right column:} final sample.  Symbols as in Fig.~\ref{fig:sel_shape_n3368}.
\label{fig:sel_shape_n4736}}
\end{figure}

\begin{figure}[ht!]
\hspace*{1.3cm}\includegraphics[scale=0.45]{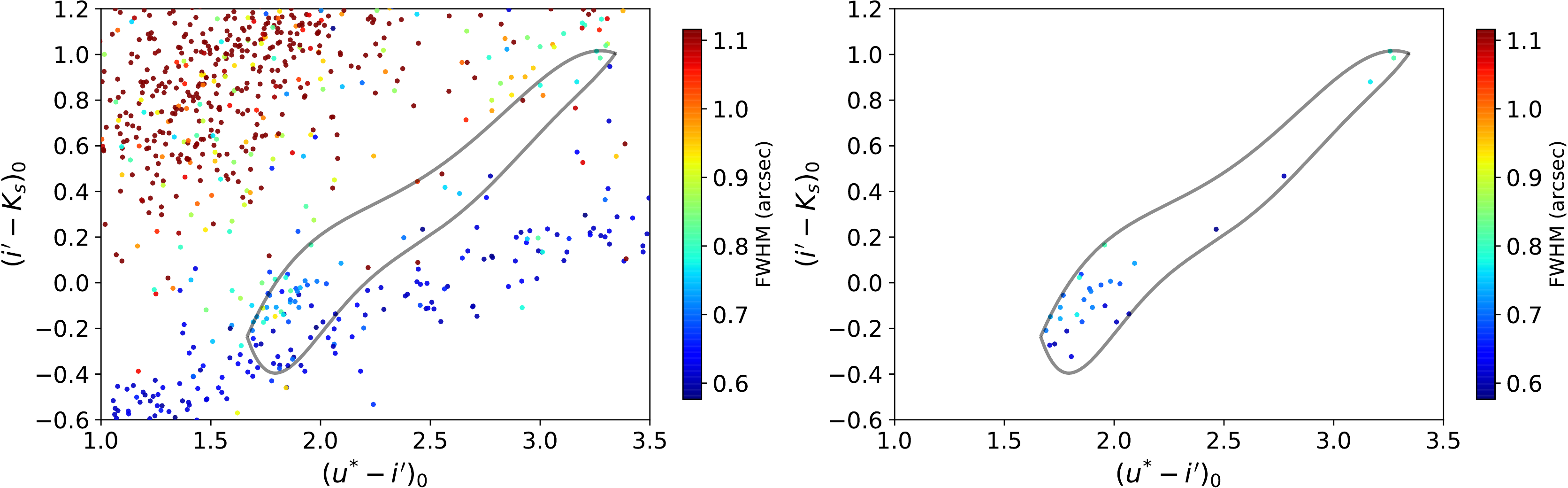}
\hspace*{1.3cm}\includegraphics[scale=0.45]{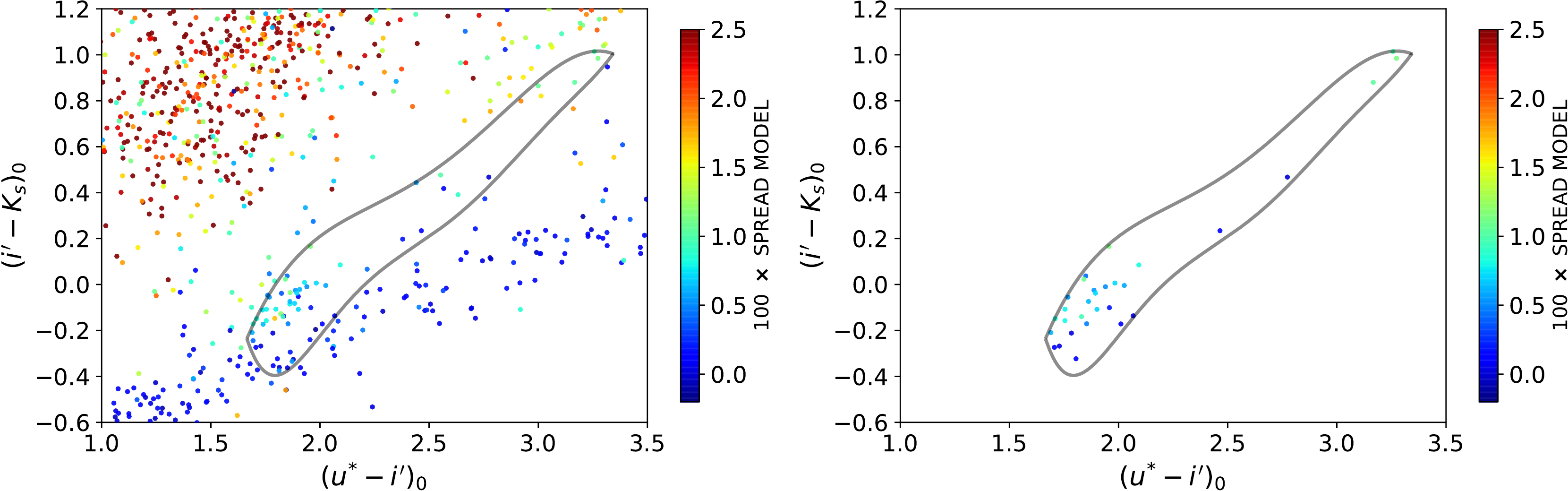}
\caption{(\ust\ - \ip) versus (\ip\ - \ks) colors and structural parameters of
sources in and closely around selection region of NGC\,4826.  {\it Top:} FWHM;
{\it bottom:} 100 $\times$ SPREAD\_MODEL.  {\it Left column:} all sources; {\it
right column:} final sample.  Symbols as in Fig.~\ref{fig:sel_shape_n3368}. 
\label{fig:sel_shape_n4826}}
\end{figure}

Figures~\ref{fig:sel_shape_n3368},~\ref{fig:sel_shape_n4395},~\ref{fig:sel_shape_n4736},
and~\ref{fig:sel_shape_n4826}, left column, show the \uiks\ diagram, FWHM (top
rows), and 100$\times$ structural parameter SPREAD\_MODEL (bottom rows) of
sources in and closely around the GCC selection region (gray contour) for
NGC\,3368, NGC\,4395, NGC\,4736 and NGC\,4826, respectively. The GCC final
samples for the galaxies are displayed in the right column.  

\begin{figure}[ht!]
\plottwo{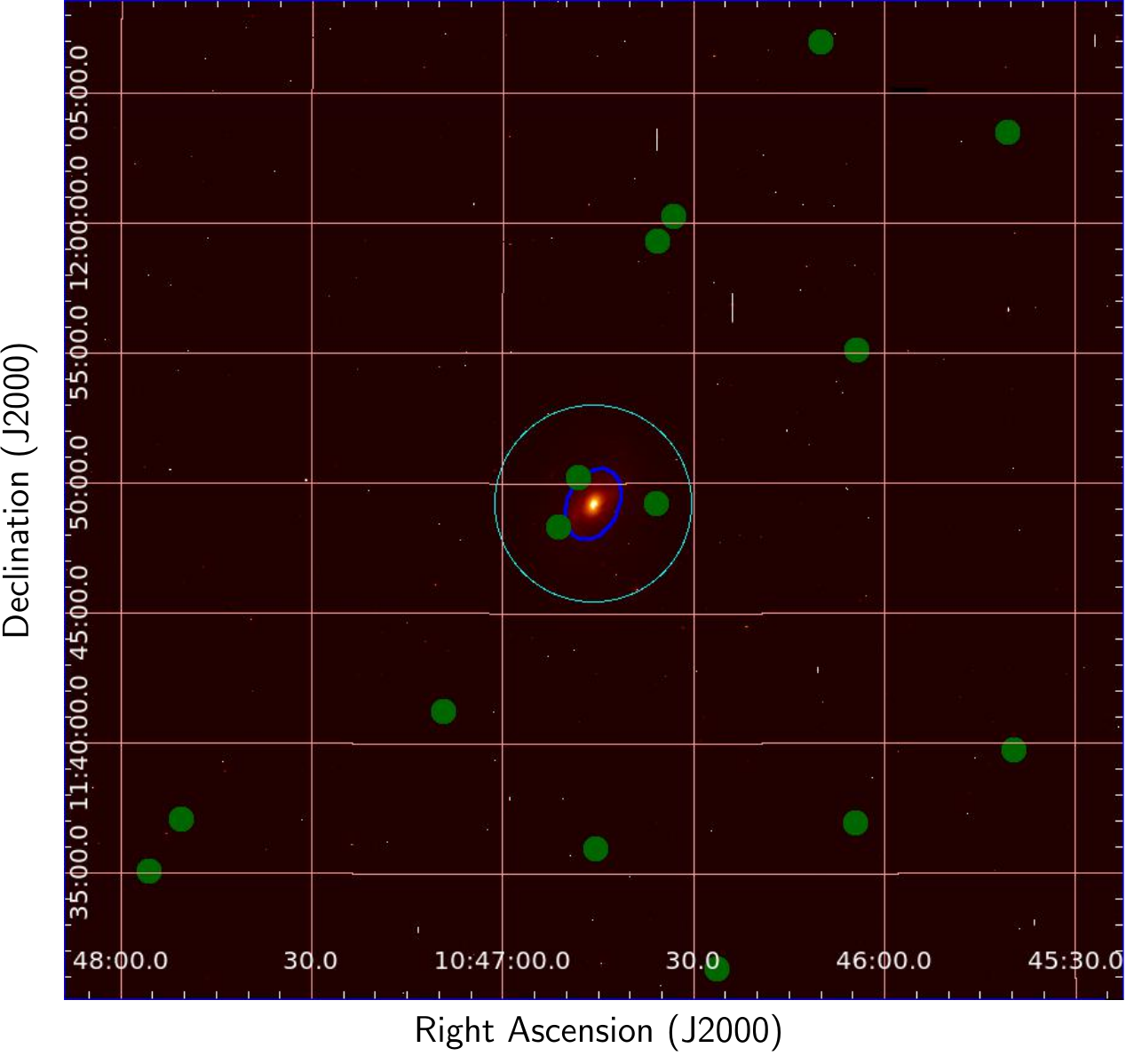}{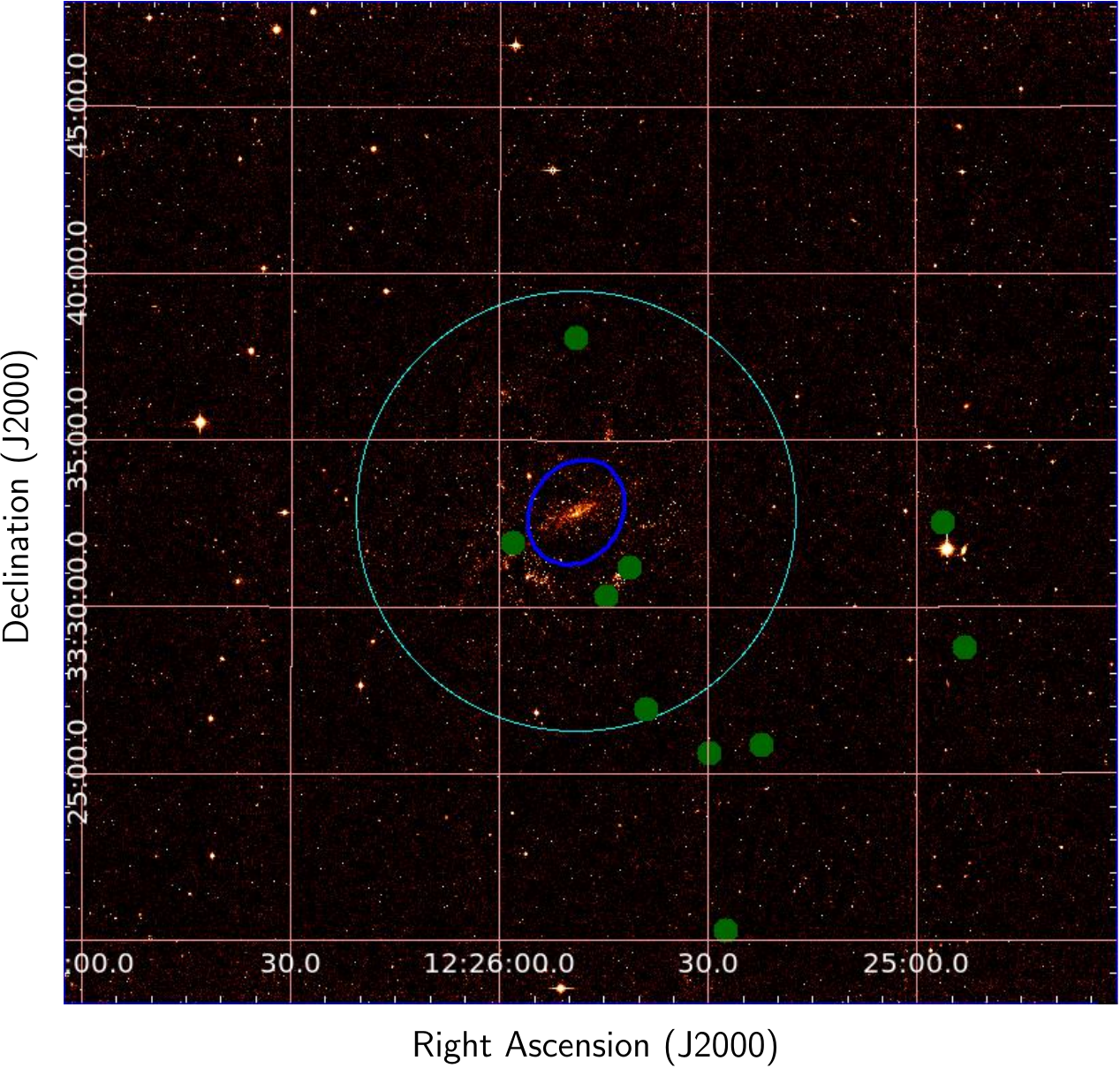}
\plottwo{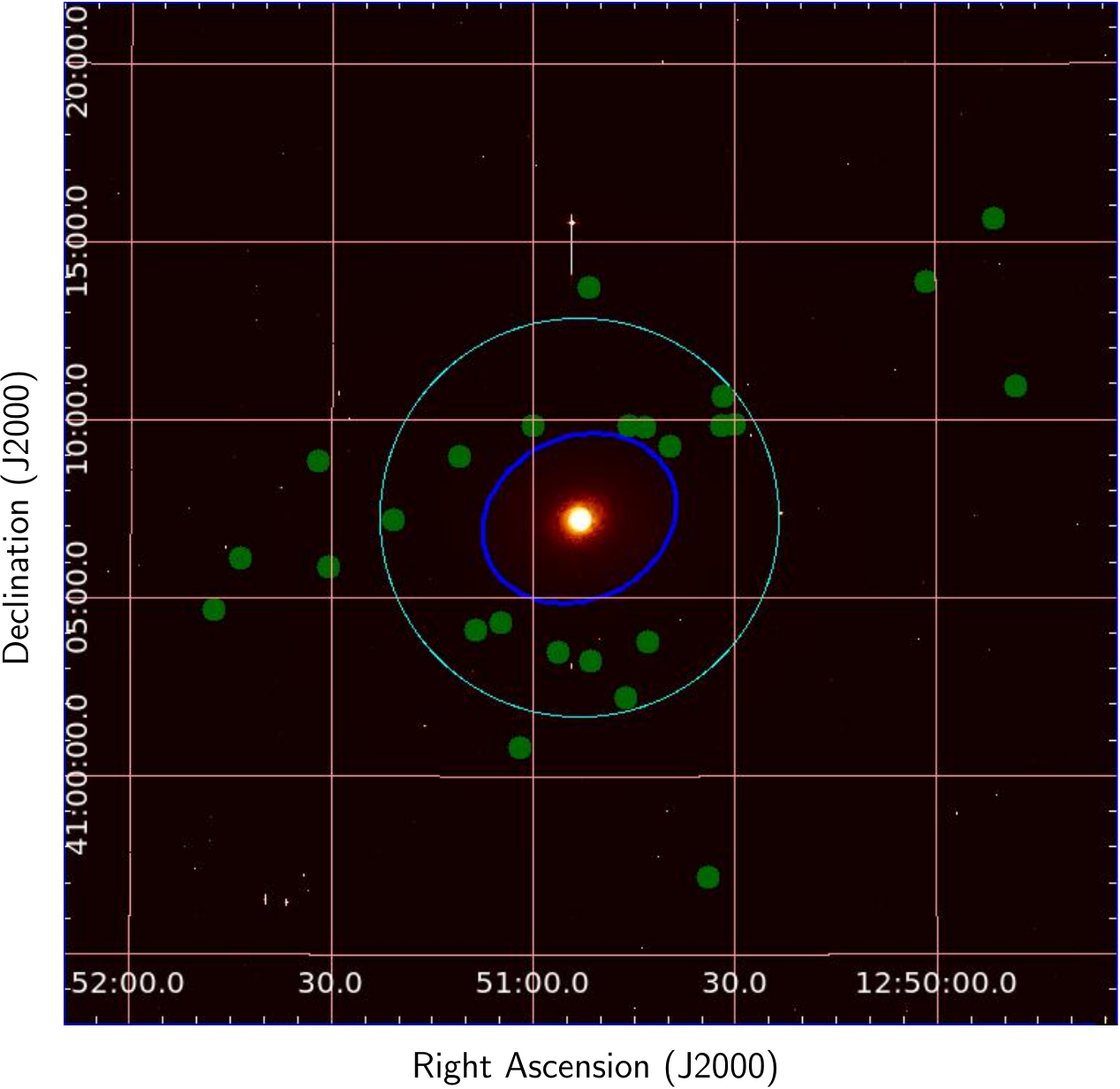}{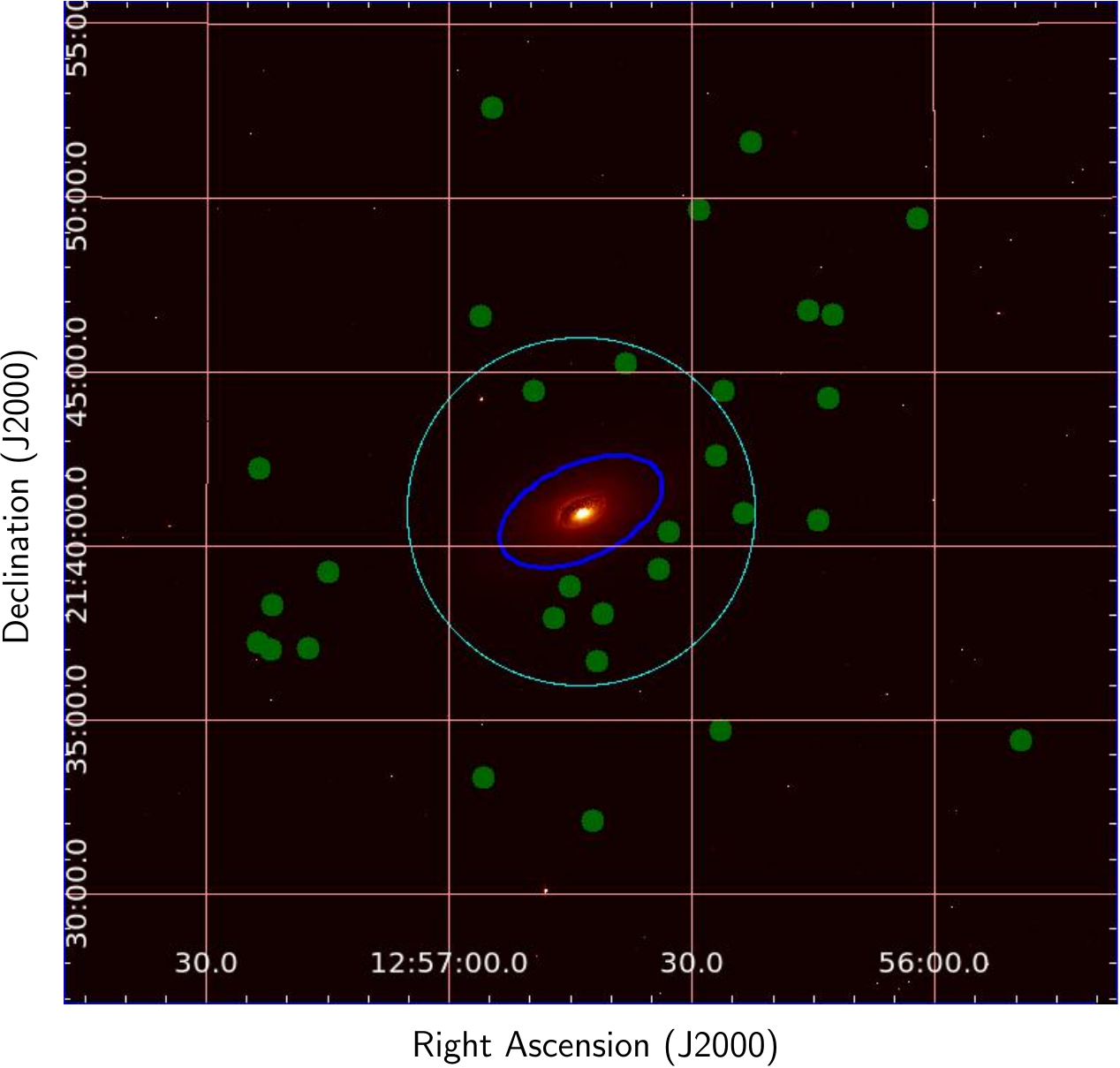}
\caption{Spatial distributions of the GCC samples. {\it Top left:} NGC\,3368;
{\it top right:} NGC\,4395; {\it bottom left:} NGC\,4736; {\it bottom right:}
NGC\,4826. The GCCs are shown as green filled circles. {\it Dark blue ellipse:}
boundary of areas where source confusion is highest (see text); {\it solid cyan
line}: $R_{25}$.
\label{fig:spatial}}
\end{figure}

The spatial distributions of the GCC samples are shown in
Figure~\ref{fig:spatial}.  The members of the samples are displayed as green
circles on the \ip-band images of NGC\,3368 (top left), NGC\,4395 (top right),
NGC\,4736 (bottom left), and NGC\,4826 (bottom right). The dark blue ellipses
outline the regions where source confusion is highest, inferred from the
completeness simulations and the dearth of GCC detections. This boundary is
located at 0.375 $R_{25}$ for NGC\,3368, 0.25 $R_{25}$ for NGC\,4395, and
0.5$R_{25}$ for both NGC\,4736 and NGC\,4826.  The cyan circles indicate
$R_{25}$.

Given the distances to the galaxies, as well as the resolution of the MegaCam
data (1 MegaCam pixel = 0$\farcs$186, i.e., $\simeq$ 3.9 pc, 4.5 pc, 6.6 pc,
and 9.4 pc, respectively, for NGC\,4395, NGC\,4736, NGC\,4826, and NGC\,3368),
we have also been able to calculate the half-light radii, $r_e$, in the
\ip-band of all the candidates.  To this end, we used the program {\sc ishape}
\citep{lars99} in the BAOLAB\footnote{\url{http://baolab.astroduo.org}.}
software package.  {\sc ishape} measures the size of a compact source by
comparing its observed light profile with models generated by convolving
different analytical profiles with the image PSF.  For data with S/N $\ga$ 30,
{\sc ishape} can measure $r_{\rm eff}$ reliably down to $\sim$ 0.1 the PSF
FWHM, or $\sim$ 1.7 pc for NGC\,4736, 2.3 pc for NGC\,4826, 2.4 pc for
NGC\,3368, and 2.5 pc for NGC\,4395; objects smaller than this are effectively
unresolved \citep{lars99,harr09}.  We fitted all the sources with King
\citep{king62,king66} and S\'ersic \citep{sers63} profiles.  For KING profiles,
we tried fixed concentration indices ($C \equiv r_{\rm tidal}/r_{\rm core}$) of
30 and 100; in the case of Sersic, we chose a fixed exponent of
2.\footnote{According to \citet{turn12}, based on the measurements of
\citet{grah09} and \citet{seth10}, a Sersic parameter of $\sim$2 represents
Galactic GCs and nuclear clusters well.} For each object, the model providing
the fit to the data with the smallest $\chi^2$ residuals was then used to
derive the effective radius, by applying the conversion factors between $r_{\rm
eff}$ and FWHM given in the {\sc ishape} manual.  For each galaxy, Tables
\ref{tab:n3368gccshapes}, \ref{tab:n4395gccshapes}, \ref{tab:n4736gccshapes},
and \ref{tab:n4826gccshapes} list shape parameters in the \ip-band: FWHM,
100$\times$SPREAD\_MODEL,  $r_e$, the type and index of the best-fitting
profile, the SNR of the data, and the $\chi^2$ of the fit.

\begin{table}[ht]
\caption{$i^\prime$-Band Shape Parameters of Globular Cluster Candidates in NGC 3368}
  \begin{tiny}
 \begin{center}
\hspace*{-1.5cm} \begin{minipage}{290mm}
  \begin{tabular}{@{\extracolsep{4pt}}cccccccrrr@{}}
\hline
\hline
\vspace*{-0.3cm}
&&&&&&&&\\
&&& 100 $\times$ (SPREAD &&&&&& \\
 Name  &     RA J2000  &   DEC J2000   & \_MODEL)$_{i^\prime}$   &   FWHM$_{i^\prime}$  &  $r_{\rm eff, i^\prime}$  &  Shape &  Index$^a$   &  SNR &  $\chi^2$ \\
       &      deg      &     deg       &              &     arcsec          &     pc              &               &       &            &            \\
\hline
J104539+113944 & 161.4148 & 11.6624 & 0.06 & 0.42 &  $<$ 8.65  &  KINGx  &   30   &      235.8  &  7.99 \\  
J104540+120330 & 161.4185 & 12.0585 & 0.04 & 0.42 &  $<$ 9.36  & KINGx   &   100   &     130.0  &  4.82   \\
J104604+115505 & 161.5179 & 11.9183 & -0.03 & 0.41&  $<$ 1.37  &  KINGx  &   30   &      241.4  & 13.27   \\
J104604+113655 & 161.5183 & 11.6154 & 0.05 & 0.42 &  $<$  8.01  & SERSICx &   2   &        74.8  &  2.33   \\
J104609+120658 & 161.5413 & 12.1161 & -0.04 & 0.42&  $<$ 3.44  &  KINGx  &   30   &      104.1  &  4.63   \\
J104626+113118 & 161.6091 & 11.5219 & -0.07 & 0.41&  $<$ 3.46  &  KINGx  &   30   &       88.9  &  3.83   \\
J104633+120015 & 161.6379 & 12.0043 & 0.02 & 0.40 &  $<$  3.59  & SERSICx &   2   &        65.4  &  2.49   \\
J104635+115918 & 161.6478 & 11.9886 & 0.10 & 0.42 &  $<$ 5.19  &  KINGx  &   30   &       86.1  &  2.20   \\
J104635+114912 & 161.6493 & 11.8201 & 0.40 & 0.46 &  $<$ 1.07  &  KINGx  &   30   &       40.4  &  1.87   \\    
J104645+113554 & 161.6890 & 11.5985 & -0.07 & 0.52&  $<$  6.15  & SERSICx &   2   &       191.0  & 10.03   \\
J104648+115012 & 161.7003 & 11.8368 & 0.30 & 0.46 &  $<$ 1.68  & KINGx   &   100   &      61.6  &  5.50   \\   
J104651+114819 & 161.7134 & 11.8053 & 0.75 & 0.54 &  $<$ 2.66   & KINGx   &  100   &       99.6  &  2.58   \\  
J104709+114112 & 161.7882 & 11.6868 & -0.00 & 0.43&  $<$  2.31  & SERSICx &   2   &        184.2 &  7.38 \\
J104750+113703 & 161.9601 & 11.6176 & 0.01 & 0.46 &  $<$ 6.00  & KINGx   &  100   &       36.9  &  1.83   \\
J104755+113503 & 161.9811 & 11.5844 & -0.00 & 0.45&  $<$  3.46  & SERSICx &   2   &        118.2 &  2.78  \\
\hline
\vspace*{-0.8cm}
\end{tabular}
\end{minipage}
\end{center}
\vspace*{0.2cm}
{{\sc Note}---
$^a$  Index denotes concentration for King profiles, exponent for Sersic profiles.

Table~\ref{tab:n3368gccshapes} is published in its entirety in the machine-readable format.
      A portion is shown here for guidance regarding its form and content.
}
\label{tab:n3368gccshapes}
\end{tiny}
\end{table}

\begin{table}[ht]
\caption{$i^\prime$-Band Shape Parameters of Globular Cluster Candidates in NGC 4395}
  \begin{tiny}
 \begin{center}
\hspace*{-1.5cm} \begin{minipage}{290mm}
  \begin{tabular}{@{\extracolsep{4pt}}cccccccrrr@{}}
\hline
\hline
\vspace*{-0.3cm}
&&&&&&&&\\
&&& 100 $\times$ (SPREAD &&&&&& \\
 Name  &     RA J2000  &   DEC J2000   & \_MODEL)$_{i^\prime}$   &   FWHM$_{i^\prime}$  &  $r_{\rm eff, i^\prime}$  &  Shape &  Index$^a$   &  SNR &  $\chi^2$ \\
       &      deg      &     deg       &              &     arcsec          &     pc              &               &       &            &            \\
\hline
J122452+332844 & 186.2204 & 33.4792 & 0.08 & 0.59 &   $<$  3.40  &   KINGx   & 30   &   78.2   &  0.57 \\ 
J122456+333230 & 186.2335 & 33.5418 & 0.11 & 0.60 &   $<$  3.39  &   SERSICx & 2    &  111.8   &  0.63  \\
J122522+332549 & 186.3426 & 33.4303 & 0.12 & 0.59 &   $<$  2.95  &   KINGx   & 100  &   46.4   &  0.48  \\
J122527+332015 & 186.3639 & 33.3376 & 0.05 & 0.59 &   $<$  3.19  &   KINGx   & 100  &   87.8   &  0.92  \\
J122529+332536 & 186.3734 & 33.4268 & 0.15 & 0.60 &   $<$  2.48  &   KINGx   & 100  &   43.6   &  0.48  \\
J122538+332654 & 186.4118 & 33.4486 & 0.49 & 0.70 &   $<$  6.33  &   SERSICx & 2    &   28.5   &  0.45  \\
J122541+333109 & 186.4210 & 33.5194 & 0.16 & 0.72 &   $<$  2.86  &   KINGx   & 30   &   19.8   &  1.00  \\
J122544+333016 & 186.4354 & 33.5046 & 0.16 & 0.61 &   $<$  2.99  &   KINGx   & 100  &   15.4   &  0.50  \\
J122548+333802 & 186.4535 & 33.6341 & 0.14 & 0.59 &   $<$  3.09  &   KINGx   & 30   &   26.9   &  0.57  \\
J122558+333154 & 186.4917 & 33.5318 & -0.13 & 0.61&   $<$  1.86  &   KINGx   & 100  &   10.1   &  0.51  \\  
\hline
\vspace*{-0.8cm}
\end{tabular}
\end{minipage}
\end{center}
\vspace*{0.2cm}
{{\sc Note}---
$^a$  Index denotes concentration for King profiles, exponent for Sersic profiles.

Table~\ref{tab:n4395gccshapes} is published in its entirety in the machine-readable format.
      A portion is shown here for guidance regarding its form and content.
}
\label{tab:n4395gccshapes}
\end{tiny}
\end{table}

\begin{table}[ht]
\caption{$i^\prime$-Band Shape Parameters of Globular Cluster Candidates in NGC 4736}
  \begin{tiny}
 \begin{center}
\hspace*{-1.5cm} \begin{minipage}{290mm}
  \begin{tabular}{@{\extracolsep{4pt}}cccccccrrr@{}}
\hline
\hline
\vspace*{-0.3cm}
&&&&&&&&\\
&&& 100 $\times$ (SPREAD &&&&&& \\
 Name  &     RA J2000  &   DEC J2000   & \_MODEL)$_{i^\prime}$   &   FWHM$_{i^\prime}$  &  $r_{\rm eff, i^\prime}$  &  Shape &  Index$^a$   &  SNR &  $\chi^2$ \\
       &      deg      &     deg       &              &     arcsec          &     pc              &               &       &            &            \\
\hline
J124947+411056 & 192.4496 & 41.1823 & 1.30 & 1.13 &   8.28$^{+0.45}_{-0.19}$   & KINGx &   30  &     70.7  &    0.57  \\ 
J124951+411538 & 192.4633 & 41.2606 & 0.12 & 0.81 &   $<$ 1.44               &  KINGx &   100   &   655.0  &    2.56  \\ 
J125001+411350 & 192.5053 & 41.2307 & 0.17 & 0.83 &   $<$ 3.75               &  KINGx &   100   &   282.2  &    1.65  \\ 
J125029+410951 & 192.6246 & 41.1644 & 0.10 & 0.84 &   $<$ 3.09               &  KINGx &   100   &    83.2  &    0.81  \\ 
J125031+411039 & 192.6314 & 41.1777 & 0.89 & 1.10 &   5.82$^{+0.34}_{-0.33}$   &SERSICx &   2   &    44.9  &    1.06  \\ 
J125031+410949 & 192.6329 & 41.1638 & 1.03 & 1.18 &   5.74$^{+0.45}_{-0.46}$   & KINGx &   30   &    29.3  &    1.04  \\ 
J125033+405708 & 192.6415 & 40.9525 & 0.04 & 0.81 &   $<$ 2.89               &  KINGx &   100   &   325.5  &    1.53  \\ 
J125039+410913 & 192.6651 & 41.1538 & 1.05 & 1.06 &   6.69$^{+0.16}_{-0.21}$   &SERSICx &   2   &    61.0  &    1.41  \\ 
J125042+410345 & 192.6786 & 41.0625 & -0.01 & 0.79 &  $<$ 4.99               &   KINGx &   30   &    92.1  &    0.95  \\ 
J125043+410947 & 192.6807 & 41.1632 & 0.72 & 0.94 &  3.81$^{+0.04}_{-0.09}$    &KINGx &   100   &   241.2  &    1.34  \\ 
J125045+410950 & 192.6902 & 41.1639 & 0.65 & 0.93 &  2.89$^{+0.01}_{-0.12}$    &KINGx &   100   &   176.8  &    1.04  \\ 
J125046+410210 & 192.6920 & 41.0361 & 0.72 & 0.94 &  3.93$^{+0.07}_{-0.02}$    &KINGx &   100   &   471.5  &    2.18  \\ 
J125051+410314 & 192.7138 & 41.0540 & 0.66 & 0.96 &  2.66$^{+0.12}_{-0.06}$    &KINGx &   100   &   140.9  &    0.90  \\ 
J125051+411341 & 192.7148 & 41.2282 & 0.57 & 0.93 &  2.54$^{+0.06}_{-0.03}$    &KINGx &   100   &   329.6  &    1.26  \\ 
J125056+410327 & 192.7340 & 41.0576 & 1.18 & 1.13 &  $<$ 7.26                &  SERSICx &   2   &    58.5  &    0.73  \\ 
J125059+410949 & 192.7500 & 41.1637 & 1.44 & 1.17 & 13.05$^{+0.51}_{-0.30}$    &KINGx &   100   &   147.9  &    1.03  \\ 
J125101+410046 & 192.7577 & 41.0128 & 0.29 & 0.91 &  1.33$^{+0.41}_{-0.41}$    & KINGx &   30   &    44.4  &    0.88  \\ 
J125104+410416 & 192.7698 & 41.0712 & 0.50 & 0.92 &  3.27$^{+0.07}_{-0.08}$    &SERSICx &   2   &   163.1  &    1.18  \\ 
J125108+410404 & 192.7848 & 41.0679 & 0.69 & 0.97 &  3.00$^{+0.35}_{-0.23}$    & KINGx &   30   &    36.7  &    0.77  \\ 
J125110+410856 & 192.7957 & 41.1491 & 0.63 & 0.94 &  3.87$^{+0.05}_{-0.09}$    & KINGx &   30   &   229.6  &    1.26  \\ 
J125120+410710 & 192.8362 & 41.1197 & 1.32 & 1.17 &  8.97$^{+0.48}_{-0.18}$    &SERSICx &   2   &    93.6  &    1.03  \\ 
J125130+410551 & 192.8767 & 41.0977 & 0.03 & 0.84 &  $<$ 3.24                &  KINGx &   100   &    90.5  &    1.05  \\ 
J125131+410849 & 192.8828 & 41.1472 & 0.58 & 0.93 &  2.42$^{+0.02}_{-0.01}$    &KINGx &   100   &   608.9  &    1.83  \\ 
J125143+410606 & 192.9315 & 41.1018 & 0.90 & 1.02 &  5.08$^{+0.22}_{-0.12}$    &KINGx &   100   &   192.4  &    0.96  \\ 
J125147+410439 & 192.9478 & 41.0776 & 1.09 & 1.08 &  8.08$^{+0.19}_{-0.38}$    &KINGx &   100   &   147.0  &    0.86 \\
\hline
\vspace*{-0.8cm}
\end{tabular}
\end{minipage}
\end{center}
\vspace*{0.2cm}
{{\sc Note}---
$^a$  Index denotes concentration for King profiles, exponent for Sersic profiles.

Table~\ref{tab:n4736gccshapes} is published in its entirety in the machine-readable format.
      A portion is shown here for guidance regarding its form and content.
}
\label{tab:n4736gccshapes}
\end{tiny}
\end{table}

\begin{table}[ht]
\caption{$i^\prime$-Band Shape Parameters of Globular Cluster Candidates in NGC 4826}
  \begin{tiny}
 \begin{center}
\hspace*{-1.5cm} \begin{minipage}{290mm}
  \begin{tabular}{@{\extracolsep{4pt}}cccccccrrr@{}}
\hline
\hline
\vspace*{-0.3cm}
&&&&&&&&\\
&&& 100 $\times$ (SPREAD &&&&&& \\
 Name  &     RA J2000  &   DEC J2000   & \_MODEL)$_{i^\prime}$   &   FWHM$_{i^\prime}$  &  $r_{\rm eff, i^\prime}$  &  Shape &  Index$^a$   &  SNR &  $\chi^2$ \\
       &      deg      &     deg       &              &     arcsec          &     pc              &               &       &            &            \\
\hline
J125549+213423 & 193.9550 & 21.5733 & 0.12 & 0.61 & $<$ 2.42               &  KINGx & 30  &   296.5   &    4.55  \\ 
J125601+214923 & 194.0083 & 21.8233 & 0.48 & 0.69 &  2.81$^{+0.11}_{-0.07}$& SERSICx & 2  &   136.4   &    2.30   \\
J125612+214638 & 194.0522 & 21.7773 & 0.12 & 0.58 & $<$ 4.72               & KINGx & 100  &    34.1   &    1.04   \\
J125613+214412 & 194.0544 & 21.7368 & 1.13 & 0.82 & 9.45$^{+0.13}_{-0.22}$ &  KINGx & 30  &   131.5   &    1.28   \\
J125614+214043 & 194.0592 & 21.6789 & 0.73 & 0.71 & 4.04$^{+0.07}_{-0.13}$ & KINGx & 100  &   136.5   &    1.44   \\
J125615+214645 & 194.0648 & 21.7794 & 0.20 & 0.62 & $<$ 4.62               & KINGx & 100  &   119.5   &    1.86   \\
J125622+215134 & 194.0944 & 21.8597 & 0.15 & 0.61 & $<$ 2.01               & KINGx & 100  &   114.8   &    1.44   \\
J125623+214054 & 194.0982 & 21.6817 & 1.10 & 0.78 & 7.21$^{+0.09}_{-0.03}$ &  KINGx & 30  &   188.4   &    2.36   \\
J125625+214427 & 194.1082 & 21.7408 & 0.72 & 0.71 & 3.37$^{+0.08}_{-0.03}$ & KINGx & 100  &   203.9   &    1.82   \\
J125626+213440 & 194.1095 & 21.5779 & 0.10 & 0.62 & $<$ 1.28               &  KINGx & 30  &   137.0   &    1.33   \\
J125626+214234 & 194.1124 & 21.7096 & 0.71 & 0.71 & 3.71$^{+0.12}_{-0.05}$ & KINGx & 100  &   173.8   &    2.65   \\
J125628+214937 & 194.1207 & 21.8272 & 1.15 & 0.81 & 8.80$_{-0.01}$         & SERSICx & 2  &   262.1   &    5.35  \\
J125632+214023 & 194.1364 & 21.6733 & 0.43 & 0.67 & 1.85$^{+0.07}_{-0.08}$ & KINGx & 100  &    80.2   &    2.13   \\
J125633+213919 & 194.1416 & 21.6554 & 0.84 & 0.74 & 4.88$^{+0.10}_{-0.10}$ & KINGx & 100  &   182.8   &    1.13   \\
J125638+214512 & 194.1583 & 21.7535 & 0.01 & 0.60 & $<$ 5.98               & KINGx & 100  &    32.3   &    0.76   \\
J125641+213802 & 194.1709 & 21.6341 & 0.64 & 0.70 & 3.90$^{+0.16}_{-0.09}$ &  KINGx & 30  &    82.2   &    0.86   \\
J125641+213641 & 194.1734 & 21.6115 & 0.58 & 0.69 & 2.86$^{+0.10}_{-0.02}$ & KINGx & 100  &   159.9   &    1.20   \\
J125642+213204 & 194.1761 & 21.5347 & 0.52 & 0.69 & 2.53$^{+0.10}_{-0.07}$ & KINGx & 100  &   109.9   &    1.00   \\
J125644+213848 & 194.1873 & 21.6468 & 0.79 & 0.73 & 4.38$^{+0.14}_{-0.05}$ & KINGx & 100  &   185.4   &    2.14   \\
J125646+213754 & 194.1954 & 21.6317 & 0.55 & 0.69 & 2.70$^{+0.18}_{-0.07}$ & KINGx & 100  &    82.2   &    1.01   \\
J125649+214425 & 194.2065 & 21.7404 & 0.81 & 0.73 & 4.88$^{+0.07}_{-0.12}$ & KINGx & 100  &   151.2   &    1.52   \\
J125654+215233 & 194.2275 & 21.8760 & 1.04 & 0.77 & 8.76$^{+0.34}_{-0.30}$ & KINGx & 100  &   109.5   &    1.16   \\
J125655+213319 & 194.2323 & 21.5554 & 0.05 & 0.60 & $<$ 5.97               &  KINGx & 30  &    64.2   &    1.07   \\
J125655+214633 & 194.2332 & 21.7761 & 0.97 & 0.77 & 6.74$^{+0.20}_{-0.15}$ & KINGx & 100  &   151.3   &    1.02   \\
J125714+213914 & 194.3117 & 21.6540 & 0.13 & 0.62 & $<$ 1.44               &  KINGx & 30  &   256.8   &    3.05   \\
J125717+213702 & 194.3225 & 21.6173 & 0.56 & 0.69 & 3.60$^{+0.12}_{-0.17}$ &  KINGx & 30  &    79.0   &    0.87   \\
J125721+213816 & 194.3410 & 21.6378 & 0.98 & 0.78 & 7.40$^{+0.18}_{-0.18}$ &  KINGx & 30  &    87.3   &    0.87   \\
J125722+213659 & 194.3418 & 21.6166 & 0.03 & 0.61 & $<$ 1.14               &  KINGx & 30  &   208.9   &    2.46   \\
J125723+214212 & 194.3472 & 21.7036 & 0.62 & 0.71 & 3.37$^{+0.12}_{-0.07}$ & KINGx & 100  &   134.4   &    1.64   \\
J125723+213711 & 194.3480 & 21.6198 & 0.80 & 0.74 & 4.72$^{+0.14}_{-0.10}$ & KINGx & 100  &   160.6   &    1.27  \\
\hline
\vspace*{-0.8cm}
\end{tabular}
\end{minipage}
\end{center}
\vspace*{0.2cm}
{{\sc Note}---
$^a$  Index denotes concentration for King profiles, exponent for Sersic profiles.

Table~\ref{tab:n4826gccshapes} is published in its entirety in the machine-readable format.
      A portion is shown here for guidance regarding its form and content.
}
\label{tab:n4826gccshapes}
\end{tiny}
\end{table}

\begin{figure}[h!]
\begin{tabular}{ll}
\raisebox{0.4cm}{\includegraphics[scale=0.45,angle=0.]{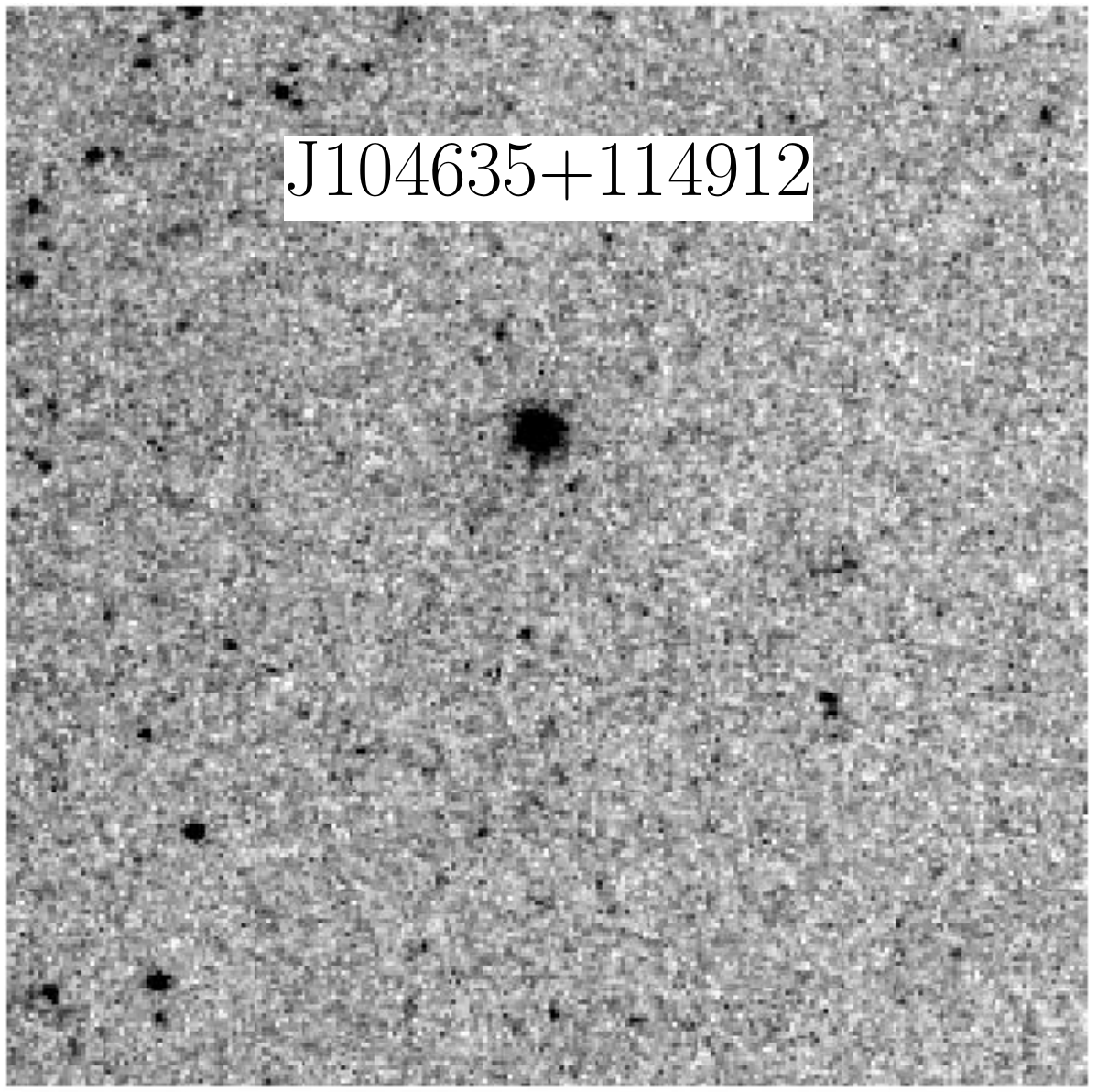}}
\hspace*{0.55cm}\raisebox{0.4cm}{\includegraphics[scale=0.45,angle=0.]{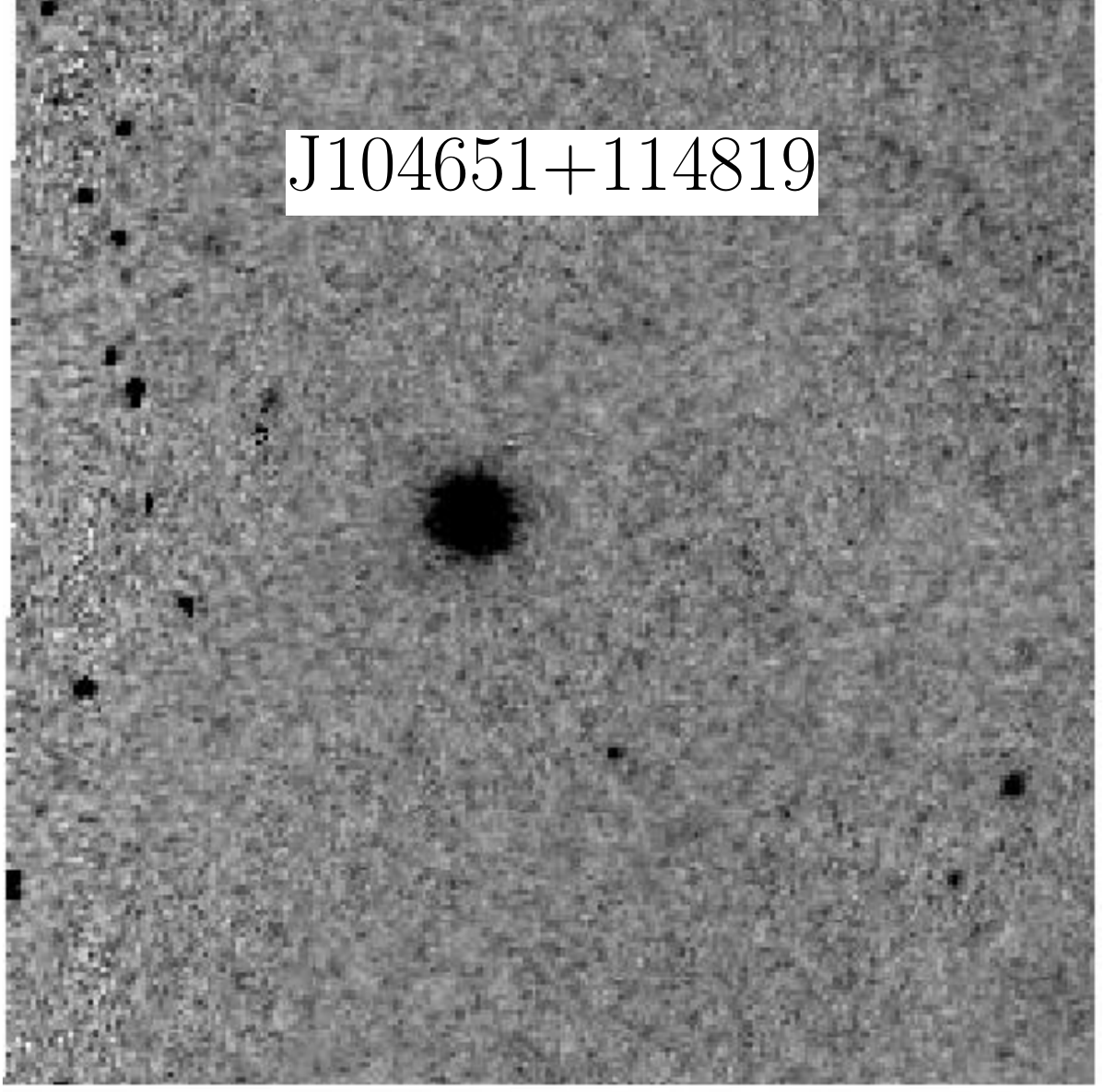}}
\end{tabular}
\begin{tabular}{lll}
\includegraphics[scale=0.45,angle=0.]{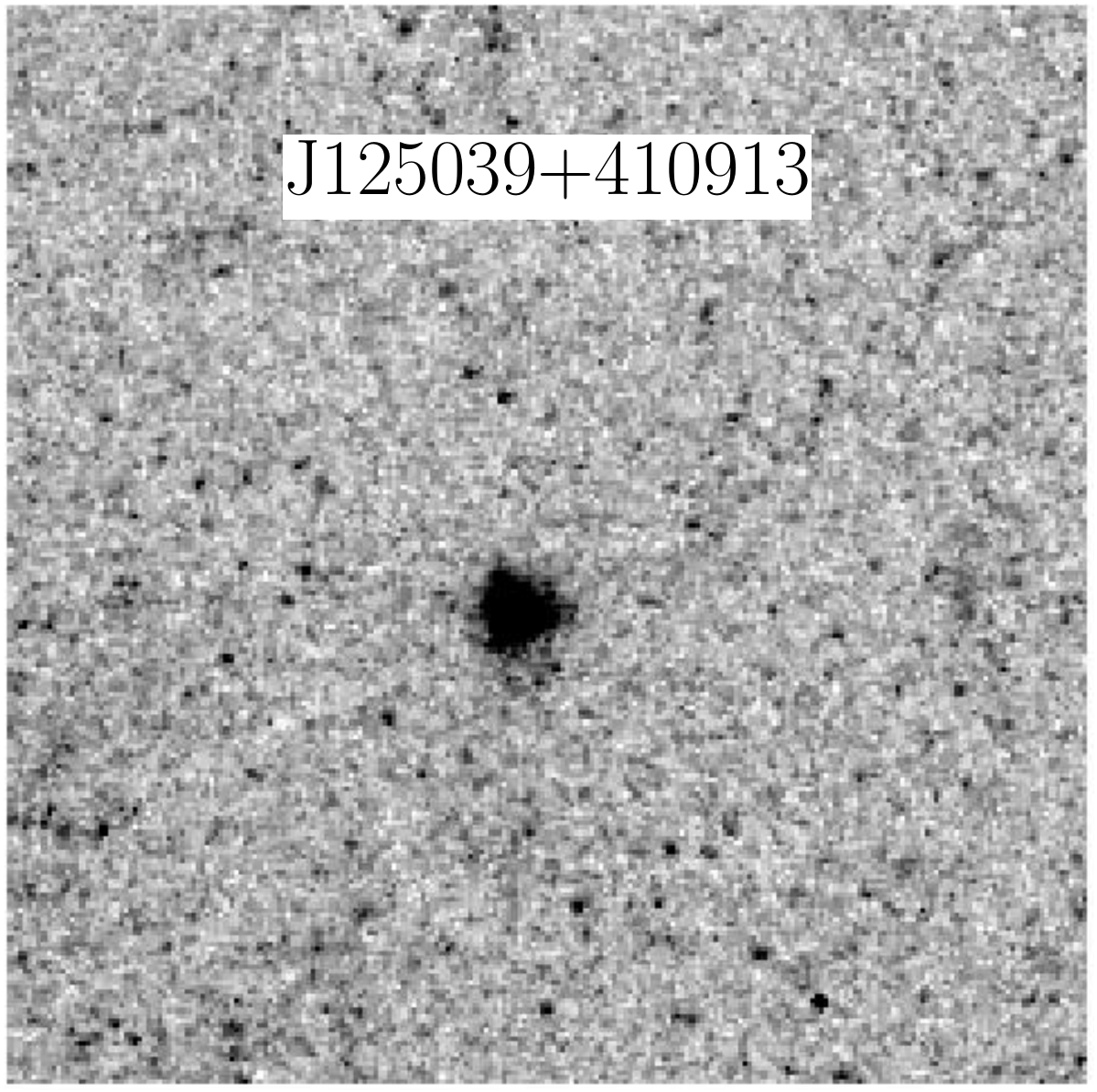}
&
\hspace*{0.3cm}\includegraphics[scale=0.45,angle=0.]{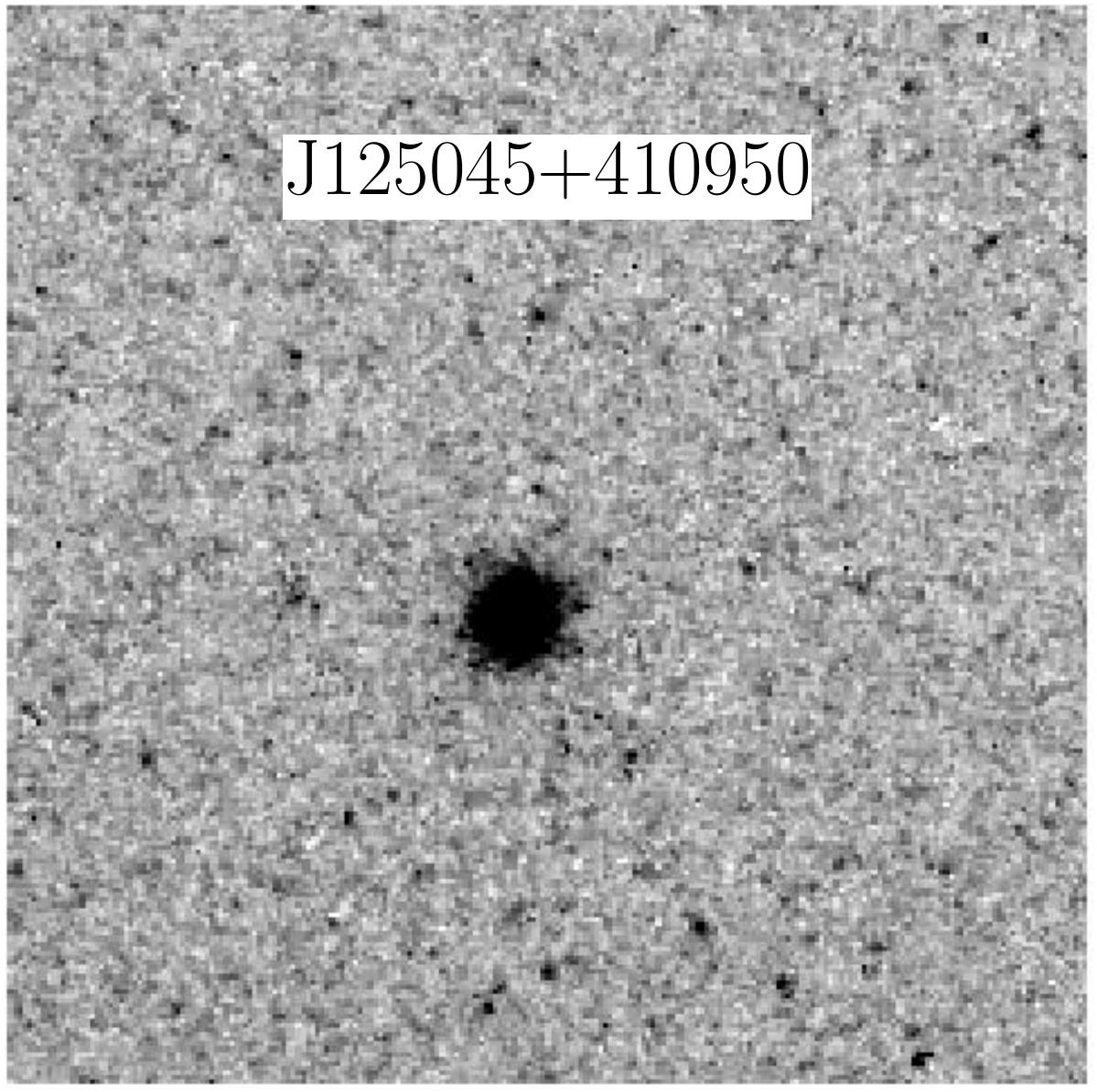}
&
\hspace*{0.3cm}\includegraphics[scale=0.45,angle=0.]{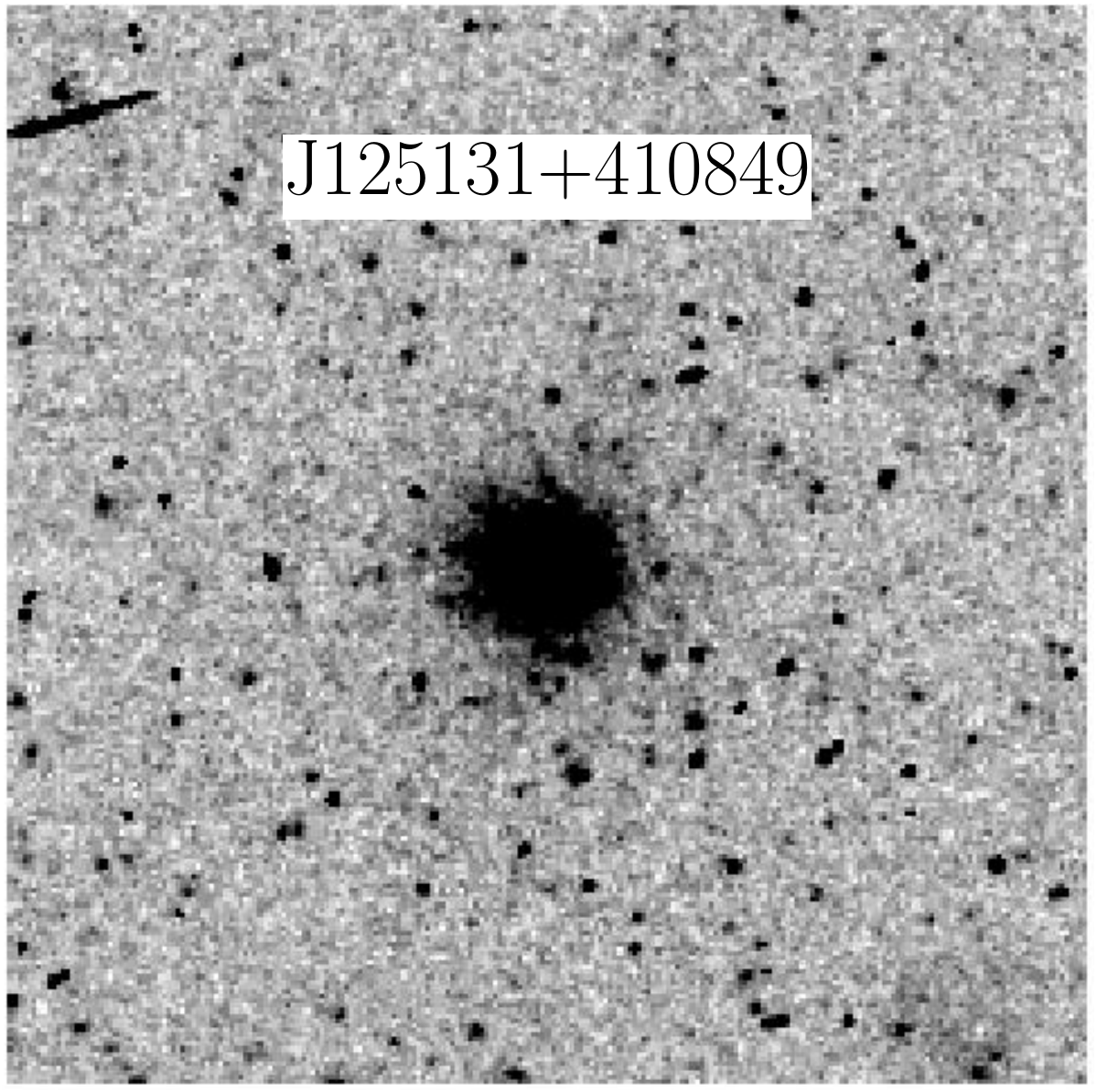}
\end{tabular}
\caption{Grayscales of GC candidates, from {\sl HST} images. {\it Top, left to
right:} J104635+114912 and J104651+114819, in NGC\,3368.  {\it Bottom:}
J125039+410913, J125045+410950, and J125131+410849, in NGC\,4736.
\label{fig:grayscands}}
\end{figure}

We present in Figure ~\ref{fig:grayscands} a few exposures of  GCCs in
NGC\,3368 and NGC\,4736 that we were able to find in the Hubble Space Telescope
({\sl HST}) Legacy Archive; their visual appearance is consistent with their
classification.  J104635+114912 was identified in the Advance Camera for
Surveys (ACS) frame hst\_10433\_02\_acs\_wfc\_f555w.fits (filter F555W,
proposal GO 10433, P.I.\ B.\ Madore); J104651+114819, in the Wide Field Camera
3 (WFC3) image hst\_13364\_07\_wfc3\_uvis\_f555w.fits (F555W, proposal GO
13364, P.I.\ D.\ Calzetti).  All the NGC\,4736 images were acquired with the
ACS.  J125039+410913 and J125045+410950 were identified in the frame
hst\_10402\_09\_acs\_wfc\_f555w\_drz.fits (proposal GO 10402, P.I.\ R.\
Chandar); J125131+410849, in hst\_9480\_u8\_acs\_wfc\_f775w\_drz.fits (filter
F775W, proposal GO/PAR 9480, P.I.\ J.\ Rhodes). 

\section{Properties of the GC systems} \label{sec:properties}

\subsection{Color distribution} \label{sec:colordist}

\begin{sidewaysfigure}[ht]
\begin{tabular}{llll}
\hspace*{-2.3cm}\includegraphics[scale=0.20]{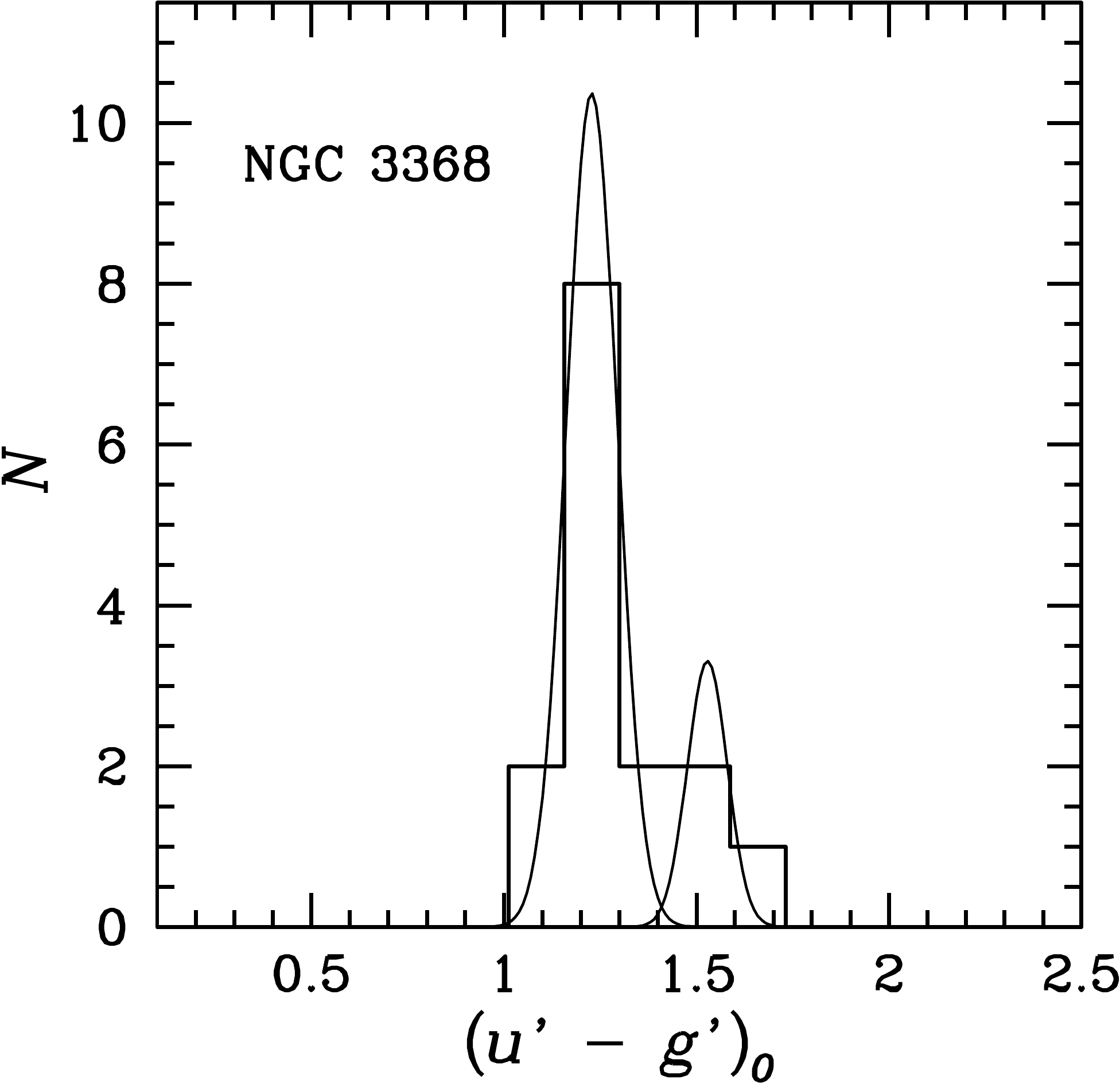}
&
\hspace*{0.05cm}\includegraphics[scale=0.20]{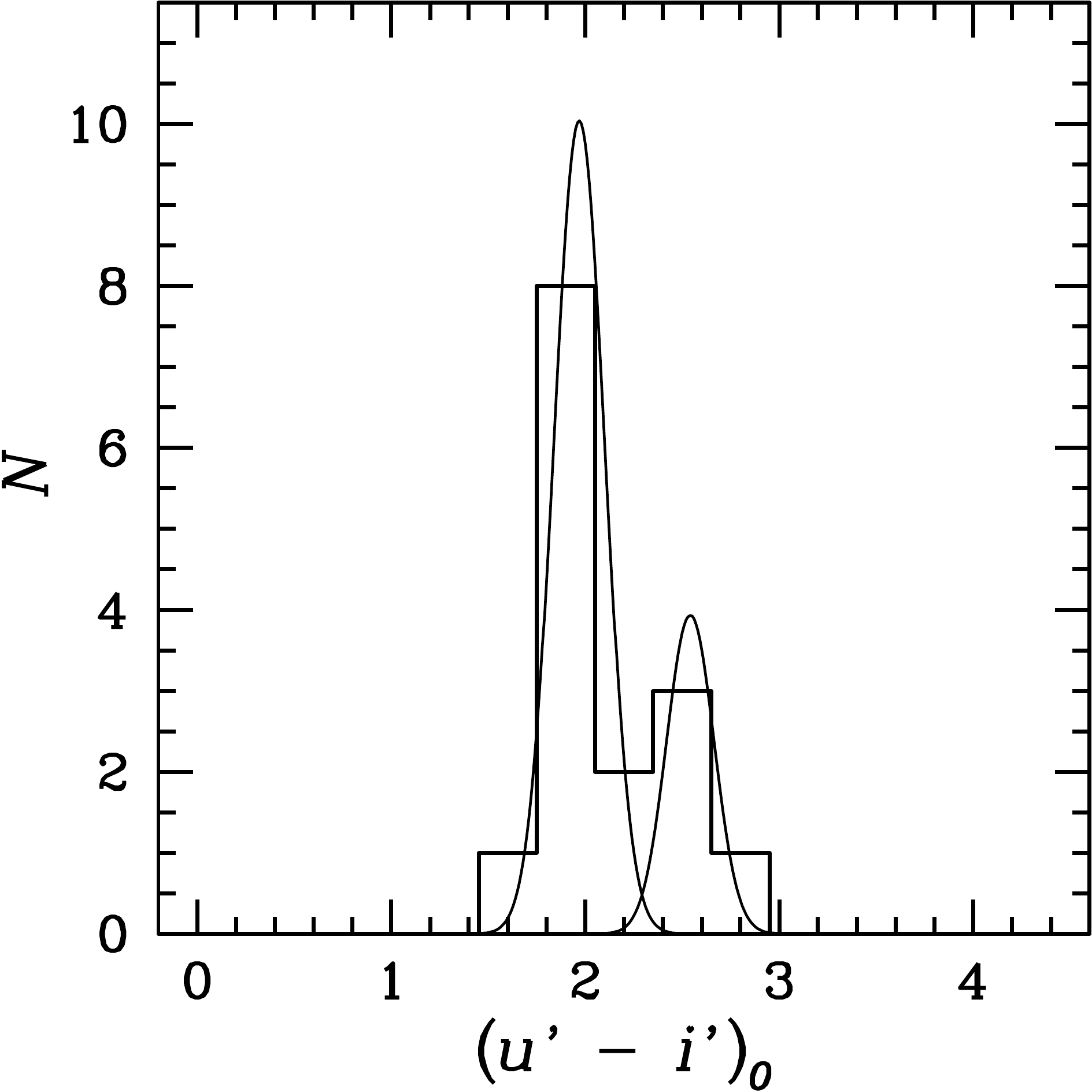}
&
\hspace*{4.35cm}\includegraphics[scale=0.20]{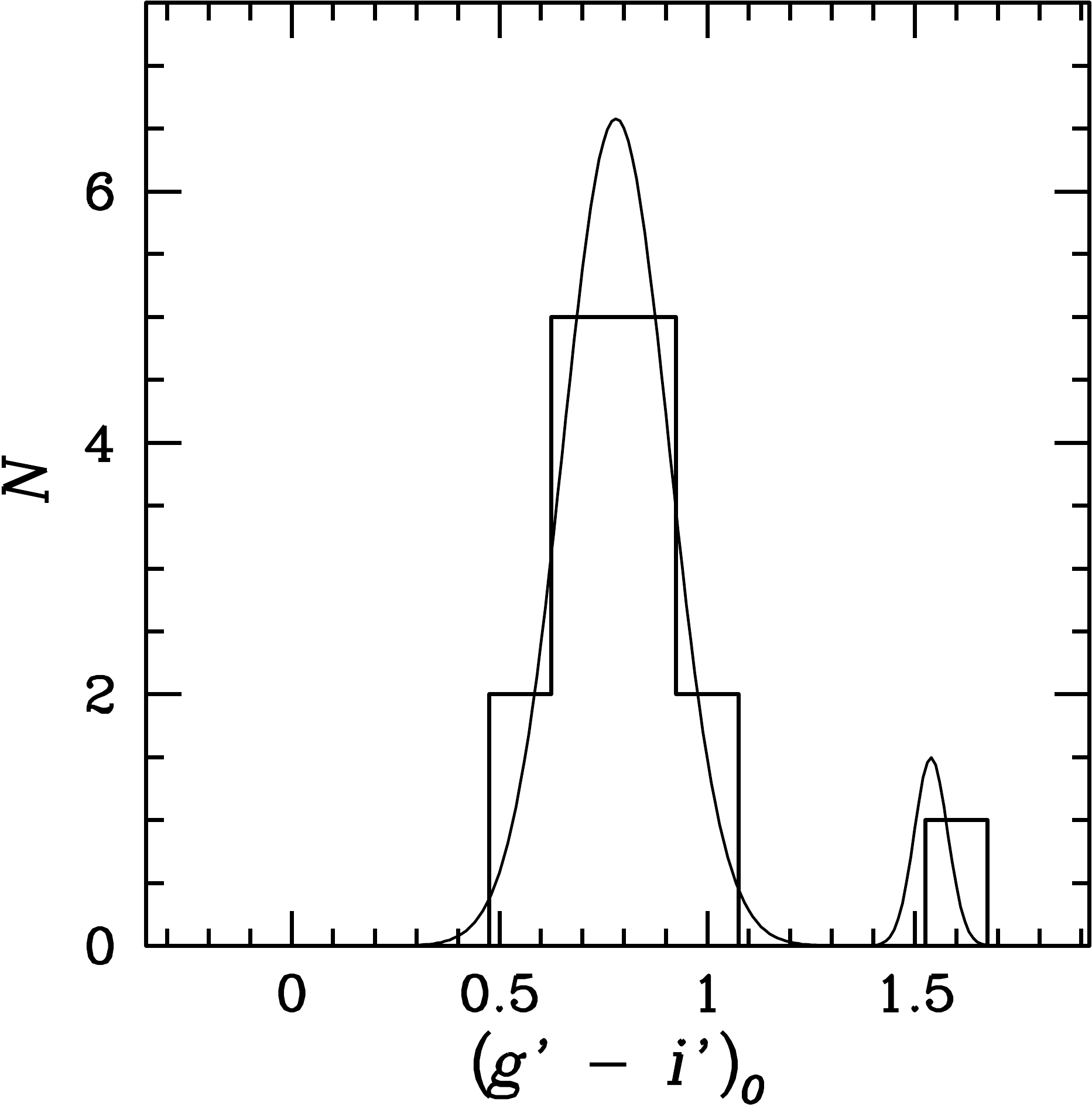}
&
\hspace*{0.2mm}\includegraphics[scale=0.20]{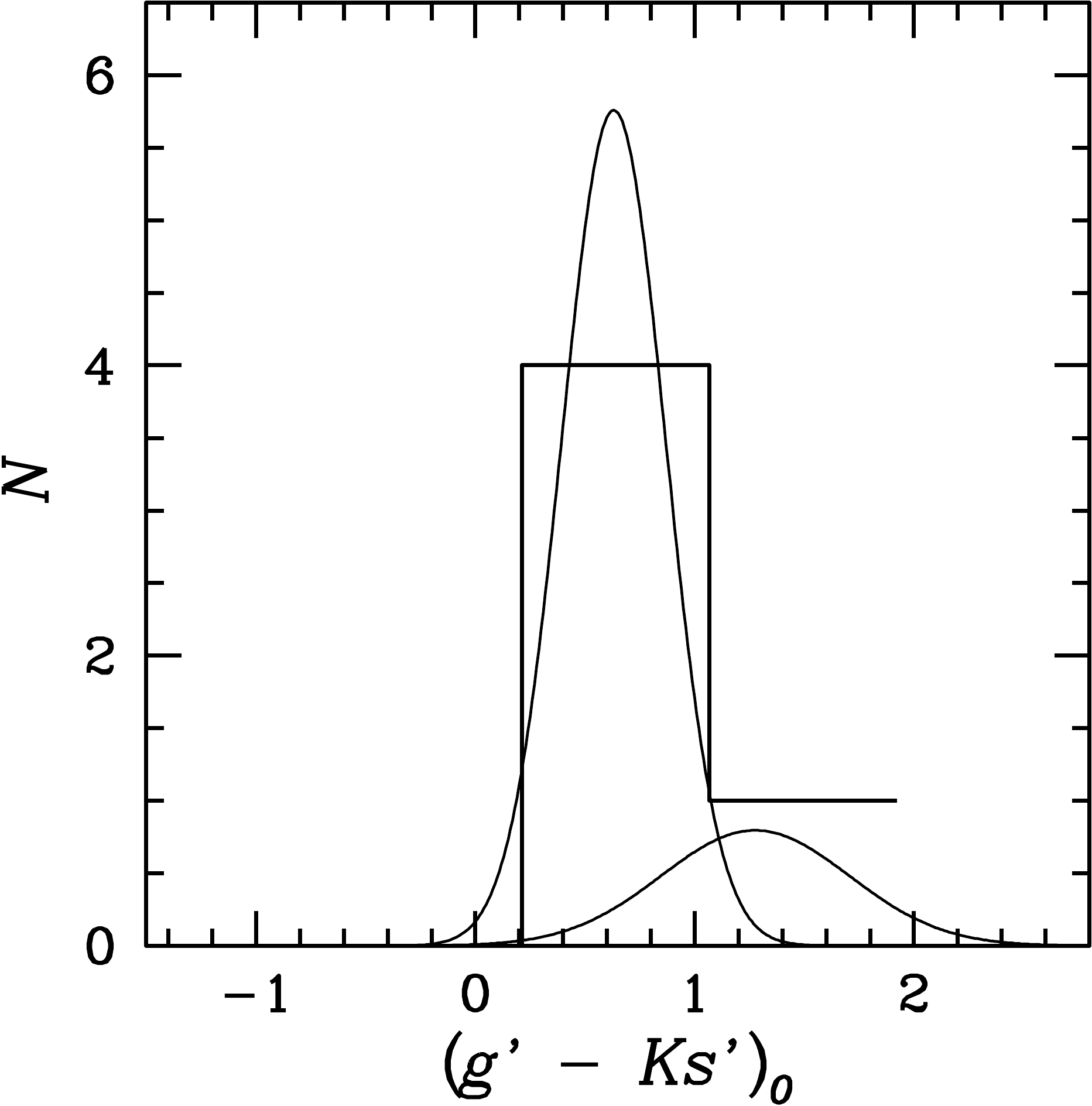}
\end{tabular}
\begin{tabular}{llll}
\hspace*{-1.1cm}\includegraphics[scale=0.20]{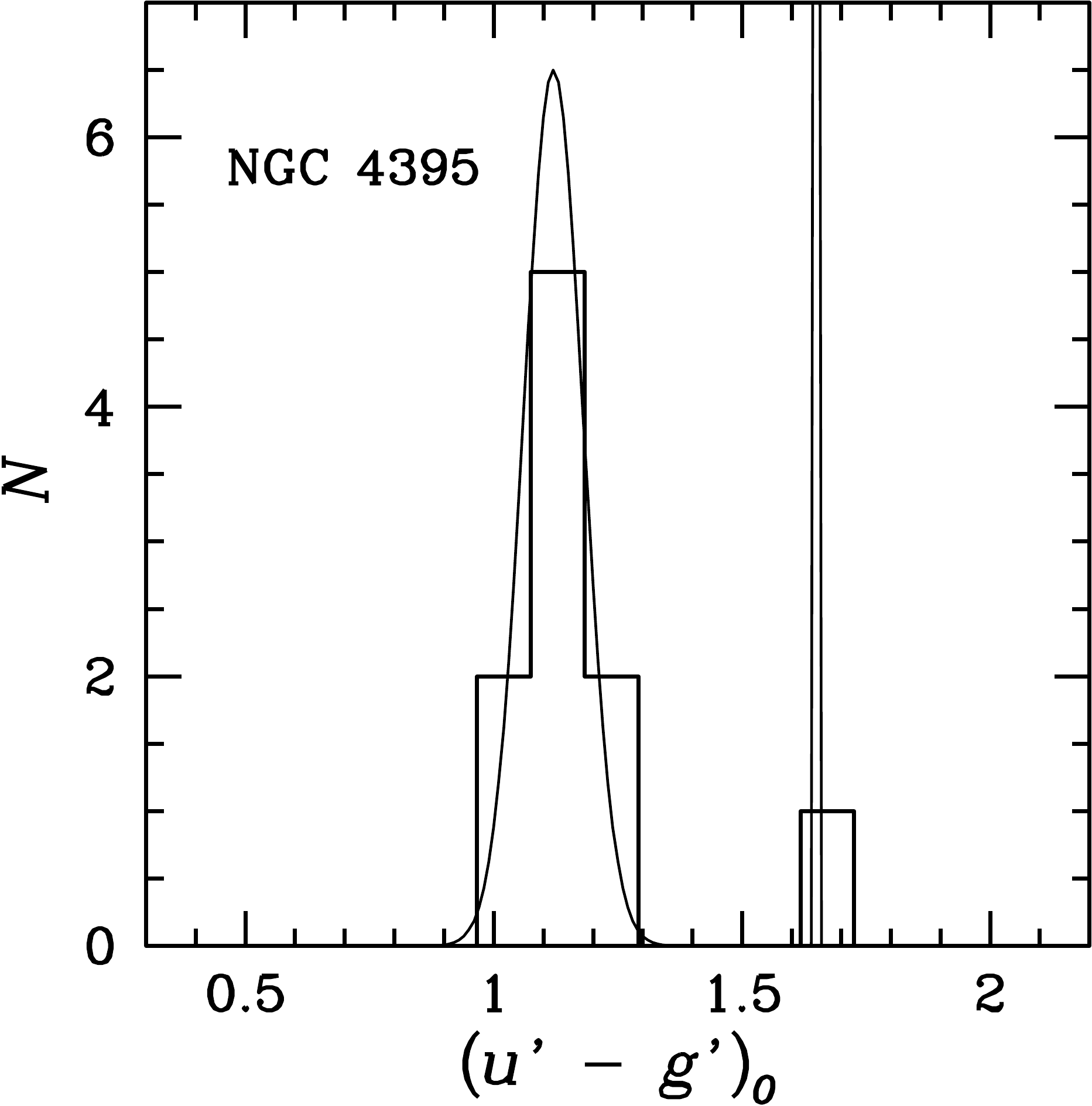}
&
\hspace*{0.05cm}\includegraphics[scale=0.20]{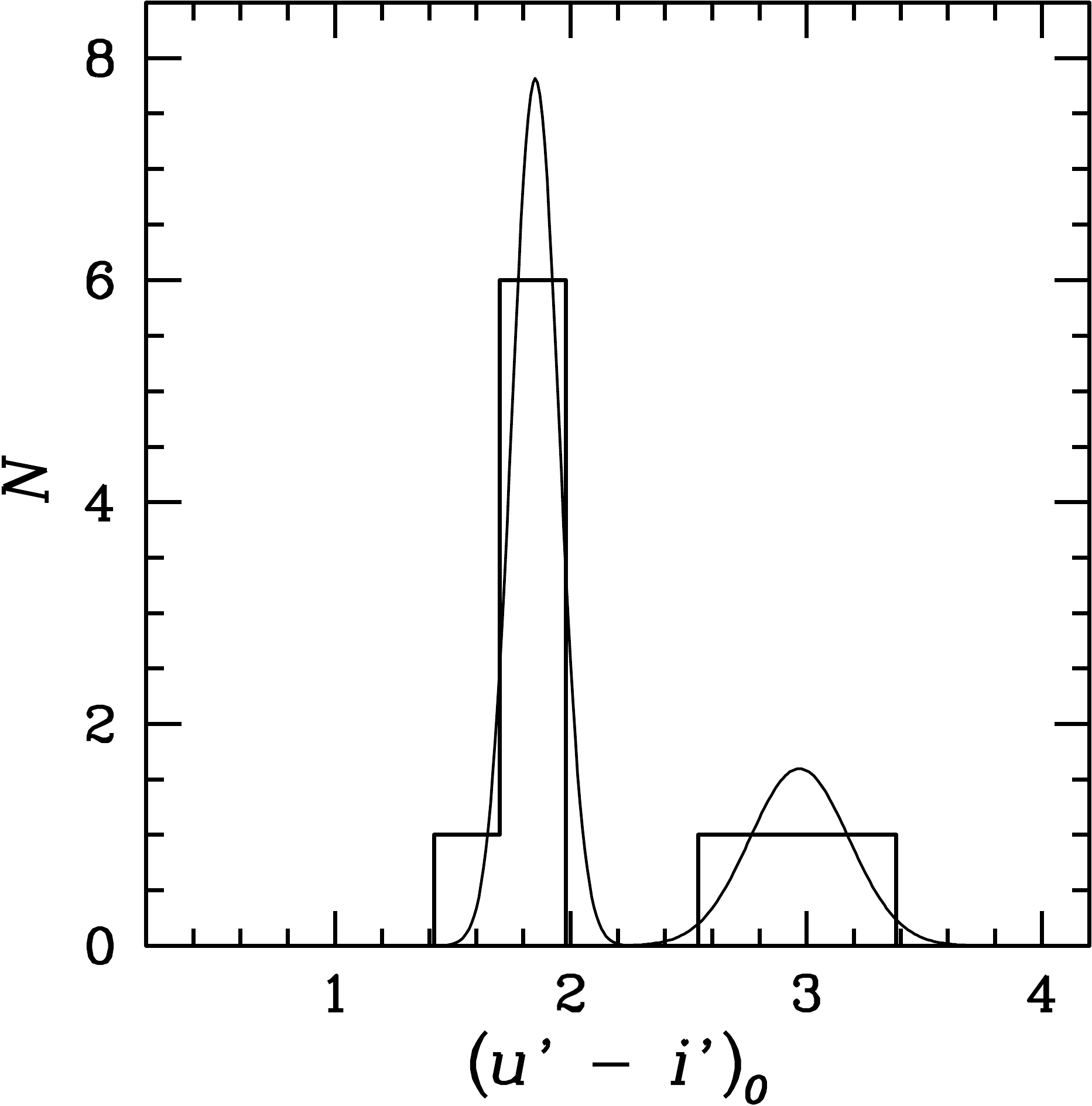}
&
\hspace*{4.35cm}\includegraphics[scale=0.20]{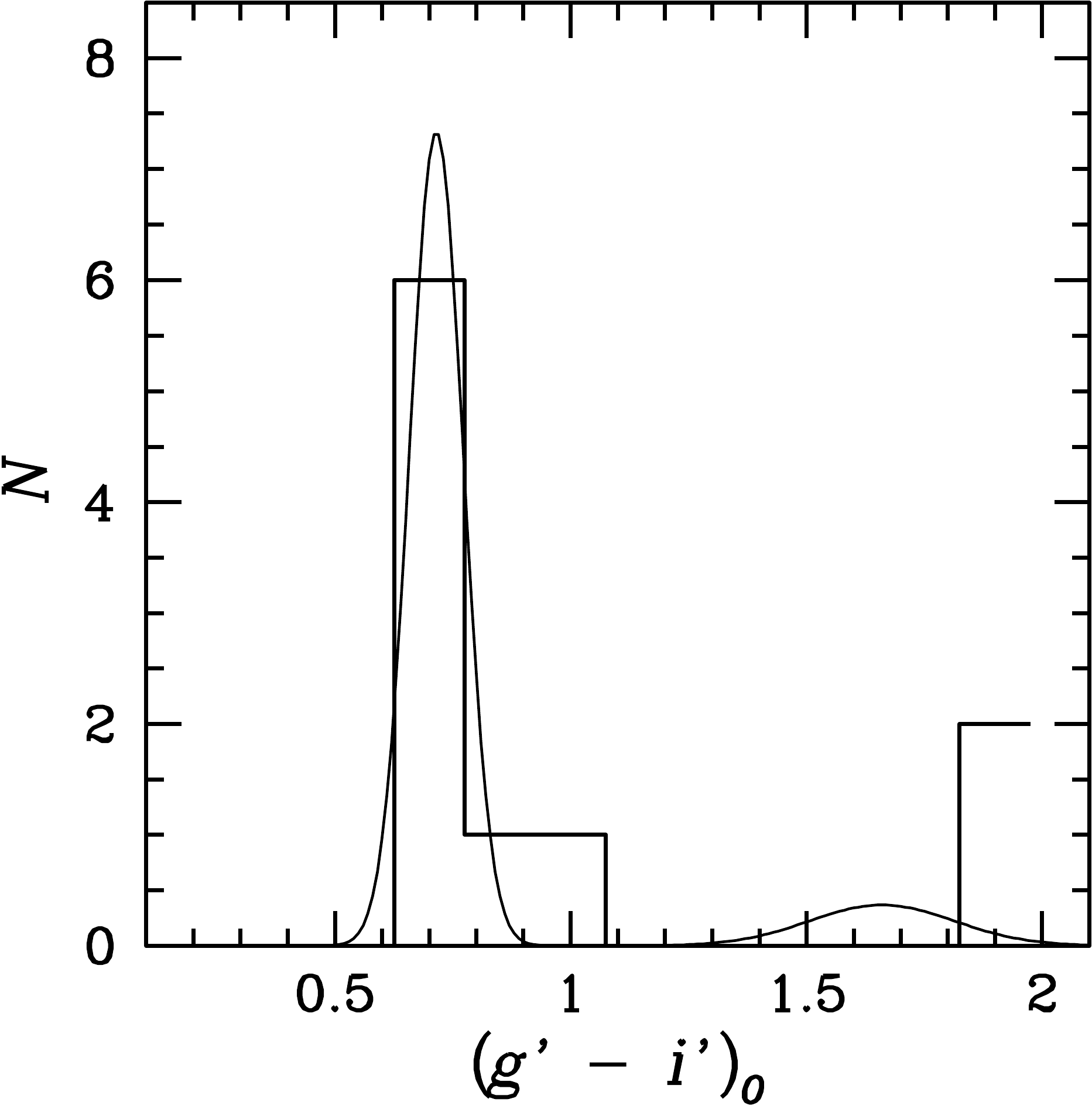}
&
\hspace*{0.2mm}\includegraphics[scale=0.20]{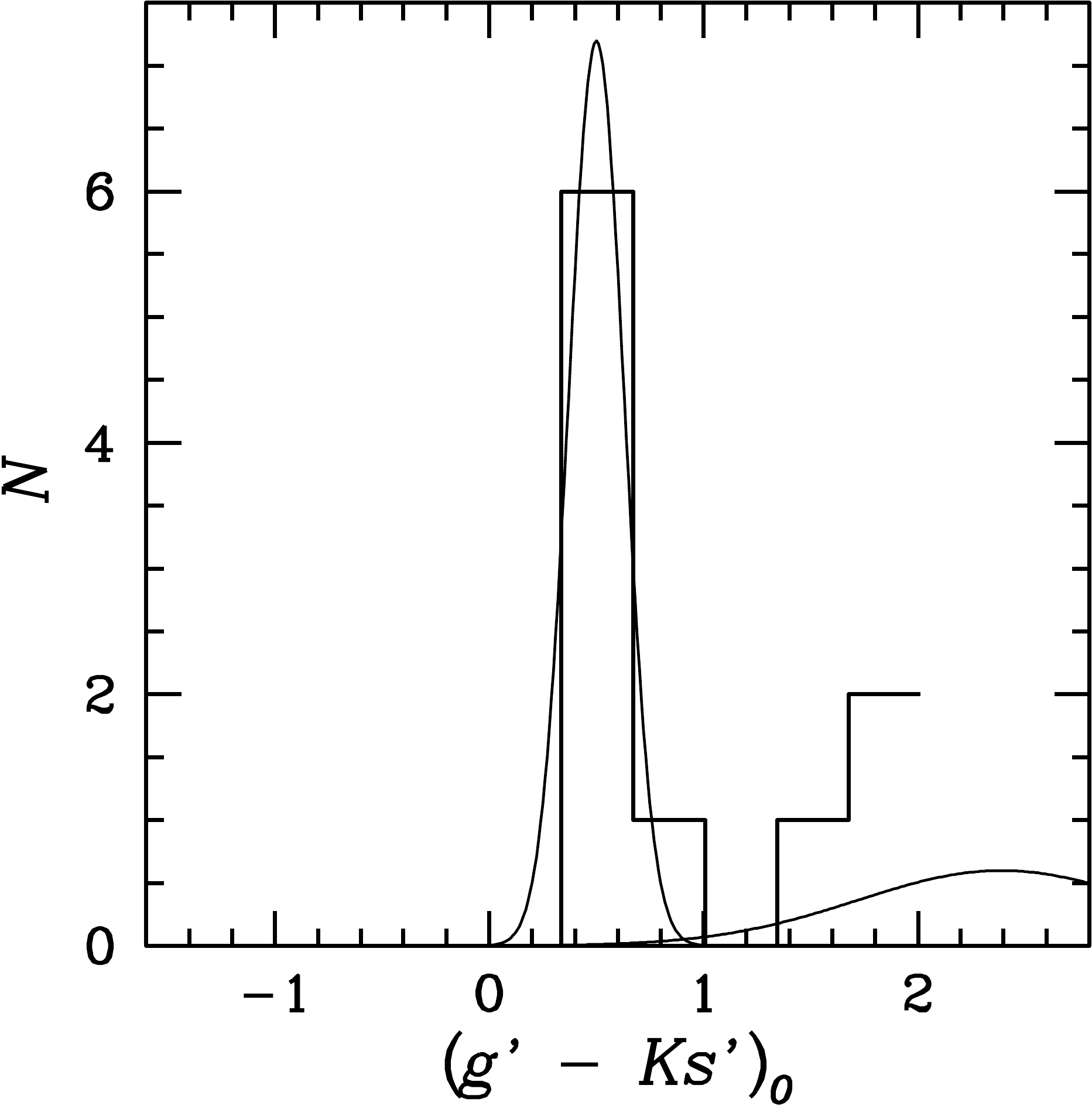}
\end{tabular}
\begin{tabular}{llllll}
\hspace*{-1.15cm}\includegraphics[scale=0.20]{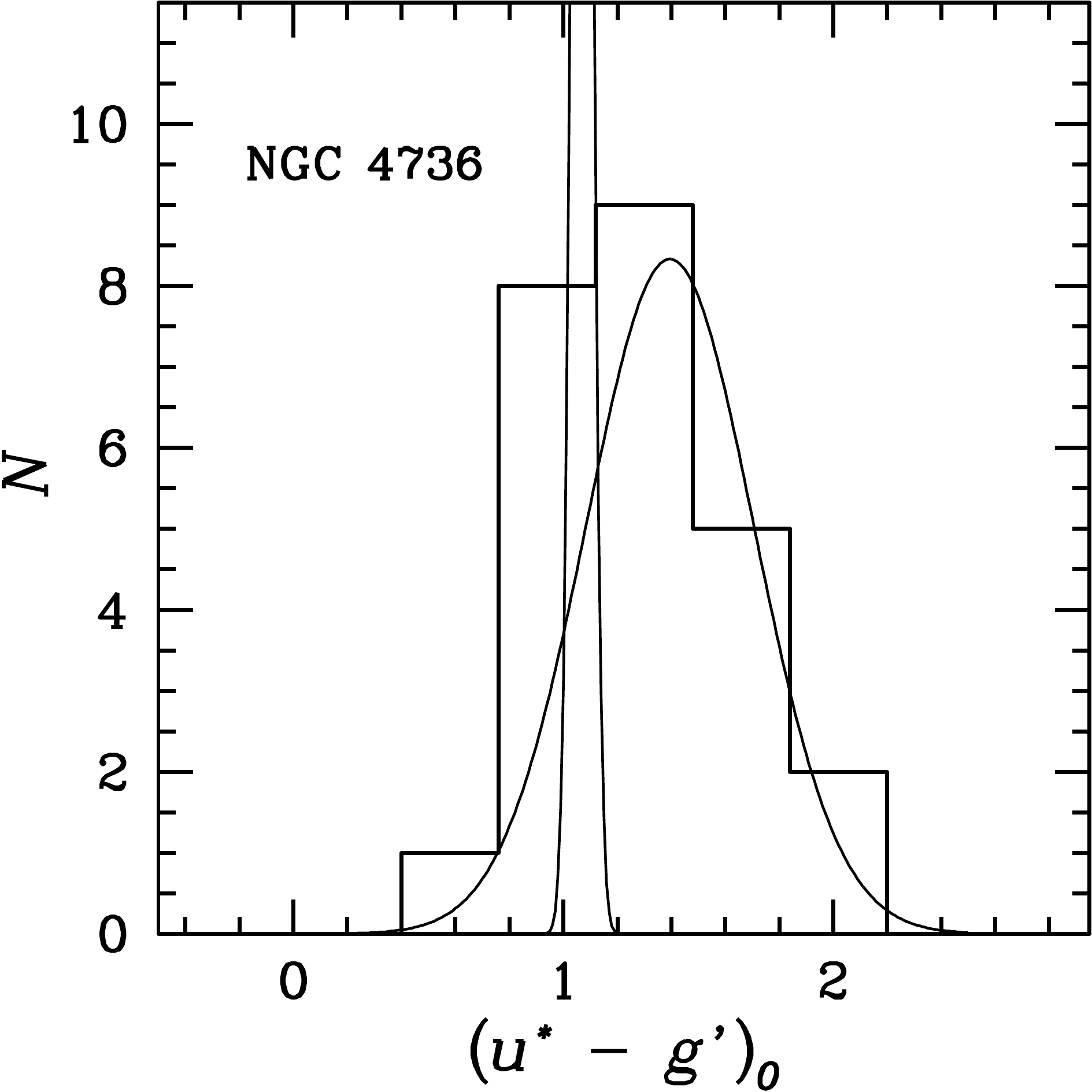}
&
\includegraphics[scale=0.20]{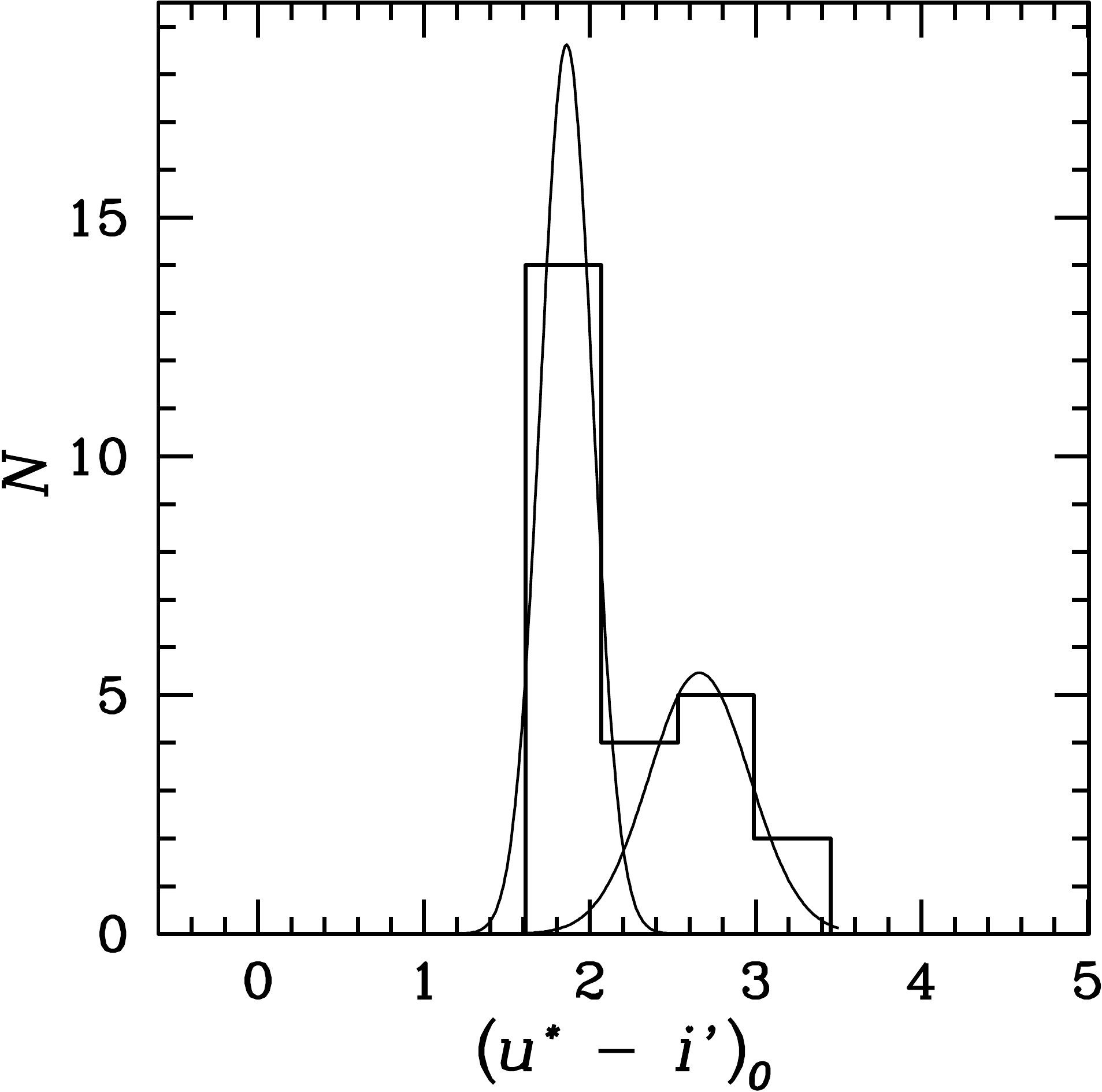}
&
\includegraphics[scale=0.20]{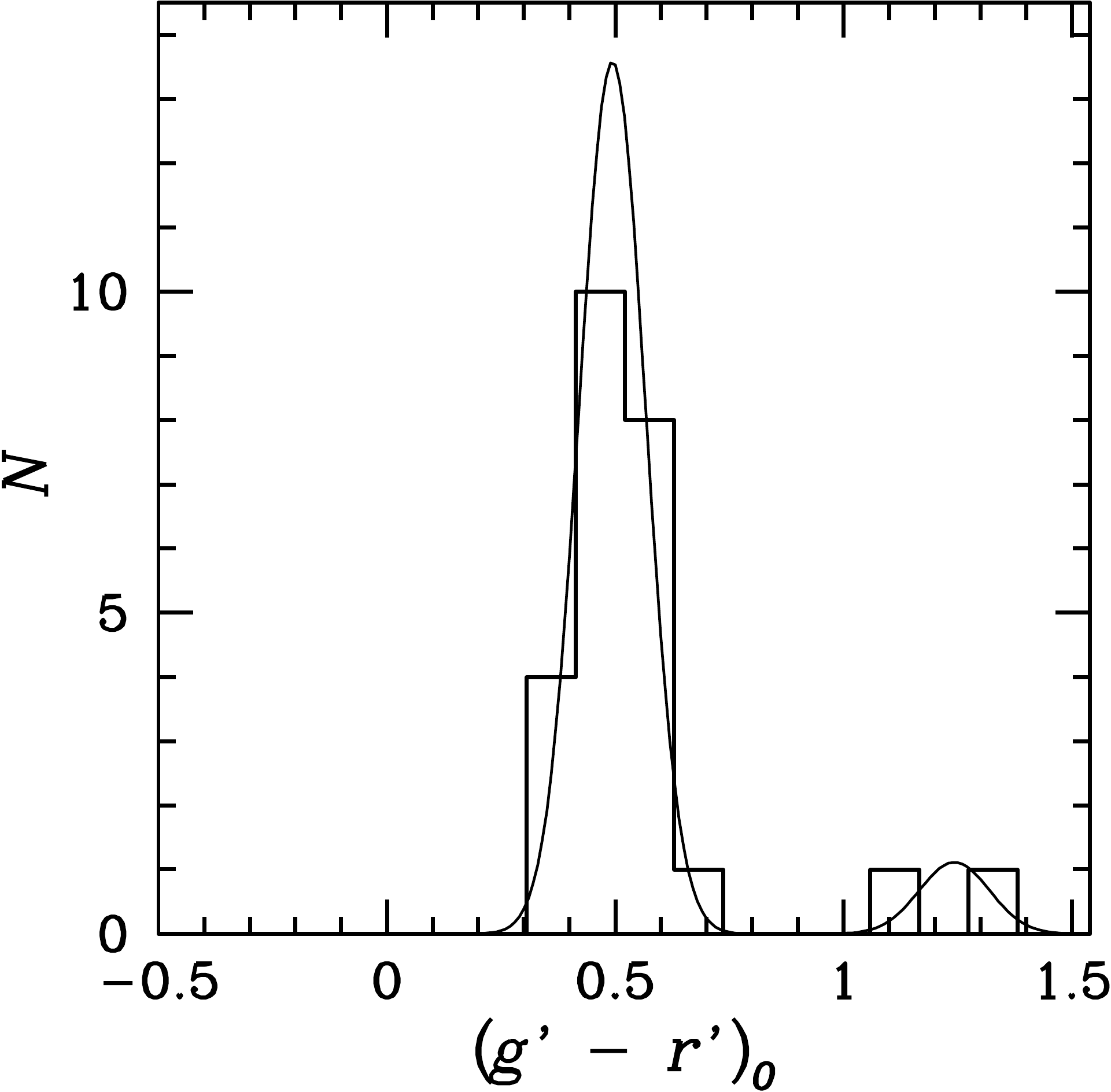}
&
\includegraphics[scale=0.20]{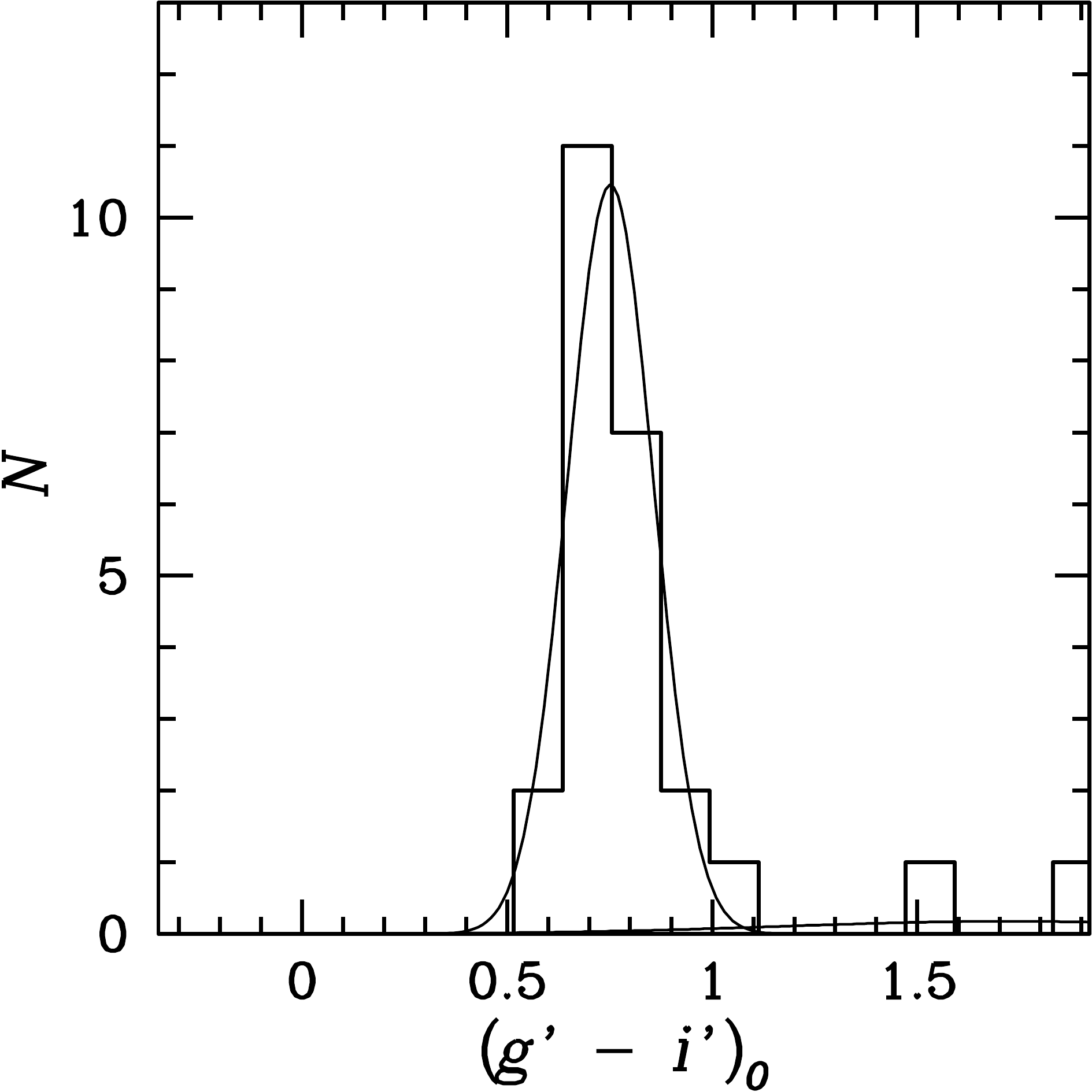}
&
\includegraphics[scale=0.20]{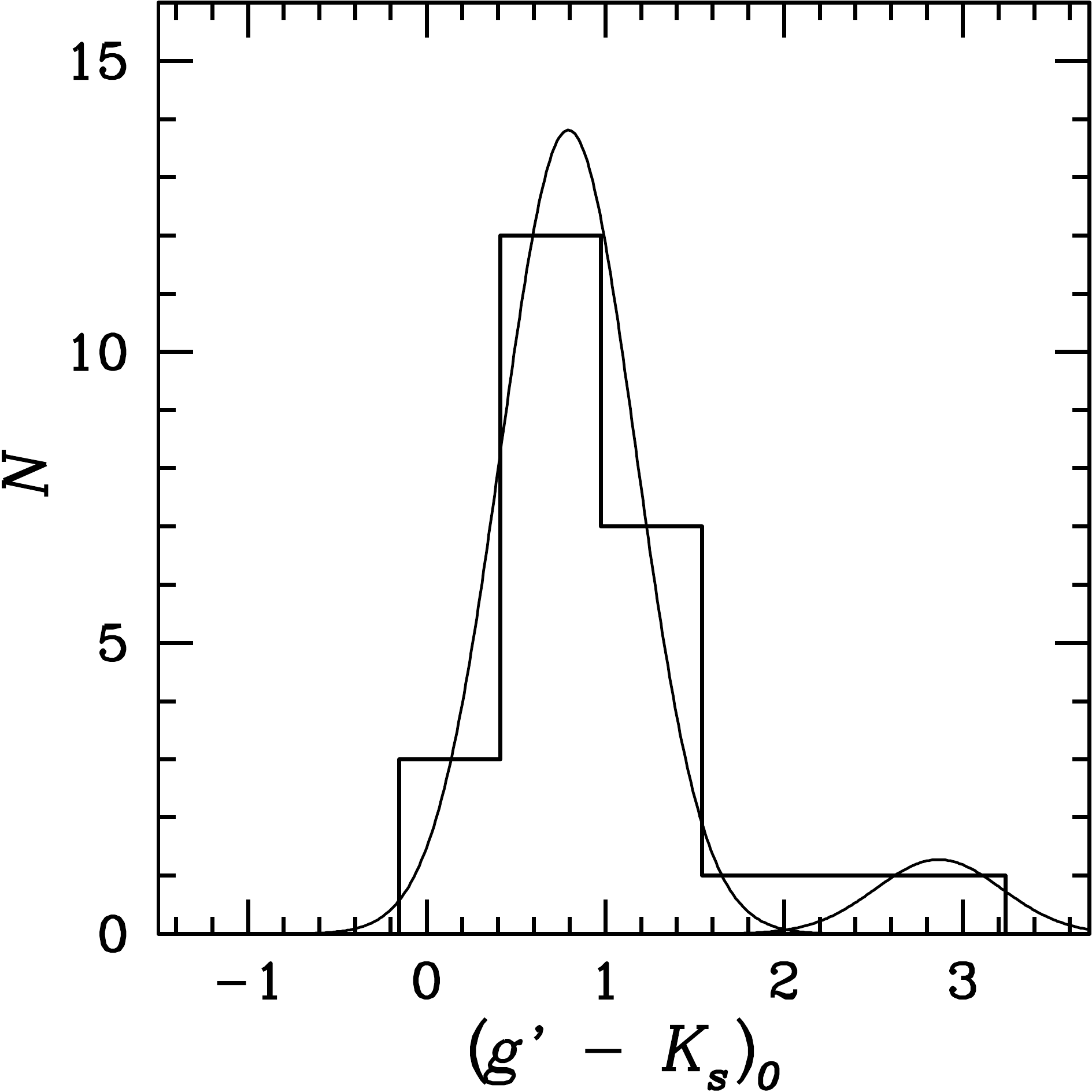}
&
\includegraphics[scale=0.20]{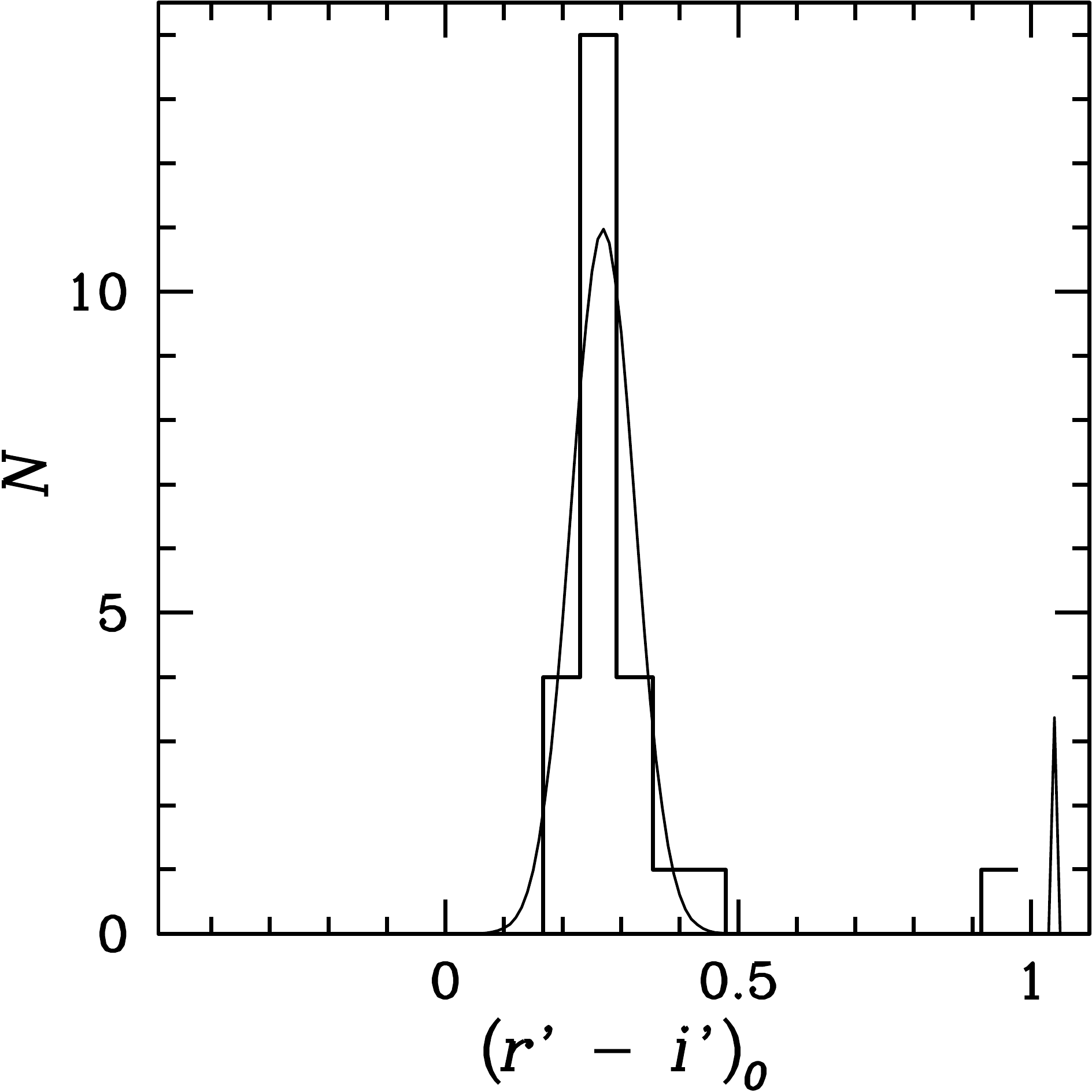}
\end{tabular}
\begin{tabular}{llllll}
\hspace*{-1.15cm}\includegraphics[scale=0.20]{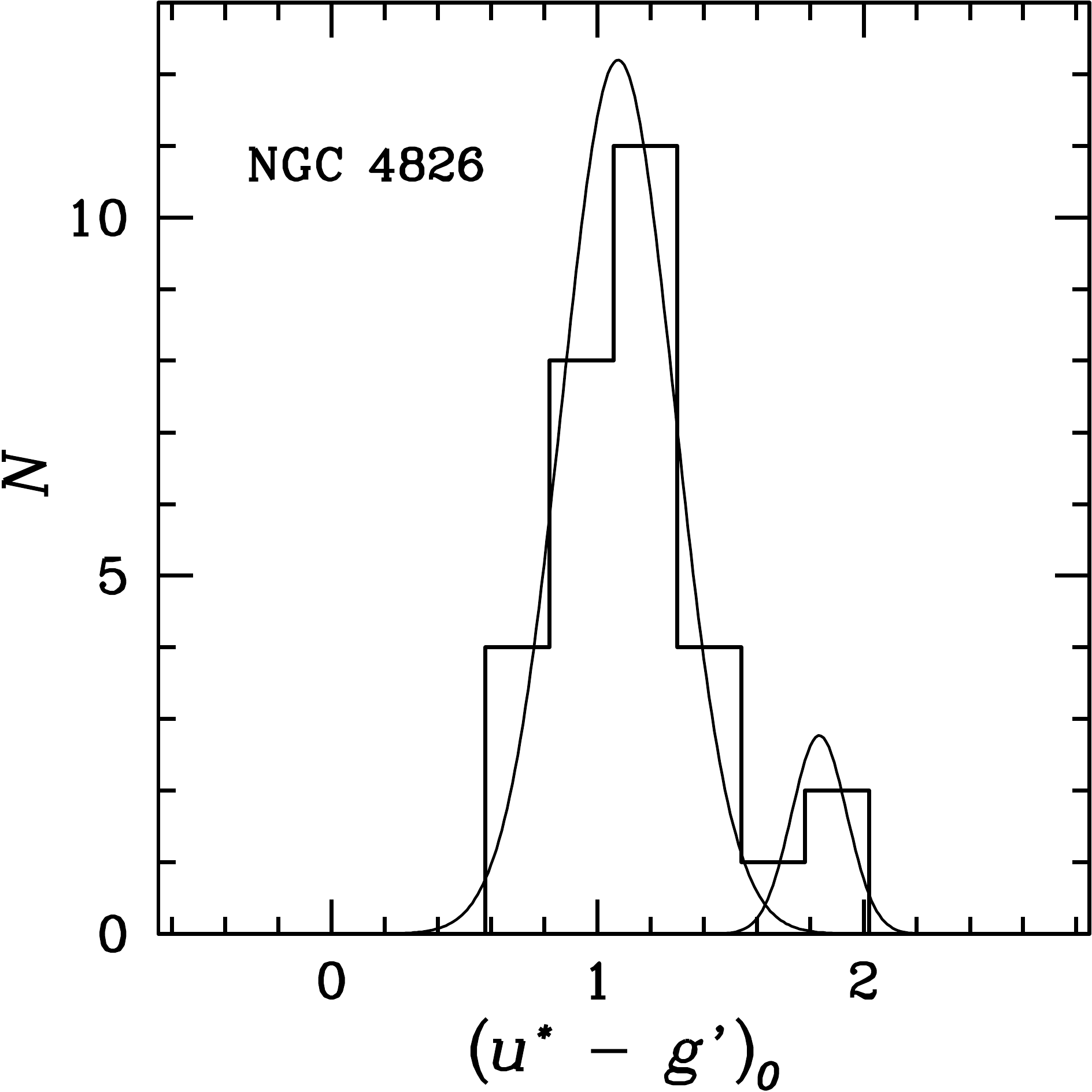}
&
\includegraphics[scale=0.20]{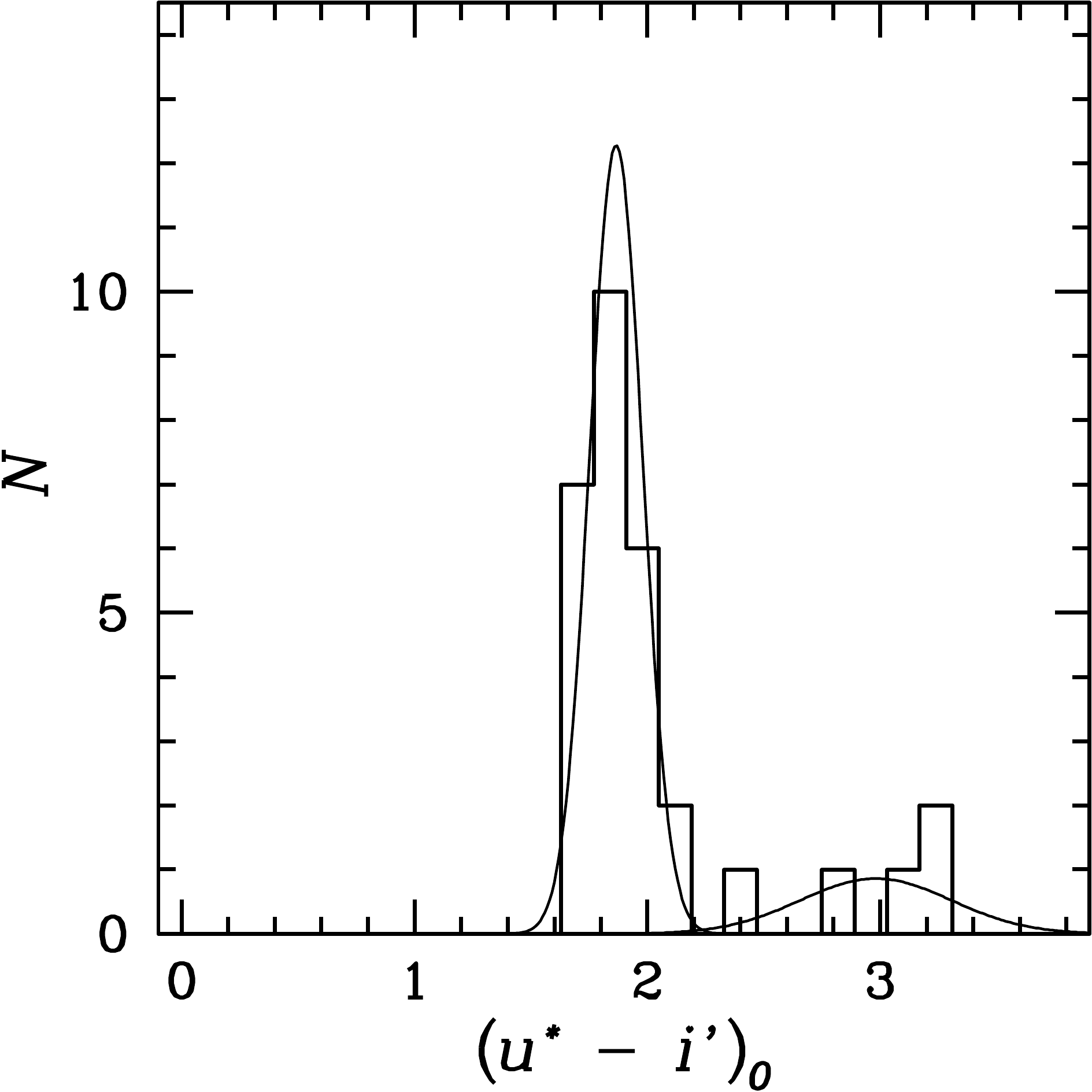}
&
\includegraphics[scale=0.20]{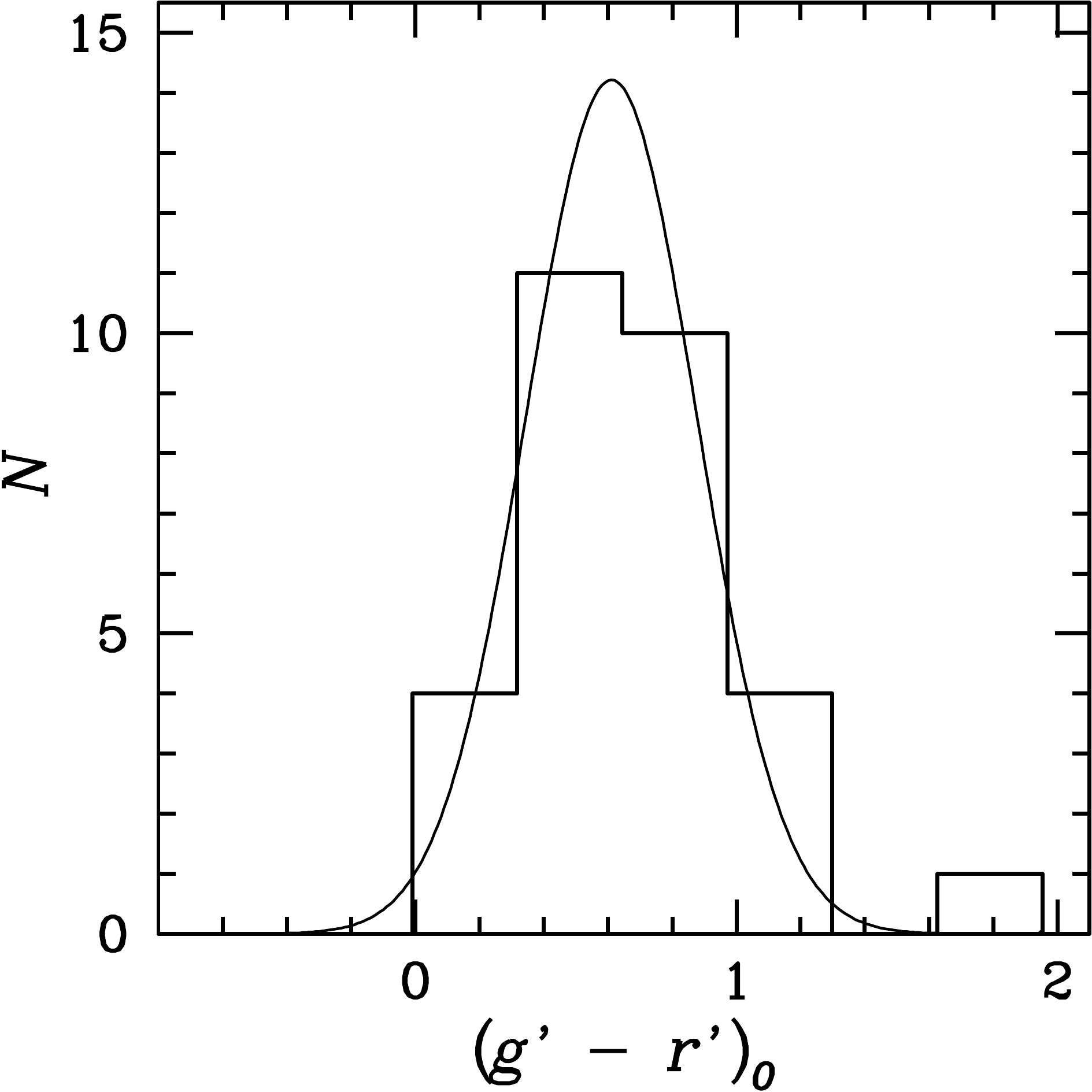}
&
\hspace*{1mm}\includegraphics[scale=0.20]{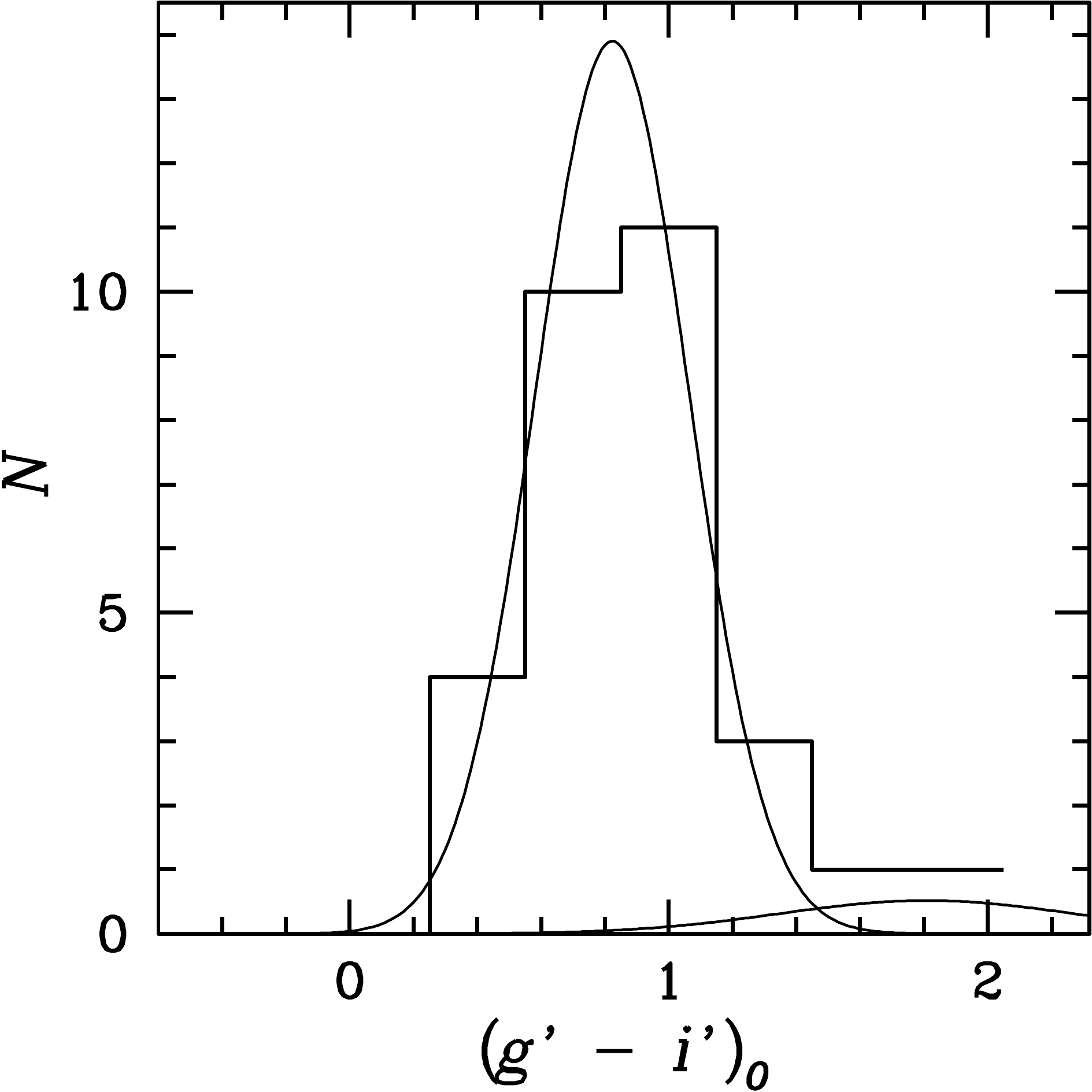}
&
\includegraphics[scale=0.20]{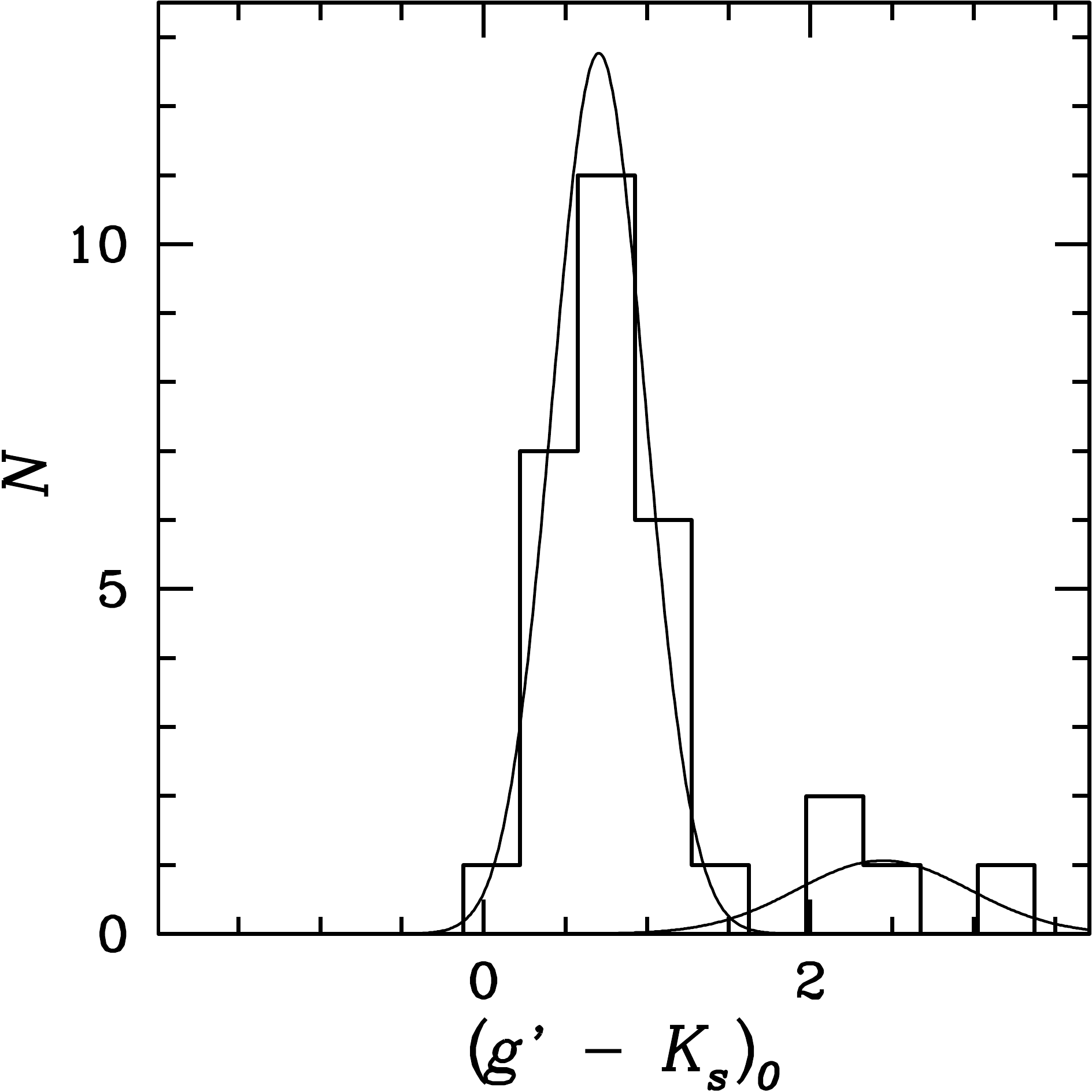}
&
\includegraphics[scale=0.20]{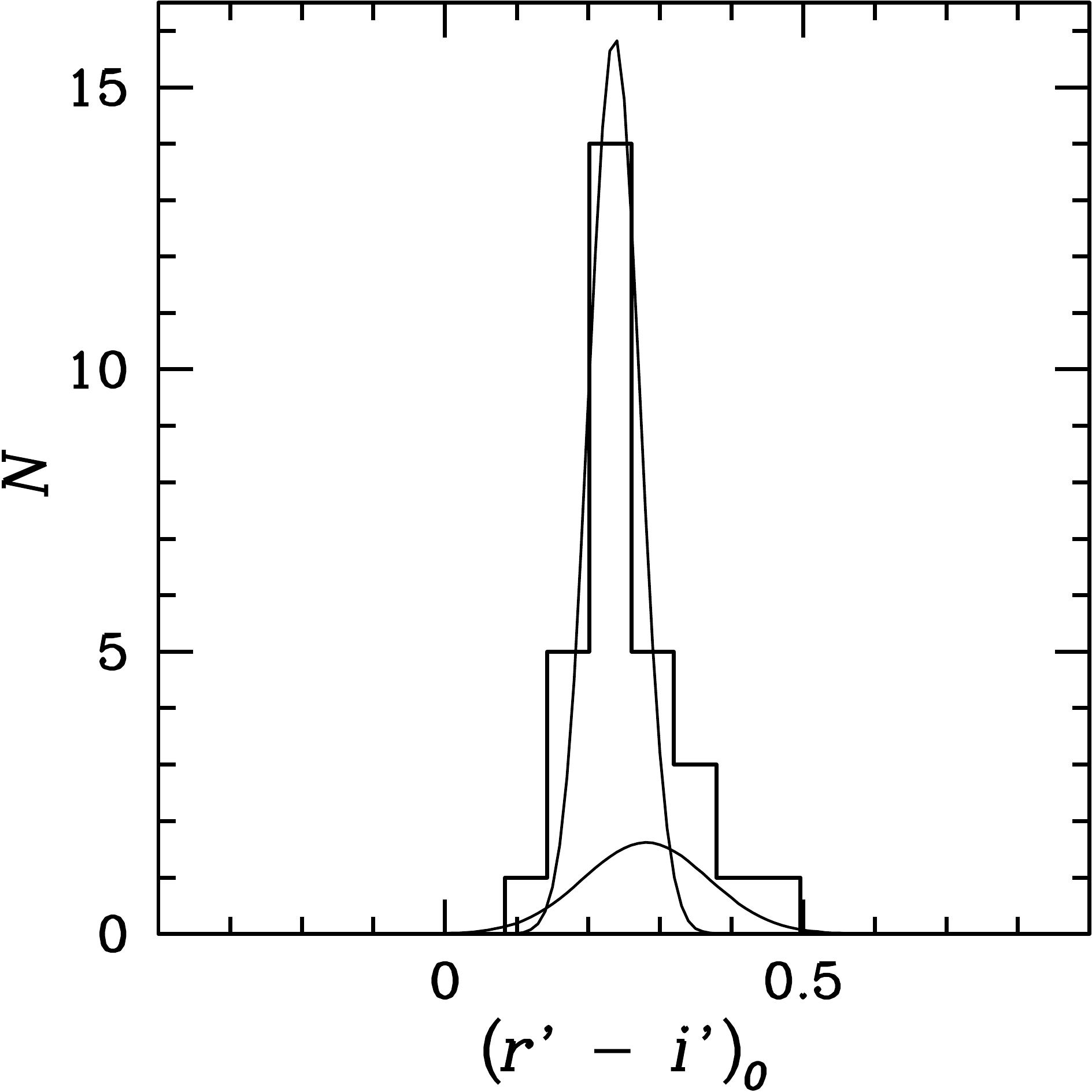}
\end{tabular}
\caption{
Color distributions of GC systems. {\it Top:} NGC\,3368 GCCs; {\it second row:}
NGC\,4395; {\it third row:} NGC\,4736; {\it bottom:} NGC\,4826.  From left to
right: ($u^\ast\ - g^\prime$), ($u^\ast\ - i^\prime$), ($g^\prime - r^\prime$),
($g^\prime - i^\prime$), ($g^\prime - K_s$), ($r^\prime - i^\prime$).
\label{fig:2gauss_colordist}}
\end{sidewaysfigure}

The distributions of our final samples of GCCs in the colors (\ust\ - \gp),
(\ust\ - \ip), (\gp\ - \rp), (\gp - \ip), (\rp - \ip), and (\gp - \ks), fit
with two heteroscedastic Gaussians, are shown in
Figure~\ref{fig:2gauss_colordist}. The means and dispersions of the Gaussian
fits are presented in Table~\ref{tab:2gauss}. The colors of the individual
candidates, shown in
tables~\ref{tab:n3368gcccolors},~\ref{tab:n4395gcccolors},~\ref{tab:n4736gcccolors}
and~\ref{tab:n4826gcccolors}, have been corrected for foreground extinction in
the Milky Way (MW), as already mentioned. 

\begin{sidewaystable}[ht]
\caption{Colors of Globular Cluster Candidates in NGC\,3368}
  \begin{tiny}
 \begin{center}
\hspace*{-0.8cm} \begin{minipage}{170mm}
\hskip-3cm\begin{tabular}{@{\extracolsep{4pt}}ccccccccccccc@{}}
\hline
\hline
\vspace*{-0.3cm}
&&&&&&&&&&&\\
 Name  &     RA J2000  &   DEC J2000   & $g_0^\prime$  & $\Delta g_0^\prime$  & $K_{s,0}$  & $\Delta K_{s,0}$ & ($u^\ast - i^\prime$)$_0$  & $\Delta$ ($u^\ast - i^\prime$)$_0$ &  ($g^\prime\ - i^\prime$)$_0$ & $\Delta$ ($g^\prime\ -i^\prime$)$_0$  & ($i^\prime\ - K_s$)$_0$ & $\Delta$ ($i^\prime\ - K_s$)$_0$  \\
     &    deg     &   deg      &    mag      &      mag           &    mag   &     mag          &          mag                  &             mag                      &     mag                    &       mag                           &           mag          &             mag                                 \\
\hline
J104539+113944 & 161.4148 & 11.6624 & 20.493 & 0.002 & 19.97 & 0.03 & 1.824 & 0.005 & 0.668 & 0.003 & -0.22 & 0.03 \\
J104540+120330 & 161.4185 & 12.0585 & 21.092 & 0.003 & 20.76 & 0.08 & 1.736 & 0.008 & 0.566 & 0.004 & -0.31 & 0.08 \\
J104604+115505 & 161.5179 & 11.9183 & 20.401 & 0.002 & 19.97 & 0.03 & 1.823 & 0.005 & 0.602 & 0.003 & -0.25 & 0.03 \\
J104604+113655 & 161.5183 & 11.6154 & 21.950 & 0.006 & 20.85 & 0.08 & 2.148 & 0.018 & 0.811 & 0.007 & 0.21 & 0.08 \\
J104609+120658 & 161.5413 & 12.1161 & 21.510 & 0.004 & 20.89 & 0.10 & 2.023 & 0.010 & 0.736 & 0.005 & -0.19 & 0.10 \\
J104626+113118 & 161.6091 & 11.5219 & 21.837 & 0.005 & 20.73 & 0.13 & 2.363 & 0.014 & 0.900 & 0.006 & 0.13 & 0.13 \\
J104633+120015 & 161.6379 & 12.0043 & 22.081 & 0.007 & 21.28 & 0.15 & 1.917 & 0.015 & 0.787 & 0.008 & -0.06 & 0.15 \\
J104635+115918 & 161.6478 & 11.9886 & 22.025 & 0.006 & 20.56 & 0.04 & 2.616 & 0.020 & 1.026 & 0.006 & 0.36 & 0.04 \\
J104635+114912 & 161.6493 & 11.8201 & 22.666 & 0.011 & 21.77 & 0.10 & 2.132 & 0.028 & 0.880 & 0.012 & -0.06 & 0.10 \\
J104645+113554 & 161.6890 & 11.5985 & 20.774 & 0.002 & 20.07 & 0.02 & 2.040 & 0.007 & 0.714 & 0.003 & -0.08 & 0.02 \\
J104648+115012 & 161.7003 & 11.8368 & 22.118 & 0.006 & 20.70 & 0.03 & 2.495 & 0.021 & 0.963 & 0.007 & 0.38 & 0.03 \\
J104651+114819 & 161.7134 & 11.8053 & 21.527 & 0.004 & 20.54 & 0.02 & 2.049 & 0.011 & 0.828 & 0.005 & 0.09 & 0.02 \\
J104709+114112 & 161.7882 & 11.6868 & 20.870 & 0.003 & 20.05 & 0.04 & 2.023 & 0.006 & 0.759 & 0.003 & -0.02 & 0.04 \\
J104750+113703 & 161.9601 & 11.6176 & 23.451 & 0.018 & 21.43 & 0.11 & 2.682 & 0.045 & 1.539 & 0.019 & 0.41 & 0.11 \\
J104755+113503 & 161.9811 & 11.5844 & 21.309 & 0.004 & 20.74 & 0.06 & 1.914 & 0.010 & 0.687 & 0.005 & -0.19 & 0.06 \\
\hline
\vspace*{-0.8cm}
\end{tabular}
\end{minipage}
\end{center}
\label{tab:n3368gcccolors}
\end{tiny}
\end{sidewaystable}

\begin{sidewaystable}[ht]
\caption{Colors of Globular Cluster Candidates in NGC\,4395}
  \begin{tiny}
 \begin{center}
\hspace*{-0.8cm} \begin{minipage}{170mm}
\hskip-3cm\begin{tabular}{@{\extracolsep{4pt}}ccccccccccccc@{}}
\hline
\hline
\vspace*{-0.3cm}
&&&&&&&&&&&\\
 Name  &     RA J2000  &   DEC J2000   & $g_0^\prime$  & $\Delta g_0^\prime$  & $K_{s,0}$  & $\Delta K_{s,0}$ & ($u^\ast - i^\prime$)$_0$  & $\Delta$ ($u^\ast - i^\prime$)$_0$ &  ($g^\prime\ - i^\prime$)$_0$ & $\Delta$ ($g^\prime\ -i^\prime$)$_0$  & ($i^\prime\ - K_s$)$_0$ & $\Delta$ ($i^\prime\ - K_s$)$_0$  \\
     &    deg     &   deg      &    mag      &      mag           &    mag   &     mag          &          mag                  &             mag                      &     mag                    &       mag                           &           mag          &             mag                                 \\
\hline
J122452+332844 & 186.2204 & 33.4792 & 20.001 & 0.004 & 19.59 & 0.03 & 1.802 & 0.007 & 0.656 & 0.006 & -0.30 & 0.03 \\
J122456+333230 & 186.2335 & 33.5418 & 19.580 & 0.003 & 19.17 & 0.02 & 1.739 & 0.005 & 0.654 & 0.004 & -0.30 & 0.02 \\
J122522+332549 & 186.3426 & 33.4303 & 20.697 & 0.006 & 20.08 & 0.04 & 1.940 & 0.012 & 0.761 & 0.009 & -0.20 & 0.04 \\
J122527+332015 & 186.3639 & 33.3376 & 20.041 & 0.004 & 19.23 & 0.03 & 1.877 & 0.008 & 0.813 & 0.005 & -0.05 & 0.03 \\
J122529+332536 & 186.3734 & 33.4268 & 20.734 & 0.007 & 20.17 & 0.04 & 1.952 & 0.013 & 0.730 & 0.009 & -0.22 & 0.04 \\
J122538+332654 & 186.4118 & 33.4486 & 21.223 & 0.009 & 20.71 & 0.16 & 1.696 & 0.016 & 0.657 & 0.013 & -0.20 & 0.16 \\
J122541+333109 & 186.4210 & 33.5194 & 21.903 & 0.016 & 20.28 & 0.05 & 2.709 & 0.035 & 1.058 & 0.020 & 0.52 & 0.06 \\
J122544+333016 & 186.4354 & 33.5046 & 23.298 & 0.053 & 20.12 & 0.05 & 3.232 & 0.069 & 2.148 & 0.055 & 0.98 & 0.05 \\
J122548+333802 & 186.4535 & 33.6341 & 21.255 & 0.010 & 20.70 & 0.08 & 1.939 & 0.016 & 0.738 & 0.013 & -0.24 & 0.08 \\
J122558+333154 & 186.4917 & 33.5318 & 23.479 & 0.062 & 20.88 & 0.10 & 2.977 & 0.079 & 1.877 & 0.066 & 0.67 & 0.10 \\
\hline
\vspace*{-0.8cm}
\end{tabular}
\end{minipage}
\end{center}
\label{tab:n4395gcccolors}
\end{tiny}
\end{sidewaystable}

\begin{sidewaystable}[ht]
\caption{Colors of Globular Cluster Candidates in NGC\,4736}
  \begin{tiny}
 \begin{center}
\hspace*{-0.8cm} \begin{minipage}{290mm}
\hskip-3cm\begin{tabular}{@{\extracolsep{4pt}}ccccccccccccccc@{}}
\hline
\hline
\vspace*{-0.3cm}
&&&&&&&&&&&&&&\\
 Name  &     RA J2000  &   DEC J2000   & $g_0^\prime$  & $\Delta g_0^\prime$  & $K_{s,0}$  & $\Delta K_{s,0}$ & ($u^\ast - i^\prime$)$_0$  & $\Delta$ ($u^\ast - i^\prime$)$_0$ & ($g^\prime\ - r^\prime$)$_0$ & $\Delta$ ($g^\prime\ - r^\prime$)$_0$ & ($g^\prime\ - i^\prime$)$_0$ & $\Delta$ ($g^\prime\ -i^\prime$)$_0$  & ($i^\prime\ - K_s$)$_0$ & $\Delta$ ($i^\prime\ - K_s$)$_0$  \\
     &    deg     &   deg      &    mag      &      mag           &    mag   &     mag          &          mag              &            mag                     &             mag            &             mag                      &     mag                    &       mag                           &           mag          &             mag                                 \\
\hline
J124947+411056 & 192.4496 & 41.1823 & 22.850 & 0.048 & 20.28 & 0.05 & 3.226 & 0.063 & 1.166 & 0.053 & 1.545 & 0.060 &  0.97 & 0.06 \\
J124951+411538 & 192.4633 & 41.2606 & 19.098 & 0.008 & 18.57 & 0.01 & 1.717 & 0.010 & 0.425 & 0.010 & 0.667 & 0.012 & -0.20 & 0.01 \\
J125001+411350 & 192.5053 & 41.2307 & 20.147 & 0.014 & 19.71 & 0.02 & 1.699 & 0.017 & 0.419 & 0.018 & 0.661 & 0.022 & -0.28 & 0.02 \\
J125029+410951 & 192.6246 & 41.1644 & 21.591 & 0.027 & 20.95 & 0.05 & 1.743 & 0.032 & 0.492 & 0.033 & 0.737 & 0.041 & -0.15 & 0.06 \\
J125031+411039 & 192.6314 & 41.1777 & 22.573 & 0.053 & 21.28 & 0.07 & 2.711 & 0.076 & 0.612 & 0.063 & 0.850 & 0.076 &  0.39 & 0.09 \\
J125031+410949 & 192.6329 & 41.1638 & 23.248 & 0.075 & 21.60 & 0.10 & 2.607 & 0.106 & 0.638 & 0.089 & 1.060 & 0.103 &  0.54 & 0.12 \\
J125033+405708 & 192.6415 & 40.9525 & 19.946 & 0.012 & 19.49 & 0.01 & 1.821 & 0.014 & 0.399 & 0.015 & 0.647 & 0.019 & -0.25 & 0.02 \\
J125039+410913 & 192.6651 & 41.1538 & 22.044 & 0.039 & 21.40 & 0.09 & 1.884 & 0.048 & 0.482 & 0.048 & 0.690 & 0.060 & -0.10 & 0.10 \\
J125042+410345 & 192.6786 & 41.0625 & 23.082 & 0.061 & 19.81 & 0.02 & 3.000 & 0.040 & 1.320 & 0.067 & 2.358 & 0.068 &  0.86 & 0.04 \\
J125043+410947 & 192.6807 & 41.1632 & 20.309 & 0.016 & 19.89 & 0.02 & 1.746 & 0.019 & 0.417 & 0.020 & 0.647 & 0.024 & -0.28 & 0.03 \\
J125045+410950 & 192.6902 & 41.1639 & 20.772 & 0.020 & 19.92 & 0.02 & 2.065 & 0.022 & 0.502 & 0.024 & 0.784 & 0.029 &  0.01 & 0.03 \\
J125046+410210 & 192.6920 & 41.0361 & 19.819 & 0.014 & 18.47 & 0.01 & 2.688 & 0.014 & 0.587 & 0.016 & 0.923 & 0.019 &  0.37 & 0.01 \\
J125051+410314 & 192.7138 & 41.0540 & 21.023 & 0.022 & 20.32 & 0.03 & 2.033 & 0.025 & 0.466 & 0.027 & 0.747 & 0.033 & -0.10 & 0.04 \\
J125051+411341 & 192.7148 & 41.2282 & 19.920 & 0.013 & 19.39 & 0.01 & 1.708 & 0.015 & 0.399 & 0.016 & 0.634 & 0.019 & -0.16 & 0.02 \\
J125056+410327 & 192.7340 & 41.0576 & 22.302 & 0.040 & 20.78 & 0.05 & 2.794 & 0.057 & 0.582 & 0.048 & 0.906 & 0.058 &  0.56 & 0.06 \\
J125059+410949 & 192.7500 & 41.1637 & 21.343 & 0.026 & 20.56 & 0.04 & 2.130 & 0.030 & 0.516 & 0.031 & 0.745 & 0.038 & -0.02 & 0.05 \\
J125101+410046 & 192.7577 & 41.0128 & 22.393 & 0.042 & 21.26 & 0.08 & 2.136 & 0.050 & 0.560 & 0.050 & 0.844 & 0.060 &  0.23 & 0.09 \\
J125104+410416 & 192.7698 & 41.0712 & 20.960 & 0.023 & 19.85 & 0.02 & 2.321 & 0.025 & 0.539 & 0.028 & 0.866 & 0.033 &  0.19 & 0.03 \\
J125108+410404 & 192.7848 & 41.0679 & 22.459 & 0.043 & 21.60 & 0.11 & 1.984 & 0.053 & 0.523 & 0.052 & 0.755 & 0.063 &  0.05 & 0.12 \\
J125110+410856 & 192.7957 & 41.1491 & 20.314 & 0.017 & 19.81 & 0.02 & 1.753 & 0.020 & 0.403 & 0.021 & 0.646 & 0.026 & -0.20 & 0.03 \\
J125120+410710 & 192.8362 & 41.1197 & 21.730 & 0.029 & 20.51 & 0.04 & 2.473 & 0.037 & 0.546 & 0.035 & 0.790 & 0.043 &  0.38 & 0.05 \\
J125130+410551 & 192.8767 & 41.0977 & 21.431 & 0.025 & 20.92 & 0.06 & 1.714 & 0.030 & 0.434 & 0.031 & 0.657 & 0.039 & -0.20 & 0.07 \\
J125131+410849 & 192.8828 & 41.1472 & 19.149 & 0.009 & 18.68 & 0.01 & 1.743 & 0.010 & 0.399 & 0.011 & 0.633 & 0.013 & -0.22 & 0.01 \\
J125143+410606 & 192.9315 & 41.1018 & 20.757 & 0.018 & 20.10 & 0.04 & 2.027 & 0.021 & 0.469 & 0.022 & 0.678 & 0.027 & -0.07 & 0.04 \\
J125147+410439 & 192.9478 & 41.0776 & 21.403 & 0.026 & 20.16 & 0.04 & 2.570 & 0.030 & 0.548 & 0.031 & 0.871 & 0.037 &  0.32 & 0.05 \\
\hline
\vspace*{-0.8cm}
\end{tabular}
\end{minipage}
\end{center}
\label{tab:n4736gcccolors}
\end{tiny}
\end{sidewaystable}

\begin{sidewaystable}[ht]
\caption{Colors of Globular Cluster Candidates in NGC\,4826}
  \begin{tiny}
 \begin{center}
\hspace*{-0.8cm} \begin{minipage}{290mm}
\hskip-3cm\begin{tabular}{@{\extracolsep{4pt}}ccccccccccccccc@{}}
\hline
\hline
\vspace*{-0.3cm}
&&&&&&&&&&&&&&\\
 Name  &     RA J2000  &   DEC J2000   & $g_0^\prime$  & $\Delta g_0^\prime$  & $K_{s,0}$  & $\Delta K_{s,0}$ & ($u^\ast - i^\prime$)$_0$  & $\Delta$ ($u^\ast - i^\prime$)$_0$ & ($g^\prime\ - r^\prime$)$_0$ & $\Delta$ ($g^\prime\ - r^\prime$)$_0$ & ($g^\prime\ - i^\prime$)$_0$ & $\Delta$ ($g^\prime\ -i^\prime$)$_0$  & ($i^\prime\ - K_s$)$_0$ & $\Delta$ ($i^\prime\ - K_s$)$_0$  \\
     &    deg     &   deg      &    mag      &      mag           &    mag   &     mag          &          mag              &            mag                     &             mag            &             mag                      &     mag                    &       mag                           &           mag          &             mag                                 \\
\hline
J125549+213423 & 193.9550 & 21.5733 & 19.982 & 0.009 & 19.36 & 0.03 & 2.011 & 0.007 & 0.424 & 0.011 & 0.665 & 0.011 & -0.17 & 0.03 \\
J125601+214923 & 194.0083 & 21.8233 & 21.238 & 0.018 & 20.26 & 0.03 & 1.854 & 0.013 & 0.811 & 0.021 & 1.027 & 0.021 & -0.17 & 0.03 \\
J125612+214638 & 194.0522 & 21.7773 & 22.550 & 0.033 & 21.38 & 0.09 & 2.465 & 0.044 & 0.484 & 0.040 & 0.810 & 0.040 &  0.23 & 0.10 \\
J125613+214412 & 194.0544 & 21.7368 & 22.804 & 0.034 & 19.36 & 0.01 & 3.273 & 0.027 & 1.946 & 0.036 & 2.331 & 0.036 &  0.98 & 0.02 \\
J125614+214043 & 194.0592 & 21.6789 & 21.255 & 0.016 & 20.17 & 0.02 & 1.983 & 0.013 & 0.730 & 0.019 & 0.956 & 0.019 &  0.01 & 0.03 \\
J125615+214645 & 194.0648 & 21.7794 & 20.949 & 0.015 & 20.30 & 0.03 & 1.958 & 0.013 & 0.349 & 0.019 & 0.625 & 0.019 & -0.10 & 0.03 \\
J125622+215134 & 194.0944 & 21.8597 & 21.335 & 0.017 & 20.39 & 0.05 & 2.067 & 0.014 & 0.716 & 0.020 & 0.952 & 0.020 & -0.14 & 0.05 \\
J125623+214054 & 194.0982 & 21.6817 & 21.091 & 0.015 & 19.85 & 0.02 & 1.841 & 0.011 & 0.972 & 0.017 & 1.096 & 0.018 &  0.02 & 0.02 \\
J125625+214427 & 194.1082 & 21.7408 & 20.446 & 0.013 & 19.72 & 0.02 & 1.894 & 0.010 & 0.355 & 0.016 & 0.638 & 0.015 & -0.04 & 0.02 \\
J125626+213440 & 194.1095 & 21.5779 & 20.705 & 0.013 & 20.34 & 0.03 & 1.706 & 0.011 & 0.266 & 0.017 & 0.516 & 0.017 & -0.27 & 0.03 \\
J125626+214234 & 194.1124 & 21.7096 & 20.816 & 0.013 & 19.95 & 0.02 & 1.901 & 0.011 & 0.530 & 0.016 & 0.851 & 0.016 & -0.11 & 0.02 \\
J125628+214937 & 194.1207 & 21.8272 & 20.872 & 0.014 & 19.38 & 0.01 & 1.956 & 0.011 & 0.954 & 0.016 & 1.198 & 0.017 &  0.17 & 0.02 \\
J125632+214023 & 194.1364 & 21.6733 & 21.338 & 0.018 & 20.55 & 0.04 & 1.850 & 0.016 & 0.335 & 0.023 & 0.624 & 0.023 &  0.04 & 0.04 \\
J125633+213919 & 194.1416 & 21.6554 & 20.733 & 0.013 & 20.01 & 0.02 & 1.710 & 0.010 & 0.538 & 0.015 & 0.743 & 0.016 & -0.15 & 0.03 \\
J125638+214512 & 194.1583 & 21.7535 & 23.490 & 0.049 & 21.22 & 0.08 & 2.774 & 0.053 & 1.230 & 0.054 & 1.678 & 0.054 &  0.47 & 0.08 \\
J125641+213802 & 194.1709 & 21.6341 & 21.815 & 0.021 & 20.74 & 0.05 & 1.862 & 0.016 & 0.762 & 0.025 & 1.020 & 0.025 & -0.07 & 0.05 \\
J125641+213641 & 194.1734 & 21.6115 & 20.810 & 0.013 & 19.93 & 0.02 & 2.026 & 0.011 & 0.494 & 0.016 & 0.760 & 0.016 & -0.00 & 0.02 \\
J125642+213204 & 194.1761 & 21.5347 & 21.075 & 0.016 & 20.35 & 0.03 & 1.940 & 0.014 & 0.352 & 0.020 & 0.612 & 0.020 & -0.01 & 0.03 \\
J125644+213848 & 194.1873 & 21.6468 & 20.897 & 0.014 & 19.90 & 0.02 & 1.755 & 0.010 & 0.808 & 0.016 & 0.979 & 0.016 & -0.11 & 0.02 \\
J125646+213754 & 194.1954 & 21.6317 & 21.352 & 0.018 & 20.72 & 0.04 & 1.768 & 0.016 & 0.314 & 0.023 & 0.562 & 0.023 & -0.05 & 0.05 \\
J125649+214425 & 194.2065 & 21.7404 & 21.212 & 0.016 & 20.01 & 0.02 & 2.093 & 0.012 & 0.787 & 0.019 & 0.993 & 0.019 &  0.09 & 0.02 \\
J125654+215233 & 194.2275 & 21.8760 & 22.006 & 0.026 & 19.70 & 0.04 & 3.167 & 0.028 & 1.013 & 0.029 & 1.297 & 0.029 &  0.88 & 0.04 \\
J125655+213319 & 194.2323 & 21.5554 & 21.993 & 0.023 & 21.18 & 0.06 & 1.729 & 0.020 & 0.707 & 0.028 & 0.960 & 0.028 & -0.27 & 0.06 \\
J125655+214633 & 194.2332 & 21.7761 & 20.895 & 0.015 & 20.32 & 0.03 & 1.830 & 0.013 & 0.391 & 0.018 & 0.587 & 0.018 & -0.14 & 0.03 \\
J125714+213914 & 194.3117 & 21.6540 & 19.931 & 0.009 & 19.66 & 0.02 & 1.805 & 0.008 & 0.271 & 0.012 & 0.466 & 0.012 & -0.32 & 0.02 \\
J125717+213702 & 194.3225 & 21.6173 & 21.816 & 0.022 & 20.73 & 0.04 & 1.887 & 0.016 & 0.789 & 0.025 & 0.990 & 0.026 & -0.03 & 0.04 \\
J125721+213816 & 194.3410 & 21.6378 & 22.215 & 0.028 & 19.73 & 0.02 & 3.257 & 0.032 & 0.998 & 0.031 & 1.343 & 0.031 &  1.01 & 0.02 \\
J125722+213659 & 194.3418 & 21.6166 & 20.674 & 0.013 & 19.80 & 0.02 & 1.784 & 0.009 & 0.705 & 0.015 & 0.961 & 0.015 & -0.21 & 0.02 \\
J125723+214212 & 194.3472 & 21.7036 & 20.805 & 0.015 & 20.36 & 0.04 & 1.689 & 0.012 & 0.301 & 0.019 & 0.528 & 0.018 & -0.21 & 0.04 \\
J125723+213711 & 194.3480 & 21.6198 & 20.644 & 0.014 & 20.18 & 0.03 & 1.754 & 0.011 & 0.324 & 0.018 & 0.493 & 0.017 & -0.16 & 0.03 \\
\hline
\vspace*{-0.8cm}
\end{tabular}
\end{minipage}
\end{center}
\label{tab:n4826gcccolors}
\end{tiny}
\end{sidewaystable}

\begin{table}
 \caption{Parameters of Double Gaussian Fits to Color Distributions}
 \begin{center}
 \begin{minipage}{290mm}
\hspace*{1.7cm}\begin{tabular}{@{}lccccc@{}}
\hline
\hline
Color           & Parameter   &  NGC\,3368  & NGC\,4395   & NGC\,4736  &     NGC\,4826  \\
\hline

\multirow{6}{*}{$(u^\ast - g')_0$} & $\mu_1$  &  1.23    &  1.12  &  1.07 &  1.08 \\
                                   & $\sigma_1$& 0.07    &  0.06  &  0.03 &  0.21 \\
                                   & $N_1$    &  12      &  9     &  7    &  27   \\
                                   & $\mu_2$  &  1.53    &  1.651 &  1.40 &  1.83 \\
                                   & $\sigma_2$& 0.05    &  0.001 &  0.31 &  0.10 \\
                                   & $N_2$    &  3       &  1     &  18   &   3   \\
\hline
\multirow{6}{*}{$(u^\ast - i')_0$} & $\mu_1$  &  1.97    &  1.85  & 1.86  &  1.87 \\
                                   & $\sigma_1$& 0.13    &  0.10  & 0.16  &  0.11 \\
                                   & $N_1$    &  11      &  7     & 16    &  25   \\
                                   & $\mu_2$  &  2.54    &  2.97  & 2.66  &  2.98 \\
                                   & $\sigma_2$& 0.12    &  0.21  & 0.30  &  0.32 \\
                                   & $N_2$    &  4       &  3     &  9    &   5   \\
\hline
\multirow{6}{*} {$(g' - r')_0$}    &$\mu_1$   &          &        & 0.49  &  0.61 \\
                                   & $\sigma_1$&         &        & 0.07  &  0.27 \\
                                   & $N_1$    &          &        & 23    &  29   \\
                                   & $\mu_2$  &          &        & 1.24  &  1.946\\
                                   & $\sigma_2$&         &        & 0.08  &  0.001\\
                                   & $N_2$    &          &        & 2     &  1    \\
\hline
\multirow{6}{*} {$(g' - i')_0$}    & $\mu_1$  &  0.78    &  0.72  & 0.75  &  0.82 \\
                                   & $\sigma_1$& 0.13    &  0.06  & 0.10  &  0.24 \\
                                   & $N_1$    &  14      &  7     &  23   &  28   \\
                                   & $\mu_2$  &  1.539   &  1.66  & 1.72  &  1.81 \\
                                   & $\sigma_2$& 0.001   &  0.49  & 0.55  &  0.46 \\
                                   & $N_2$    &  1       &  3     &  2    &  2    \\
\hline
\multirow{6}{*}{$(g' - K_s)_0$}    & $\mu_1$  &  0.63    &  0.50  & 0.79  &  0.71 \\
                                   & $\sigma_1$& 0.24    &  0.13  & 0.38  &  0.28 \\
                                   & $N_1$    &  12      &  7     &  23   &  26   \\
                                   & $\mu_2$  &  1.28    &  2.39  & 2.87  &  2.45 \\
                                   & $\sigma_2$& 0.43    &  0.67  & 0.35  &  0.53 \\
                                   & $N_2$    &  3       &  3     & 2     &  4    \\
\hline
\multirow{6}{*}{$(r' - i')_0$}     & $\mu_1$  &          &        & 0.27  &  0.24 \\
                                   & $\sigma_1$&         &        &  0.05 &  0.04 \\
                                   & $N_1$    &          &        &  24   &  24   \\
                                   & $\mu_2$  &          &        & 1.038 &  0.28 \\
                                   & $\sigma_2$&         &        & 0.001 &  0.09 \\
                                   & $N_2$    &          &        & 1     &    6  \\
\hline
\vspace*{-0.5cm}
\end{tabular}
\end{minipage}
\end{center}
\label{tab:2gauss}
\end{table}


\citet{gonz17} found that the colors of GCs and GCCs in the Milky Way, M31, and
NGC\,4258 were remarkably consistent, once internal extinction was taken into
account. The same is true for NGC\,3368, NGC\,4395, NGC\,4736 and NGC\,4826.
Their GCC colors have not been corrected for internal extinction, and the bluer
Gaussians fit to their color distributions are bracketed by the values of
single Gaussian fits to the GC system of M\,31, with and without such
correction. Both these and the colors of the MW GC system are shown in
Table~\ref{tab:M31MW}. 

\begin{table}[ht]
 \caption{Single Gaussian Fits to the Colors of the GC Systems of the MW and M\,31}
 \begin{center}
 \begin{minipage}{280mm}
  \begin{tiny}
\hspace*{-5.6cm}  \begin{tabular}{@{}lcccccccccccccccccccccccc@{}}
\hline
\hline
System & \multicolumn{2}{c}{$(u^\ast - g')$} & \multicolumn{2}{c}{$(u^\ast - g')_0$} & \multicolumn{2}{c}{$(u^\ast - i')$} & \multicolumn{2}{c}{$(u^\ast - i')_0$} &
                   \multicolumn{2}{c}{$(g' - r')$} & \multicolumn{2}{c}{$(g' - r')_0$} & \multicolumn{2}{c}{$(g' - i')$} & \multicolumn{2}{c}{$(g' - i')_0$} &
                   \multicolumn{2}{c}{$(g' - K_s)$} & \multicolumn{2}{c}{$(g' - K_s)_0$} & \multicolumn{2}{c}{$(r' - i')$} & \multicolumn{2}{c}{$(r' - i')_0$} \\
 & $\mu$ & $\sigma$ & $\mu$ & $\sigma$ & $\mu$ & $\sigma$ & $\mu$ & $\sigma$ &
           $\mu$ & $\sigma$ & $\mu$ & $\sigma$ & $\mu$ & $\sigma$ & $\mu$ & $\sigma$ &
           $\mu$ & $\sigma$ & $\mu$ & $\sigma$ & $\mu$ & $\sigma$ & $\mu$ & $\sigma$  \\
\hline
M\,31                   & 1.30 & 0.26 & 1.05 &      & 2.29 & 0.53 & 1.72 &      & 0.64 & 0.17 & 0.45 &      & 1.00 & 0.28 & 0.68 &      & 1.24 & 0.61 & 0.56 &        & 0.36 & 0.12 & 0.23 &      \\
MW                      &      &      & 1.01 & 0.15 &      &      & 1.67 & 0.26 &      &      & 0.45 & 0.09 &      &      & 0.66 & 0.12 &      &      &      &        &      &      & 0.21 & 0.03 \\
\hline
\vspace*{-0.5cm}
\end{tabular}
\end{tiny}
\end{minipage}
\end{center}
We remind the reader that colors using the  new \ust\ filter are $\sim$ 0.2 mag
redder. The colors of the M\,31 GCs were transformed by \citet{gonz17} to the
previous CFHT system.
\label{tab:M31MW}
\end{table}

\subsection{Luminosity functions}\label{subsec:gclf}

\begin{figure}[ht!]
\plottwo{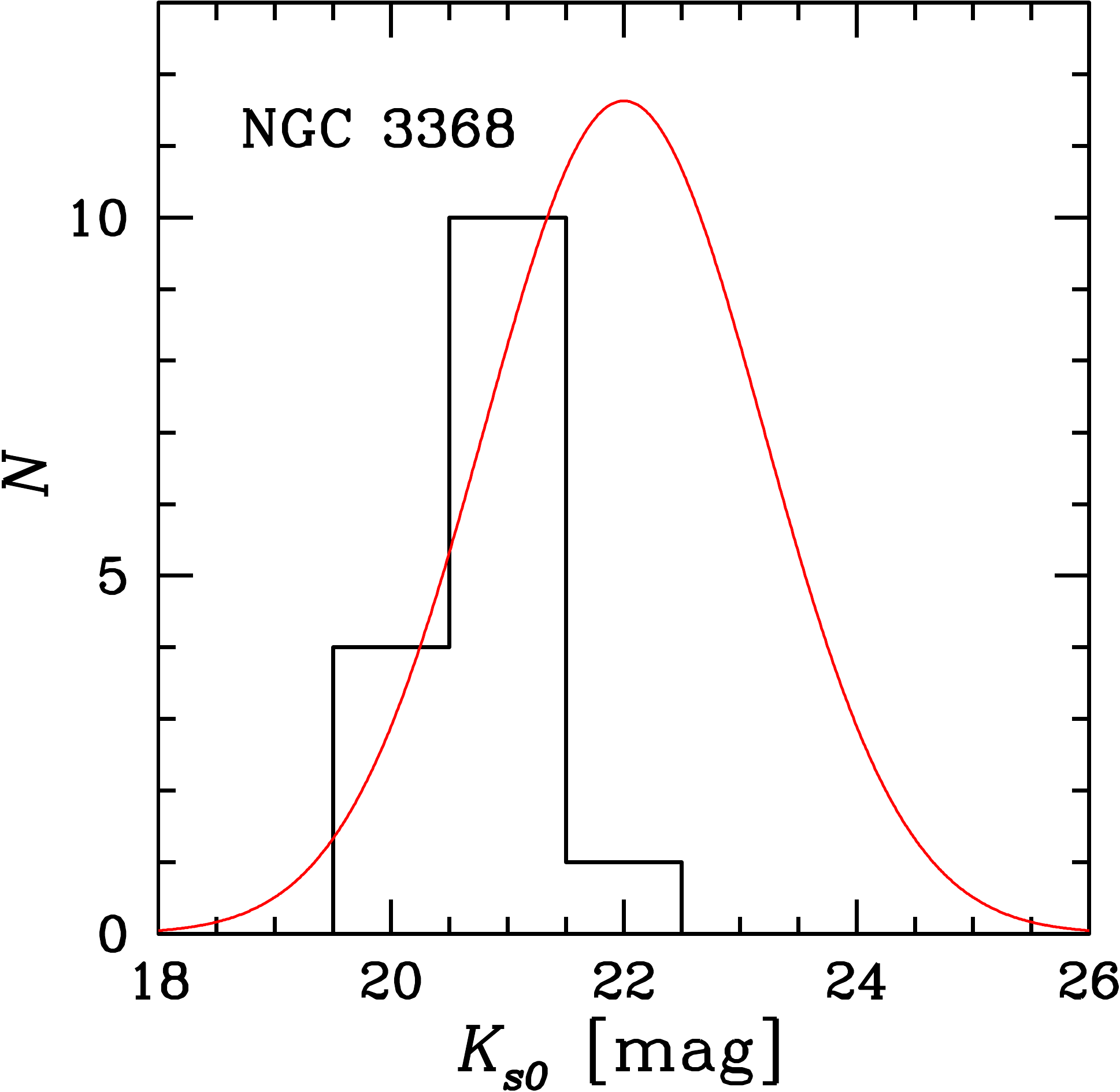}{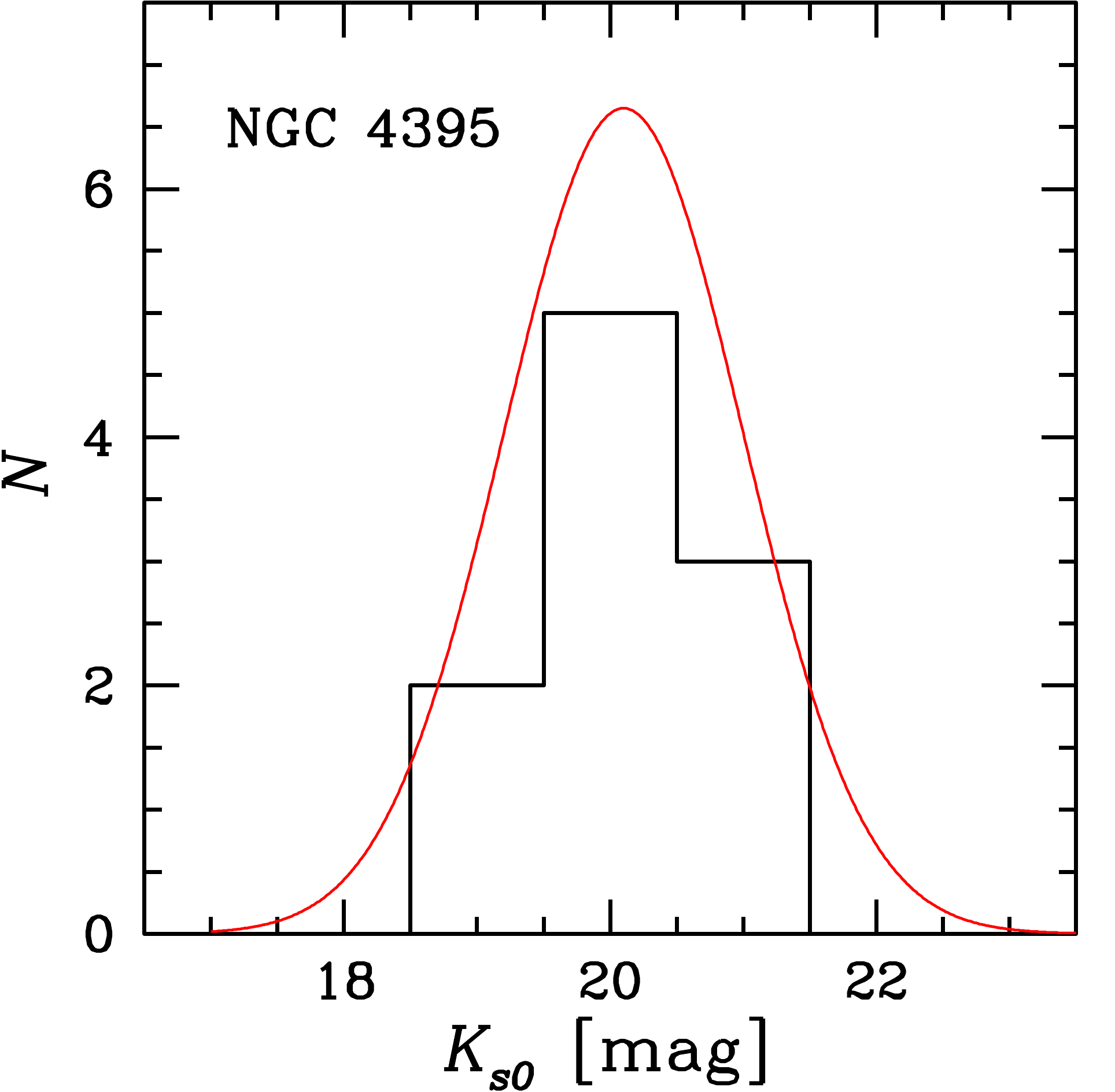}
\plottwo{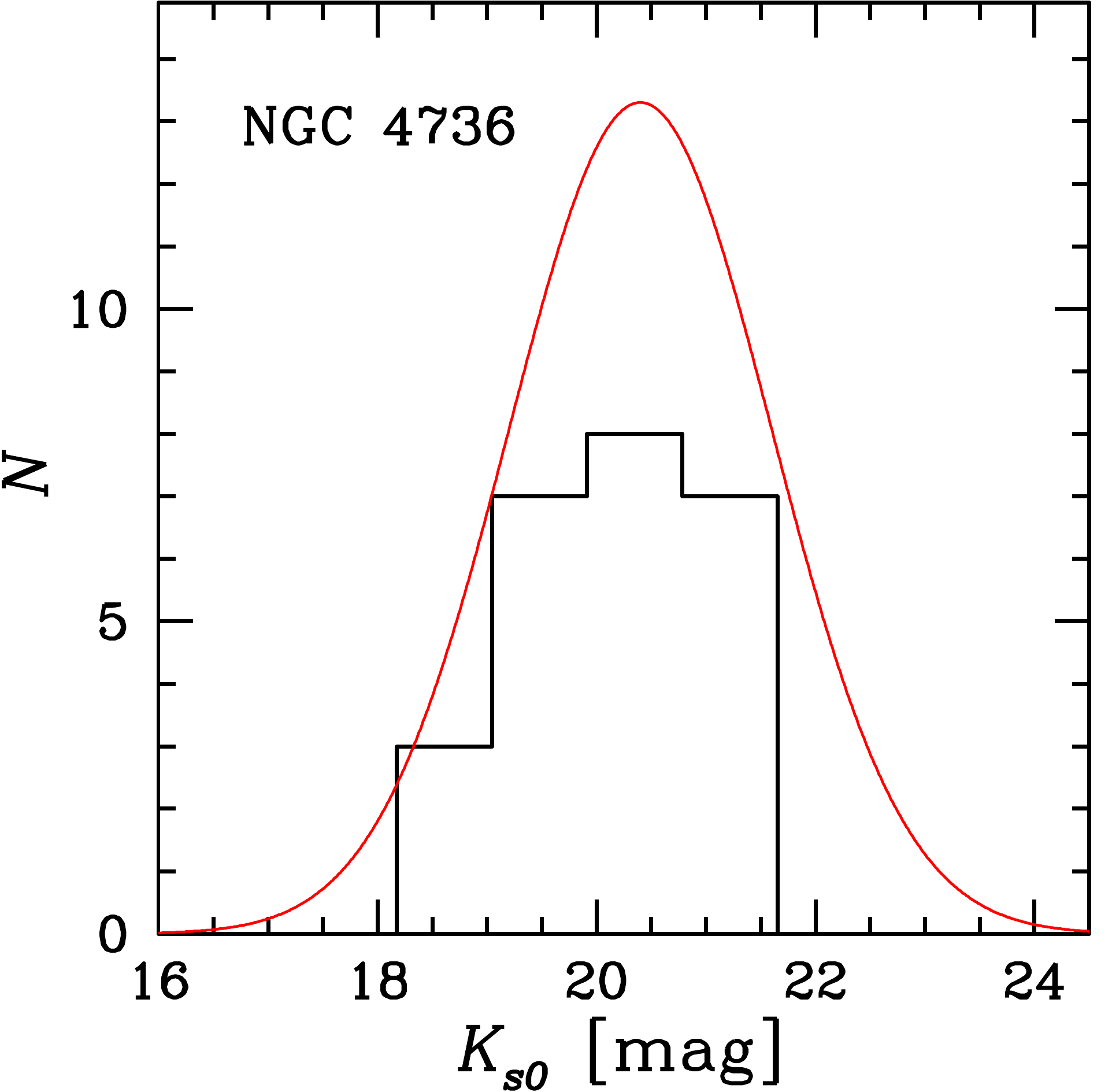}{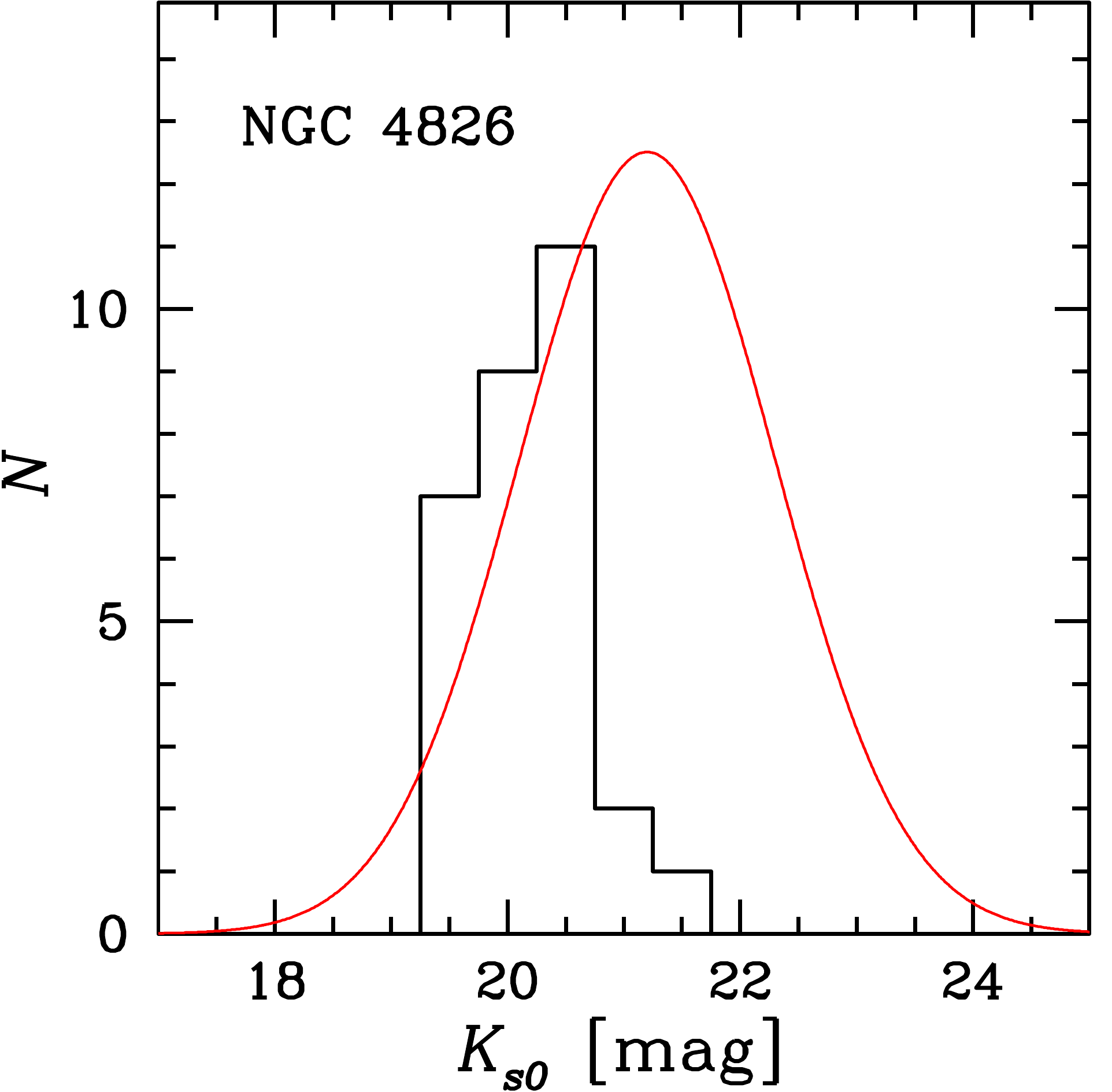}
\caption{\ks-band luminosity functions of the GCCs. {\it Top left:} NGC\,3368;
{\it top right:} NGC\,4395; {\it bottom left:} NGC\,4736; {\it bottom right:}
NGC\,4826.  {\it Histograms:} numbers of detected objects; {\it solid red
lines:} Expected Gaussian GCLFs (see Table~\ref{tab:TOdat} and
Section~\ref{sec:totalnum}. 
\label{fig:gclf}}
\end{figure}

The \ks-band luminosity functions of the final samples of GCCs of the four
galaxies are shown in Figure~\ref{fig:gclf}.  Given the exclusion of the
regions close to the centers of the galaxies, i.e., the areas most affected by
crowding and confusion, the corrections for incompleteness would be barely
significant.  The solid red lines are Gaussians with the expected
characteristics of the luminosity functions: mean = 22.0 mag, $\sigma = 1.2$
mag for NGC\,3368; mean = 20.1 mag, $\sigma = 0.9$ mag for NGC\,4395; mean =
20.4 mag, $\sigma = 1.2$ mag for NGC\,4736; mean = 21.2 mag, $\sigma = 1.1$ mag
for NGC\,4826.  The curves have been scaled such that their integrals are equal
to the expected GCCs for each galaxy, respectively, 35, 15, 46, and 69, for
NGC\,3368, NGC\,4395, NGC\,4736, and NGC\,4826 (see Section~\ref{sec:totalnum}
and Table~\ref{tab:fitgal}.)

We also take the opportunity here to stress that we do not use the
extrapolation over the luminosity function to derive the total number of
globular clusters in the system. We perform the completeness and GCLF
corrections implicitly, as we explain in the next section.

\begin{figure}[ht!]
\hspace*{-1cm}\plotone{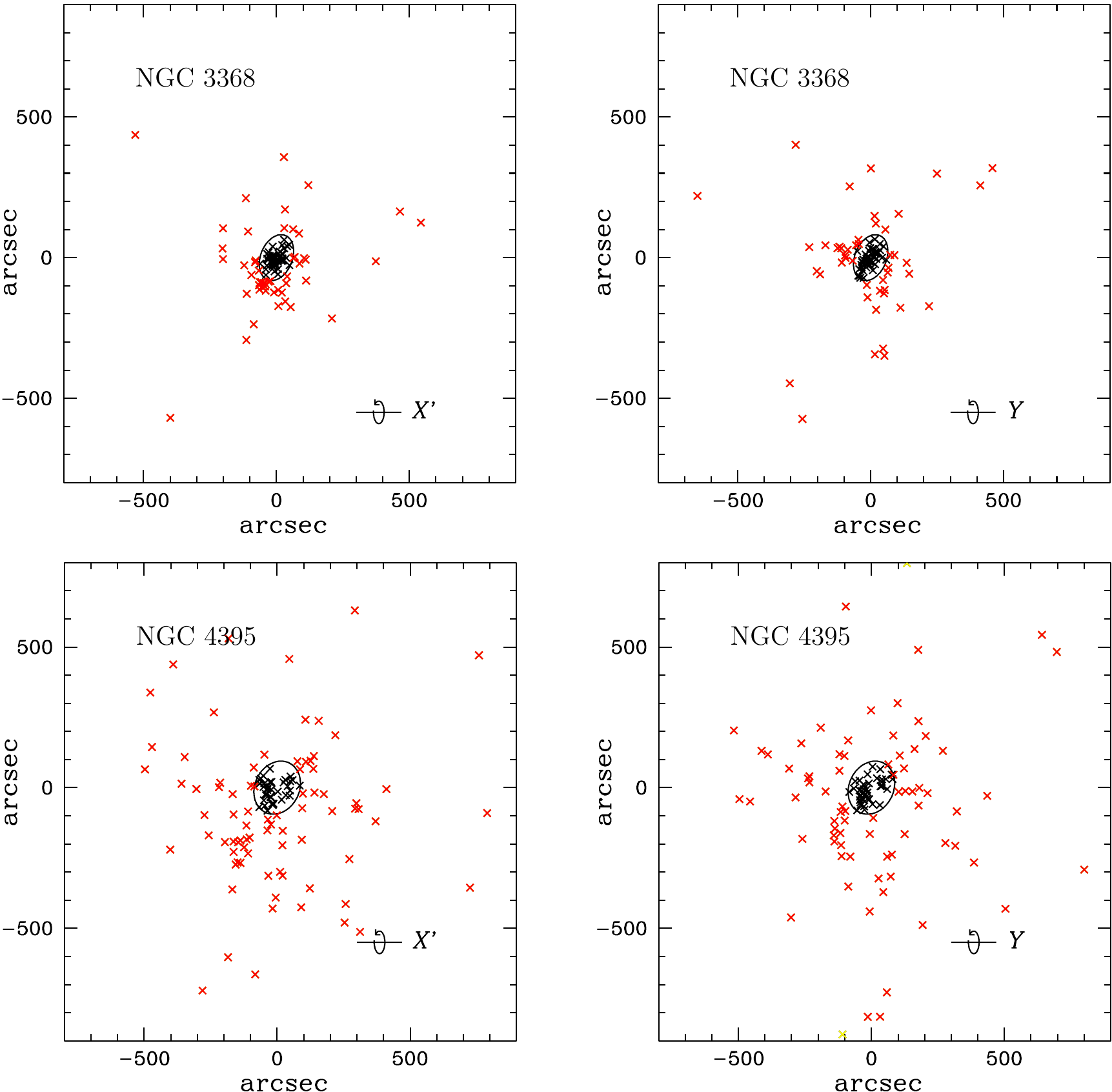}
\caption{
MW GC system, at the distance and orientation of studied galaxies, projected
coordinates $+X_{\rm proj},+Y_{\rm proj}$. {\it Top:} NGC\,3368; {\it bottom:}
NGC\,4395. {\it Left:} 3-D rotation about Galactic $X^\prime$ axis before
projection on the plane of the sky.  {\it Right:} 3-D rotation about Galactic
$Y$ axis. {\it Solid black line:} mask; {\it red crosses:} sources visible in
the WIRCam FOV; {\it yellow crosses:} recovered in the simulation, but outside
of the WIRCam FOV; {\it black crosses:} masked-out GCs.
\label{fig:mask}}
\end{figure}

\begin{figure}[ht!]
\hspace*{-1cm}\plotone{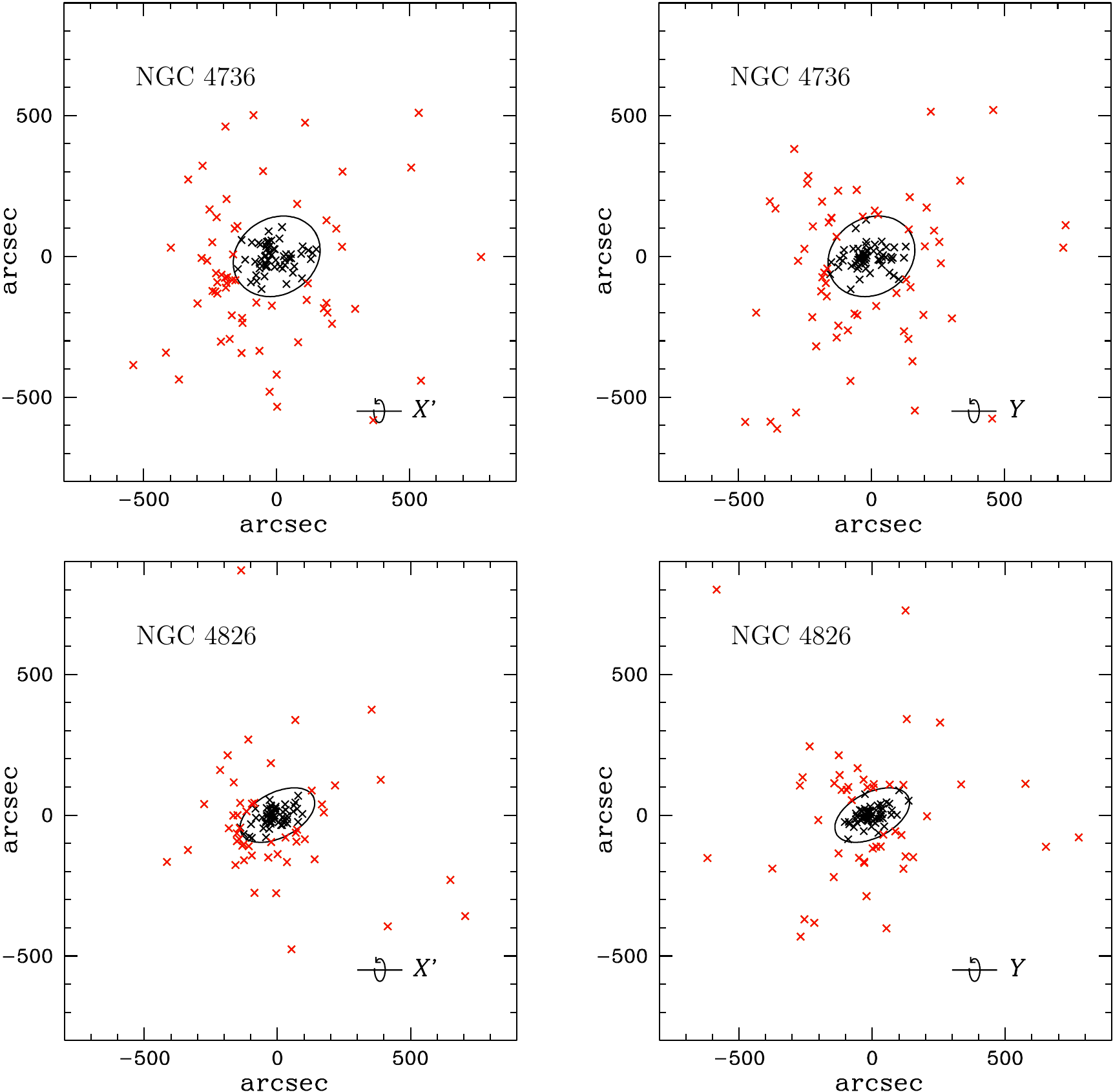}
\caption{
MW GC system, at the distance and orientation of studied galaxies, projected
coordinates $+X_{\rm proj},+Y_{\rm proj}$. {\it Top:} NGC\,4736; {\it bottom:}
NGC\,4826.  Symbols as in Figure~\ref{fig:mask}.
\label{fig:mask2}}
\end{figure}

\section{Total number of globular clusters} \label{sec:totalnum}

\subsection{Correction for incomplete spatial coverage}\label{subsec:spatcov}

Given the small number of clusters detected, and the lack of constraints in the
centermost regions of the galaxies due to confusion, here we derive the total
number of GCs for each galaxy through a comparison with the GC system of the
Milky Way. The procedure was introduced by \citet{kiss99}, who applied it to
edge-on spirals; its results are consistent with those from a fit to the radial
number density, when both methods can be applied to the same object
\citep{kiss99,goud03}.  \citet{gonz17} adapted it for use on any galaxy,
regardless of orientation. With this method, the correction for incomplete
spatial coverage implicitly takes care of the completeness correction and the
extrapolation over the LF.

In brief, the method consists in estimating the number of GCs that one would
detect if the MW were at the distance and orientation of the galaxy of
interest, and were observed with the same instrument and to the same depth. By
comparison, the total number of GCs would be:

\begin{equation}
N_{\rm GC} = N_{\rm GC} {\rm (MW)} \frac{N_{\rm obs}}{N_{\rm FOV}}.
\label{eq:kissler}
\end{equation}

\noindent
$N_{\rm GC}$(MW) is the total number of clusters in the MW, including objects
invisible to us behind the bulge; $N_{\rm obs}$ is the number of clusters
observed in the target galaxy, and $N_{\rm FOV}$ is the number of objects
recovered in the artificial observation of the MW.  We take $N_{\rm GC}$ (Milky
Way) = 160$\pm$10 \citep{harr14}.

In order to estimate $N_{\rm FOV}$ we use the catalog by \citet{harr96}, which
provides  $X,Y,Z$ coordinates for 144 MW GCs. The origin of this coordinate
system is at the position of the Sun; the $X$ axis points towards the Galactic
center, the $Y$ axis in the direction of Galactic rotation, and $Z$ is
perpendicular to the Galactic plane. We define a new coordinate system,
$X^{\prime}YZ$, with origin in the Galactic center, assuming a Galactocentric
distance for the Sun R${_0}$ = 8.34 kpc \citep{reid14}.  For galaxies that are
not edge-on, we first rotate the MW GC system in 3-D by the inclination angle
$i$ of the target galaxy (i.e., 46$\fdg$2 for NGC\,3368, 33$\fdg$7 for
NGC\,4395, 35$\fdg$6 for NGC\,4736 and 57$\fdg$5 for NGC\,4826) with respect to
either the $X^\prime$ or $Y$ axis, then project it on the plane of the sky; the
rotated pairs ($X^\prime_{{\rm rot},X^\prime}, Y_{{\rm rot},X^\prime}$),
($X^\prime_{{\rm rot},Y}, Y_{{\rm rot},Y}$), where the subscript indicates the
Galactic axis of 3-D rotation, are directly the projected coordinates. The
easiest way to place the projected system on the WIRCam FOV is to then apply a
rotation around the line of sight equal to the PA of the target galaxy
(5$\degr$ for NGC\,3368, 147$\degr$ for NGC\,4395, and 115$\degr$ for both
NGC\,4736 and NGC\,4826, always North through East). 

To ensure a fair comparison with the MW, we then mask out elliptical regions in
the centers of the galaxies.  For NGC\,4395, NGC\,4736, and NGC\,4826, they
have the same axis ratios [i.e., cos($i$)] and PA as the galaxies. Their
semimajor axes are equal to, respectively, 98$\farcs9$, 168$\farcs3$, and
150$\arcsec$, or 0.25 $R_{25}$ for NGC\,4395, and 0.5 $R_{25}$ for both
NGC\,4736 and NGC\,4826. For NGC\,3368, we exclude an ellipse with axis ratio
0.69 and PA=155$\degr$, like the region in the galaxy with the highest surface
brightness, and semimajor axis equal to 85$\farcs6$, or 0.375 $R_{25}$
\citep{rc3}.  These are the areas most affected by confusion (see
Section~\ref{sec:completeness} and Figures~\ref{fig:artificial}
and~\ref{fig:cmpltnss}).  We also need to set detection limits for these
simulated observations. To this end, we combine the real limiting magnitude of
the $K_s$ images, with the difference between the $K_s$ and $V$ bands TO
magnitudes; the results are $V$ = 22.5, 21.6, 23.4, and 23.1 mag, respectively,
for NGC\,3368, NGC\,4395, NGC\,4736, and NGC\,4826.  Finally, the $V$ mag of
each MW cluster is corrected for the Galactic extinction given in the
\citet{harr96} catalog, and subsequently dimmed by the foreground extinction in
the direction of each galaxy, i.e., 0.069 mag (NGC\,3368), 0.047 mag
(NGC\,4395), 0.049 mag (NGC\,4736), and 0.113 mag (NGC\,4826), respectively. 

For every galaxy, there are four possible orientations of the FOV and mask that
preserve its alignment; they can be seen as mirror reflections of the projected
coordinates ($+X_{\rm proj},+Y_{\rm proj}$;$+X_{\rm proj},-Y_{\rm proj}$;
$-X_{\rm proj},+Y_{\rm proj}$; $-X_{\rm proj},-Y_{\rm proj}$, with two sets of
four pairs, i.e., one for each of the two possible rotation axes.) 

We show in Figures~\ref{fig:mask} and~\ref{fig:mask2} the results of the
artificial observations of the MW GC system, for the rotation around the
$X^\prime$ ({\it left}) and $Y$ ({\it right}) axes, respectively, and $+X_{\rm
proj},+Y_{\rm proj}$. Results for NGC\,3368 and NGC\,4395  are displayed,
respectively, in the top and bottom panels of Figure~\ref{fig:mask}, whereas
outcomes for NGC\,4736 and NGC\,4826 are presented in the top and bottom panels
of Figure~\ref{fig:mask2}.  Angular distances in arcsec are measured relative
to the center of the galaxies, with the horizontal axis increasing in the
direction of decreasing RA.  The ellipses delineated with the black solid line
mark the outer edges of the masks; the red crosses are clusters visible in the
WIRCam FOV, whilst sources that have been masked out are represented by black
crosses.

In the case of NGC\,3368, for the 4 realizations rotating around the $X^\prime$
axis we detect, respectively, 50 ($+X_{\rm proj},+Y_{\rm proj}$), 50 ($+X_{\rm
proj},-Y_{\rm proj}$), 50 ($-X_{\rm proj},+Y_{\rm proj}$), and 50 ($-X_{\rm
proj},-Y_{\rm proj}$) sources; for rotating around the $Y$ axis, the numbers
are 46, 46, 46, 46.  With these numbers, there would be, on average, 48$\pm$2
simulated clusters visible in NGC\,3368, and eq.~\ref{eq:kissler} gives a total
$N_{\rm GC} = 50\pm13$. This error is statistical only; it includes errors in
the assumed number of total GCs in the MW, Poisson errors in the observed
number of GCCs in NGC\,3368, and Poisson errors in the number of simulated
clusters in the WIRCam FOV. To this error, we add potential systematics. Errors
in the distance to NGC\,3368 result in uncertainties in the detection limiting
magnitude, and in the effective areas of the galaxy covered by the FOV and
excluded by the mask. For the given uncertainty in the distance to NGC\,3368
($^{+1.1}_{-1.0}$ Mpc), $N_{\rm GC}$ would vary by +21/-14 if the random and
systematic errors are added in quadrature. Another potential concern is that
the MW and NGC\,3368 likely have a different number of obscured clusters. To
account for this possibility, we include an additional uncertainty of 25\%
($\pm$ 12). We end up with $N_{\rm GC} = 50^{+24}_{-19}$, or
50$\pm13^{+20}_{-13}$, with the first error statistical and the second
systematic (from distance and differential obscuration).
 
For NGC\,4395, rotations around $X^\prime$ yield 80, 80, 80, 79; for the $Y$
axis, the results are: 73, 71, 75, 71. Therefore, the average simulated
clusters visible would be 76$\pm$4, and a total $N_{\rm GC} = 21\pm5$ from
eq.~\ref{eq:kissler}.  The distance uncertainty of $\pm$ 0.4 Mpc results in an
additional error of $\pm$1 in  $N_{\rm GC}$. Including a 25\% ($\pm$6)
uncertainty for differential obscuration, the census for NGC\,4395 is $N_{\rm
GC} = 21\pm8$, or 21$\pm5\pm6$. 

For NGC\,4736, we find, respectively, 62, 62, 62, and 61 for rotations around
$X^\prime$; for the $Y$ axis, the results are: 60, 60, 61, and 60. Hence,  the
average simulated clusters visible would be 61$\pm$1, and a total $N_{\rm GC} =
66\pm14$ from eq.~\ref{eq:kissler}.  The distance uncertainty of $\pm$ 0.4 Mpc
results in an additional error of +6/-3 in  $N_{\rm GC}$. Including a 25\%
($\pm$16) uncertainty for differential obscuration, the tally for NGC\,4736 is
$N_{\rm GC} = 66^{+22}_{-21}$, or 66$\pm 14^{+17}_{-16}$. 

Finally, for NGC\,4826, rotations around $X^\prime$ give 49, 49, 50, 49
detections; for the $Y$ axis, the results are: 48, 47, 47, 48. Accordingly, the
average simulated clusters visible would be 48$\pm$1, and a total $N_{\rm GC} =
99\pm19$ from eq.~\ref{eq:kissler}.  The distance uncertainty of $\pm$ 0.7 Mpc
results in an additional error of +20/-10 in  $N_{\rm GC}$. Including a 25\%
($\pm$25) uncertainty for differential obscuration, the count for NGC\,4826 is
$N_{\rm GC} = 99^{+37}_{33}$, or 99$\pm 19^{+32}_{21}$. 

We make a final, downward, correction of 30\% to these numbers, based on the
percentage contamination found for NGC\,4258 by \citet{gonz19}, from
spectroscopic observations of its GC candidates. Hence, we get $N_{\rm GC} =
35^{+20}_{-16}$ for NGC\,3368, $N_{\rm GC} = 15\pm7$ for NGC\,4395, $N_{\rm GC}
= 46\pm18$ for NGC\,4736, and $N_{\rm GC} = 69^{+31}_{-28}$ for NGC\,4826. 

The correlation of $N_{\rm GC}$ with virial mass provides an additional sanity
check, at least for NGC\,4736 and NGC\,4826. From their rotation curves
(respectively, Jalocha et al.\ 2008 and Saburova et al.\ 2009), both galaxies
are quite light, with total halo masses $\sim$ 4$\times 10^{10} M_\odot$; their
log $N_{\rm GC} \sim $ 1.7 is fully consistent with the larger dispersion of
the correlation at low masses (see Burkert \& Forbes 2020, their Figure
1).\footnote{We note here that, from their analysis of the NGC\,4736 rotation
curve, \citet{jalo08} conclude that it violates the sphericity condition at
large radii (i.e., mass ceases to increase with distance). Hence, the galaxy
would not have a massive spherical halo and its mass distribution would be best
fit with a disk model.} 

\section{Specific frequency, and the $N_{\rm GC}$ vs.\ $M_\bullet$ relation} \label{sec:results}

From the distance to NGC\,3368 and the values of $B_{\rm T,0} = 9.80\pm 15$ mag
and ($B\ - V$)$_{\rm T,0} = 0.79 \pm 0.01$  given in \citet{rc3}, we find $M_V
= -21.08 \pm 0.36$ mag, and a specific frequency $S_N = N_{\rm GC} \times
10^{0.4 \times [M_V + 15]} = 0.13 \pm 0.06$, if we consider random errors only,
and $S_N = 0.13\pm 0.07$, if we include the uncertainty in the number of
obscured clusters.\footnote{The systematic error in the distance basically
cancels out, because the absolute magnitude of the galaxy also changes.} For
NGC\,4395, \citet{rc3} report $B_{\rm T,0} = 10.57 \pm 0.54$ mag and ($B\ -
V$)$_{\rm T,0} = 0.46 \pm 0.08$ mag; this results in $M_V = -18.06 \pm 0.59$
mag and $S_N = 0.90\pm 0.55$, with statistical errors only, and $S_N = 0.90 \pm
0.65$ if the uncertainty in the number of obscured clusters is included.  For
NGC\,4736, the values are $B_{\rm T,0} = 8.75 \pm 0.13$ mag and ($B\ -
V$)$_{\rm T,0} = 0.72\pm0.01$ mag, from which $M_V = -20.46\pm 0.21$ mag, and
$S_N = 0.30 \pm 0.10$, with random errors only, and $S_N = 0.30 \pm 0.13$, if
we include systematics.  For NGC\,4826, \citet{rc3} give $B_{\rm T,0} = 8.82
\pm 0.24$ mag and ($B\ - V$)$_{\rm T,0} = 0.71\pm0.01$ mag; $M_V = -21.21 \pm
0.32$ mag, and $S_N = 0.23\pm0.09$ or $S_N = 0.23 \pm 0.11$, respectively,
without and with the uncertainty in the number of obscured clusters. For
comparison, the Milky Way has $S_N = 0.5\pm0.1$ \citep{ashm98}.  

Before placing our four galaxies on the $N_{\rm GC}$ -- $M_\bullet$
correlation, we redetermine it in the Appendix for the sample in
\citet{harr14}, plus NGC\,4258 and the spirals in this paper, albeit with
updated galaxy classifications from \citet{sahu19b}.  


\renewcommand{\arraystretch}{3}

\begin{sidewaysfigure}[ht!]
\begin{tabular}{lll}
\hspace*{-1.5cm}\includegraphics[scale=0.60]{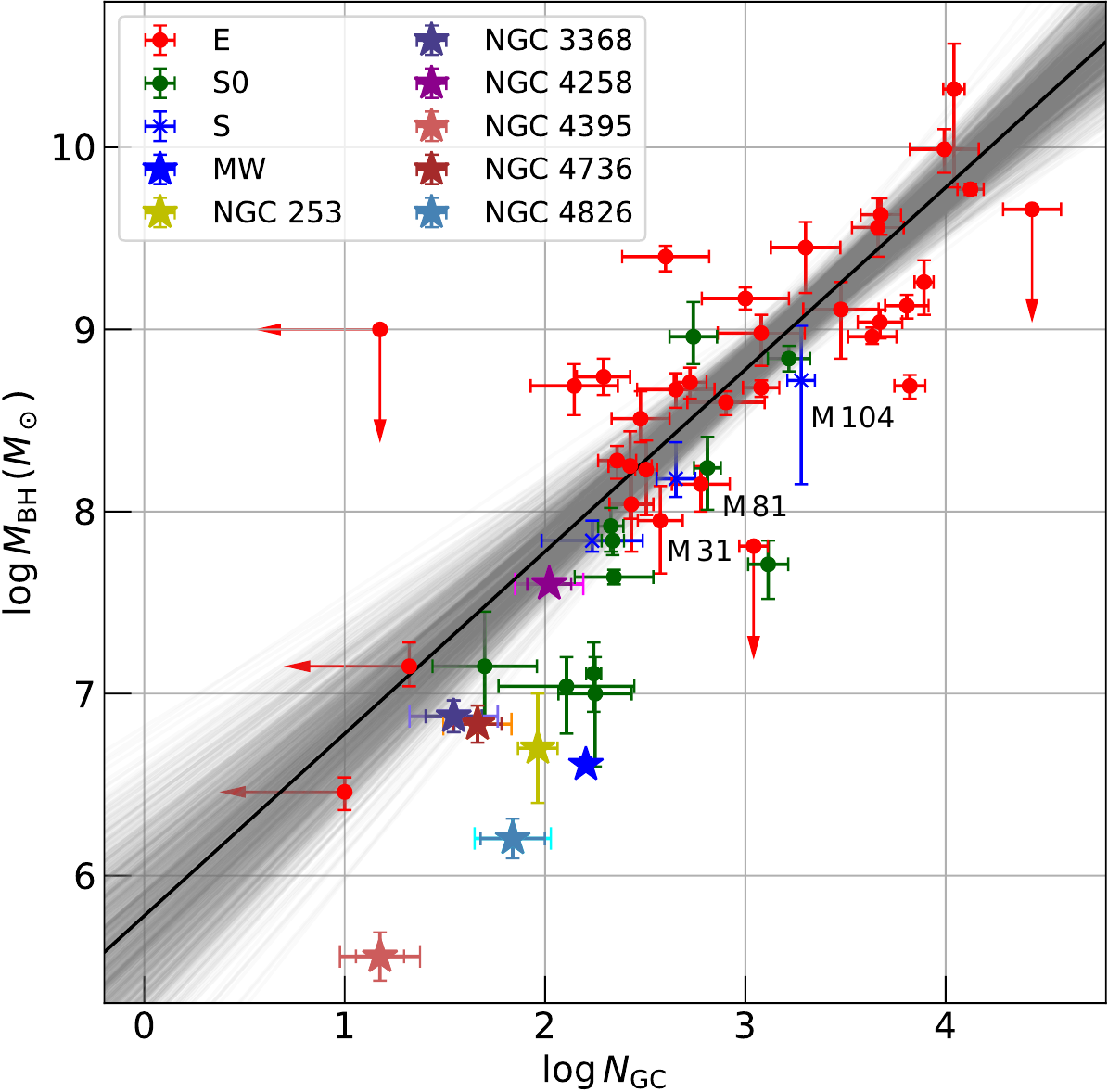}
&
\hspace{0.36cm}\includegraphics[scale=0.60]{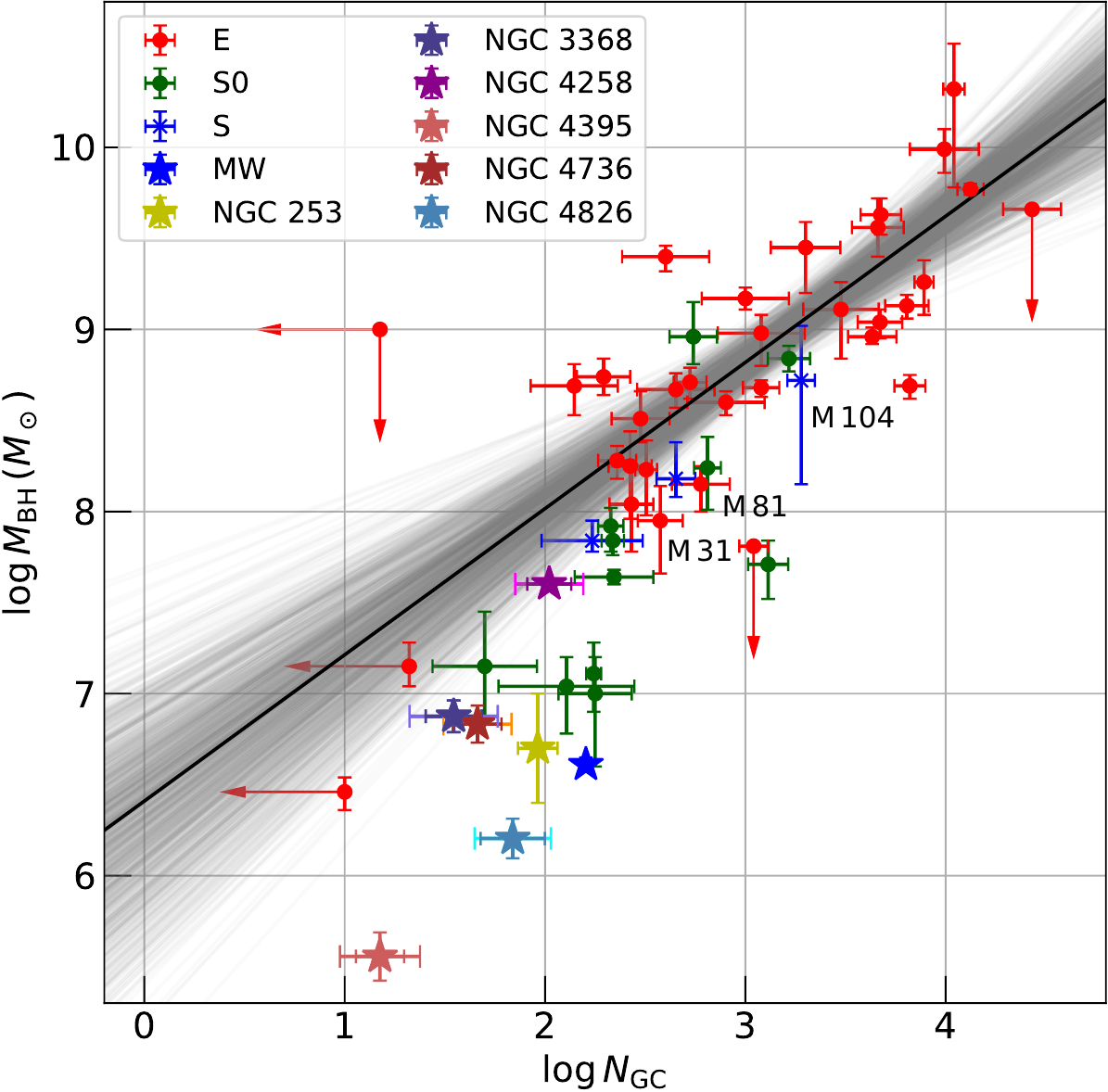}
&
\hspace{0.36cm}\includegraphics[scale=0.60]{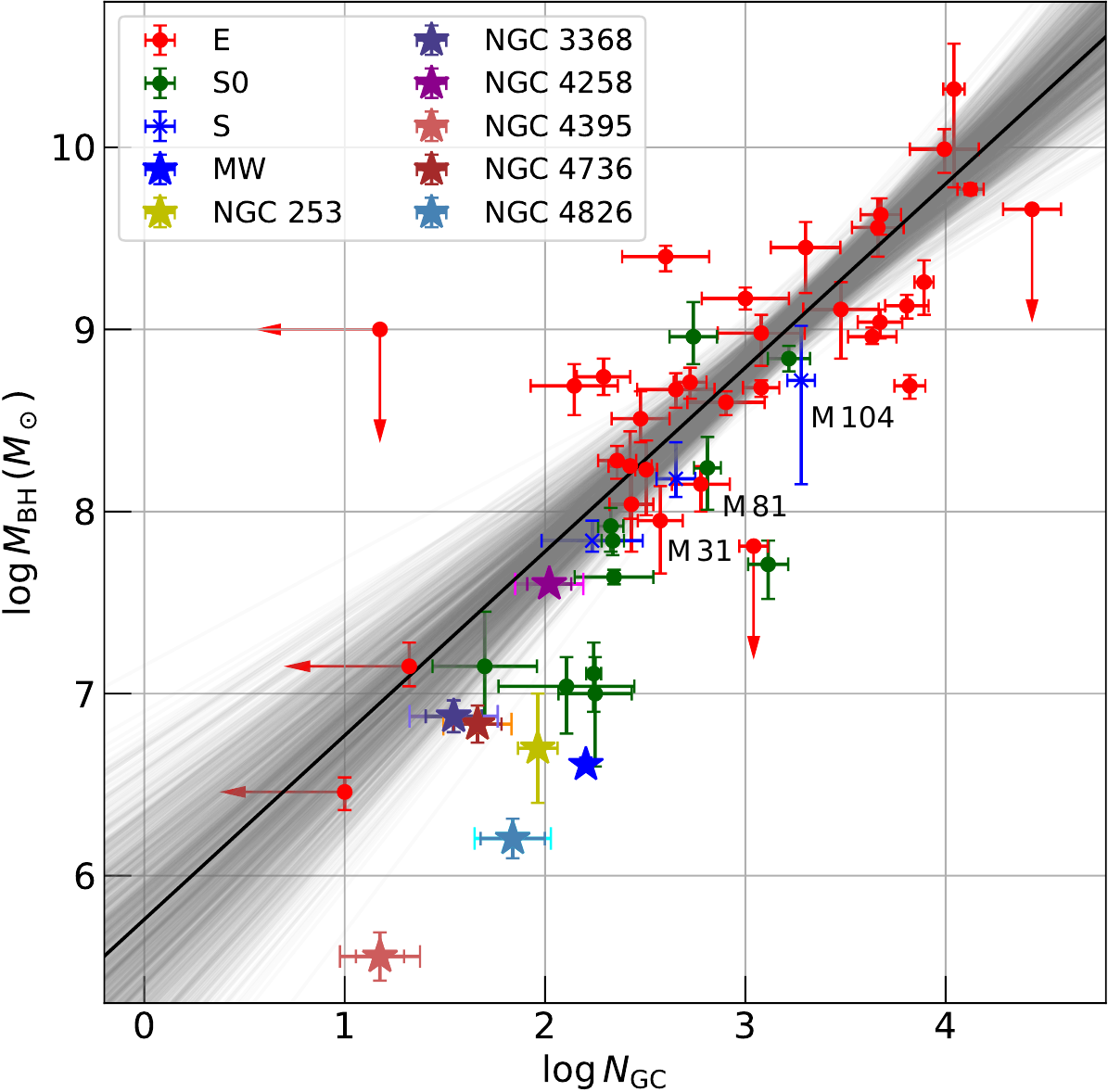}
\end{tabular}
\hspace*{7.83cm}\includegraphics[scale=0.586]{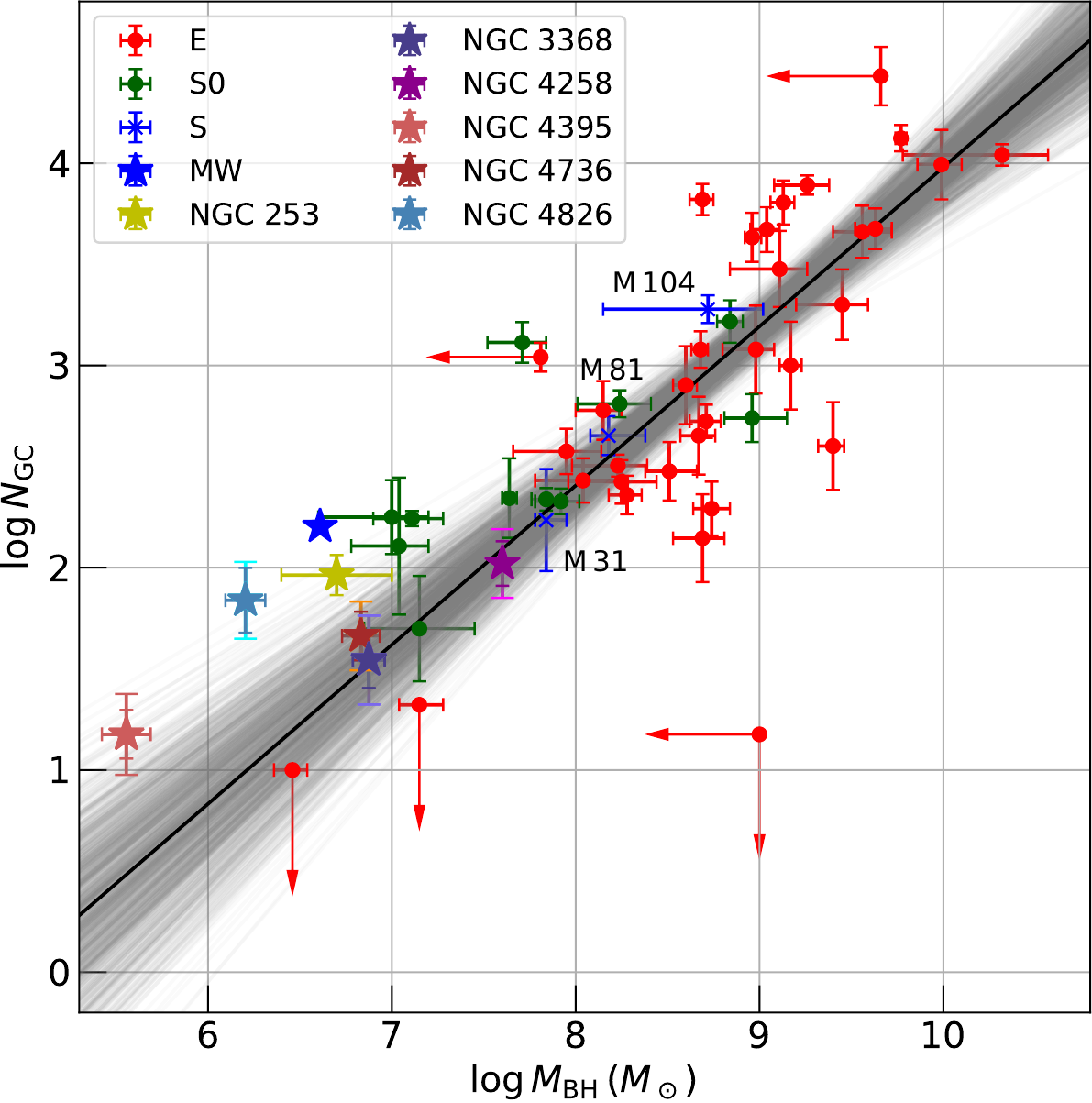}
\caption{BCES fits to elliptical galaxies. {\it Top:} log $M_\bullet$ versus log $N_{\rm GC}$; 
{\it bottom:} log $N_{\rm GC}$ versus log $M_\bullet$. {\it Left:} no additional scatter;
{\it center:} additional $\epsilon_y$; {\it right:} $\epsilon_y$ and $\epsilon_x$.
{\it Solid black lines:} best fits; {\it translucent gray lines:} range of possible solutions.
Solid ({\it red}) and open ({\it green}) circles
represent, respectively, elliptical and lenticular galaxies in the sample of \citet{harr14}, reclassified
as per \citet{sahu19b};
the blue crosses are the spiral galaxies M\,31, M\,81, and M\,104.
{\it Solid blue star:} Milky Way. {\it Solid light olive star:} NGC\,256, with
correct number of GCs. 
{\it Solid purple star:} NGC\,4258 \citep{gonz17,gonz19};
{\it Solid violet, light brown, dark brown,} and {\it light blue stars:} NGC\,3368, 
NGC\,4395, NGC\,4736, and NGC\,4826, respectively (this work).
For NGC\,3368, NGC\,4395, NGC\,4258, NGC\,4736, and NGC\,4826, random errors of $N_{\rm GC}$ are shown by small bars, 
while systematic errors, added in quadrature, are illustrated by the large bars.
Errors for the MW are smaller than the blue star.
\label{fig:bces_ell}}
\end{sidewaysfigure}


\renewcommand{\arraystretch}{3}

\begin{sidewaysfigure}[ht!]
\begin{tabular}{lll}
\hspace*{-1.5cm}\includegraphics[scale=0.60]{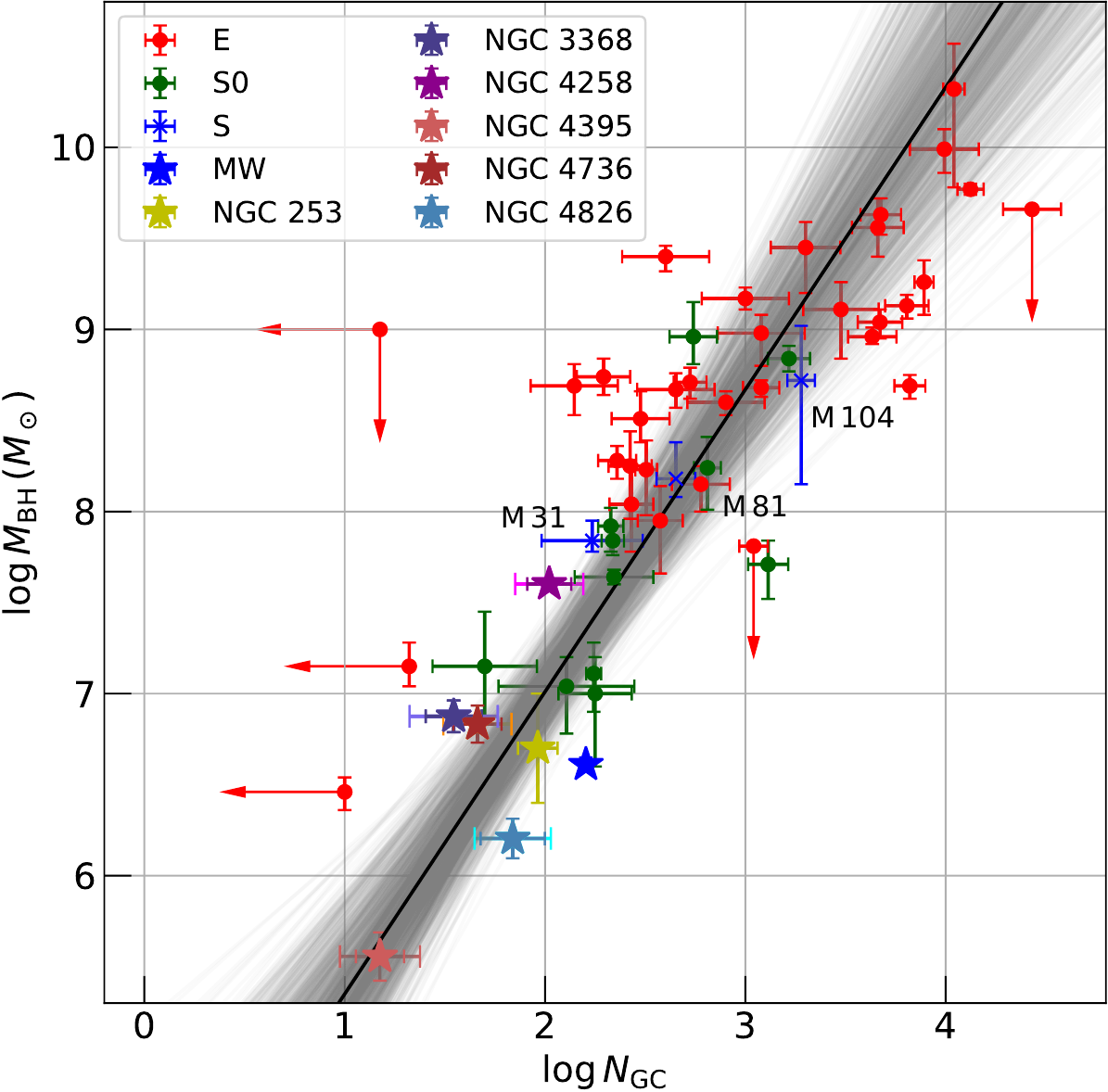}
&
\hspace{0.36cm}\includegraphics[scale=0.60]{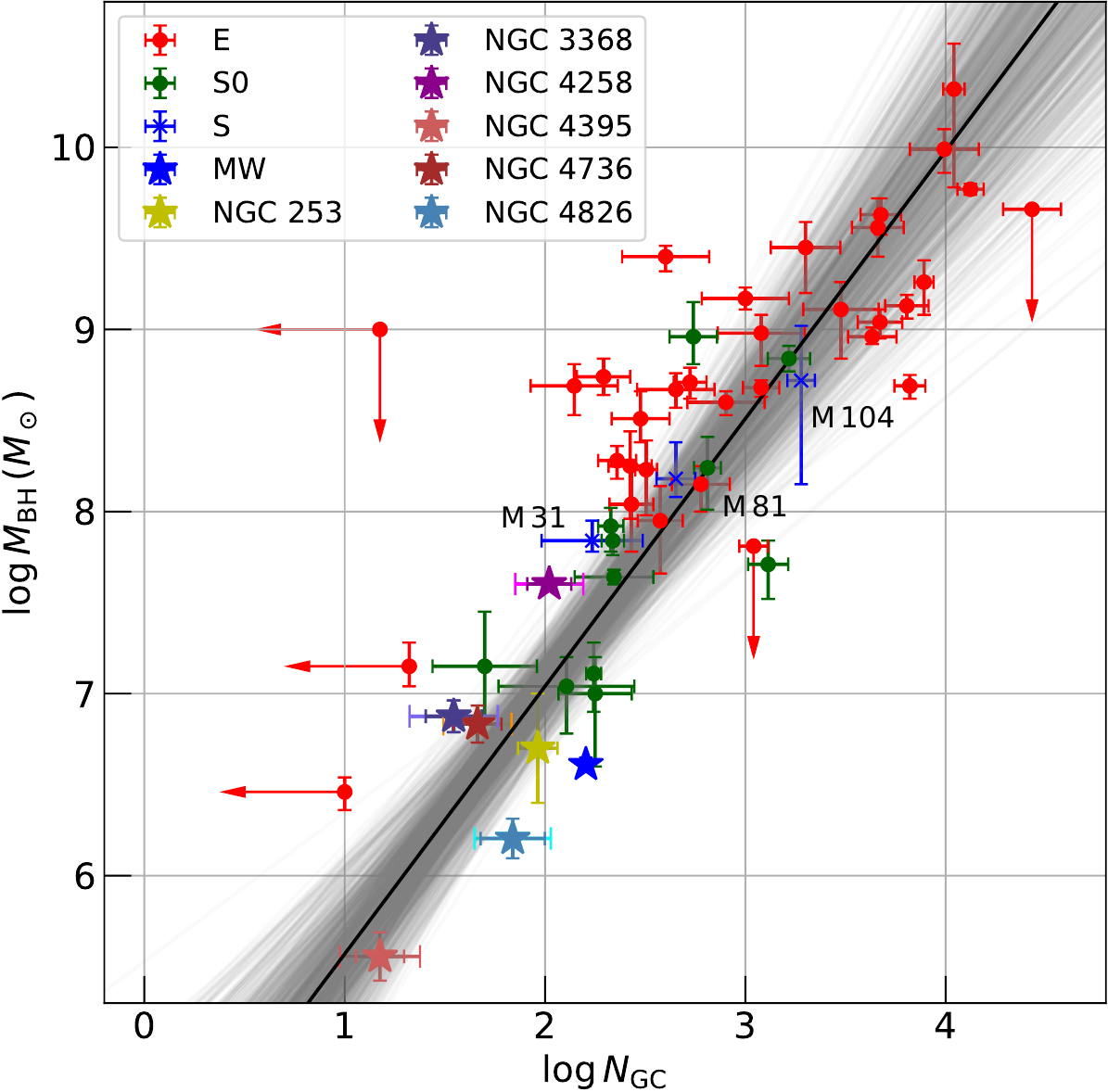}
&
\hspace{0.36cm}\includegraphics[scale=0.60]{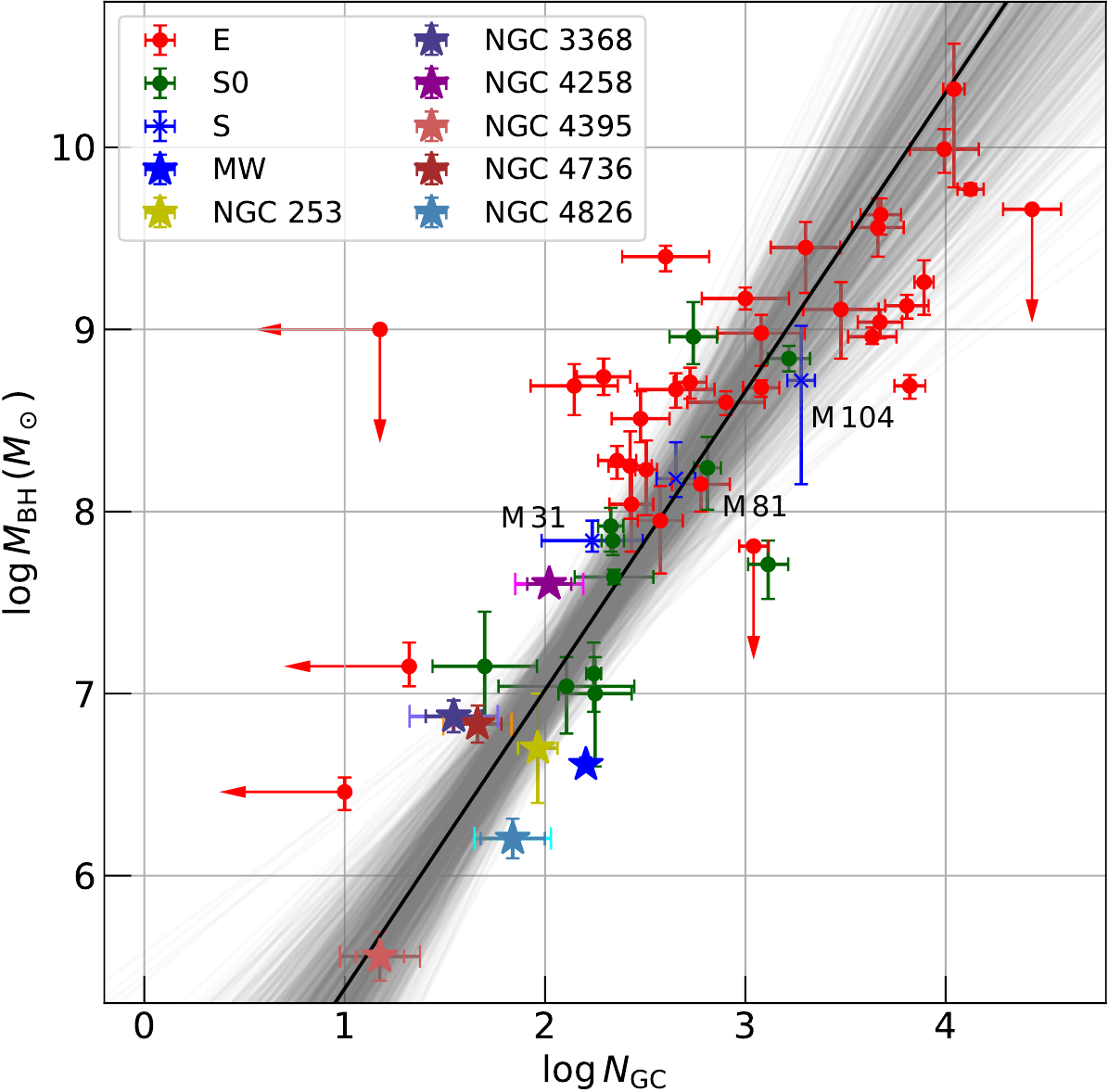}
\end{tabular}
\hspace*{7.83cm}\includegraphics[scale=0.586]{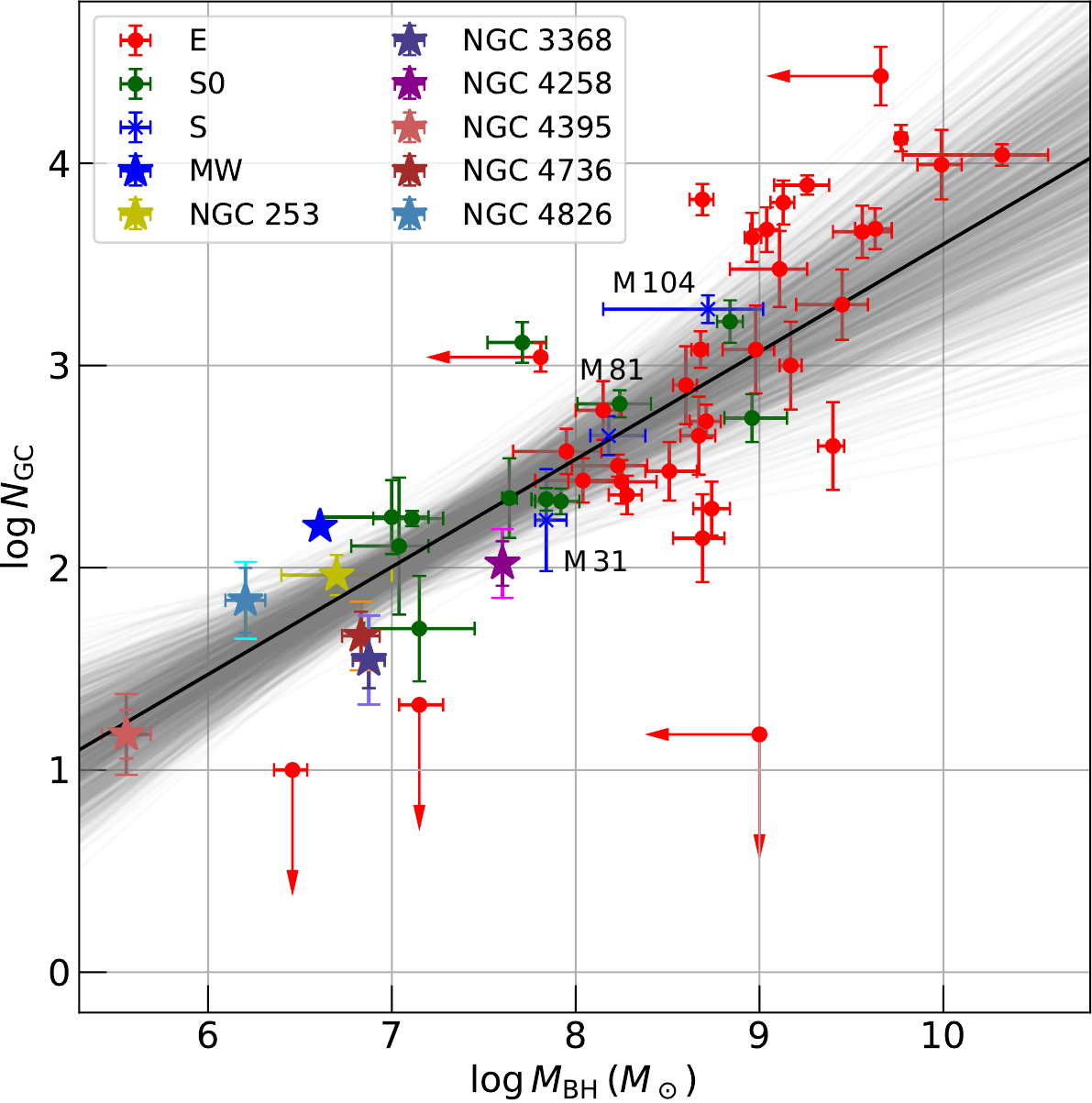}
\caption{BCES fits to spiral galaxies. {\it Top:} log $M_\bullet$ versus log $N_{\rm GC}$; 
{\it bottom:} log $N_{\rm GC}$ versus log $M_\bullet$. {\it Left:} no additional scatter;
{\it center:} additional $\epsilon_y$; {\it right:} $\epsilon_y$ and $\epsilon_x$.
Lines and symbols as in Figure~\ref{fig:bces_ell}.
\label{fig:bces_spir}}
\end{sidewaysfigure}

\renewcommand{\arraystretch}{1}

Figures~\ref{fig:bces_ell} and \ref{fig:bces_spir}, and Table~\ref{tab:bces}
display the Bivariate Correlated Errors and Intrinsic Scatter
\citep[BCES,][]{akri96} fits for, respectively, ellipticals and spirals. The
black solid lines represents the best fit, and the translucent gray lines
illustrate the range of possible solutions. Each gray line corresponds to a
realization of the fit, obtained by adding a random variable to each the slope
$\beta$ and the intercept $\alpha$ of the best fit, drawn from the normal
distributions of errors,  $\sigma_\beta$ and $\sigma_\alpha$. Since 1,000
realizations were obtained for each fit, roughly 50 gray lines ($5\%$ of the
total 1,000) belong to fits separated by $2\sigma$ or more from the best
values.  In the figures, ellipticals appear as solid red circles and
lenticulars as solid dark green ones; the spirals M\,31, M\,81, and M\,104 are
the blue crosses. The locations of the MW, NGC\,253, NGC\,3378, NGC\,4258,
NGC\,4395, NGC\,4736, and NGC\,4826 are shown as solid stars, respectively,
blue, light olive, violet, purple, light brown, dark brown, and light blue.  We
correct the position of NGC\,253, whose number of GC had been read erroneously
by \citet{harr14}. Indeed, \citet{olse04} report $N_{\rm GC} = 21$, {\em but}
with a completeness of 0.21, i.e., they really implied $N_{\rm GC} \sim 100$.
They also give a specific frequency in the $K$-band $S_{N,K} = N_{\rm GC}
\times 10^{0.4 \times [M_K + 18]} = 0.30 \pm 0.07$, or $N_{\rm GC} = 92\pm21$.
Previously, \citet{blec86} had found $N_{\rm GC} = 80 \pm 40$ for NGC\,253. 

The Figures show $M_\bullet$ versus $N_{\rm GC}$ in the top row; from left to
right, fits without extra ``cosmic" scatter, fits with additional $\epsilon_y$,
and fits with additional intrinsic scatter in both $x$ and $y$ ( see Appendix).
The bottom row illustrates $N_{\rm GC}$ versus $M_\bullet$, only for fits with
additional $\epsilon_y$, which are not symmetrical.  We find that the slope of
the correlation is roughly 60\% steeper for spirals than for ellipticals (see
Table~\ref{tab:bces}).

\section{The $M_\bullet$ versus $M_\ast$ correlation from $N_{\rm GC}$}
\label{mgaltot} 

In order to investigate whether the discrepant slopes of the $M_\bullet$ versus $N_{\rm
GC}$ correlations for elliptical and spiral galaxies reflect the slopes between
$M_\bullet$ and $M_\ast$, we derive $M_\ast$ from its correlation with $N_{\rm GC}$, for the ellipticals 
and LTGs in our sample. \citet{huds14} have obtained log $M_{\rm GC}$ versus
log $M_\ast$ for the 422 galaxies (248 ellipticals, 93 lenticulars, and 81
spirals and irregulars) in the \citet{harr13} sample, where $M_{\rm GC}$ is
the total mass of the GC system of a galaxy. In actuality, $N_{\rm GC}$ is the
observable, while $M_{\rm GC}$ has the advantage of a smaller sensitivity to
undetected faint GCs, since 99\% of the total GC system mass is roughly
contained in the clusters brighter than 1 mag below the GCLF turnover
\citep{gonz17}.\footnote{The trends of $N_{\rm GC}$ with total $K$-band
luminosity, a good tracer of mass, are remarkably similar for all galaxy
types.} \citet{huds14} assume a relation between total number and total mass of
GC log $N_{\rm GC} = 0.90 \times M_{\rm GC}$ - 4.65.  The relations between log
$M_{\rm GC}$ or $N_{\rm GC}$ and $M_\ast$ are broken power laws, with log
$M_{\rm GC} = 0.53\ \times$ log $M_\ast$ + 2.00, log $M_\ast = 2.11\ \times$
log $N_{\rm GC}$ + 5.99, for lower galaxy masses, and $M_{\rm GC} = 1.60
\times$ log $M_\ast$ - 9.22, log $M_\ast =  0.69\ \times$ log $N_{\rm GC}$ +
8.96, for higher galaxy masses, with the inflection point at  log
$M_\ast/M_\odot \sim$ 10.4, log $N_{\rm GC} \sim$  2.1. 

\begin{sidewaysfigure}[ht!]
\begin{tabular}{lll}
\hspace*{-1.5cm}\includegraphics[scale=0.60]{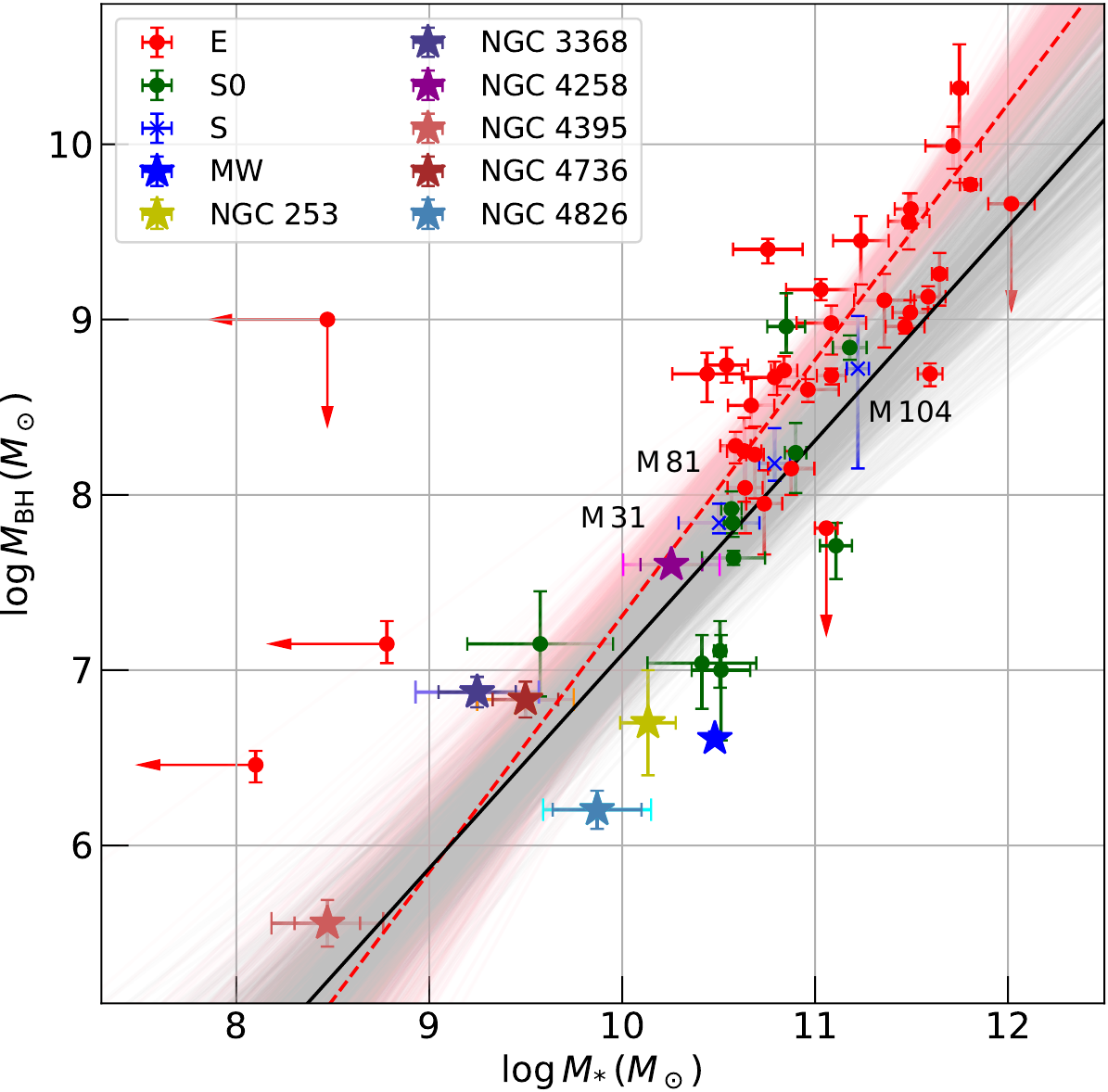}
&
\hspace{0.36cm}\includegraphics[scale=0.60]{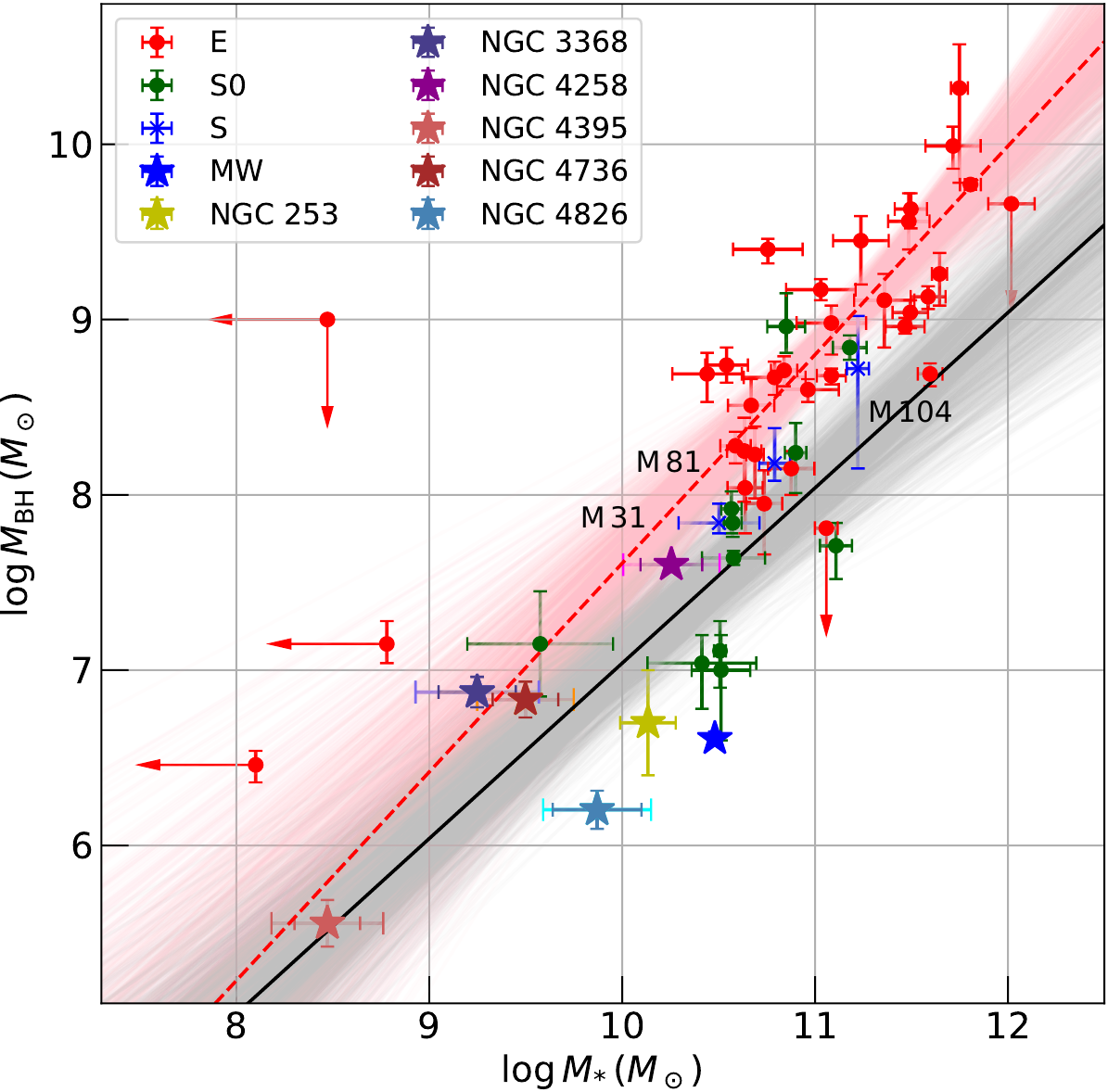}
&
\hspace{0.36cm}\includegraphics[scale=0.60]{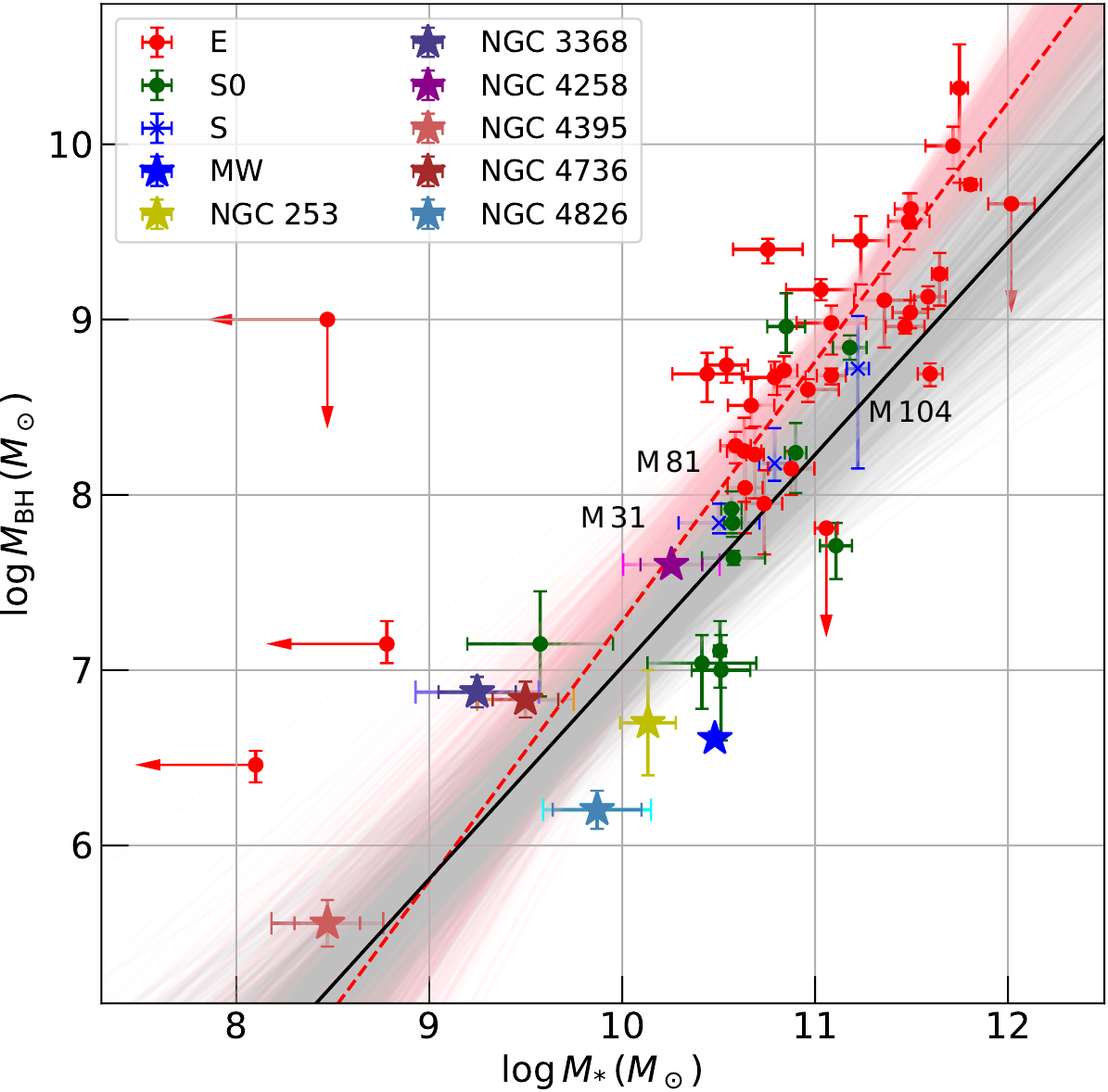}
\end{tabular}
\hspace*{7.63cm}\includegraphics[scale=0.602]{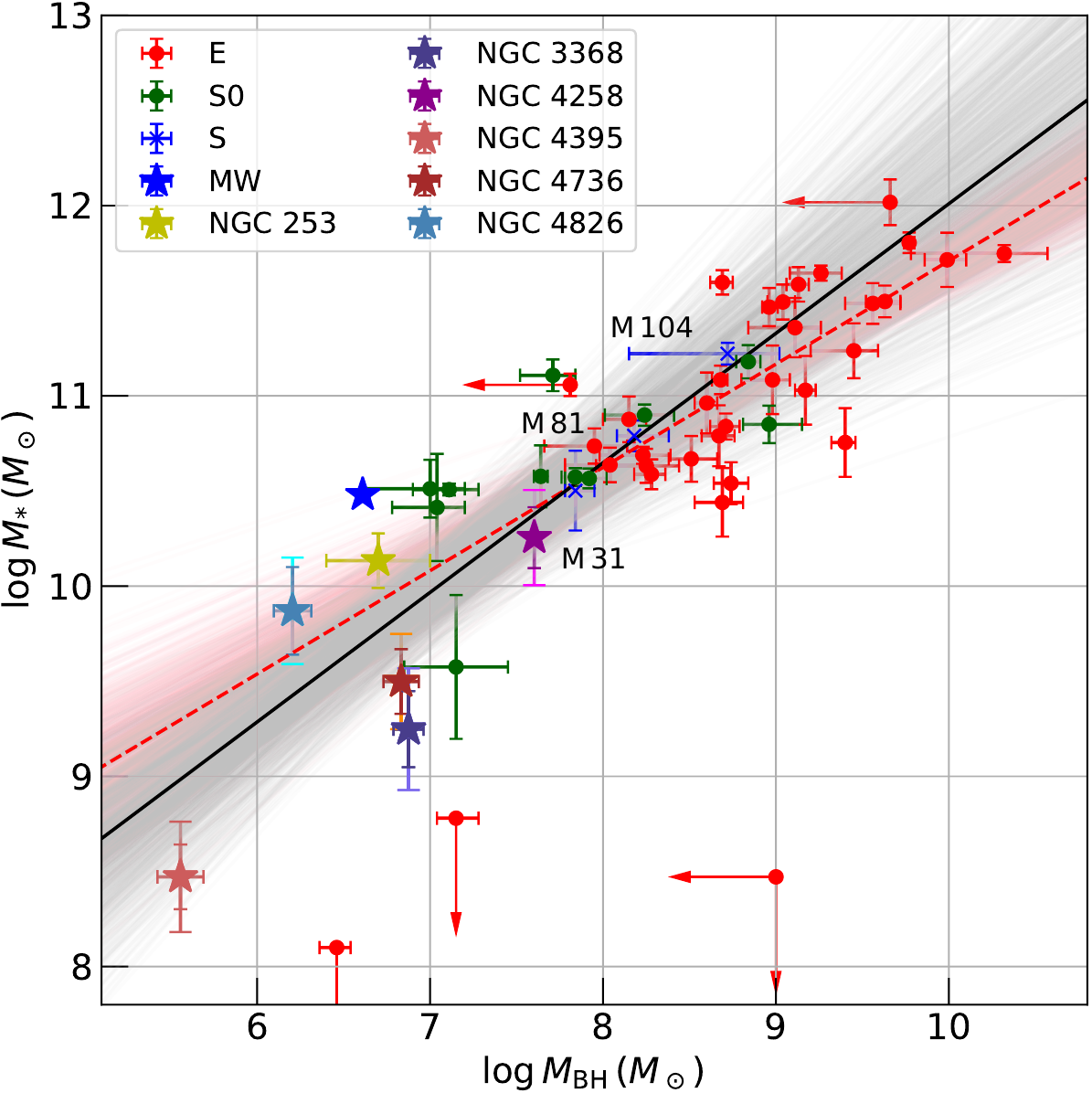}
\caption{BCES fits to elliptical and spiral galaxies. {\it Top:} log $M_\bullet$ versus log $M_\ast$; 
{\it bottom:} log $M_\ast$ versus log $M_\bullet$. {\it Left:} no additional scatter;
{\it center:} additional $\epsilon_y$; {\it right:} $\epsilon_y$ and $\epsilon_x$.
{\it Dashed red lines:} best fits for ellipticals; {\it translucent pink lines:} range of possible solutions for ellipticals.
{\it Solid black lines:} best fits for spirals; {\it translucent gray lines:} range of possible solutions for spirals.
Symbols as in Figure~\ref{fig:bces_ell}.
\label{fig:bces_mgal_ell_spir}}
\end{sidewaysfigure}

\begin{sidewaysfigure}[ht!]
\begin{tabular}{lll}
\hspace*{-1.5cm}\includegraphics[scale=0.60]{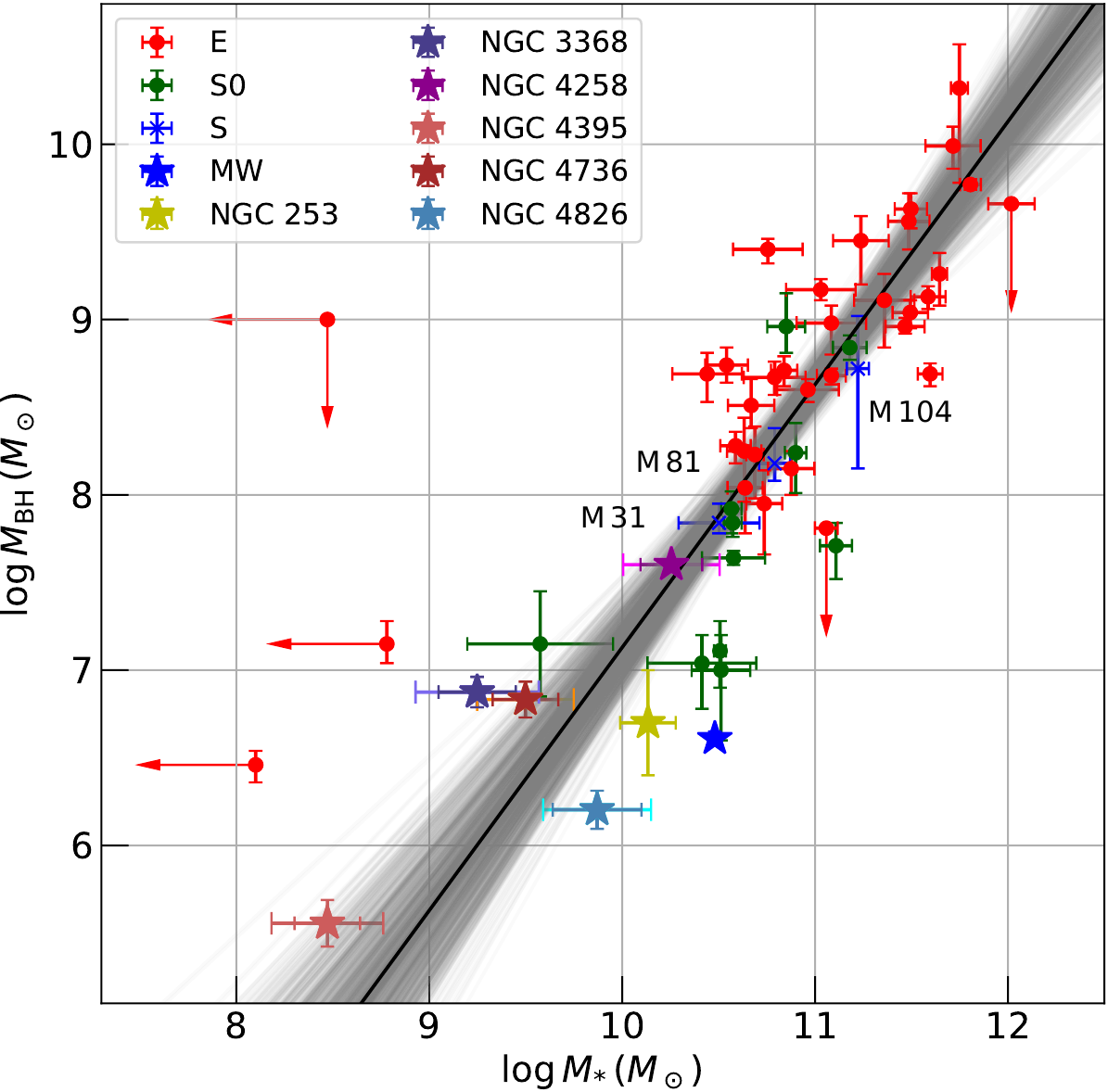}
&
\hspace{0.36cm}\includegraphics[scale=0.60]{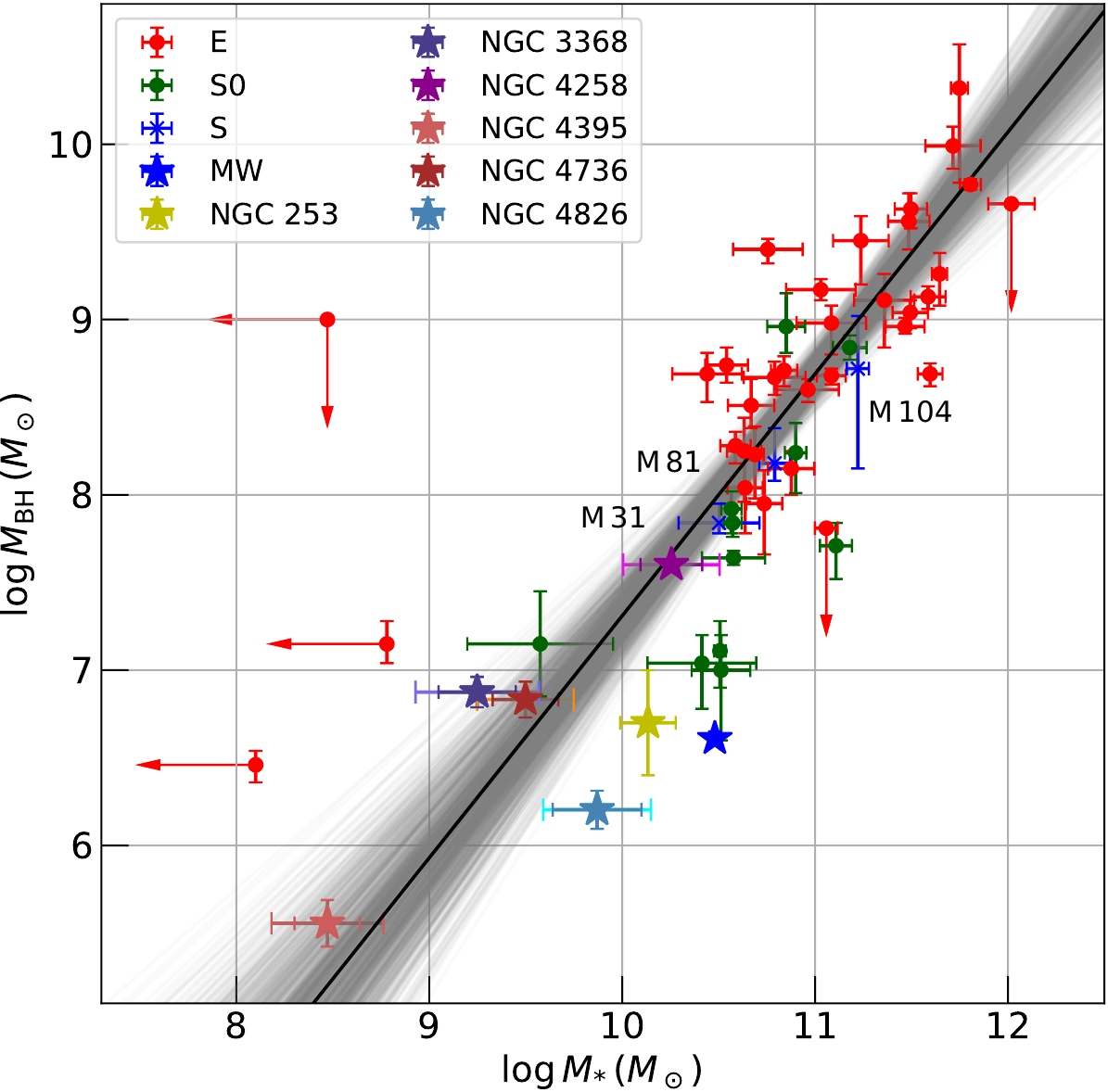}
&
\hspace{0.36cm}\includegraphics[scale=0.60]{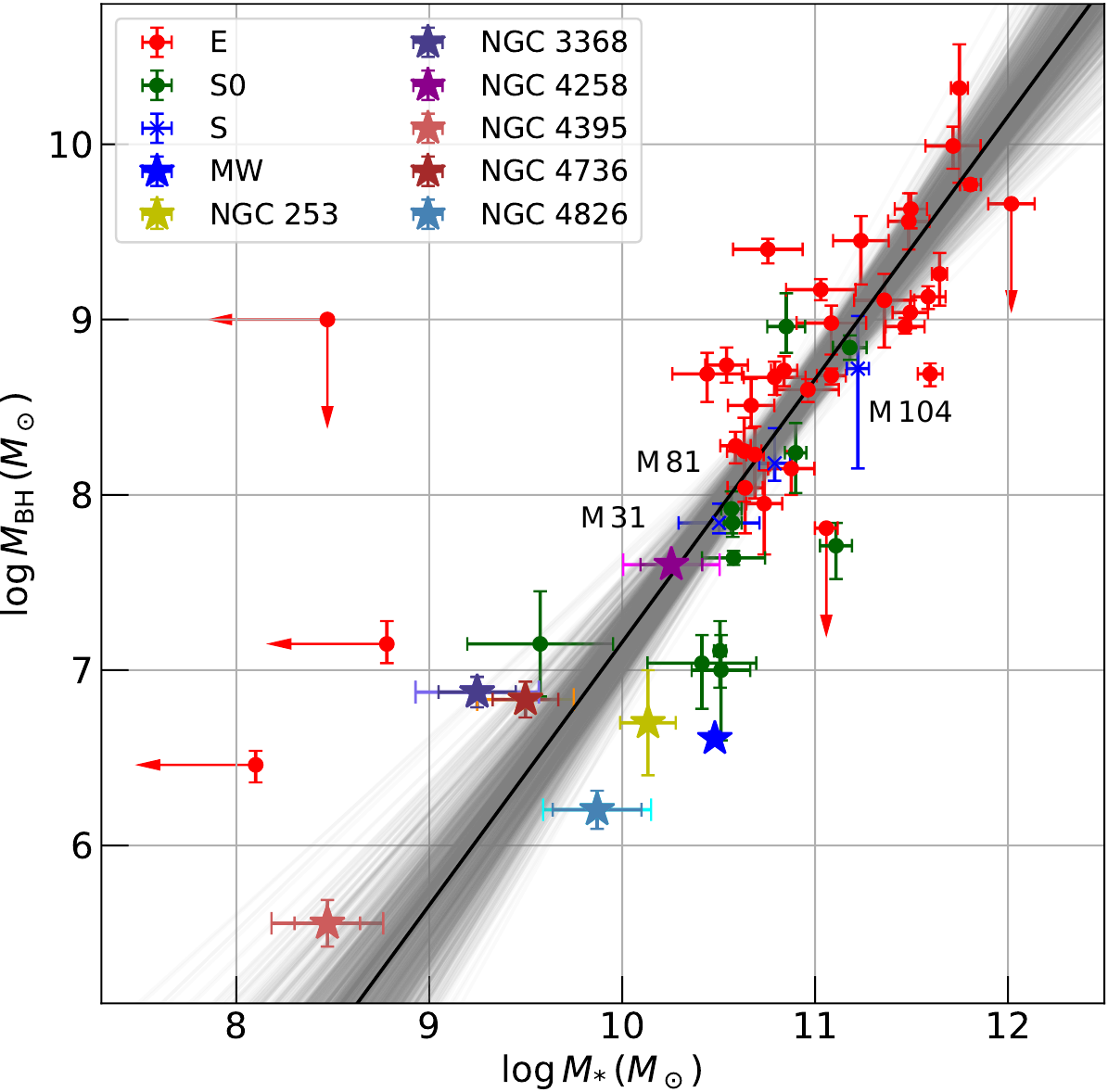}
\end{tabular}
\hspace*{7.63cm}\includegraphics[scale=0.603]{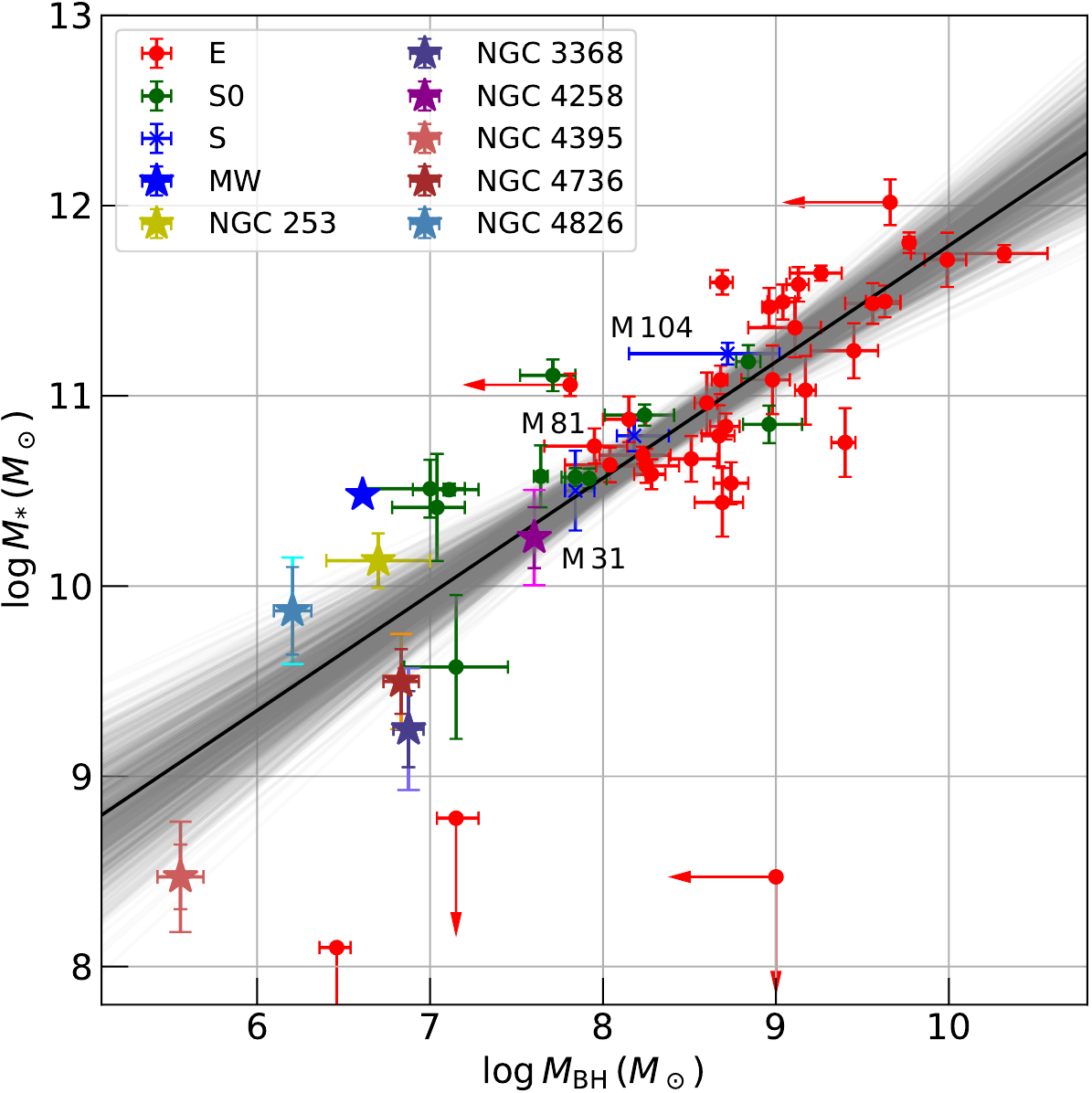}
\caption{BCES fits to E+S galaxies. {\it Top:} log $M_\bullet$ versus log $M_\ast$; 
{\it bottom:} log $M_\ast$ versus log $M_\bullet$. {\it Left:} no additional scatter;
{\it center:} additional $\epsilon_y$; {\it right:} $\epsilon_y$ and $\epsilon_x$.
Lines and symbols as in Figure~\ref{fig:bces_ell}.
\label{fig:bces_mgal_all}}
\end{sidewaysfigure}

Figure~\ref{fig:bces_mgal_ell_spir} and Table~\ref{tab:galmass_bces} show the
BCES fits to $M_\bullet$ versus $M_\ast$ for, respectively, ellipticals ({\it
red-dashed} and {\it pink translucent} lines) and spirals ({\it black solid}
and {\it gray translucent} lines). Their slopes are, roughly, 1.47$\pm$0.16 and
1.22$\pm$15, respectively, i.e., the same within the errors, with the ``best"
fits as close or closer than those found by \citet{rein15}, also less steep for
spirals than for ellipticals, and with an offset of $\sim$ 0.2-0.3 dex at log
$M_\ast/M_\odot$ = 10. A fit to both ellipticals and spirals together
(Figure~\ref{fig:bces_mgal_all}, Table~\ref{tab:galmass_bces}) yields a slope
$\sim 1.5\pm 0.15$, and a cosmic scatter $\epsilon \sim$ 0.4, of the order of
half that found by \citet{vand16}. 

%
\renewcommand{\arraystretch}{3}

%
\renewcommand{\arraystretch}{3}

\renewcommand{\arraystretch}{1}

\section{Discussion and conclusions} \label{sec:concl}

We have now successfully applied the \uiks\ diagram GC selection technique
\citep{muno14} to five spiral galaxies: NGC\,4258 \citep{gonz17}, NGC\,3368,
NGC\,4395, NGC\,4736, and NGC\,4826 (this work). We complement the \uiks\ plot
with GCC structural parameters; this combination of techniques constitutes a
very efficient tool to study GC systems of nearby galaxies. The sample of
spirals with measurements of both the GC system and the SMBH mass has now been
increased to 10, or 100\% more than before we embarked on this project. Our
full spiral working sample comprises all galaxies observable from the northern
hemisphere with a direct measurement of their central SMBH mass, and within 16
Mpc, regardless of orientation in the sky, i.e., a total of 9; so far, we have
analyzed the 5 for which we have complete \ust, \ip, and \ks\ data. We have
also taken the opportunity to correct the position of NGC\,253 in the
$M_\bullet$ versus $N_{\rm GC}$ diagram.  

The spiral galaxies for which we have determined $N_{\rm GC}$, as well as
NGC\,253, seem to have undergone few major mergers.  Like the MW, they all
belong to small groups, i.e., Leo I, Canes I, Canes II, and Sculptor. They also
have pseudobulges, except for NGC\,4395, which is a bulgeless, purely disk
galaxy. Additionally, the GC systems of both NGC\,4258 and NGC\,253 seem to lie
in rotating disks \citep{gonz19,olse04}.  Together with simulations
\citep[e.g.,][]{krav05,ager09,krui15} and observations of clumpy star-forming
disks at $z \sim 2$ \citep[e.g.,][]{cowi95,guo15}, the discovery of GC disks in
the local universe supports the idea that GCs probably formed in disks, and
came to populate halos through multiple mergers
\citep[e.g.,][]{schw87,ashm92,cote98}. 

We have investigated the correlation between $M_\bullet$ and $N_{\rm GC}$, for
both ellipticals and LTG.  We find log $M_\bullet \propto$ (1.01
$\pm$ 0.13) log $N_{\rm GC}$ for ellipticals, and log $M_\bullet \propto$ (1.64 $\pm$
0.24) log $N_{\rm GC}$ for LTG, in both cases including cosmic scatter in both variables. 
Hence, they are not consistent with each other,
and the slope of the correlation for LTG, in contrast with that for ellipticals, is not
linear. This, together with the likely scarcity of major interactions for LTG,
may argue in favor of statistical convergence through mergers as the cause of
the correlation in elliptical galaxies. 

On the other hand, we have derived $M_\ast$, for the ellipticals and LTG in our sample,
from its relation with $N_{\rm GC}$, and measured the correlations with $M_\bullet$. Far from
mirroring the discrepant slopes of the $M_\bullet$ versus $N_{\rm GC}$ fits, they are
quite similar, i.e., $M_\bullet \propto$ (1.48 $\pm$
0.18) log $M_\ast$ for ellipticals, and $M_\bullet \propto$ (1.21 $\pm$ 0.16) log
$M_\ast$ for LTG, with a small offset of $\sim$ 0.2-0.3 dex at log
$M_\ast/M_\odot = 10$, in both cases including cosmic scatter in both variables.  These 
results are in line with those obtained by \citet{rein15}
and \citet{simm17}, and argue for a mostly non-merger driven $M_\bullet$
growth, and galaxy-black hole co-evolution through secular processes, like calm
accretion and feedback.  \citep{simm17,mart18,smet19}. 

Researching GC systems and scaling relations in spirals is very important.
While these relations in ellipticals seem to be showing the end result of star
formation and galaxy assembly, in lower mass galaxies we may be witnessing the
fingerprints of the individual steps of the process.

\begin{acknowledgements}
R.A.G.L. acknowledges the financial support of DGAPA, UNAM, project IN108518.
R.A.G.L. and L.L.N. acknowledge the support of CONACYT, Mexico, project
A1-S-8263.  L.L. acknowledges the support of DGAPA PAPIIT grants IN112417 and
IN112820, CONACYT-AEM grant 275201, and CONACYT-CF grant 263356.  Y.O.-B.
acknowledges support from FONDECYT Postdoctoral Fellowship project No.\
3210442.  We would like to thank the anonymous referee, for their pertinent
criticisms and suggestions, and Daniel D\'{\i}az-Gonz\'alez, for his help with
the routine we used to calculate two-Gaussian fits. 

This research has made use of the VizieR catalogue access tool, CDS,
Strasbourg, France (DOI: 10.26093/cds/vizier). The original description of the
VizieR service was published in A\&AS 143, 23.  \end{acknowledgements}

\vspace{5mm}
\facilities{CFHT}

\software{
           daophot \citep{ 1987PASP...99..191S,2011ascl.soft04011S},
           galsim \citep{ 2015A&C....10..121R},
           iraf \citep{ 1986SPIE..627..733T, 1993ASPC...52..173T},
           ishape \citep{lars99},
           matplotlib \citep{ 2007CSE.....9...90H},
           numpy \citep{ 2020Natur.585..357H},
           PSFEx \citep{bert11},
           saoimage \citep{2003ASPC..295..489J},
           scipy \citep{2020NatMe..17..261V},
           SExtractor \citep{bert96},
           supermongo \citep{ 1991supe.book.....L}
          }


\appendix 

\begin{table}[ht]
\begin{scriptsize}
\begin{center}
\caption{Properties of galaxies used in the fits}
\begin{minipage}{80mm}
\hskip-3cm\begin{tabular}{@{\extracolsep{4pt}}lccrrrl@{}}
\hline
\hline
\vspace*{-0.3cm}
&&&&&&\\
Galaxy  &Type1 &  Type2  & $d$ (Mpc) & $N_{\rm GC}$ &$\Delta\ N_{\rm GC}$&  log  $M_{\bullet}/M_\odot$\\ 
\hline
NGC\,221   & E2  &            &   0.8   &   10  &        &   6.46(+0.08,-0.10)  \\  
NGC\,4486A & E2  &   E        &  15.6   &   21  &        &   7.15(+0.13,-0.11)  \\  
NGC\,821   & E4  &   E        &  24.1   &  320  &    40  &   8.23(+0.16,-0.25)  \\
NGC\,1332  & S0  &   ES       &  22.9   & 1000  &   500  &   9.17(+0.06,-0.06)  \\
NGC\,1399  & E1  &   E        &  20.0   & 6625  &  1180  &   8.69(+0.06,-0.07)  \\  
NGC\,3377  & E5  &   E        &  11.2   &  266  &    66  &   8.25(+0.19,-0.29)  \\
NGC\,3379  & E1  &   E        &  10.6   &  270  &    68  &   8.04(+0.22,-0.26)  \\
NGC\,3414  & S0  &   E        &  25.2   &  400  &   200  &   9.40(+0.06,-0.08)  \\
NGC\,3585  & S0  &   E        &  20.0   &  300  &   100  &   8.51(+0.15,-0.13)  \\
NGC\,3607  & E   &   E        &  22.8   &  600  &   200  &   8.15(+0.10,-0.15)  \\
NGC\,3608  & E2  &   E        &  22.9   &  450  &   200  &   8.67(+0.09,-0.10)  \\
NGC\,3842  & E   &   E        &  98.4   & 9850  &  3900  &   9.99(+0.11,-0.13)  \\
NGC\,4261  & E2  &   E        &  31.6   &  530  &   100  &   8.71(+0.08,-0.09)  \\
NGC\,4291  & E2  &   E        &  26.2   & 1200  &   600  &   8.98(+0.10,-0.18)  \\
NGC\,4350  & S0  &   EBS      &  15.4   &  196  &    60  &   8.74(+0.10,-0.10)  \\
NGC\,4374  & E1  &   E        &  18.4   & 4300  &  1200  &   8.96(+0.05,-0.04)  \\
NGC\,4472  & E2  &   E        &  16.3   & 7800  &   850  &   9.26(+0.12,-0.18)  \\
NGC\,4473  & E5  &   E        &  15.7   &  376  &    97  &   7.95(+0.19,-0.29)  \\
NGC\,4486  & E0  &   E        &  16.1   &13300  &  2000  &   9.77(+0.03,-0.03)  \\
NGC\,4552  & E1  &   E        &  15.3   & 1200  &   250  &   8.68(+0.04,-0.05)  \\
NGC\,4621  & E4  &   E        &  18.3   &  800  &   355  &   8.60(+0.06,-0.07)  \\
NGC\,4649  & E2  &   E        &  16.8   & 4745  &  1100  &   9.63(+0.09,-0.11)  \\
NGC\,4697  & E6  &   E        &  11.7   &  229  &    50  &   8.28(+0.08,-0.10)  \\
NGC\,4889  & E   &   E        & 103.2   &11000  &  1340  &  \hspace*{-0.15cm}10.32(+0.25,-0.54)  \\
NGC\,5846  & E0  &   E        &  24.9   & 4700  &  1200  &   9.04(+0.07,-0.09)  \\
NGC\,6086  & E   &   E        & 132.0   & 4584  &  1362  &   9.56(+0.16,-0.16)  \\
NGC\,7768  & E   &   E        & 112.8   & 3000  &  1300  &   9.11(+0.15,-0.27)  \\
IC\,1459   & E4  &   E        &  29.2   & 2000  &   800  &   9.45(+0.14,-0.25)  \\
IC\,4296   & BCG &   E        &  50.8   & 6400  &  1600  &   9.13(+0.06,-0.07)  \\
NGC\,5845  & E3  &   ES       &  25.9   &  140  &    70  &   8.69(+0.12,-0.16)  \\
NGC\,4382  & E2  &            &  18.4   & 1100  &   181  &   7.81               \\
A\,2052    & BCG &            &  141.0  &27000  &  9000  &   9.66               \\
NGC\,4486B & E1  &   E        &  16.3   &   15  &        &   9.00               \\
NGC\,5128  & E0p &   S0merg   &   3.8   & 1300  &   300  &   7.71(+0.13,-0.19)  \\ 
NGC\,1023  & SB0 &   SB0      &  11.4   &  221  &   100  &   7.64(+0.04,-0.04)  \\
NGC\,1316  & E   &   SAB0merg &  21.5   &  647  &   100  &   8.24(+0.17,-0.23)  \\
NGC\,2778  & E2  &   S0       &  22.9   &   50  &    30  &   7.15(+0.30,-0.30)  \\
NGC\,3115  & S0  &   S0       &   9.7   &  550  &   150  &   8.96(+0.19,-0.15)  \\
NGC\,3384  & SB0 &   S0       &  11.6   &  128  &   100  &   7.04(+0.16,-0.26)  \\
NGC\,4459  & E2  &   S0       &  16.1   &  218  &    28  &   7.84(+0.08,-0.08)  \\
NGC\,4564  & S0  &   S0       &  15.0   &  213  &    31  &   7.92(+0.10,-0.14)  \\
NGC\,5813  & E2  &   S0       &  32.2   & 1650  &   400  &   8.84(+0.07,-0.07)  \\
NGC\,7332  & S0  &   SB0      &  23.0   &  175  &    15  &   7.11(+0.17,-0.21)  \\
NGC\,7457  & S0  &   S0       &  13.2   &  178  &    75  &   7.00(+0.20,-0.40)  \\
NGC\,224   & Sb  &   SBb      &   0.7   &  450  &   100  &   8.18(+0.20,-0.10)  \\ 
NGC\,253   & SBc &   SABc     &   3.4   &   92  &    21  &   6.70(+0.30,-0.30)  \\ 
NGC\,3031  & Sb  &   SABab    &   3.6   &  172  &   100  &   7.84(+0.11,-0.06)  \\
NGC\,3368$^a,b$& &   SABa     &  10.4   &  35   &    18  &   6.88(+0.09,-0.09)  \\
NGC\,4258$^a,c$& &   SABb     &   7.6   &  105  &    40  &   7.60(+0.01,-0.01)  \\ 
NGC\,4395$^a,d$& &   SBm      &   4.3   &  15   &     7  &   5.56(+0.13,-0.13)  \\
NGC\,4594& Sa    &   Sa       &   9.8   & 1900  &   300  &   8.72(+0.30,-0.57)  \\
NGC\,4736$^a,b$& &   SABab    &   5.0   &  46   &   18   &   6.83(+0.10,-0.10)  \\
NGC\,4826$^a,b$& &   Sab      &   7.3   &  69   &   30   &   6.20(+0.11,-0.11)  \\
MilkyWay & Sbc   &   SBbc     &   0.0   &  160  &    10  &   6.61(+0.04,-0.04)  \\
\hline
\vspace*{-0.8cm}
\end{tabular}
\end{minipage}
\end{center}
\vspace*{0.3cm}
{{\sc Note}--- Galaxy sample from \cite{harr14}; data from references therein. Col.\ 3: galaxy types from \citet{sahu19b}, based on the classification grid of \citet{grah19}. \\ 
$^a$ These galaxies were not in the original sample. $N_{\rm GC}$ from \citet{gonz17,gonz19}, this work. 
$^b$ $M_\bullet$ from \citet{korm13a}.\\
$^c$ $M_\bullet$ from \citet{hump13}.\\
$^d$ $M_\bullet$ from \citet{pete05}.\\
Galaxies NGC\,221, NGC\,4486A, NGC\,4486B, NGC\,4382, A\,2052 (upper limits), NGC\,5845, NGC\,5128 (outliers and/or large uncertainties) are excluded from the fits.}
\label{tab:fitgal}
\end{scriptsize}
\end{table}


We determine the $M_\bullet$ versus $N_{\rm GC}$ correlation for the sample in
Table~\ref{tab:fitgal}. We use the galaxy types in \cite{sahu19b} (Col.\ 3),
which are based on the classification grid of \cite{grah19}.  We obtain
separate fits for ellipticals (E), lenticulars (L) and spirals (S), with the
FITEXY regression routines \citep{pres92}, as modified by \citet{trem02} and
\citet{nova06} to allow for the addition of intrinsic (or ``cosmic") scatter in
the parameters.  Summarizing, if data pairs ($x_i,y_i$), with individual
measurement uncertainties ($\sigma_{x,i},\sigma_{y,i}$) are related by the
linear equation $y = \alpha + \beta x$, the expression to minimize is 

\begin{equation}
\chi^2 = \Sigma \frac{y_i - \alpha - \beta (x_i - \langle x \rangle)^2}{(\sigma^2_{y,i} + \epsilon^2_y) + \beta^2(\sigma^2_{x,i} + \epsilon^2_x)}, 
\label{eq:chisq}
\end{equation}

\noindent
where $\langle x \rangle \equiv N^{-1} \Sigma^N_{i=1} x_i$ is the sample mean,
and $\epsilon_x$ and $\epsilon_y$ represent the additional intrinsic scatter
over the formal measurement errors.  If additional ``cosmic" scatter indeed
exists, as implied by a reduced $\chi^2_\nu > 1$, the epsilons can be increased
until $\chi^2_\nu = 1$.  We searched for solutions both assuming $\epsilon_x =
0$, with all additional scattering in $y$, and distributing the potential
supplemental scattering between $\epsilon_x$ and $\epsilon_y$. In the absence
of other constraints, when epsilons are added simultaneously to the $M_\bullet$
and $N_{\rm GC}$ values, their squares are kept in the same ratio observed when
only $\epsilon_y$ is appended to either variable. 

Correlation solutions with the FITEXY routines are shown in
Table~\ref{tab:fitexy}.  We also carry out fits with the Bivariate Correlated
Errors and Intrinsic Scatter (BCES) method by \citet{akri96}, with results for
the bisector line of the forward ($Y|X$) and inverse ($X|Y$) regressions
presented in Table~\ref{tab:bces}.  The solutions have the form $Y = \alpha +
\beta (X - \langle x \rangle)$.  To note: lenticular galaxies do show a
correlation.

We perform analogous fits to $M_\bullet$ versus $M_\ast$, for E, S, and E+S
galaxies. The solutions are listed in Tables~\ref{tab:galmass_fitexy}
and~\ref{tab:galmass_bces}.  



\floattable
\begin{deluxetable}{cccccccccc}
\tablecaption{Correlation solutions with FITEXY\label{tab:fitexy}}
\tablewidth{0pt}
\tablehead{
\colhead{Solutions for: }& \colhead{Sample      }& \colhead{$N$}&\colhead{$\alpha$}& \colhead {$\beta$} & \colhead{$\langle x \rangle$ }& \colhead{$\epsilon_y$}&\colhead{$\epsilon_x$} & \colhead{$\chi^2_\nu$} &\colhead {rms}\\ 
\colhead{      }& \colhead{      }& \colhead{}&\colhead{dex}& \colhead {} & \colhead{ dex }& \colhead{dex}&\colhead{dex} &\colhead{}  &\colhead{dex}
}
\decimals
\startdata
log $M_\bullet$ versus log $N_{\rm GC}$  &  E   &   27     & 8.871 $\pm$ 0.029 & 0.915 $\pm$ 0.051 & 3.2 & 0    &  0    & 6.9  & 0.403 \\
                                         &      &          & 8.963 $\pm$ 0.074 & 0.771 $\pm$ 0.130 & 3.2 & 0.35 &  0    & 1.0  & 0.378 \\
                                         &      &          & 8.967 $\pm$ 0.080 & 0.972 $\pm$ 0.158 & 3.2 & 0.26 &  0.27 & 1.0  & 0.396 \\
                                         &  S   &   10     & 6.455 $\pm$ 0.080 & 2.394 $\pm$ 0.261 & 2.1 & 0    &  0    & 6.2  & 0.964 \\
                                         &      &          & 7.086 $\pm$ 0.164 & 1.515 $\pm$ 0.310 & 2.1 & 0.42 &  0    & 1.0  & 0.444 \\
                                         &      &          & 7.077 $\pm$ 0.170 & 1.673 $\pm$ 0.329 & 2.1 & 0.32 &  0.17 & 1.0  & 0.464 \\
                                         &  L   &   10     & 7.802 $\pm$ 0.072 & 1.617 $\pm$ 0.243 & 2.4 & 0    &  0    & 2.6  & 0.374 \\
                                         &      &          & 7.767 $\pm$ 0.127 & 1.617 $\pm$ 0.342 & 2.4 & 0.28 &  0    & 1.0  & 0.372 \\
                                         &      &          & 7.752 $\pm$ 0.133 & 1.776 $\pm$ 0.388 & 2.4 & 0.21 &  0.12 & 1.0  & 0.387 \\
log $N_{\rm GC}$ versus log $M_\bullet$  &  E   &   27     & 3.232 $\pm$ 0.032 & 1.093 $\pm$ 0.061 & 8.9 & 0    &  0    & 6.9  & 0.440 \\
                                         &      &          & 3.138 $\pm$ 0.077 & 0.815 $\pm$ 0.136 & 8.9 & 0.36 &  0    & 1.0  & 0.378 \\
                                         &      &          & 3.132 $\pm$ 0.082 & 1.029 $\pm$ 0.166 & 8.9 & 0.27 &  0.26 & 1.0  & 0.407 \\
                                         &  S   &   10     & 2.334 $\pm$ 0.030 & 0.418 $\pm$ 0.045 & 7.1 & 0    &  0    & 6.2  & 0.403 \\
                                         &      &          & 2.083 $\pm$ 0.098 & 0.534 $\pm$ 0.103 & 7.1 & 0.25 &  0    & 1.0  & 0.273 \\
                                         &      &          & 2.080 $\pm$ 0.101 & 0.598 $\pm$ 0.118 & 7.1 & 0.17 &  0.32 & 1.0  & 0.277 \\
                                         &  L   &   10     & 2.398 $\pm$ 0.045 & 0.618 $\pm$ 0.093 & 7.8 & 0    &  0    & 2.6  & 0.231 \\
                                         &      &          & 2.434 $\pm$ 0.071 & 0.509 $\pm$ 0.116 & 7.8 & 0.16 &  0    & 1.0  & 0.213 \\
                                         &      &          & 2.427 $\pm$ 0.073 & 0.564 $\pm$ 0.123 & 7.8 & 0.12 &  0.21 & 1.0  & 0.218 \\
\enddata
\end{deluxetable}


\floattable
\begin{deluxetable}{cccccccccc}
\tablecaption{Correlation solutions with BCES\label{tab:bces}}
\tablewidth{0pt}
\tablehead{
\colhead{Solutions for: }& \colhead{Sample      }& \colhead{$N$}&\colhead{$\alpha$}& \colhead {$\beta$} & \colhead{$\langle x \rangle$ }& \colhead{$\epsilon_y$}&\colhead{$\epsilon_x$} & \colhead{$\chi^2_\nu$} &\colhead {rms}\\
\colhead{      }& \colhead{      }& \colhead{}&\colhead{dex}& \colhead {} & \colhead{ dex }& \colhead{dex}&\colhead{dex} &\colhead{}  &\colhead{dex}
}
\decimals
\startdata
log $M_\bullet$ versus log $N_{\rm GC}$  &  E   &   27     & 8.980 $\pm$ 0.078 & 1.000 $\pm$ 0.098  & 3.2 & 0    &  0    & 7.8  & 0.400 \\
                                         &      &          & 8.980 $\pm$ 0.074 & 0.803 $\pm$ 0.126  & 3.2 & 0.35 &  0    & 1.0  & 0.378 \\
                                         &      &          & 8.992 $\pm$ 0.079 & 1.010 $\pm$ 0.126  & 3.2 & 0.26 &  0.27 & 1.0  & 0.403 \\
                                         &  S   &   10     & 7.110 $\pm$ 0.146 & 1.660 $\pm$ 0.218  & 2.1 & 0    &  0    & 20.9 & 0.460 \\
                                         &      &          & 7.128 $\pm$ 0.139 & 1.470 $\pm$ 0.182  & 2.1 & 0.42 &  0    & 1.0  & 0.441 \\
                                         &      &          & 7.118 $\pm$ 0.144 & 1.640 $\pm$ 0.237  & 2.1 & 0.32 &  0.17 & 1.0  &  0.457 \\
                                         &  L   &   10     & 7.760 $\pm$ 0.126 & 1.850  $\pm$ 0.447  & 2.4 & 0    &  0    & 2.7  & 0.398 \\
                                         &      &          & 7.758 $\pm$ 0.119 & 1.670  $\pm$ 0.418  & 2.4 & 0.28 &  0    & 1.0  & 0.376 \\
                                         &      &          & 7.750 $\pm$ 0.126 & 1.850  $\pm$ 0.496  & 2.4 & 0.21 &  0.12 & 1.0  & 0.398 \\
log $N_{\rm GC}$ versus log $M_\bullet$  &  E   &   27     & 3.114 $\pm$ 0.072 & 0.787 $\pm$ 0.089  & 8.9 & 0.36 &  0    & 1.0  & 0.377 \\
                                         &  S   &   10     & 2.063 $\pm$ 0.086 & 0.532 $\pm$ 0.101  & 7.1 & 0.25 &  0    & 1.0  & 0.272 \\
                                         &  L   &   10     & 2.413 $\pm$ 0.067 & 0.485 $\pm$ 0.142  & 7.8 & 0.16 &  0    & 1.0  & 0.212 \\
\enddata
\end{deluxetable}




\floattable
\begin{deluxetable}{cccccccccc}
\tablecaption{Correlation solutions with FITEXY for $M_\bullet$ and Total Galaxy Mass \label{tab:galmass_fitexy}}
\tablewidth{0pt}
\tablehead{
\colhead{Solutions for: }& \colhead{Sample      }& \colhead{$N$}&\colhead{$\alpha$}& \colhead {$\beta$} & \colhead{$\langle x \rangle$ }& \colhead{$\epsilon_y$}&\colhead{$\epsilon_x$} & \colhead{$\chi^2_\nu$} &\colhead {rms}\\ 
\colhead{      }& \colhead{      }& \colhead{}&\colhead{dex}& \colhead {} & \colhead{ dex }& \colhead{dex}&\colhead{dex} &\colhead{}  &\colhead{dex}
}
\decimals
\startdata
log $M_\bullet$ versus log $M_\ast$      &  E   &   27     & 8.812 $\pm$ 0.035 & 1.369 $\pm$ 0.086 &11.1 & 0    &  0    & 5.5  & 0.411 \\
                                         &      &          & 8.915 $\pm$ 0.075 & 1.141 $\pm$ 0.190 &11.1 & 0.34 &  0    & 1.0  & 0.380 \\
                                         &      &          & 8.906 $\pm$ 0.080 & 1.416 $\pm$ 0.228 &11.1 & 0.25 &  0.18 & 1.0  & 0.399 \\
                                         &  S   &   10     & 5.472 $\pm$ 0.203 & 3.065 $\pm$ 0.418 &10.0 & 0    &  0    &10.0  & 2.34  \\
                                         &      &          & 7.092 $\pm$ 0.183 & 1.028 $\pm$ 0.262 &10.0 & 0.51 &  0    & 1.0  & 0.502 \\
                                         &      &          & 7.077 $\pm$ 0.196 & 1.238 $\pm$ 0.310 &10.0 & 0.37 &  0.30 & 1.0  & 0.535 \\
                                         & E+S  &   37     & 7.865 $\pm$ 0.040 & 2.425 $\pm$ 0.082 &10.8 & 0    &  0    & 8.8  & 1.082 \\
                                         &      &          & 8.425 $\pm$ 0.082 & 1.390 $\pm$ 0.129 &10.8 & 0.43 &  0    & 1.0  & 0.475 \\
                                         &      &          & 8.407 $\pm$ 0.085 & 1.556 $\pm$ 0.146 &10.8 & 0.31 &  0.20 & 1.0  & 0.322 \\
log $M_\ast$ versus log $M_\bullet$      &  E   &   27     & 11.224 $\pm$ 0.024 & 0.730 $\pm$ 0.046 & 8.9 & 0    &  0    & 5.5  & 0.300 \\
                                         &      &          & 11.153 $\pm$ 0.053 & 0.567 $\pm$ 0.094 & 8.9 & 0.24 &  0    & 1.0  & 0.263 \\
                                         &      &          & 11.154 $\pm$ 0.056 & 0.706 $\pm$ 0.113 & 8.9 & 0.18 &  0.25 & 1.0  & 0.282 \\
                                         &  S   &   10     & 10.585 $\pm$ 0.027 & 0.327 $\pm$ 0.045 & 7.1 & 0    &  0    &10.0  & 0.763 \\
                                         &      &          & 10.084 $\pm$ 0.150 & 0.678 $\pm$ 0.168 & 7.1 & 0.41 &  0    & 1.0  & 0.422 \\
                                         &      &          & 10.076 $\pm$ 0.157 & 0.807 $\pm$ 0.202 & 7.1 & 0.30 &  0.37 & 1.0  & 0.432 \\
                                         & E+S  &   37     & 11.077 $\pm$ 0.016 & 0.413 $\pm$ 0.014 & 8.4 & 0    &  0    & 8.8  & 0.446 \\
                                         &      &          & 10.869 $\pm$ 0.053 & 0.576 $\pm$ 0.054 & 8.4 & 0.27 &  0    & 1.0  & 0.320 \\
                                         &      &          & 10.861 $\pm$ 0.054 & 0.643 $\pm$ 0.060 & 8.4 & 0.20 &  0.31 & 1.0  & 0.500 \\
\enddata
\end{deluxetable}



\floattable
\begin{deluxetable}{cccccccccc}
\tablecaption{Correlation solutions with BCES for $M_\bullet$ and Total Galaxy Mass \label{tab:galmass_bces}}
\tablewidth{0pt}
\tablehead{
\colhead{Solutions for: }& \colhead{Sample      }& \colhead{$N$}&\colhead{$\alpha$}& \colhead {$\beta$} & \colhead{$\langle x \rangle$ }& \colhead{$\epsilon_y$}&\colhead{$\epsilon_x$} & \colhead{$\chi^2_\nu$} &\colhead {rms}\\ 
\colhead{      }& \colhead{      }& \colhead{}&\colhead{dex}& \colhead {} & \colhead{ dex }& \colhead{dex}&\colhead{dex} &\colhead{}  &\colhead{dex}
}
\decimals
\startdata
log $M_\bullet$ versus log $M_\ast$      &  E   &   27     & 8.960 $\pm$ 0.078 & 1.460 $\pm$ 0.149 &11.1 & 0    &  0    & 5.5  & 0.405 \\
                                         &      &          & 8.955 $\pm$ 0.073 & 1.190 $\pm$ 0.183 &11.1 & 0.34 &  0    & 1.0  & 0.381 \\
                                         &      &          & 8.952 $\pm$ 0.078 & 1.480 $\pm$ 0.184 &11.1 & 0.25 &  0.18 & 1.0  & 0.407 \\
                                         &  S   &   10     & 7.151 $\pm$ 0.167 & 1.220 $\pm$ 0.135 &10.0 & 0    &  0    &10.0  & 0.530 \\
                                         &      &          & 7.090 $\pm$ 0.158 & 1.000 $\pm$ 0.131 &10.0 & 0.51 &  0    & 1.0  & 0.501 \\
                                         &      &          & 7.081 $\pm$ 0.166 & 1.210 $\pm$ 0.163 &10.0 & 0.37 &  0.30 & 1.0  & 0.527 \\
                                         & E+S  &   37     & 8.390 $\pm$ 0.080 & 1.500 $\pm$ 0.140 &10.8 & 0    &  0    & 8.8  & 0.491 \\
                                         &      &          & 8.469 $\pm$ 0.078 & 1.380 $\pm$ 0.131 &10.8 & 0.43 &  0    & 1.0  & 0.475 \\
                                         &      &          & 8.420 $\pm$ 0.080 & 1.500 $\pm$ 0.153 &10.8 & 0.31 &  0.20 & 1.0  & 0.488 \\
log $M_\ast$ versus log $M_\bullet$      &  E   &   27     & 11.134 $\pm$ 0.050 & 0.543 $\pm$ 0.062 & 8.9 & 0.24 &  0    & 1.0  & 0.262 \\
                                         &  S   &   10     & 10.042 $\pm$ 0.133 & 0.681 $\pm$ 0.128 & 7.1 & 0.41 &  0    & 1.0  & 0.420 \\
                                         & E+S  &   37     & 10.837 $\pm$ 0.052 & 0.611 $\pm$ 0.070 & 8.4 & 0.27 &  0    & 1.0  & 0.318 \\
\enddata
\end{deluxetable}

\bibliographystyle{aasjournal}

\begin{thebibliography}

\bibitem[Agertz et al.(2009)]{ager09} Agertz, O., Teyssier, R., \& Moore, B.\ 2009, \mnras, 397, L64. doi:10.1111/j.1745-3933.2009.00685.x
\bibitem[Akritas \& Bershady(1996)]{akri96} Akritas, M.~G. \& Bershady, M.~A.\ 1996, \apj, 470, 706. doi:10.1086/177901
\bibitem[Ashman \& Zepf(1992)]{ashm92} Ashman, K.~M. \& Zepf, S.~E.\ 1992, \apj, 384, 50. doi:10.1086/170850
\bibitem[Ashman \& Zepf(1998)]{ashm98} \samename\ 1998, Globular cluster systems / Keith M. Ashman, Stephen E. Zepf.  Cambridge, U. K. ; New York : Cambridge University Press, 1998. (Cambridge astrophysics series ; 30) 
\bibitem[Athanassoula(2005)]{atha05} Athanassoula, E.\ 2005, \mnras, 358, 1477. doi:10.1111/j.1365-2966.2005.08872.x
\bibitem[Baes et al.(2003)]{baes03} Baes, M., Buyle, P., Hau, G.~K.~T., et al.\ 2003, \mnras, 341, L44. doi:10.1046/j.1365-8711.2003.06680.x
\bibitem[Bertin(2011)]{bert11} Bertin, E.\ 2011, Astronomical Data Analysis Software and Systems XX, 442, 435
\bibitem[Bertin \& Arnouts(1996)]{bert96} Bertin, E., \& Arnouts, S.\ 1996, \aaps, 117, 393
\bibitem[Blakeslee et al.(2010)]{blak10} Blakeslee, J.~P., Cantiello, M., Mei, S., et al.\ 2010, \apj, 724, 657
\bibitem[Blakeslee et al.(1997)]{blak97} Blakeslee, J.~P., Tonry, J.~L., \& Metzger, M.~R.\ 1997, \aj, 114, 482. doi:10.1086/118488
\bibitem[Blecha(1986)]{blec86} Blecha, A.\ 1986, \aap, 154, 321
\bibitem[Bogd{\'a}n et al.(2018)]{bogd18} Bogd{\'a}n, {\'A}., Lovisari, L., Volonteri, M., et al.\ 2018, \apj, 852, 131. doi:10.3847/1538-4357/aa9ab5
\bibitem[Boulade et al.(2003)]{boul03} Boulade, O., Charlot, X., Abbon, P., et al.\ 2003, \procspie, 4841, 72
\bibitem[Bruzual \& Charlot(2003)]{bruz03} Bruzual, G., \& Charlot, S.\ 2003, \mnras, 344, 1000
\bibitem[Burkert \& Forbes(2020)]{burk20} Burkert, A. \& Forbes, D.~A.\ 2020, \aj, 159, 56. doi:10.3847/1538-3881/ab5b0e
\bibitem[Burkert \& Tremaine(2010)]{burk10} Burkert, A., \& Tremaine, S.\ 2010, \apj, 720, 516
\bibitem[Capuzzo-Dolcetta \& Donnarumma(2001)]{capu01} Capuzzo-Dolcetta, R., \& Donnarumma, I.\ 2001, \mnras, 328, 645
\bibitem[Capuzzo-Dolcetta \& Mastrobuono-Battisti(2009)]{capu09} Capuzzo-Dolcetta, R., \& Mastrobuono-Battisti, A.\ 2009, \aap, 507, 183
\bibitem[Capuzzo-Dolcetta \& Vicari(2005)]{capu05} Capuzzo-Dolcetta, R., \& Vicari, A.\ 2005, \mnras, 356, 899 
\bibitem[Chabrier(2003)]{chab03} Chabrier, G.\ 2003, \pasp, 115, 763
\bibitem[C{\^o}t{\'e} et al.(1998)]{cote98} C{\^o}t{\'e}, P., Marzke, R.~O., \& West, M.~J.\ 1998, \apj, 501, 554. doi:10.1086/305838
\bibitem[Cowie et al.(1995)]{cowi95} Cowie, L.~L., Hu, E.~M., \& Songaila, A.\ 1995, \nat, 377, 603. doi:10.1038/377603a0
\bibitem[Davis et al.(2018)]{davi18} Davis, B.~L., Graham, A.~W., \& Cameron, E.\ 2018, \apj, 869, 113. doi:10.3847/1538-4357/aae820 
\bibitem[Davis et al.(2019a)]{davi19a} \samename\ 2019a, \apj, 873, 85. doi:10.3847/1538-4357/aaf3b8 
\bibitem[Davis et al.(2019b)]{davi19b} Davis, B.~L., Graham, A.~W., \& Combes, F.\ 2019b, \apj, 877, 64. doi:10.3847/1538-4357/ab1aa4
\bibitem[Desai et al.(2012)]{desa12} Desai, S., Armstrong, R., Mohr, J.~J., et al.\ 2012, \apj, 757, 83
\bibitem[de Vaucouleurs et al.(1991)]{rc3} de Vaucouleurs, G., de Vaucouleurs, A., Corwin, H.~G., Jr., et al.\ 1991, Third Reference Catalogue of Bright Galaxies.~Volume I: Explanations and references.~ Volume II: Data for galaxies between 0$^{h}$ and 12$^{h}$.~ Volume III: Data for galaxies between 12$^{h}$ and 24$^{h}$., by de Vaucouleurs, G.; de Vaucouleurs, A.; Corwin, H.~G., Jr.; Buta, R.~J.; Paturel, G.; Fouqu{\'e}, P..~Springer, New York, NY (USA), 1991, 2091 p. (https://vizier.u-strasbg.fr/viz-bin/VizieR?-source=VII/155)
\bibitem[Dressler(1989)]{dres89} Dressler, A.\ 1989,  in IAU Symp. 134, Active Galactic Nuclei, ed. D. E. Osterbrock \& J. S. Miller (Dordrecht: Kluwer), 217
\bibitem[Fabian(2012)]{fabi12} Fabian, A.~C.\ 2012, \araa, 50, 455
\bibitem[Ferrarese(2002)]{ferr02} Ferrarese, L.\ 2002, \apj, 578, 90. doi:10.1086/342308
\bibitem[Ferrarese \& Merritt(2000)]{ferr00} Ferrarese, L., \& Merritt, D.\ 2000, \apj, 539, L9
\bibitem[Filippenko \& Ho(2003)]{fili03} Filippenko, A.~V. \& Ho, L.~C.\ 2003, \apjl, 588, L13. doi:10.1086/375361
\bibitem[Forbes et al.(2018)]{forb18} Forbes, D.~A., Read, J.~I., Gieles, M., et al.\ 2018, \mnras, 481, 5592. doi:10.1093/mnras/sty2584
\bibitem[Gebhardt et al.(2000)]{gebh00} Gebhardt, K., et al.\ 2000, \apj, 539, L13
\bibitem[Georgiev et al.(2010)]{geor10} Georgiev, I.~Y., Puzia, T.~H., Goudfrooij, P., et al.\ 2010, \mnras, 406, 1967. doi:10.1111/j.1365-2966.2010.16802.x
\bibitem[Gnedin et al.(2014)]{gned14} Gnedin, O.~Y., Ostriker, J.~P., \& Tremaine, S.\ 2014, \apj, 785, 71
\bibitem[Gonz{\'a}lez-L{\'o}pezlira et al.(2017)]{gonz17} Gonz{\'a}lez-L{\'o}pezlira, R.~A., Lomel{\'\i}-N{\'u}{\~n}ez, L., {\'A}lamo-Mart{\'\i}nez, K., et al.\ 2017, \apj, 835, 184
\bibitem[Gonz{\'a}lez-L{\'o}pezlira et al.(2019)]{gonz19} Gonz{\'a}lez-L{\'o}pezlira, R.~A., Mayya, Y.~D., Loinard, L., et al.\ 2019, \apj, 876, 39
\bibitem[Goudfrooij et al.(2003)]{goud03} Goudfrooij, P., Strader, J., Brenneman, L., et al.\ 2003, \mnras, 343, 665
\bibitem[Graham(2012)]{grah12} Graham, A.~W.\ 2012, \apj, 746, 113. doi:10.1088/0004-637X/746/1/113 
\bibitem[Graham(2019)]{grah19} \samename\ 2019, \mnras, 487, 4995. doi:10.1093/mnras/stz1623
\bibitem[Graham \& Scott(2013)]{grah13} Graham, A.~W. \& Scott, N.\ 2013, \apj, 764, 151. doi:10.1088/0004-637X/764/2/151 
\bibitem[Graham \& Spitler(2009)]{grah09} Graham, A.~W. \& Spitler, L.~R.\ 2009, \mnras, 397, 2148. doi:10.1111/j.1365-2966.2009.15118.x
\bibitem[Guo et al.(2015)]{guo15} Guo, Y., Ferguson, H.~C., Bell, E.~F., et al.\ 2015, \apj, 800, 39. doi:10.1088/0004-637X/800/1/39
\bibitem[Gwyn(2008)]{gwyn08} Gwyn, S.~D.~J.\ 2008, \pasp, 120, 212
\bibitem[Gwyn(2014)]{gwyn14} \samename\ 2014, Astronomical Data Analysis Software and Systems XXIII, 485, 387
\bibitem[H{\"a}ring \& Rix(2004)]{hari04} H{\"a}ring, N., \& Rix, H.-W.\ 2004, \apj, 604, L89
\bibitem[Harris et al.(2020)]{2020Natur.585..357H} Harris, C.~R., Millman, K.~J., van der Walt, S.~J., et al.\ 2020, \nat, 585, 357. doi:10.1038/s41586-020-2649-2
\bibitem[Harris \& Harris(2011)]{harr11} Harris, G.~L.~H., \& Harris, W.~E.\ 2011, \mnras, 410, 2347
\bibitem[Harris et al.(2014)]{harr14} Harris, G.~L.~H., Poole, G.~B., \& Harris, W.~E.\ 2014, \mnras, 438, 2117
\bibitem[Harris 1996 (2010 edition) ]{harr96} Harris, W.~E.\ 1996, \aj, 112, 1487
\bibitem[Harris et al.(2017)]{harr17} Harris, W.~E., Blakeslee, J.~P., \& Harris, G.~L.~H.\ 2017, \apj, 836, 67. doi:10.3847/1538-4357/836/1/67
\bibitem[Harris et al.(2013)]{harr13} Harris, W.~E., Harris, G.~L.~H., \& Alessi, M.\ 2013, \apj, 772, 82. doi:10.1088/0004-637X/772/2/82
\bibitem[Harris et al.(2009)]{harr09} Harris, W.~E., Kavelaars, J.~J., Hanes, D.~A., Pritchet, C.~J., \& Baum, W.~A.\ 2009, \aj, 137, 3314
\bibitem[Hudson et al.(2014)]{huds14} Hudson, M.~J., Harris, G.~L., \& Harris, W.~E.\ 2014, \apjl, 787, L5. doi:10.1088/2041-8205/787/1/L5
\bibitem[Humphreys et al.(2013)]{hump13} Humphreys, E.~M.~L., Reid, M.~J., Moran, J.~M., Greenhill, L.~J., \& Argon, A.~L.\ 2013, \apj, 775, 13
\bibitem[Hunter(2007)]{2007CSE.....9...90H} Hunter, J.~D.\ 2007, Computing in Science and Engineering, 9, 90. doi:10.1109/MCSE.2007.55
\bibitem[Jahnke \& Macci{\`o}(2011)]{jahn11} Jahnke, K., \& Macci{\`o}, A.~V.\ 2011, \apj, 734, 92
\bibitem[Ja{\l}ocha et al.(2008)]{jalo08} Ja{\l}ocha, J., Bratek, {\L}., \& Kutschera, M.\ 2008, \apj, 679, 373. doi:10.1086/533511 
\bibitem[Jord{\'a}n et al.(2007)]{jord07} Jord{\'a}n, A., McLaughlin, D.~E., C{\^o}t{\'e}, P., et al.\ 2007, \apjs, 171, 101
\bibitem[Joye \& Mandel(2003)]{2003ASPC..295..489J} Joye, W.~A. \& Mandel, E.\ 2003, Astronomical Data Analysis Software and Systems XII, 295, 489
\bibitem[King(1962)]{king62} King, I.\ 1962, \aj, 67, 471
\bibitem[King(1966)]{king66} King, I.~R.\ 1966, \aj, 71, 64
\bibitem[Kissler-Patig et al.(1999)]{kiss99} Kissler-Patig, M., Ashman, K.~M., Zepf, S.~E., \& Freeman, K.~C.\ 1999, \aj, 118, 197
\bibitem[Kormendy(1993)]{korm93} Kormendy, J.\ 1993, Galactic Bulges, 153, 209
\bibitem[Kormendy(2013)]{korm13b} Kormendy, J.\ 2013, Secular Evolution of Galaxies, 1
\bibitem[Kormendy \& Bender(2011)]{korm11b} Kormendy, J. \& Bender, R.\ 2011, \nat, 469, 377. doi:10.1038/nature09695 
\bibitem[Kormendy et al.(2011)]{korm11a} Kormendy, J., Bender, R., \& Cornell, M.~E.\ 2011, \nat, 469, 374. doi:10.1038/nature09694 
\bibitem[Kormendy et al.(2010)]{korm10} Kormendy, J., Drory, N., Bender, R., et al.\ 2010, \apj, 723, 54. doi:10.1088/0004-637X/723/1/54
\bibitem[Kormendy \& Ho(2013)]{korm13a} Kormendy, J. \& Ho, L.~C.\ 2013, \araa, 51, 511
\bibitem[Kormendy \& Kennicutt(2004)]{korm04} Kormendy, J. \& Kennicutt, R.~C.\ 2004, \araa, 42, 603
\bibitem[Kormendy \& Richstone(1995)]{korm95} Kormendy, J., \& Richstone, D.\ 1995, \araa, 33, 581
\bibitem[Kravtsov \& Gnedin(2005)]{krav05} Kravtsov, A.~V. \& Gnedin, O.~Y.\ 2005, \apj, 623, 650. doi:10.1086/428636
\bibitem[Kron(1980)]{kron80} Kron, R.~G.\ 1980, \apjs, 43, 305
\bibitem[Kruijssen(2015)]{krui15} Kruijssen, J.~M.~D.\ 2015, \mnras, 454, 1658. doi:10.1093/mnras/stv2026
\bibitem[Larsen(1999)]{lars99} Larsen, S.~S.\ 1999, \aaps, 139, 393
\bibitem[Lupton \& Monger(1991)]{1991supe.book.....L} Lupton, R. \& Monger, P.\ 1991, Unpublished paper, 1991
\bibitem[Magorrian et al.(1998)]{mago98} Magorrian, J., et al.\ 1998, \aj, 115, 2285
\bibitem[Martin et al.(2018)]{mart18} Martin, G., Kaviraj, S., Volonteri, M., et al.\ 2018, \mnras, 476, 2801. doi:10.1093/mnras/sty324
\bibitem[McLaughlin et al.(1994)]{mcla94} McLaughlin, D.~E., Harris, W.~E., \& Hanes, D.~A.\ 1994, \apj, 422, 486
\bibitem[Mu{\~n}oz et al.(2014)]{muno14} Mu{\~n}oz, R.~P., Puzia, T.~H., Lan{\c c}on, A., et al.\ 2014, \apjs, 210, 4
\bibitem[Novak et al.(2006)]{nova06} Novak, G.~S., Faber, S.~M., \& Dekel, A.\ 2006, \apj, 637, 96. doi:10.1086/498333
\bibitem[Oke(1974)]{oke74} Oke, J.~B.\ 1974, \apjs, 27, 21
\bibitem[Olsen et al.(2004)]{olse04} Olsen, K.~A.~G., Miller, B.~W., Suntzeff, N.~B., et al.\ 2004, \aj, 127, 2674. doi:10.1086/383297
\bibitem[Peng(2007)]{peng07} Peng, C.~Y.\ 2007, \apj, 671, 1098
\bibitem[Peterson et al.(2005)]{pete05} Peterson, B.~M., Bentz, M.~C., Desroches, L.-B., et al.\ 2005, \apj, 632, 799. doi:10.1086/444494
\bibitem[Powalka et al.(2016)]{powa16} Powalka, M., Lan{\c c}on, A., Puzia, T.~H., et al.\ 2016, \apjs, 227, 12
\bibitem[Press et al.(1992)]{pres92} Press, W.~H., Teukolsky, S.~A., Vetterling, W.~T., et al.\ 1992, Cambridge: University Press, |c1992, 2nd ed.
\bibitem[Puget et al.(2004)]{puge04} Puget, P., Stadler, E., Doyon, R., et al.\ 2004, \procspie, 5492, 978
\bibitem[Reid et al.(2014)]{reid14} Reid, M.~J., Menten, K.~M., Brunthaler, A., et al.\ 2014, \apj, 783, 130
\bibitem[Reines \& Volonteri(2015)]{rein15} Reines, A.~E. \& Volonteri, M.\ 2015, \apj, 813, 82. doi:10.1088/0004-637X/813/2/82
\bibitem[Rowe et al.(2015)]{2015A&C....10..121R} Rowe, B.~T.~P., Jarvis, M., Mandelbaum, R., et al.\ 2015, Astronomy and Computing, 10, 121. doi:10.1016/j.ascom.2015.02.002
\bibitem[Sahu et al.(2019a)]{sahu19a} Sahu, N., Graham, A.~W., \& Davis, B.~L.\ 2019a, \apj, 876, 155. doi:10.3847/1538-4357/ab0f32
\bibitem[Sahu et al.(2019b)]{sahu19b} \samename\ 2019b, \apj, 887, 10. doi:10.3847/1538-4357/ab50b7 
\bibitem[Sauvaget et al.(2018)]{sauv18} Sauvaget, T., Hammer, F., Puech, M., et al.\ 2018, \mnras, 473, 2521. doi:10.1093/mnras/stx2453
\bibitem[Savorgnan et al.(2016)]{savo16} Savorgnan, G.~A.~D., Graham, A.~W., Marconi, A., et al.\ 2016, \apj, 817, 21. doi:10.3847/0004-637X/817/1/21
\bibitem[Schlafly \& Finkbeiner(2011)]{schlaf11} Schlafly, E.~F., \& Finkbeiner, D.~P.\ 2011, \apj, 737, 103
\bibitem[Schweizer(1987)]{schw87} Schweizer, F.\ 1987, Nearly Normal Galaxies. From the Planck Time to the Present, 18
\bibitem[Scott et al.(2013)]{scot13} Scott, N., Graham, A.~W., \& Schombert, J.\ 2013, \apj, 768, 76. doi:10.1088/0004-637X/768/1/76  
\bibitem[S{\'e}rsic(1963)]{sers63} S{\'e}rsic, J.~L.\ 1963, Boletin de la Asociacion Argentina de Astronomia La Plata Argentina, 6, 41
\bibitem[Seth et al.(2010)]{seth10} Seth, A.~C., Cappellari, M., Neumayer, N., et al.\ 2010, \apj, 714, 713. doi:10.1088/0004-637X/714/1/713
\bibitem[Silk \& Rees(1998)]{silk98} Silk, J., \& Rees, M.~J.\ 1998, \aap, 331, L1
\bibitem[Simmons et al.(2017)]{simm17} Simmons, B.~D., Smethurst, R.~J., \& Lintott, C.\ 2017, \mnras, 470, 1559. doi:10.1093/mnras/stx1340
\bibitem[Smethurst et al.(2019)]{smet19} Smethurst, R.~J., Simmons, B.~D., Lintott, C.~J., et al.\ 2019, \mnras, 489, 4016. doi:10.1093/mnras/stz2443
\bibitem[Spitler \& Forbes(2009)]{spit09} Spitler, L.~R., \& Forbes, D.~A.\ 2009, \mnras, 392, L1
\bibitem[Stetson(1987)]{1987PASP...99..191S} Stetson, P.~B.\ 1987, \pasp, 99, 191. doi:10.1086/131977
\bibitem[Stetson(2011)]{2011ascl.soft04011S} \samename\ 2011, Astrophysics Source Code Library. ascl:1104.011
\bibitem[Thim et al.(2004)]{thim04} Thim, F., Hoessel, J.~G., Saha, A., et al.\ 2004, \aj, 127, 2322. doi:10.1086/382244
\bibitem[Tody(1986)]{1986SPIE..627..733T} Tody, D.\ 1986, \procspie, 627, 733. doi:10.1117/12.968154
\bibitem[Tody(1993)]{1993ASPC...52..173T} \samename\ 1993, Astronomical Data Analysis Software and Systems II, 52, 173
\bibitem[Tonry et al.(2001)]{tonr01} Tonry, J.~L., Dressler, A., Blakeslee, J.~P., et al.\ 2001, \apj, 546, 681
\bibitem[Tremaine et al.(2002)]{trem02} Tremaine, S., et al.\  2002, \apj, 574, 740
\bibitem[Turner et al.(2012)]{turn12} Turner, M.~L., C{\^o}t{\'e}, P., Ferrarese, L., et al.\ 2012, \apjs, 203, 5. doi:10.1088/0067-0049/203/1/5
\bibitem[van den Bosch(2016)]{vand16} van den Bosch, R.~C.~E.\ 2016, \apj, 831, 134. doi:10.3847/0004-637X/831/2/134
\bibitem[Virtanen et al.(2020)]{2020NatMe..17..261V} Virtanen, P., Gommers, R., Oliphant, T.~E., et al.\ 2020, Nature Methods, 17, 261. doi:10.1038/s41592-019-0686-2
\bibitem[Wang et al.(2014)]{wang14} Wang, S., Ma, J., Wu, Z., \& Zhou, X.\ 2014, \aj, 148, 4

\end{thebibliography}

\end{document}